\documentclass{aa}

\usepackage{graphicx}
\usepackage{textcomp}
\usepackage{gensymb}
\usepackage{subfigure}
\usepackage{cleveref}    
\usepackage{upgreek}
\usepackage{enumitem}
\usepackage{epstopdf}
\usepackage{stfloats}
\usepackage{longtable,multirow}
\usepackage[para,online,flushleft]{threeparttable}
\usepackage{multirow,bigdelim}
\usepackage[version=4]{mhchem}

\usepackage{color}
\usepackage{xcolor}
\usepackage{ulem}
\usepackage{cancel}
\usepackage[varg]{txfonts}
\usepackage{microtype}
\usepackage{hyperref}
\usepackage{booktabs}
\usepackage{xcolor}
\usepackage{hyperref}
\usepackage{graphicx}
\usepackage{multirow}
\usepackage{float}

\hypersetup{%
  colorlinks=true,% hyperlinks will be black
  linkcolor=blue,% hyperlink borders will be red
  citecolor=blue
}

\usepackage{etoolbox}

\let\oldcitet=\citet

\renewcommand{\citet}[1]{\textcolor[rgb]{0,0,1}{\oldcitet{#1}}}

\newcommand{\OI}{[O\,{\sc i}]}
\newcommand{\CI}{[C\,{\sc i}]}

\newcommand{\LCI}{\ensuremath{ L_{[\rm{C}\,{\textsc{i}}]}}}
\newcommand{\LNII}{\ensuremath{ L_{[\rm{N}\,{\textsc{ii}}]}}}

\newcommand{\CII}{[C\,{\sc ii}]}
\newcommand{\HII}{H\,{\sc ii}}

\newcommand{\NII}{[N\,{\sc ii}]}

\newcommand{\NIII}{N\,{\sc iii}}   
\newcommand{\OIII}{O\,{\sc iii}}
\def\kms{{\hbox {km\thinspace s$^{-1}$}}} 
 
\def\Ls{{\hbox {$L_{\odot}$}}} 

\def\Ms{{\hbox {$M_{\odot}$}}}

\begin{document} 

\title{Submillimeter imaging of the Galactic Center starburst Sgr~B2\thanks{\textit{Herschel} is an ESA space observatory with science instruments provided by European-led Principal Investigator consortia and with important participation from NASA.}$^{,}$\thanks{Includes IRAM 30m observations. IRAM is supported by INSU/CNRS (France), MPG (Germany), and IGN (Spain).}}

\subtitle{Warm molecular, atomic, and ionized gas far from massive star-forming cores }

 \titlerunning{Submm line mapping of Sgr~B2} 
\authorrunning{Santa-Maria et al.} 
          
 \author{M. G. Santa-Maria\inst{1}
          \and 
         J. R. Goicoechea\inst{1}
          \and
         M. Etxaluze\inst{2} 
          \and
         J. Cernicharo\inst{1}
          \and
         S.~Cuadrado\inst{1}
          }

 \institute{Instituto de F\'{\i}sica Fundamental
     (CSIC). Calle Serrano 121-123, E-28006, Madrid, Spain.
              \email{miriam.g.sm@csic.es}
         \and
         RAL Space, Rutherford Appleton Laboratory, Oxfordshire, OX11 0QX, UK.          
             }
   
   \date{Received December 23, 2020; accepted March 26, 2021}

% \abstract{}{}{}{}{} 
% 5 {} token are mandatory
 
  \abstract
{Star-forming galaxies emit bright molecular and atomic  lines in the submillimeter (submm) and far-infrared (FIR) domains. \mbox{However}, it is not always clear which  gas heating mechanisms dominate and which feedback processes drive their excitation.} 
{The Sgr~B2  complex  is an excellent template to spatially resolve the  main OB-type star-forming cores from the extended cloud environment and to study the properties of the  warm molecular gas  in conditions likely prevailing in distant extragalactic nuclei. }
{We present \mbox{168\,arcmin$^2$} spectral images of  Sgr~B2 taken with \mbox{\textit{Herschel}/SPIRE-FTS}  in the  complete \mbox{$\sim$\,450$-$1545\,GHz} band. We detect ubiquitous emission from  \mbox{mid-$J$ CO} \mbox{(up to  $J$\,=\,12$-$11)}, 
 \mbox{\ce{H2O} $2_{1,1}-2_{0,2}$}, \mbox{\CI\,492, 809\,GHz}, and  \NII\,205\,$\upmu$m lines. We also present \mbox{velocity-resolved} maps of the \mbox{SiO (2$-$1)}, \ce{N2H+}, \mbox{HCN}, and \mbox{\ce{HCO+} (1$-$0)}  emission obtained with the IRAM\,30m telescope.}
{The cloud environment ($\sim$\,1000\,pc$^2$ around the main cores) dominates the emitted 
FIR ($\sim$\,80\,\%), \mbox{H$_2$O\,752\,GHz ($\sim$\,60\,\%)} \mbox{mid-$J$\,CO ($\sim$\,91\%)}, \mbox{\CI~($\sim$\,93\,\%)}, and \mbox{\NII\,205\,$\upmu$m~($\sim$\,95\,\%)} luminosity. The region shows very extended \NII\,205\,$\upmu$m emission (spatially correlated with the  24\,$\upmu$m and 70\,$\upmu$m dust emission) that traces an  extended component of diffuse ionized gas of low ionization  parameter ($U$\,$\simeq$\,10$^{-3}$)  and  low  \mbox{$L_{\rm FIR}$\,/\,$M_{\rm H_2}$\,$\simeq$\,4$-$11\,\Ls\,\Ms$^{-1}$} ratios (scaling as  \mbox{$\propto$\,$T_{\rm dust}^{6}$}). The observed FIR luminosities imply a flux of nonionizing photons equivalent to  \mbox{$G_0$\,$\approx$\,10$^3$}. All these diagnostics suggest that the complex is clumpy and this allows UV photons from young massive stars to escape from their natal molecular cores. The extended \CI~emission arises from a pervasive component of neutral gas with \mbox{$n_{\rm{H}}$\,$\simeq$\,10$^3$\,cm$^{-3}$}. The high ionization rates in the region, produced by enhanced cosmic-ray (CR) fluxes, drive the gas heating in this  component  to \mbox{$T_{\rm k}$\,$\simeq$\,40$-$60\,K}. The \mbox{mid-$J$ CO} emission arises from a similarly extended but more pressurized  gas component (\mbox{$P_{\rm th}$\,/\,$k$\,$\simeq$\,10$^7$\,K\,cm$^{-3}$}): spatially unresolved clumps, thin sheets, or filaments of UV-illuminated compressed  gas ($n_{\rm{H}}$\,$\simeq$\,10$^6$\,cm$^{-3}$).  Specific regions of enhanced SiO emission and high \mbox{CO-to-FIR}  intensity ratios (\mbox{$I_{\rm CO}$\,/\,$I_{\rm FIR}$\,$\gtrsim$\,10$^{-3}$}) show mid-$J$~CO emission compatible with $C$-type shock models. A major difference compared to  more quiescent  star-forming clouds in the disk of our Galaxy  is the extended  nature of the SiO and \ce{N2H+} emission in Sgr~B2. This can  be explained by the presence of cloud-scale  shocks, induced by cloud-cloud collisions and stellar feedback, and the much higher CR ionization rate ($>$\,10$^{-15}$\,s$^{-1}$) leading to overabundant  H$_{3}^{+}$ and  \ce{N2H+}.}
{Sgr~B2 hosts a more extreme  \mbox{environment} than  star-forming regions in the disk of the Galaxy.  As a usual template for extragalactic comparisons,  Sgr~B2 shows more similarities to  nearby ultra luminous infrared galaxies such as Arp\,220, including a ``deficit'' in the \CI\,/\,FIR and \NII\,/\,FIR intensity ratios, than to pure starburst galaxies such as M82.
However, it is the extended cloud environment, rather than the cores, that serves as a useful template when telescopes do not resolve such extended regions in galaxies.
}


   \maketitle
%
%-------------------------------------------------------------------

%%%%%%%%%%%%%%%%%%%%%%
\section{Introduction}\label{sect:intro}
%%%%%%%%%%%%%%%%%%%%%%

The interstellar medium (ISM) of galactic nuclei  can host extreme conditions driven by the presence of numerous OB-type massive stars \citep[i.e., strong UV radiation fields;][]{Clark2018}, cloud-cloud collisions \citep[e.g.,][]{Tsuboi2015}, stellar winds and supernova (SNe) explosions \citep[producing expanding bubbles and widespread shocks;][]{Maeda2002}, enhanced X-ray emission (from accretion onto a super-massive black hole, SMBH), and increased cosmic-ray fluxes  \citep[e.g.,][]{Zhang_2015}. Characterizing the physical conditions and heating mechanisms of the ISM in galactic nuclei (especially the dense molecular gas component, the reservoir that may form new stars)
is critical to understand how star formation proceeds in such extreme conditions and how galaxies evolve. %The advent of ALMA has made \textcolor{orange}{it} possible to study the warm molecular gas \mbox{($T\rm{_{k}}$\,$>$\,50\,K)} in galaxies at increasingly larger redshifts.
With ALMA, it is now possible to study the emission from warm molecular gas in galaxies at increasingly larger redshift. ALMA's sensitivity  allows the detection of redshifted submillimeter (submm) and far-infrared (FIR) rotationally excited molecular lines (e.g., from  CO, H$_2$O, HCN, and \ce{HCO+}) and  also the fine-structure atomic lines \mbox{(e.g., \CI, \OI, \CII, \NII, etc.)}, which are the main gas coolants. 
 
Far-infrared and submm emission lines are very good tracers of the gas physical conditions and
chemistry in extreme star-forming environments, such as those prevailing in merger galaxies \citep{Schirm2014}, starburst (SB) galaxies, and (ultra) luminous infrared galaxies \citep[(U)LIRG;][]{Fischer2010,vanderwerf2010,Meijerink2013,Mashian2015}. Ideally, these lines can be used to disentangle the different feedback processes and heating mechanisms operating in their ISM. However, when it comes to star-forming regions of the Milky Way and nearby galaxies, these rest frame frequencies cannot be easily  observed with ground-based telescopes due to the poor or null atmospheric transmission.

Giant molecular clouds (GMCs) and their star-forming cores are typically spatially unresolved in extragalactic observations ($1''$\,$\approx$\,15\,pc at a distance of 3\,Mpc). Hence, it is
not easy to link cloud-scale structure, to galactic structure, to global star formation
and evolution. This is clearly a multi-scale problem that benefits from having 
local \mbox{templates} in which the  emission from well characterized environments:
photodissociation regions (PDRs), shocks, dense cores, diffuse gas halos, etc. can be spatially resolved. 
In general, molecular-line observations of local massive star-forming cores (where OB stars are born) should not be directly extrapolated to interpret extragalactic observations because the former ones are not sensitive to the large-scale (often colder and less dense) cloud emission that likely dominates extragalactic observations \mbox{\citep[e.g.,][]{Indriolo17}}. Until very recently, the study of the ISM and star-forming regions targeted either small Galactic ($\lesssim$\,1\,pc) or large extragalactic ($\gtrsim$\,1\,kpc) spatial scales.  The main limitation has been the difficulty to obtain wide-field spectroscopic maps of entire GMCs (scales of tens of pc) in the Milky Way, and to spatially resolve individual GMCs in nearby galaxies. In addition, the FIR and submm windows can only be fully covered from airbone or, preferably, space telescopes.

There is an increasing amount of new extragalactic studies that focus on the smaller spatial  scales \citep[{$\gtrsim$\,10\,pc}; e.g.,][]{Schinnerer2013,Schirm2014,Wu18,Lee2019}. In addition, the implementation of fast mapping techniques and  use of broadband spectrometers have made possible to simultaneously map multiple molecular lines over square degree areas of sky \mbox{\citep[e.g., $\gtrsim$\,100\,pc$^2$ in Orion~B,][]{Pety17}}. Also, the development of multi-beam receivers reduces the mapping time considerably \citep[for square degree maps of Orion~A in \CII\,158\,$\upmu$m and CO  see e.g.,][]{Pabst2019,Goico20}.

Our aim is to bridge the gap between the small and the intermediate spatial scales in the study of the warm molecular gas in massive star-forming regions of galactic nuclei. We expect to provide a useful template that may help to better interpret the submm spectrum of spatially unresolved star-forming galaxies. Our goal is to determine the properties of the extended cloud environment  at spatial scales and conditions close to those prevailing in extragalactic nuclei. In particular, we  try to answer the question of which feedback mechanisms dominate in the Galactic Center (GC), with emphasis on the heating mechanisms of the extended molecular gas and the observational signature of  X-rays, cosmic rays (CRs), shocks, and stellar radiation\footnote{Nonionizing far-UV (FUV) photons have energies $h\nu$\,$<$\,13.6\,eV and extreme-UV  (EUV)  ionizing photons have energies $h\nu$\,$>$\,13.6\,eV.}.
A related question we  study is whether the topology of GMCs affects the  propagation of stellar EUV and FUV photons.

%\textcolor{red}{So, we wonder which feedback mechanisms dominate in Galactic Center (GC) %clouds; which are the heating mechanisms of the extended and warm molecular gas; what is the %specific role and signature of X-rays, cosmic rays (CRs), shocks, and stellar %radiation\footnote{Nonionizing far-UV (FUV) photons have energies $h\nu$\,$<$\,13.6\,eV  and %extreme-UV  (EUV)  ionizing photons have energies $h\nu$\,$>$\,13.6\,eV.}, and if we are able %to distinguish them observationally; as well as what is the topology of GMCs and how does this %affect the  propagation of EUV and FUV photons.}
%%and shed light to the following questions: Which feedback mechanisms dominate in Galactic Center (GC) clouds? Which are the heating mechanisms  of the extended and warm molecular gas? What is the specific role and signature of X-rays, cosmic rays (CRs), shocks, and stellar radiation\footnote{Nonionizing far-UV (FUV) photons have energies $h\nu$\,$<$\,13.6\,eV  and extreme-UV  (EUV)  ionizing photons have energies $h\nu$\,$>$\,13.6\,eV.}? Can we distinguish them observationally? What is the topology of GMCs and how does this affect the  propagation of EUV and FUV photons? 

In this study we present and analyze \mbox{$168$\,arcmin$^2$} \mbox{($\sim$\,$31\,\rm{pc}\times31\,\rm{pc}$)} spectral-images  of multiple submm lines across the extended cloud environment of the most active high-mass star-forming region of the GC, \mbox{Sagittarius~B2} (Sgr~B2). These observations were carried out with the \mbox{SPIRE--FTS} spectrometer \citep{Griffin10} on board \mbox{\textit{Herschel}} \citep{Pilbratt10}. \mbox{We expand} previous \mbox{SPIRE--FTS} observations of the main massive star-forming cores \mbox{Sgr~B2(N, M, and S)} that we introduced in \cite{Etxaluze13}. We also present new velocity-resolved maps of the \mbox{SiO~($J$\,=\,2$-$1)}, HCN, \ce{HCO+}, and \mbox{\ce{N2H+}~($J$\,=\,1$-$0)} emission obtained with the \mbox{IRAM\,30m} telescope in the 3\,mm band. We complement these line maps with photometric images of the dust emission obtained with \textit{Spitzer}  \citep{Carey09} and \textit{Herschel} \citep{Molinari11}. We also make use of 3$-$79\,keV X-ray images taken with \textit{NuSTAR} \citep[][and Shuo Zhang priv. communication]{Zhang_2015}.

The paper is organized as follows: in Sect.\;\ref{sect:observations} we give specific  details of the Sgr~B2 complex and of the observational dataset. In Sect.\;\ref{sect:results} we present the molecular and atomic emission maps. In Sect.\;\ref{sect:Analysis} we analyze the dust continuum emission, the excitation of the  CO and \CI\;lines and  discuss the results in the context of PDR and shock models.
 We also analyze the extended \ce{N2H+} emission in relation with the
 high CR ionization rates.  In  Sect.\;\ref{sect:discussion} we discuss 
 particular  tracers of different gas heating mechanisms and compare our results with similar observations of prototypical extragalactic sources, and we discuss specific regions in Sgr~B2. In Sect.\;\ref{sect:conclusions} we summarize our work and conclude.

% -----------------------------------------------------------------------------
\begin{figure}[t]
    \centering
    \includegraphics[width=0.49\textwidth]{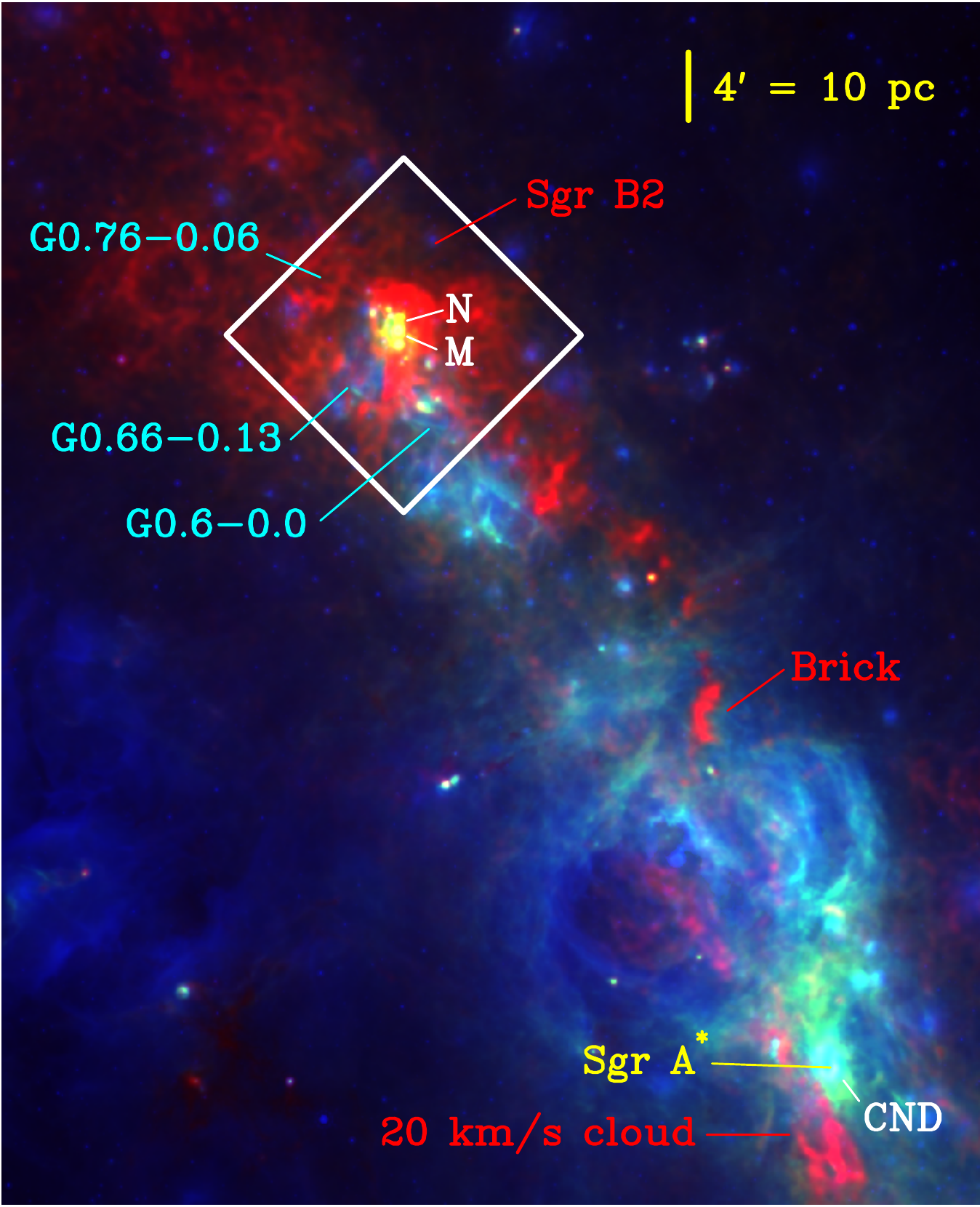}
    \caption{RGB view of about half of the CMZ ($\sim$50$’\,\times\,$60$’$) in the GC. \mbox{Red: SPIRE~350\,$\upmu$m} tracing cold dust from the most prominent molecular clouds. Green: PACS~70\,$\upmu$m tracing warm dust, mostly in extended PDR-like environments. Blue: MIPS~24\,$\upmu$m tracing hot dust, mostly from  ionized regions. The rhombus marks the $\sim$13$’\times$13$'$ area mapped with SPIRE-FTS around Sgr B2(M,N)  massive star-forming cores.}
    \label{fig:Sgrb2-sketch}
\end{figure}
% -----------------------------------------------------------------------------

%%%%%%%%%%%%%%%%%%%%%%
\section{Observations of the Sgr B2 complex} \label{sect:observations}
%%%%%%%%%%%%%%%%%%%%%%

\subsection{Sgr B2, a starburst in the Galactic Center}\label{subsec:sgrb2}

Compared to the disk of the Galaxy, GMCs in the GC are between 10 and 100 times denser and more turbulent \citep[e.g.,][]{Morris1996} but star formation is currently restricted to a few regions like Sgr~B2 \citep[e.g.,][]{Hatchfield2020}. \mbox{Sagittarius~B2} is the most massive  molecular cloud with ongoing  high-mass star formation \citep[with lines of sight having gas column densities above $N_{\rm{H}_2}$\,$\gtrsim$\,$10^{24}$\,cm$^{-2}$; e.g.,][]{Lis1990,Etxaluze13,Schmiedeke16}. At a distance of $\sim$8.18\,kpc \citep[e.g.,][]{Gravity2019}, Sgr~B2 is located in the so called central molecular zone (CMZ, see Fig.~\ref{fig:Sgrb2-sketch}). This is an area of high molecular gas fraction  within $\sim$\,200\,pc of the very center \citep[][]{Morris1996}. \mbox{Sagittarius~B2} is one of the most luminous star-forming regions in the Galaxy 
\citep[$\sim$\,10$^7\,\Ls$;][]{Goldsmith1992}. Together with Sgr~B1 and \mbox{G0.6$-$0.0}, also in the Sgr~B complex, Sgr~B2 is located at a projected distance of about 100\,pc from the dynamical center of the Milky Way (Sgr~A*; see Fig.~\ref{fig:Sgrb2-sketch}). The Sgr~B2 complex lays in the semi-major axis of the 100\,pc ring of gas and dust that rotates around the GC \citep{Molinari11}. This ring is likely located in the $x_2$ orbit system \citep{Binney1991}, and Sgr~B2 seems to be placed where orbits $x_1$ and $x_2$  intersect. 

\mbox{Sagittarius~B2} contains three main high-mass star-forming cores, Sgr~B2(N), Sgr~B2(M) and Sgr~B2(S), with no less than 49 compact \HII\;regions \citep{Gaume1995,Ginsburg2018, Ginsburg2018b}.
These cores are embedded in a moderate-density \mbox{($n(\rm{H_2})$\,$\sim$\,$10^5$\,$-$\,$10^6\,\rm{cm}^{-3}$)} molecular cloud of about \mbox{5$-$10~pc} \citep{Huttemeister1993,Etxaluze13,Schmiedeke16}. The entire complex is surrounded by a larger envelope ($\sim$\,45~pc) \citep{Huttemeister1995,Goicoechea_2004}, hereafter the Sgr~B2  ``envelope'', that shows diverse physical conditions. Indeed, given its location in the GC, the whole extended region is a good laboratory to study the properties of the ISM in the nucleus of ULIRG and SB galaxies.

In this study we consider that most, if not all, the submm line and continuum emission
inside the white rhombus of Fig.~\ref{fig:Sgrb2-sketch} arises from  Sgr~B2 envelope;
that is to say, that all sources in the mapped area are physically associated
(we justify this further in the text). The Sgr~B2 envelope has a complex structure. %Figure~\ref{fig:Sgrb2-sketch} shows  the entire region and the main features.
 The kinematic structure presents a cone shape in the $\{l,b,\rm{v_{LSR}}\}$ space, increasing from $\sim$\,$+$20\,\kms\;at the edges to $\sim$\,$+$65\,\kms\;at the center \citep{Henshaw16}. Sgr~B2(R) and (V) are two additional \HII\;regions ionized by at least one massive O6 star in each region \citep{Martin1972,Mehringer1993}. 
Sgr~B2 Deep South \citep[DS;][]{Meng2019} is located about 2.8$\arcmin$ south of Sgr~B2(M). This is a peculiar \HII~region that shows nonthermal emission  \citep[][]{Meng2019,Padovani2019} and it hosts a $\gamma$-ray source \mbox{\citep{YusefZadeh2013}}. In the southwest of the envelope we find the cloud \mbox{G0.6$-$0.0}. This region contains no less than four ultracompact \HII\;regions, and is surrounded by an arc, observed in the 3.6\,cm radio-continuum, that seems to bridge the southern part of Sgr~B2 with the northern part of Sgr~B1 \citep{Mehringer1992}. Further to the east we find the \mbox{G0.66--0.13} region, which was first reported in hard \mbox{X-ray} observations with \textit{NuSTAR} \citep{Zhang_2015}. These authors suggest that this region is a molecular clump and a local column density peak. In addition, this region, which seems to have a hollow hemispherical structure, it is likely a site of ongoing cloud-cloud collision \citep{Tsuboi2015}.%\cite{Tsuboi2015} concluded that this region, which has a hollow hemispherical structure,  it is a site of ongoing cloud-cloud collision.

Signatures of ongoing star-formation are found all over the region:  very high FIR luminosity, molecular maser emission, \mbox{X-ray} sources, extended, compact, ultra-compact, and hypercompact \HII\;regions, young stellar objects (YSOs), and massive hot cores. These attributes make the complex one of the most prolific star-forming regions in our Galaxy \citep[e.g.,][]{Martin1972,Whiteoak1983,Benson1984,Mehringer1992,Mehringer1993,YusefYadeh2009,Ginsburg2018}. From a general perspective, it has been proposed that shocks produced by cloud-cloud collisions triggered the star formation in Sgr~B2 starburst \citep{Hasegawa1994,Sato2000}. Indeed, Sgr~B2 could be experiencing an extended ($3\,\rm{pc}\,\times\,12\,\rm{pc}$) star-formation event, not just an isolated starburst within the main star-forming cores \citep[see e.g.,][]{Ginsburg2018}.

As in many galactic nuclei, it is not clear what dominates the molecular gas heating in the Sgr~B2 envelope.  Shocks produced by the peculiar location of Sgr~B2 in a region where the $x_1$ and $x_2$ orbits are tangential, a position favorable for cloud-cloud collisions \citep{Hasegawa1994,Tsuboi2015,Armijos2020b} must play a role.  Other studies suggest that both low-velocity shocks and stellar UV radiation permeating a clumpy medium, contribute to the gas heating at large scales \citep[][]{Goicoechea_2004}. Previous \mbox{X-ray} observations suggest that the complex is an X-ray reflection nebula  because it shows  diffuse \mbox{X-ray} emission in the  Fe$^0$ K$\alpha$ line at 6.4~keV \citep[e.g.,][]{Koyama1996,Murakami2000}. In addition, the  Fe$^0$ K$\alpha$ line correlates with the SiO emission, advocating to a relation between shocks and the source of X-rays \citep{MartinPintado2000}. Several studies conclude that the region was irradiated with the X-ray emission originated during brief outburst periods of Sgr~A$^{*}$ that ended a few hundred years ago \citep[e.g.,][]{Zhang_2015,Terrier2018}. In addition, bombardment of low-energy cosmic ray protons (LECRp), from SNe, accretion onto Sgr~A$^*$, or local LECR sources, can be related to the X-ray emission as well \citep{Zhang_2015}. However,
several studies exclude the low-energy cosmic ray electrons (LECRe)  as the dominant origin of the Fe$^0$ K$\alpha$ and hard X-ray continuum emission in Sgr~B2 \citep[e.g.,][]{Revnivtsev2004,Terrier2010,Zhang_2015}. For a more detailed discussion on the possible contribution of LECRe and LECRp to the hard X-ray emission in the CMZ, see for example: \cite{Zhang_2015,Hess16,Padovani2020}.

\subsection{Herschel/SPIRE-FTS spectroscopic maps of Sgr B2}

%-----------------------------------------------------------------------------------------
\begin{figure*}[t]
\centering
\includegraphics[height=0.44\textwidth]{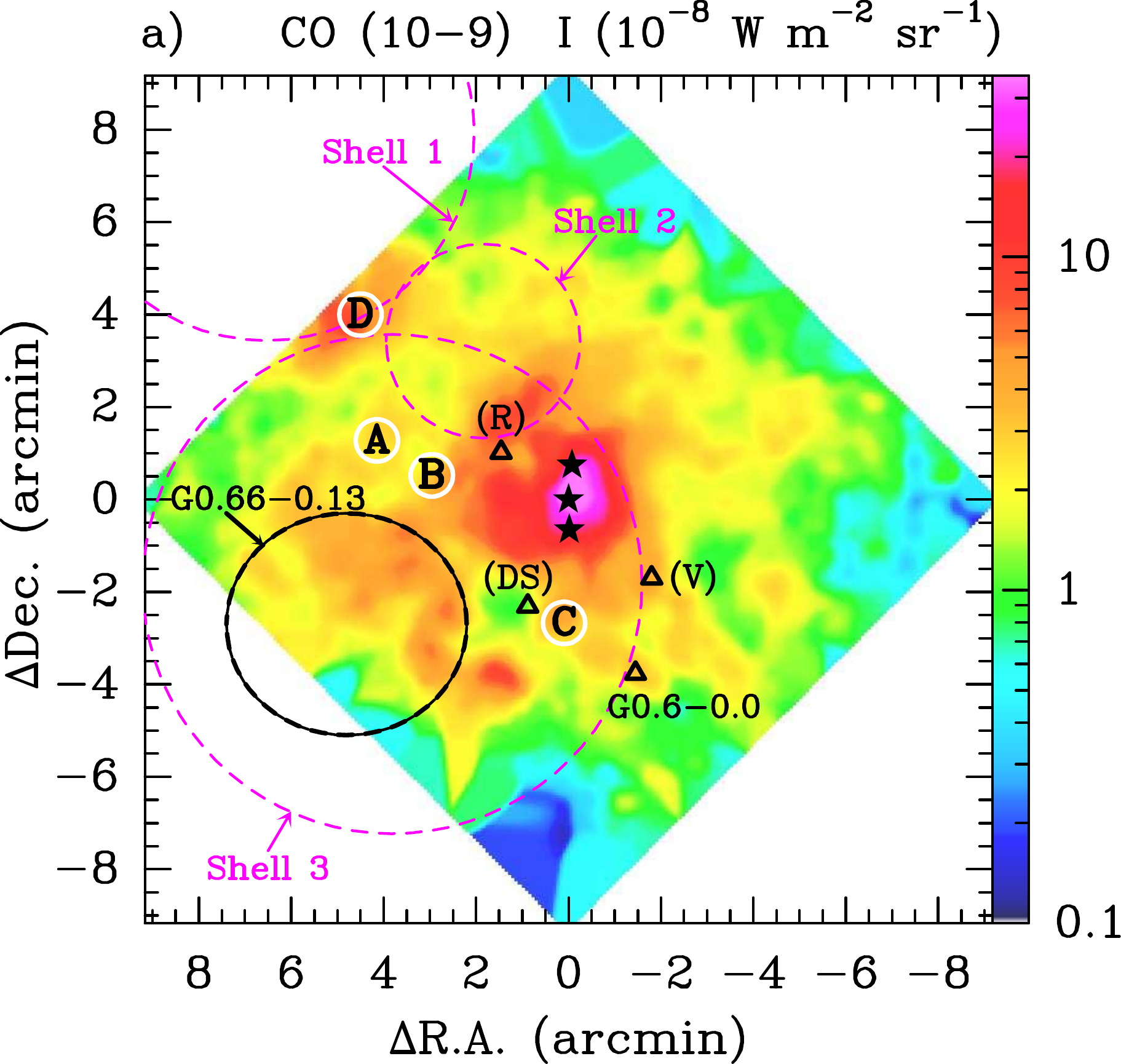}\hspace{0.8cm} 
\includegraphics[height=0.44\textwidth]{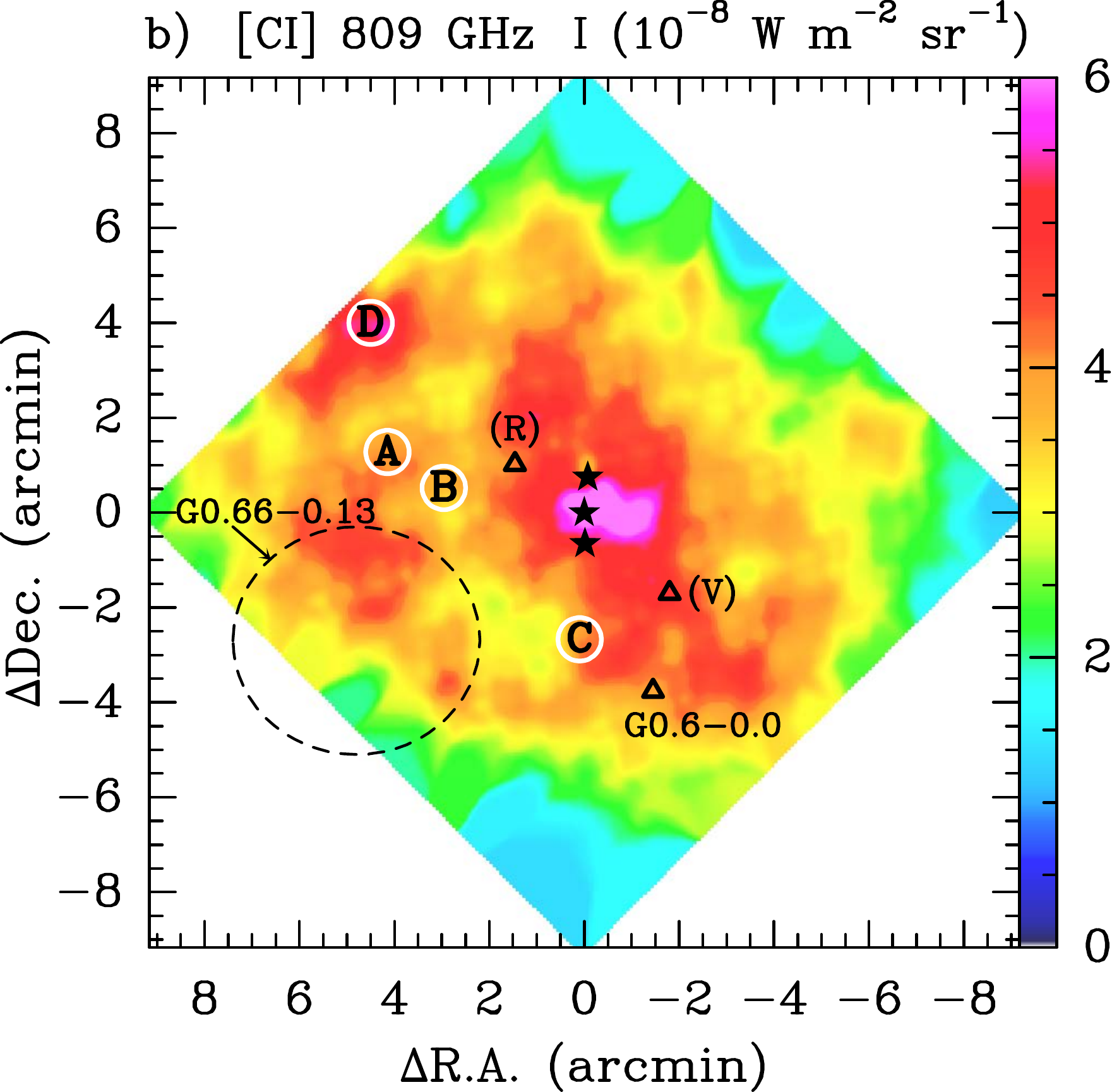}\\
\vspace{0.4cm}
\hspace{0.1cm}\includegraphics[height=0.44\textwidth]{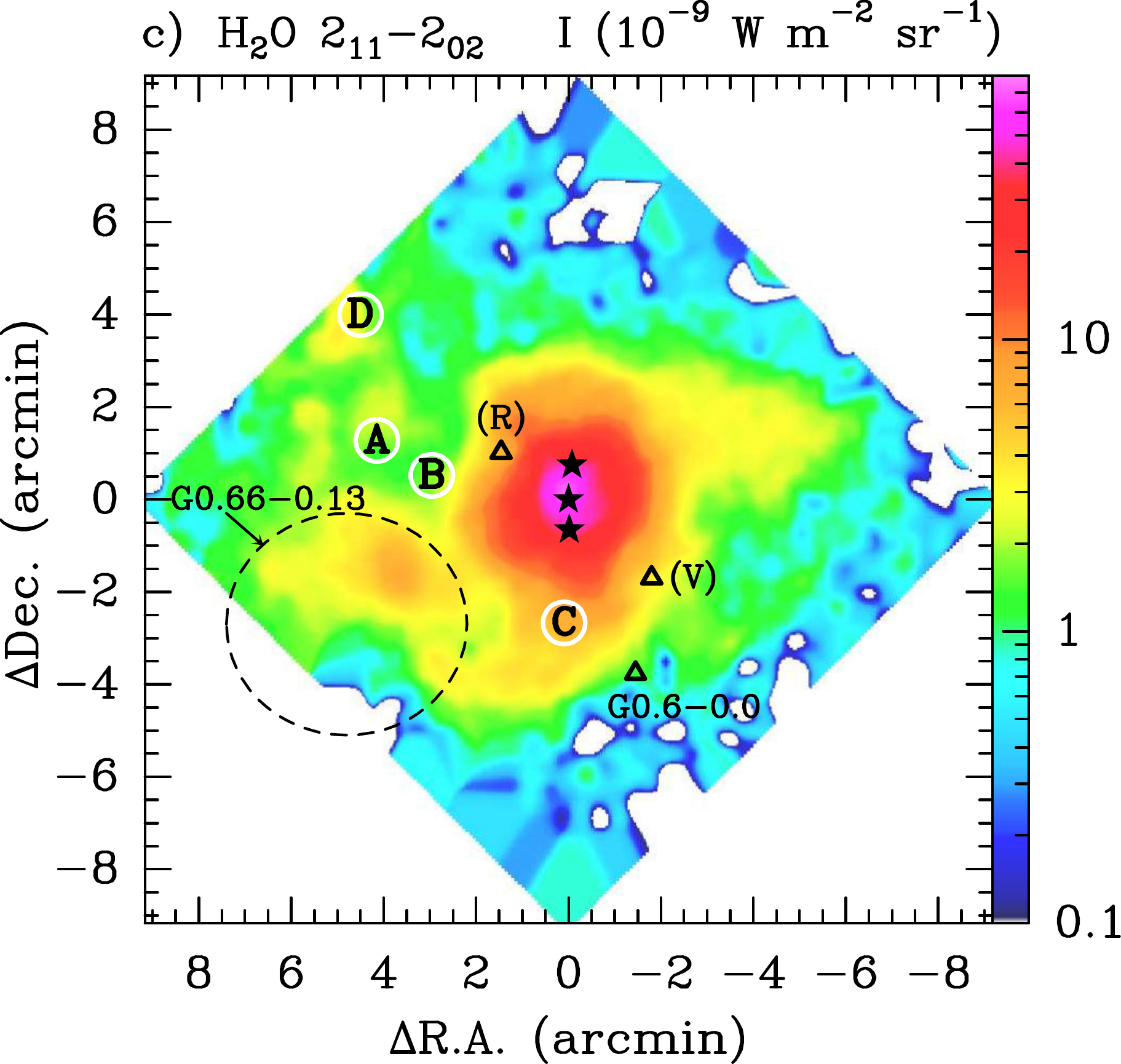} \hspace{0.6cm}
\includegraphics[height=0.44\textwidth]{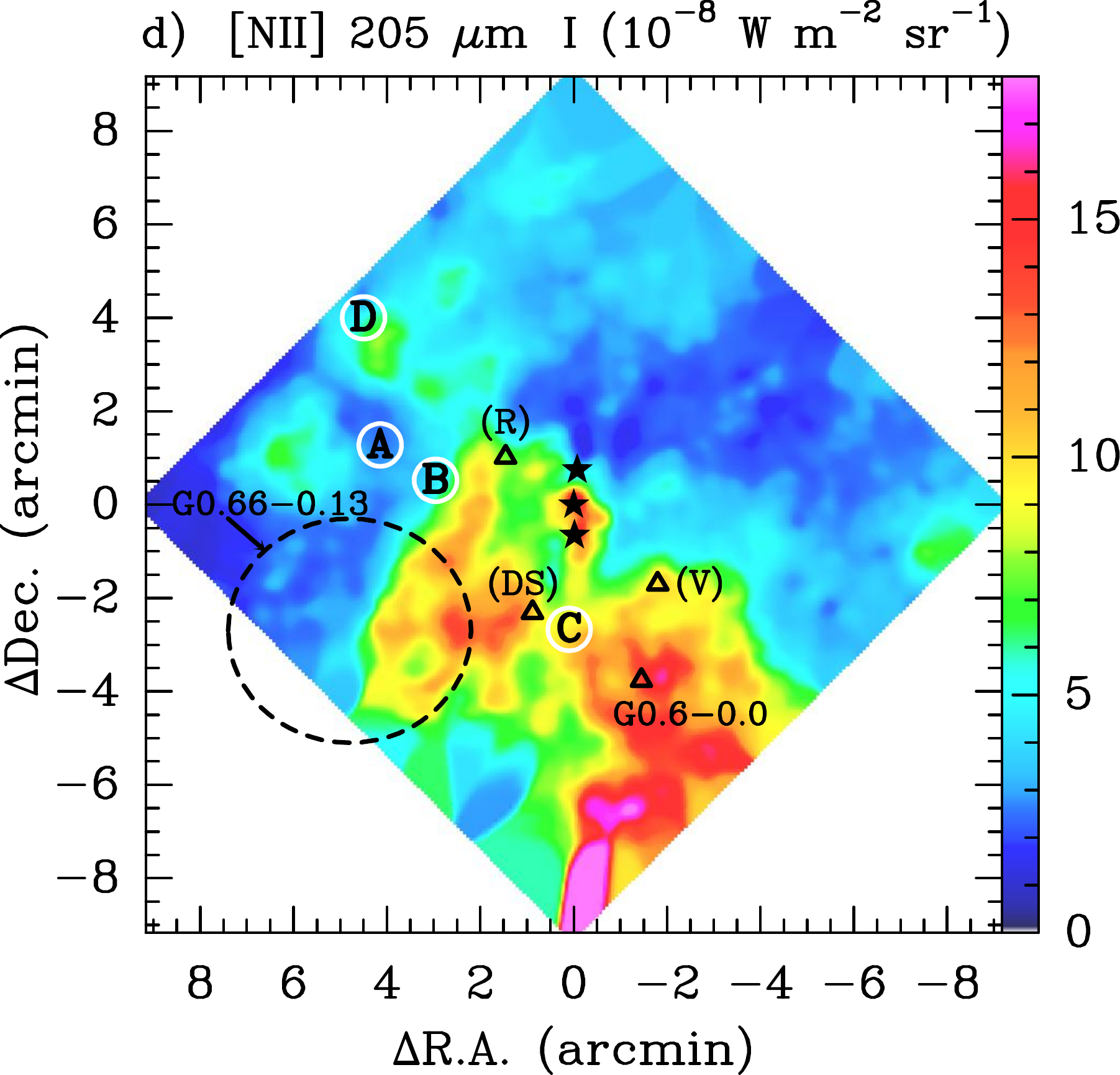}
\vspace{0.4cm}
\caption{Selection of SPIRE-FTS integrated line intensity maps. The map center coordinates are those of  \mbox{Sgr B2(M) core}: \mbox{R.A., Dec.\,=\,17$^{\rm h}$\,47$^{\rm m}$\,20.5$^{\rm s}$, $-$28$\degree$\,23$\arcmin$\,06$\arcsec$}. Black stars mark the location of the three massive star-forming cores Sgr~B2(N, M, and S). Black  triangles mark other \HII\, regions \citep{Mehringer1992,Mehringer1993,Meng2019}. The black dashed circle of radius 2.5$\arcmin$ ($\sim$5.0\,pc), centered at \mbox{R.A., Dec.\,=\,17$^{\rm h}$\,47$^{\rm m}$\,39.7$^{\rm s}$, $-$28$\degree$\,25$\arcmin$\,48$\arcsec$}  marks a specific X-ray irradiated region \citep{Zhang_2015}. The pink dashed circles in panel $a$ show Shells 1, 2, and 3 of \citet{Tsuboi2015}. White circles show  positions A, B, C, and D.}
\label{fig:SPIRE}
\end{figure*}
%-----------------------------------------------------------------------------------------

%------------------------------------  
\begin{figure*}[t]
    \hspace{-0.8cm}\includegraphics[width=1.05\textwidth]{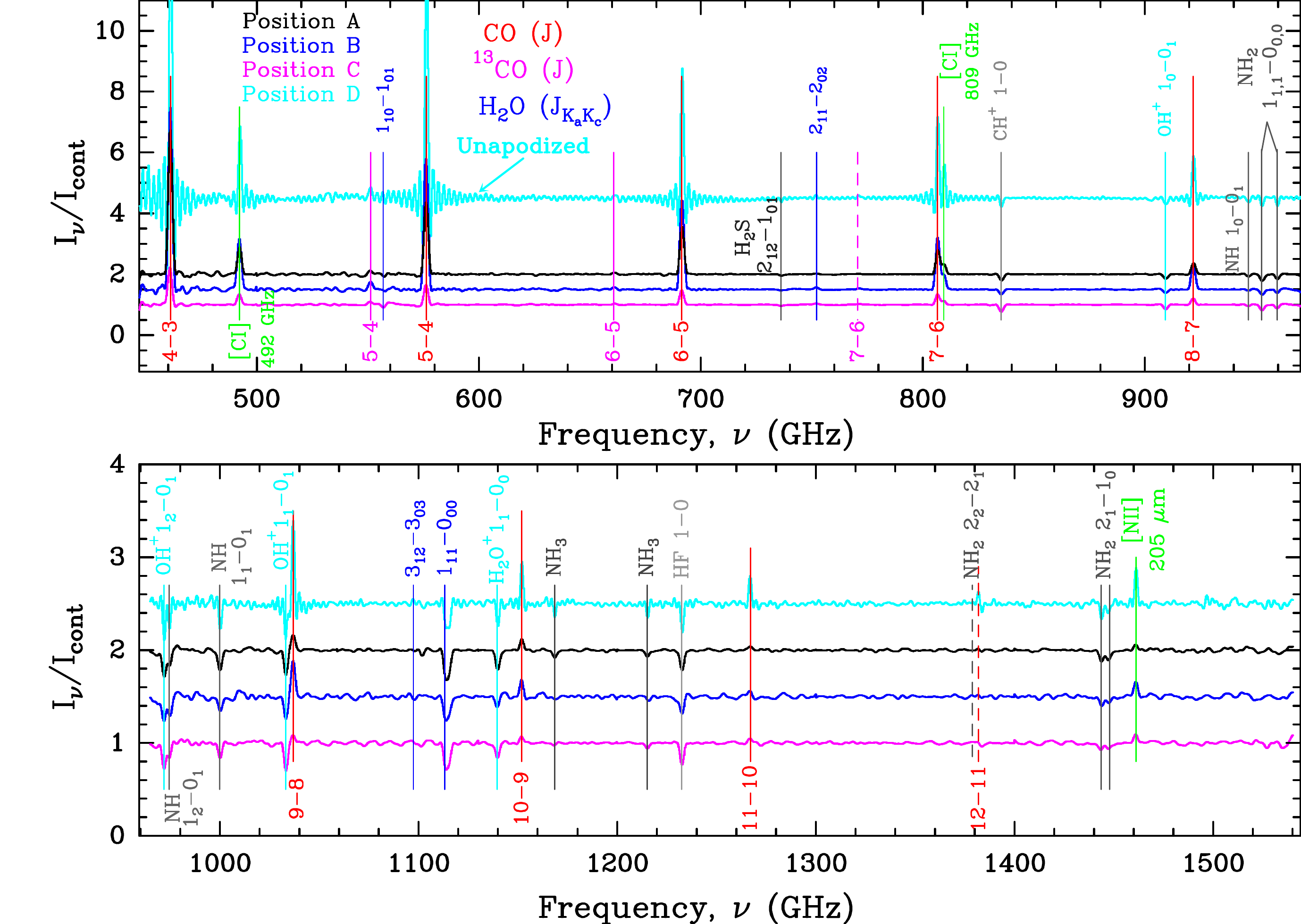}
    \vspace{-0.2cm}
    \caption{SPIRE-FTS continuum-divided spectra  toward four representative positions of Sgr~B2 envelope: A in black, B in blue, C in magenta, and D in cyan
(the $I_{\nu}/I_{\rm cont}$ scale applies to the C spectrum, all the others are shifted). The D spectrum is unapodized, the other three are apodized.}
    \label{fig:spectra}
\end{figure*}
%------------------------------------  
We obtained \mbox{13$'$\,$\times$\,13$'$} submm spectroscopic maps
 with the SPIRE-FTS instrument  \citep{Griffin10}, and centered them to 
the coordinates of Sgr~B2(M): \mbox{R.A.\,$=$\,17$^{\rm h}$47$^{\rm m}$20.5$^{\rm s}$},~\mbox{Dec.\,$=$\,$-$28$\degree$23$\arcmin$06$\arcsec$}. The spectrometer used two bolometer arrays, the Short Wavelength Spectrometer array (SSW), with 37 detectors covering the wavelength range \mbox{194$-$313\,$\upmu$m} \mbox{(1545$-$958\,GHz)} and the Long Wavelength Spectrometer array (SLW), with 19 detectors covering the range \mbox{303$-$671\,$\upmu$m}  \mbox{(989$-$447\,GHz)}. The SLW and SSW detectors sampled an unvignetted  field  of view of about 2$\arcmin$ in diameter. The major advantage of this multi-beam FTS was that it obtained a complete $\sim$\,450 to 1545\,GHz spectrum
at every detector of the array. This allowed to map  nine rotational  lines of $^{12}$CO simultaneously. These spectra have the maximum spectral resolution allowed by the FTS scanning mechanism \mbox{($\sim$\,0.04\,cm$^{-1}$\,$=$\,1.2\,GHz)}. Figure~\ref{fig:SPIRE} shows a selection of these maps and \mbox{Figs.~\ref{fig:apendSPIRECO} and \ref{fig:apendSPIRE13COCI}  show all the others.}

These maps combine observations from an open time program (\mbox{$OT2\_jgoicoec\_5$}; 10.8\,h of observing time) and a \textit{Must-Do}  program (\mbox{$DDT\_mustdo\_1$}; 4.5\,h). In program \mbox{$OT2\_jgoicoec\_5$} \mbox{(obsID$=$1342265842)}, we mapped Sgr~B2 at intermediate spatial sampling (one beam spacing) using the bright mode to avoid saturation. The mapping mode was  raster scanning  with 29 pointings and 4 jiggle positions each. In \mbox{$DDT\_mustdo\_1$} \mbox{(obsID$=$1342252288)}, we covered the region  with sparse sampling
(a single pointing of the array per position, 36 in total at roughly two beam spacing).
The combined data provide nearly fully spatially sampled maps.
We calibrated the data using Herschel Interactive Processing Environment 
\citep[HIPE;][]{Ott10}. We adopted the extended emission calibration mode taking into account the diffraction losses and coupling efficiency to a point source model \citep[][]{Swinyard14}. The resulting integrated line intensities were extracted from the unapodized  interferograms after baseline subtraction and  classical $sinc$ function fits using the line-fitting tool available in HIPE. The FTS beam (the half power beam width or HPBW) depends on the frequency. It ranges from $\sim$\,17$''$ in the SSW detectors to $\sim$\,42$''$ in the SLW detectors. However, the FTS HPBW significantly departs from basic diffraction theory \citep[it does not simply  scale as the inverse of the frequency;][]{Swinyard14}. To compare the different line maps (among them and also with  photometric images)  we convolved all SPIRE-FTS  spectral-images to a uniform  resolution (HPBW) of 42$''$ (that of the CO $J$\,=\,4$-$3 line) using gaussian kernels. The typical 1$\sigma$ line sensitivity in these maps is about \mbox{10$^{-9}$\,W\,m$^{-2}$\,sr$^{-1}$}.

\subsection{Herschel and Spitzer photometric images}
%---------------------------------------  

We made use of archival photometric images 
of the FIR and submm dust continuum emission taken with \textit{Herschel} PACS
and SPIRE cameras. These observations are part of the \mbox{Hi-GAL \textit{Herschel}} Key Project \citep[the Herschel infrared Galactic Plane Survey;][]{Molinari2010}. We retrieved them from the Hi-GAL repository at \mbox{\url{https://vialactea.iaps.inaf.it}}. We made use of fully calibrated PACS\,70\,$\upmu$m and SPIRE\,250, 350, and 500\,$\upmu$m \mbox{images} in MJy\,sr$^{-1}$ units (we discarded the PACS\,160\,$\upmu$m data because they are saturated toward several positions of Sgr~B2).
These data include a correction of the zero-intensity level using  offsets derived from a comparison  with  \textit{Planck}/IRAS images \mbox{\citep{Bernard10}}. The nominal angular resolution (HPBW) of the \mbox{Hi-GAL \textit{Herschel}} images are \mbox{$\sim$\,6$''$ (70\,$\upmu$m)}, \mbox{$\sim$\,18$''$ (250\,$\upmu$m)}, \mbox{$\sim$\,25$''$ (350\,$\upmu$m)}, and \mbox{$\sim$\,36$''$ (500\,$\upmu$m)}. \mbox{In addition}, we also made use of the  \textit{Spitzer}/MIPS~24\,$\upmu$m image ($\sim$\,6$''$ resolution) taken as part of the MIPSGAL Legacy Survey \citep{Carey09}.

For the dust spectral energy distribution (SED) analysis \mbox{(see Sect.\,\ref{Sect:SED})} and to compare with the SPIRE-FTS spectral-images, we smoothed these photometric images to a uniform resolution of $\sim$\,42$''$ by convolving with gaussian kernels.

%-------------------------------------------------------------------------
\begin{table*}[th]
\centering
\caption{Spectroscopic and observational parameters of the lines discussed in this work.} 
\label{tab:linesall}
%\resizebox{1\textwidth}{!}{
\begin{tabular}{@{\vrule height 7pt depth 3pt width 0pt}lcccccccc}
\hline\hline
 {Species} & {Transition}  & {$\nu$}&$E_{\rm u}/k$  & $n_{\rm cr}(\rm H_2)^{a}$ &  $I(\rm{A})$ & $I(\rm{B})$ & $I(\rm{C})$ &   $I(\rm{D})$ \\
   &   & {[GHz]}    &   [K]      & {[cm$^{-3}$]} & {[W\,m$^{-2}$\,sr$^{-1}$]} & {[W\,m$^{-2}$\,sr$^{-1}$]} & {[W\,m$^{-2}$\,sr$^{-1}$]} &{[W\,m$^{-2}$\,sr$^{-1}$]}\\ 
\hline
{\NII}& \mbox{$^3P_1$\,$-$\,$^3P_0$}	& 1461.13 &  \phantom{0}70.1          &  173      & 3.0E-8  &   5.1E-8  &	9.5E-8  &   5.1E-8 	 \\
{\CI}&	\mbox{$^3P_2$\,$-$\,$^3P_1$}	& \phantom{0}809.23 & \phantom{0}62.5 &  2.0E+3   & 4.0E-8  &	3.6E-8 	&	4.0E-8  &   5.3E-8   \\	  
{\CI}&	\mbox{$^3P_1$\,$-$\,$^3P_0$}  & \phantom{0}492.15 & \phantom{0}23.6 &  1.0E+3  & 1.2E-8  &	1.1E-8	&	1.1E-8  &   1.4E-8   \\	   
\hline
CO  &	$J$\,=\,12$-$11	&   1381.99             & 431.3              &   1.8E+6  &   1.1E-8     &   1.3E-8 	&	1.2E-8 	&   2.2E-8   \\ 
CO	&	$J$\,=\,11$-$10	&	1267.01             & 365.0              &   1.4E+6  &   1.7E-8     &	2.0E-8 	&   2.1E-8 	&   3.6E-8   \\ 
CO	&	$J$\,=\,10$-$9	&	1151.90             & 304.2              &   1.0E+6  &   2.5E-8     &	3.2E-8  &	3.3E-8  &	5.2E-8   \\  
CO	&	$J$\,=\,9$-$8	    &	1036.91             & 248.0              &   8.4E+5  &   3.2E-8     &   4.6E-8  &   3.9E-8  &	7.0E-8   \\ 
CO	&	$J$\,=\,8$-$7	    &	\phantom{0}921.80   & 199.1              &   6.3E+5  &   5.6E-8     &	7.3E-8 	&	7.5E-8 	&   1.2E-7   \\
CO	&	$J$\,=\,7$-$6	    &	\phantom{0}806.65   & 154.9              &   4.4E+5  &   8.1E-8 	&	1.0E-7 	&	8.9E-8  &   1.3E-7 	 \\
CO	&	$J$\,=\,6$-$5	    &	\phantom{0}691.47   & 116.2              &   2.9E+5  &   8.0E-8 	&	8.7E-8 	&	6.3E-8  &   1.1E-7   \\	   
CO	&	$J$\,=\,5$-$4	    &	\phantom{0}576.26   &  \phantom{0}83.0   &   1.7E+5  &   6.8E-8	    &	6.3E-8 	&	4.3E-8  &   9.4E-8 	 \\	   
CO	&	$J$\,=\,4$-$3     &	\phantom{0}461.04   &  \phantom{0}55.3   &   8.7E+4  &   4.4E-8	    &	3.8E-8	&	3.2E-8  &   6.8E-8   \\	
\hline
$^{13}$CO	&   $J$\,=\,7$-$6	&   \phantom{0}771.18 & 148.1  &   3.9E+5      &   1.1E-9  &	\phantom{0}9.3E-10 &	4.0E-9  &   1.8E-9     \\
$^{13}$CO	&	$J$\,=\,6$-$5	&	\phantom{0}661.07 & 111.1  &   2.5E+5      &   \phantom{0}7.0E-10 &	1.2E-9 	&	2.3E-9  &   2.1E-9	\\	 
$^{13}$CO	&	$J$\,=\,5$-$4	&	\phantom{0}550.92 & \phantom{0}79.3   &  1.5E+5     &   2.4E-9	&	3.2E-9	&	5.5E-9  &	4.6E-9	\\  
\hline
H$_2$O	  &	\mbox{$J_{K_{\rm{a}},K_{\rm{c}}}$\,=\,$2_{1,1}$\,$-$\,$2_{0,2}$}   & \phantom{0}752.03   & 136.9  &       &   1.8E-9 	&	1.7E-9 	&	6.4E-9     &  2.3E-9 	\\	   
\hline
\ce{N2H+} & $J$\,=\,1$-$0 & \phantom{00}93.17  & \phantom{00}4.5 & 1.8E+5 & \phantom{0}4.1E-11 & \phantom{0}3.5E-11 & \phantom{0}8.0E-11 & \phantom{0}5.6E-11 \\
\ce{HCO+} & $J$\,=\,1$-$0    &  \phantom{00}89.19 & \phantom{00}4.3 & 2.1E+5  & \phantom{0}7.0E-11 & \phantom{0}5.6E-11 & \phantom{0}5.2E-11 & \phantom{0}1.1E-10 \\
HCN & $J$\,=\,1$-$0       & \phantom{00}88.63  & \phantom{00}4.3 & 2.0E+6 & \phantom{0}1.1E-10 & \phantom{0}7.8E-11 & \phantom{0}8.7E-11 & \phantom{0}1.8E-10  \\
SiO & $J$\,=\,2$-$1        & \phantom{00}86.85  & \phantom{00}6.3 & 1.6E+5 & \phantom{0}1.6E-11 & \phantom{0}1.3E-11 & \phantom{0}1.8E-11 & \phantom{0}3.8E-11  \\
\hline
\end{tabular}
%}
\vspace{0.1cm}
\tablefoot{Line intensities of the SPIRE-FTS observations extracted from maps convolved at a uniform resolution of $\sim$\,42$''$. Line intensity uncertainty of $\sim$\,20\,$\%$.$^a$ Critical density, 
$n_{\rm cr}$\,$=$\,$A_{\rm ul}$\,/\,$\gamma_{\rm ul}(T_{\rm k})$, in collisions with \textit{o}-H$_2$ at 50\,K. For $n_{\rm cr}$(\NII) in collisions with $e^-$ at 8000\,K.}
\end{table*}
%-------------------------------------------------------------------------

\subsection{IRAM\,30m line maps}
%-------------------------------  

We complemented  the \textit{Herschel} spectral- and photometric-images with  velocity-resolved maps of the  3\,mm band ($\sim$\,90\,GHz) molecular emission obtained by us  with the IRAM\,30m telescope in Pico Veleta (Spain). The size of these maps is \mbox{12$'$\,$\times$\,12$'$}. We used the E090 receiver, providing an instantaneous bandwidth of 16\,GHz per polarization, in combination with the 200\,kHz resolution FFT backend. We employed two local oscillator tunings to cover the \mbox{[84-92]+[100-108]\,GHz} and \mbox{[92-100]+[108-116]\,GHz} windows.
Each map consists of 9 sub-maps observed in the \mbox{``on-the-fly''} mode, scanning the sky in two orthogonal directions. The reference  position was taken 20$'$ away in right ascension. This position was calibrated by observing several positions outside the galactic plane. Here we focused only in the \mbox{SiO\;($J$\,=\,2$-$1)}, \mbox{\ce{N2H+}}, \mbox{HCN}, and \mbox{\ce{HCO+} ($J$\,=\,1$-$0)} emission (including the H$^{13}$CN and H$^{13}$CO$^+$ isotopologues). The angular resolution at the target frequencies is $\sim$\,27$''$. We carried out these  IRAM\,30m observations  in \mbox{August 2014} and 2015 during average summer weather conditions (up to 9\,mm of precipitable water vapor). Due to the low declination of the Galactic Center, Sgr~B2 was always below 25 degrees of elevation from Pico Veleta.
The total observing time was about 28\,h and we reached a 1$\sigma$ sensitivity of $\sim$\,60\,mK per  1\,\kms\;resampled velocity channel. 
We reduced this data with the GILDAS software and subtracted a polynomial baseline of order 1 or 2 avoiding the velocities with molecular emission. We finally gridded the spectra into a data cube through a convolution with a Gaussian kernel of about one third of the HPBW.  
%%All maps were projected to a common spatial grid.

%-----------------------------------------------------------------------------------------
\begin{figure*}[!ht]
\centering
\includegraphics[height=0.44\textwidth]{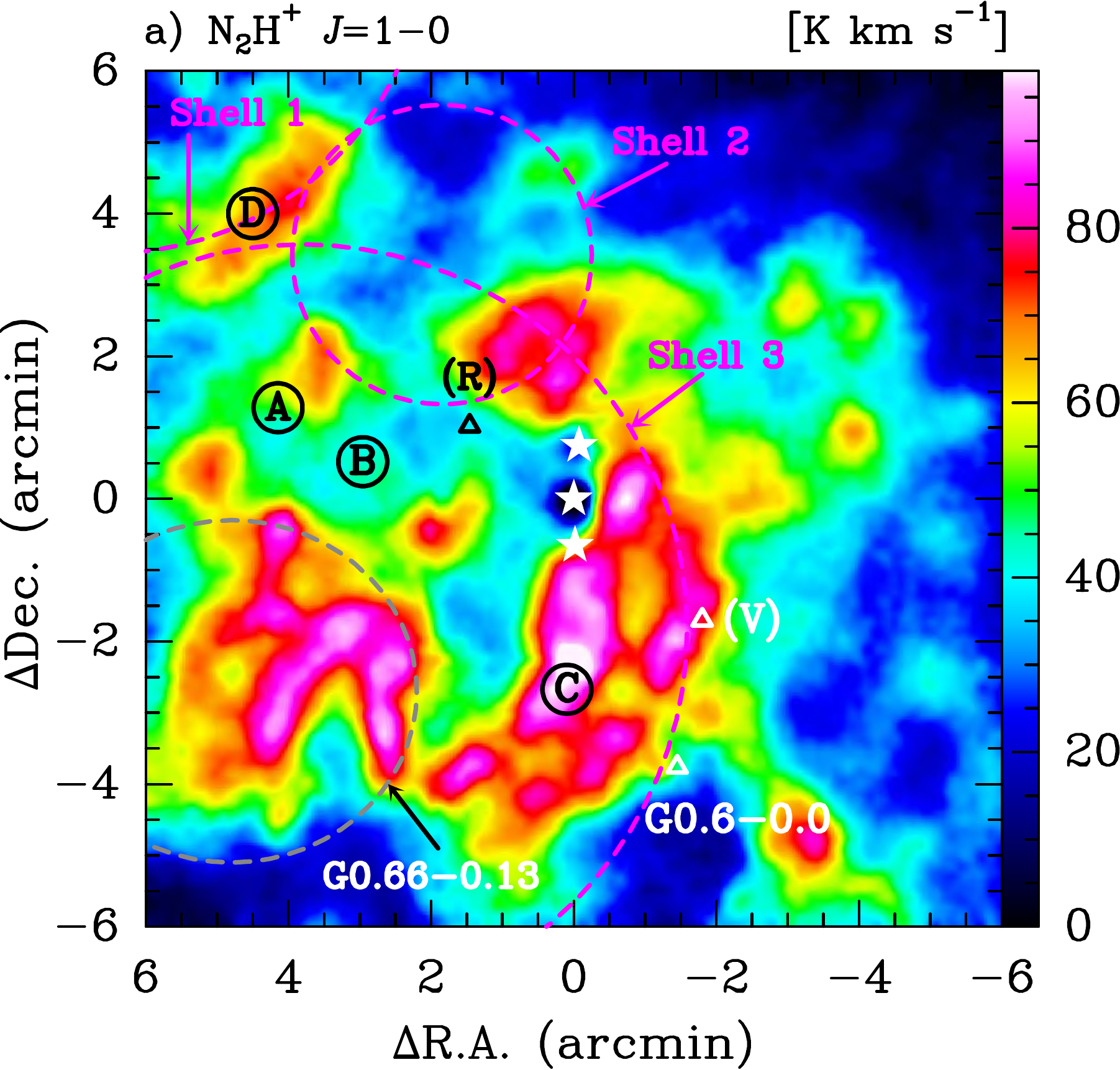} 
\hspace{0.4cm}
\includegraphics[height=0.44\textwidth]{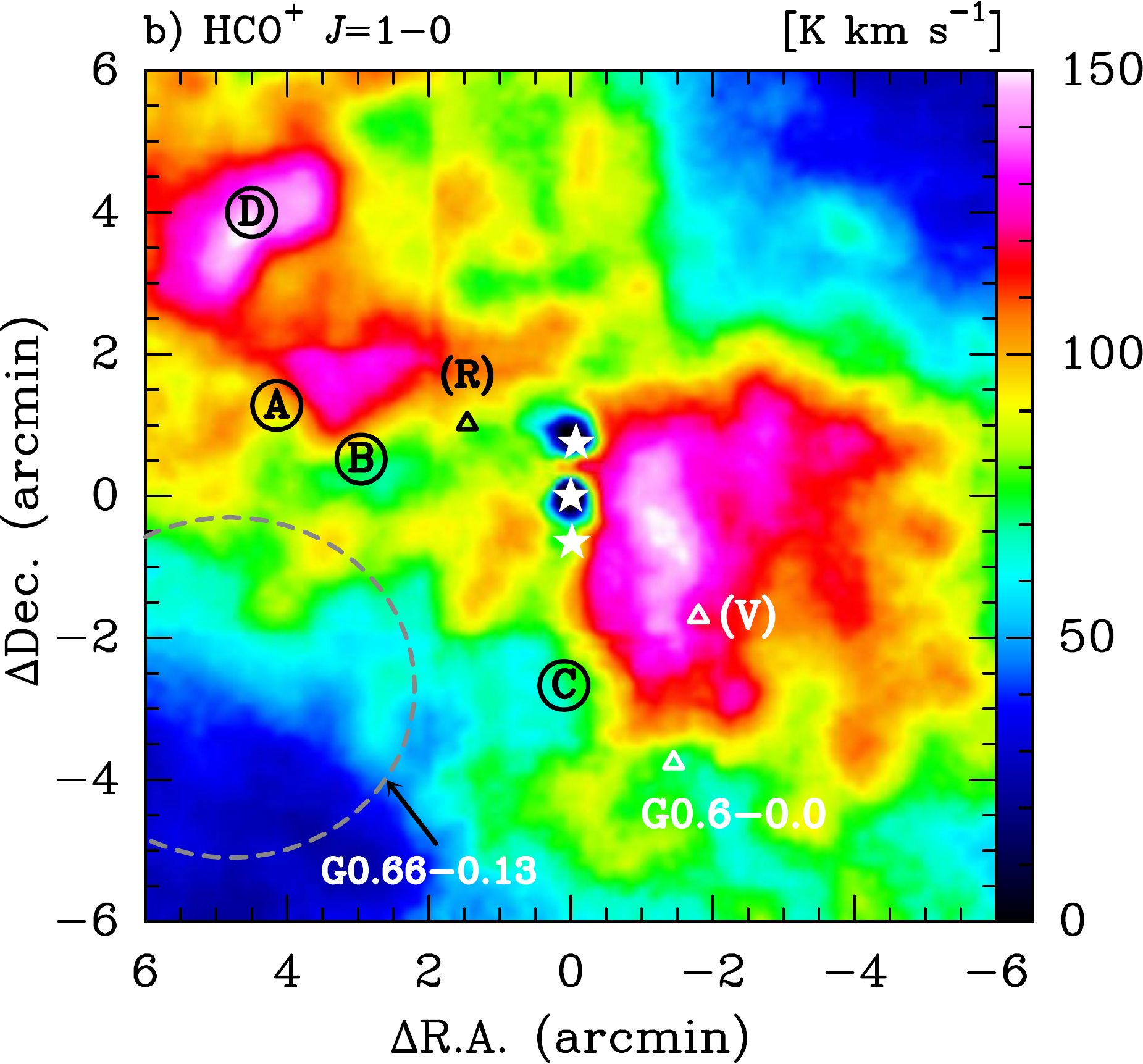}\\ \vspace{0.3cm} 
%\vspace{0.6cm}
\includegraphics[height=0.44\textwidth]{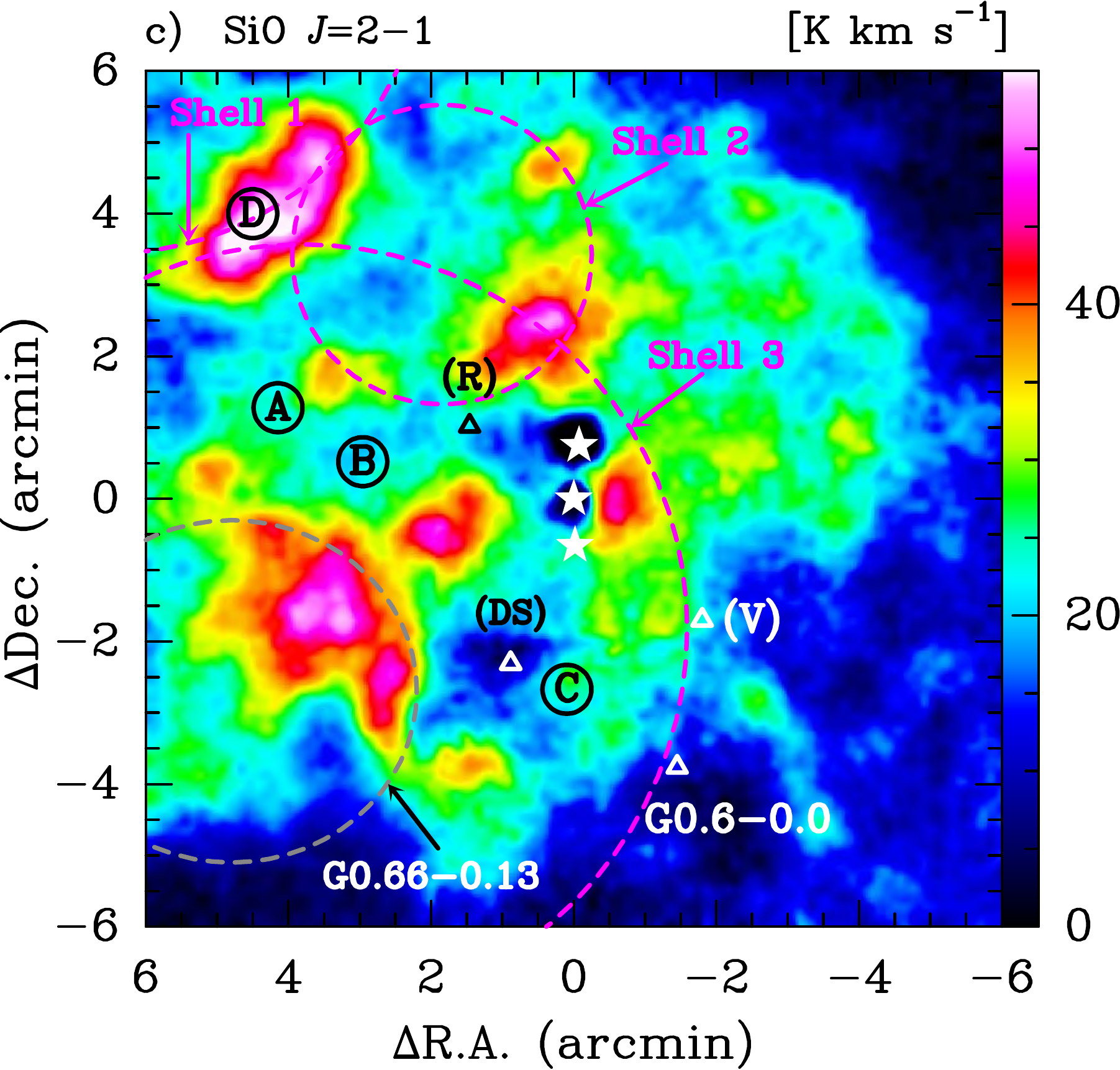} 
\hspace{0.4cm}
\includegraphics[height=0.44\textwidth]{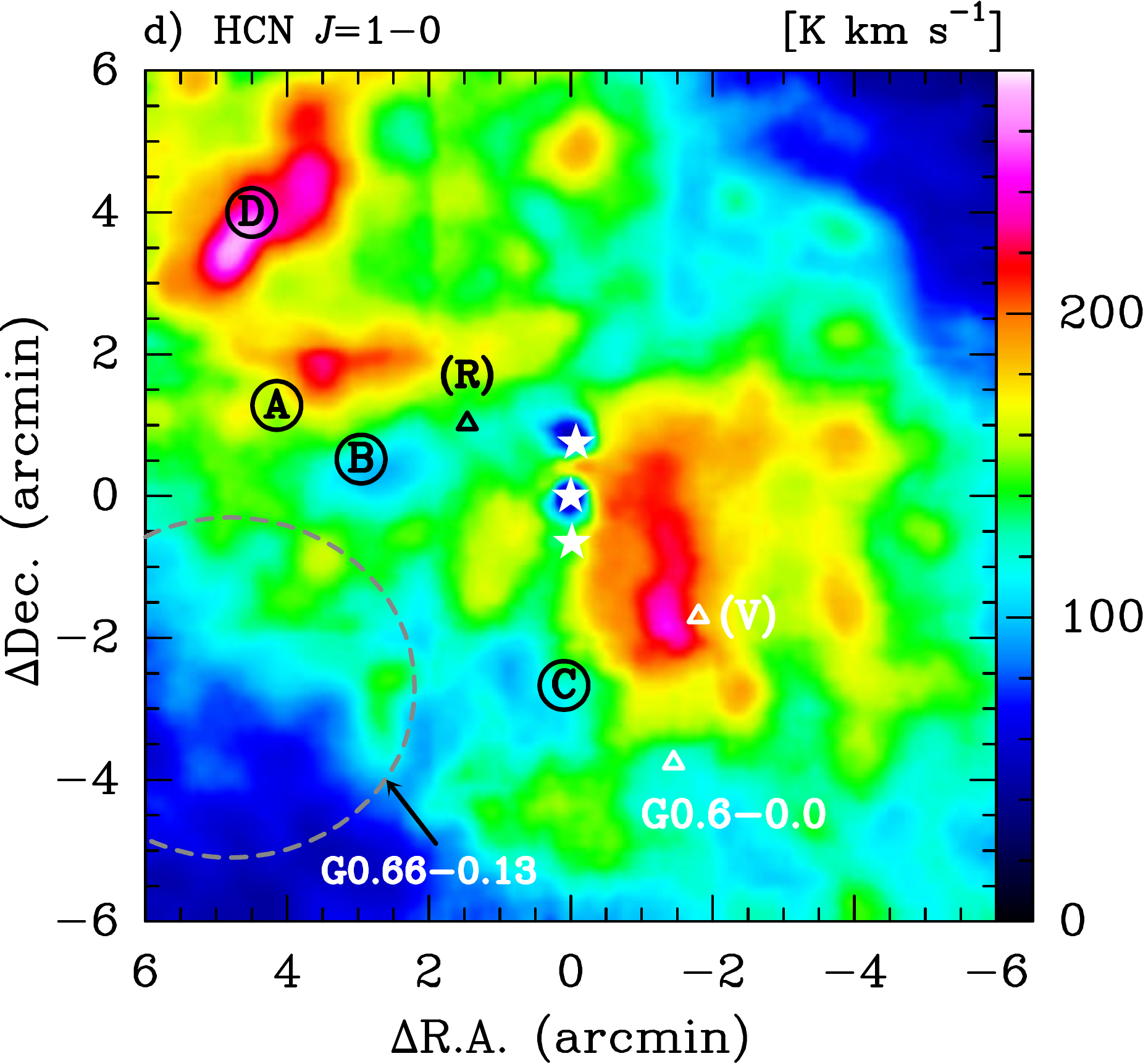}
\vspace{0.3cm}
\caption{Total integrated line intensity maps (in units K \kms) of different molecular lines  observed with the IRAM\,30m telescope toward the Sgr~B2 molecular complex and integrated in the LSR velocity range [$-20, 120$]\,\kms. The gray dashed circle marks a specific X-ray irradiated region \citep{Zhang_2015}. The pink dashed circles show  Shells 1, 2, and 3 of \citet{Tsuboi2015}.}
\vspace{-0.4cm}
\label{fig:30mtotal}
%\vspace{0.1cm}
\end{figure*}
%-----------------------------------------------------------------------------------------

%%%%%%%%%%%%%%%%%%%%%%
\section{Results} \label{sect:results}
%%%%%%%%%%%%%%%%%%%%%%

\subsection{The submm spectrum of the Sgr B2 envelope}
%-----------------------------------------------------  

The mapped area of Sgr~B2 envelope extends $\sim$1000\,pc$^2$ around the main star-forming cores Sgr~B2(N, M, S). To simplify the analysis, we chose four positions at different distances from Sgr~B2(M): \mbox{A ($\Delta$R.A, $\Delta$Dec.)\,=\,(4.15$'$,\,1.28$'$)}, \mbox{B (2.95$'$,\,0.52$'$)}, and \mbox{C (0$'$,\,$-$2.67$'$)} as representative positions of the physical conditions in the extended cloud environment, and \mbox{D (4.5$'$,\,4$'$)}, a peculiar and remarkably bright position at the edge of the mapped area.
Figure~\ref{fig:spectra} shows the  continuum-divided \mbox{SPIRE-FTS} spectra from 450 to 1545\,GHz obtained toward the four positions (A, B, C, and D) located at 10.4, 7.2, 6.4, and 15\,pc from \mbox{Sgr B2(M)}, respectively. Their submm spectra are dominated by rotationally excited CO emission lines, often called \mbox{``mid-$J$ CO} lines''. The brightest ones, those contributing more to the gas cooling, are the \mbox{$J$\,=\,6$-$5}, 7$-$6, and 8$-$7 lines. The spectra also show the emission from the less abundant $^{13}$CO lines (from \mbox{$J$\,=\,5$-$4 to 7$-$6}). Other lines also detected in emission  include the \mbox{\ce{H2O}\:$J_{K_{\rm{a}},K_{\rm{c}}}$\,=\,$2_{1,1}-2_{0,2}$} ($\sim$\,752\,GHz) line as well as  the atomic fine-structure lines \mbox{\CI\,492, 809\,GHz} and \NII\,205\,$\upmu$m (\mbox{Table~\ref{tab:linesall}}). In addition, the spectra  display  absorption lines produced by \mbox{low-energy}  rotational lines of hydride molecules: CH$^+$, OH$^+$, H$_2$O, HF, H$_2$S, NH, NH$_2$, and NH$_3$ \citep[for a review on hydrides see][]{Gerin16}. These absorption lines   are produced by diffuse clouds  in the line of sight toward the GC (in the spiral arms of the Galaxy), by gas in the CMZ, and by gas in the Sgr~B2 envelope itself. \mbox{\textit{Herschel}/HIFI} detected and spectrally resolved these absorption lines toward Sgr~B2(N) \citep[e.g.,][]{Neill2014,Indriolo15}.
 In these high-resolution HIFI spectra, the ground-state lines of abundant species such as
\mbox{$o$--H$_2$O ($\sim$557\,GHz)}, \mbox{CH$^+$ ($\sim$835\,GHz)}, or \mbox{$p$--H$_2$O ($\sim$1113\,GHz)}, are saturated to \mbox{$I_{\nu}/I_{\rm cont}$\,$\simeq$\,0}  (i.e., they absorb 100\,$\%$ of continuum) over a broad velocity range: \mbox{$\Delta v$\,$\simeq$150$-$200\,km\,s$^{-1}$}. At the low velocity resolution of \mbox{SPIRE-FTS}, inversely proportional to the frequency, these spectrally unresolved lines  absorb $\sim$30, $\sim$40, and $\sim$50\,$\%$ of the continuum emission from Sgr~B2(N), respectively.  However, their absorption depth across the envelope is smaller ($\sim$10, $\sim$20, and $\sim$30\,$\%$) which  means either that they are not saturated, or that they are saturated over a narrower velocity range. Lines absorbing less than  the above percentages (ground-state lines from less abundant species or rotationally excited lines) are certainly not saturated, and thus have absorption depths roughly proportional to $e^{-\tau}$.
Even if the \mbox{continuum} level changes from one position to another in the envelope,  the absorption depth of these lines do not show remarkable differences from one position to another in these maps. This implies a rather  uniform column density of foreground absorbing material.

\subsection{Distribution of mid-$J$ CO, \CI\, and \NII\,205\,$\upmu$m emission}
%---------------------------------------------------------------------------- 

Figure~\ref{fig:SPIRE} shows the spatial distribution of the 
\mbox{CO $J$\,=\,10$-$9} (Fig.~\ref{fig:SPIRE}a), \mbox{\CI\,809\,GHz}~(Fig.~\ref{fig:SPIRE}b), \mbox{\ce{H2O} $J_{K_{\rm{a}},K_{\rm{c}}}$\,=\,2$_{1,1}$\,$-$\,$2_{0,2}$} (Fig.~\ref{fig:SPIRE}c), and \mbox{\NII\,205\,$\upmu$m} (Fig.~\ref{fig:SPIRE}d) integrated line emission  (i.e.,~their surface brightness map). The spatial distribution of each CO emission line depends on the rotational number $J$. Compared to the extended emission of the lower $J$ lines  (see Fig.~\ref{fig:apendSPIRECO}a and b), the  higher $J$ lines show a more compact distribution around the main star-forming cores (Fig.~\ref{fig:apendSPIRE13COCI}g and h). The \mbox{CO (10$-$9)}  surface luminosity around  Sgr~B2(M, and N) is very high, \mbox{$L_{\rm 10-9}$\,$\simeq$\,150\,\Ls\,pc$^{-2}$} (over an area of $\sim$\,1.3\,pc$^{2}$). For comparison, this is much more luminous than toward a similar area around the Trapezium cluster (that hosts only a few late-type O stars) in Orion\,A star-forming region \citep[\mbox{$L_{\rm 10-9}$\,$\simeq$\,4\,\Ls\,pc$^{-2}$},][]{Goicoechea2019}. In Sgr~B2, the \mbox{CO (10$-$9)} line luminosity remains high through the $\sim$\,1000\,pc$^2$ area of the mapped regions (\mbox{$L_{\rm 10-9}$\,$\simeq$\,0.8\,\Ls\,pc$^{-2}$}).
Interestingly, the \mbox{CO (6$-$5)} emission does not peak toward the main cores but toward position D. This specific region is discussed in Sect.~\ref{subsubsec:P}. 

Figure~\ref{fig:SPIRE}b shows the spatial distribution of the \mbox{\CI\,$^3P_2$\,$-$\,$^3P_1$} fine-structure emission at 809\,GHz. This line\footnote{The \CI\,809\,GHz and CO~$J$\,=\,7$-$6 lines are blended in the apodized FTS spectra, but not in the unapodized spectra. We independently extracted their integrated line intensities from the  unapodized spectra.} from  neutral atomic carbon is very widespread through out the envelope (the distribution of the \CI\,$^3P_1-^3P_0$ emission at 492\,GHz is similar, see Fig.~\ref{fig:apendSPIRE13COCI}l).
Figure~\ref{fig:SPIRE}c shows that the emission from rotationally excited water vapor \mbox{($p$-H$_2$O\;$2_{1,1}-2_{0,2}$} line) is also extended over scales of tens of pc. Detailed excitation and radiative transfer models already predicted that this line would be observed in emission over a broad range of physical conditions and background FIR illuminations \citep{Cernicharo_2006}.

Finally, Fig.~\ref{fig:SPIRE}d shows a map of the \mbox{\NII\,$^3P_1-^3P_0$} fine-structure line emission at 205\,$\upmu$m. This line stems from  ionized gas. The ionization potential of nitrogen is 14.5\,eV, thus the detection of  \mbox{\NII\,205\,$\upmu$m}  emission implies the presence of ionizing EUV photons, perhaps also \mbox{X-rays}, electron collisional ionization, or proton charge exchanges \citep[e.g.,][]{Langer2015}. This line is easily excited by electron collisions because of its  low critical density,  \mbox{173\,cm$\rm ^{-3}$} at 8000\,K \citep[from collisional rates of ][]{Tayal11}, and because of the small energy difference of the fine-structure levels \mbox{($\Delta E$\,/\,$k$\,$=$\,70\,K)}. Hence, this map traces the spatial distribution of relatively low-density ionized gas. \cite{Goicoechea_2004} determined electron densities down to \mbox{$n_e$\,$\simeq$\,50$-$100\,cm$^{-3}$}  in the southern  envelope (with an average of 240\,cm$^{-3}$)  from \mbox{low-angular} resolution ($\sim$80$''$)  \mbox{\textit{Infrared Space Observatory}}  observations of the \mbox{[\OIII]\,52, and 88\,$\upmu$m} lines. We suspect that this is the same ionized gas component, which is much more extended than the dense \HII~regions in the vicinity of young massive stars typically and traced by mm-wave hydrogen radio recombination lines \mbox{\citep[e.g.,][]{Jones2012}}. The \mbox{\NII\,205\,$\upmu$m}  emission is brighter in the southern envelope and displays a remarkably different spatial distribution than that of the \mbox{mid-$J$ CO} and \CI\;lines. 
Table~\ref{tab:corr30m} shows the derived correlation coefficients of the spatial distribution from several gas lines and dust emission wavelengths mapped
in Sgr~B2.

%---------------------------------------------------------------
\begin{figure*}[!ht]
\centering
\includegraphics[height=0.44\textwidth]{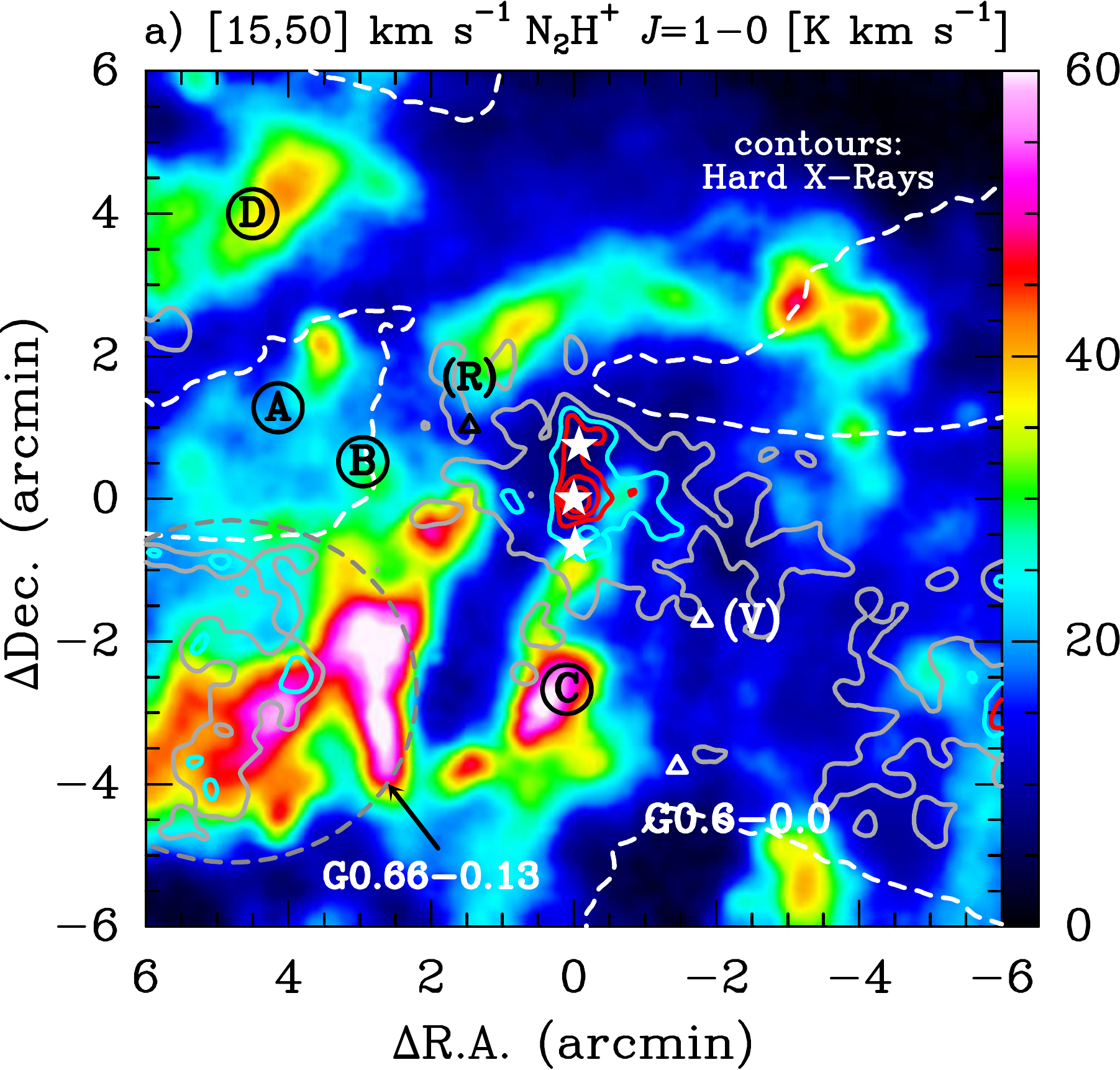} \hspace{0.4cm}
\includegraphics[height=0.44\textwidth]{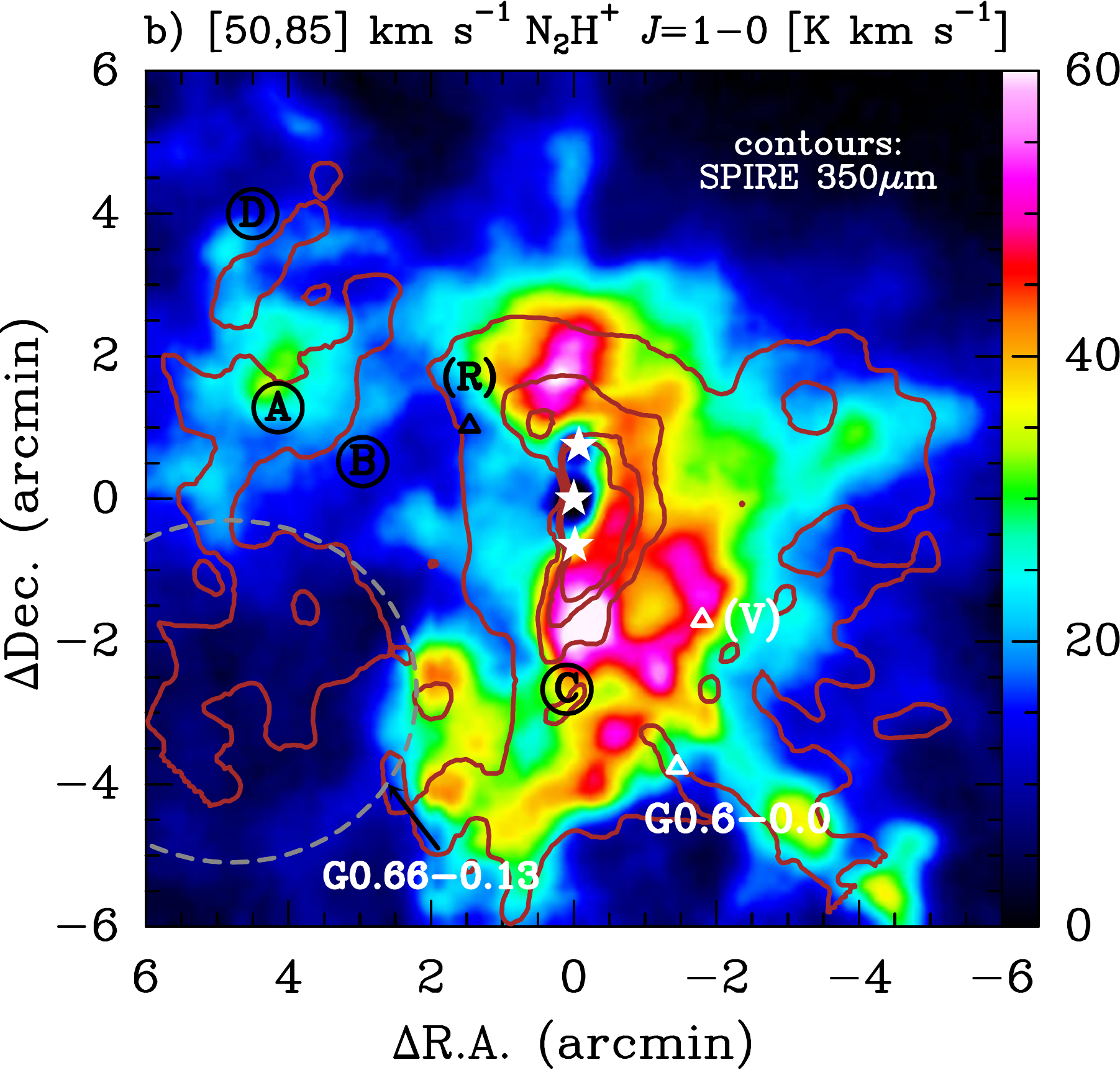}\\ \vspace{0.4cm}
\includegraphics[height=0.44\textwidth]{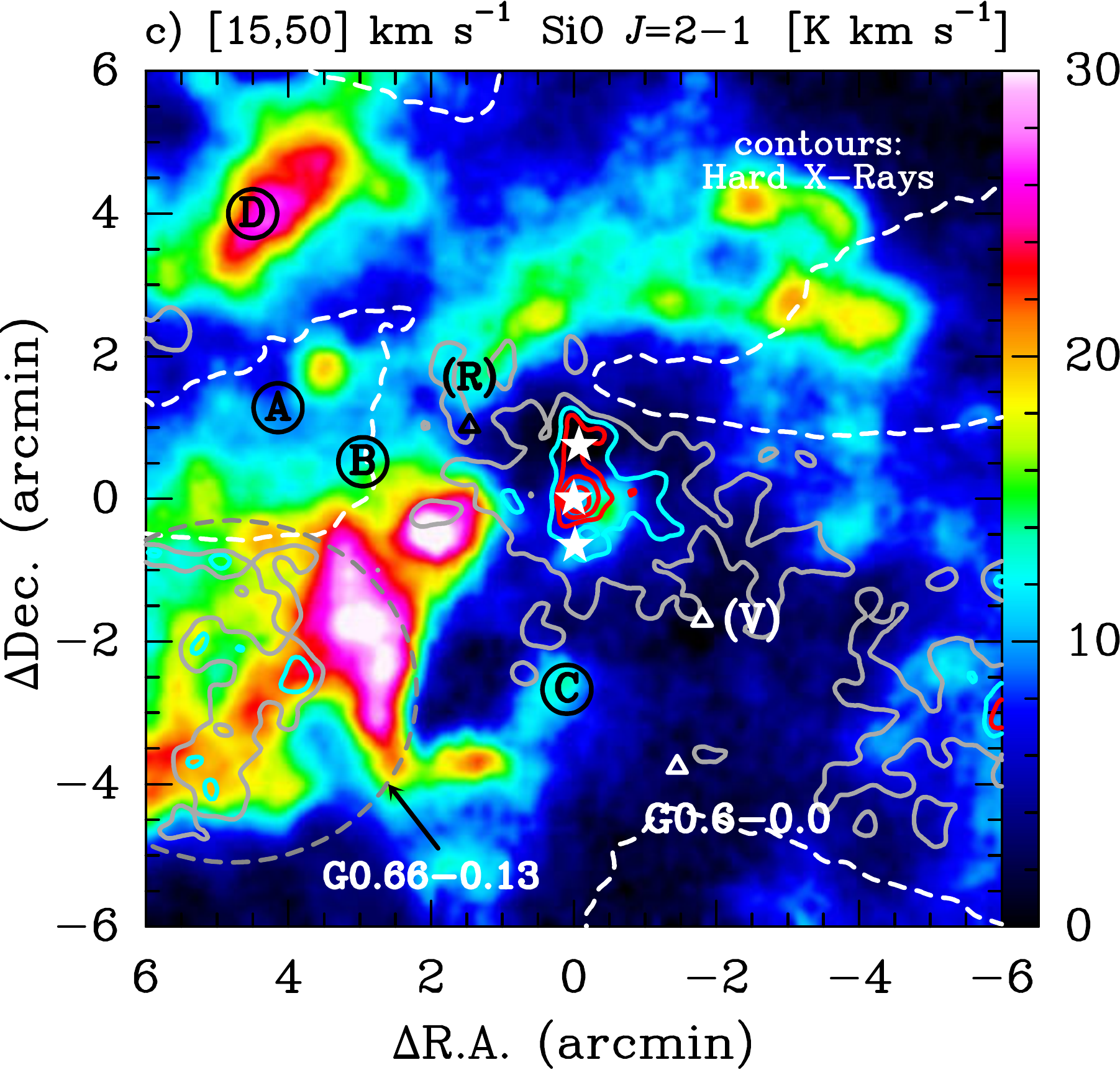} \hspace{0.4cm} 
\includegraphics[height=0.44\textwidth]{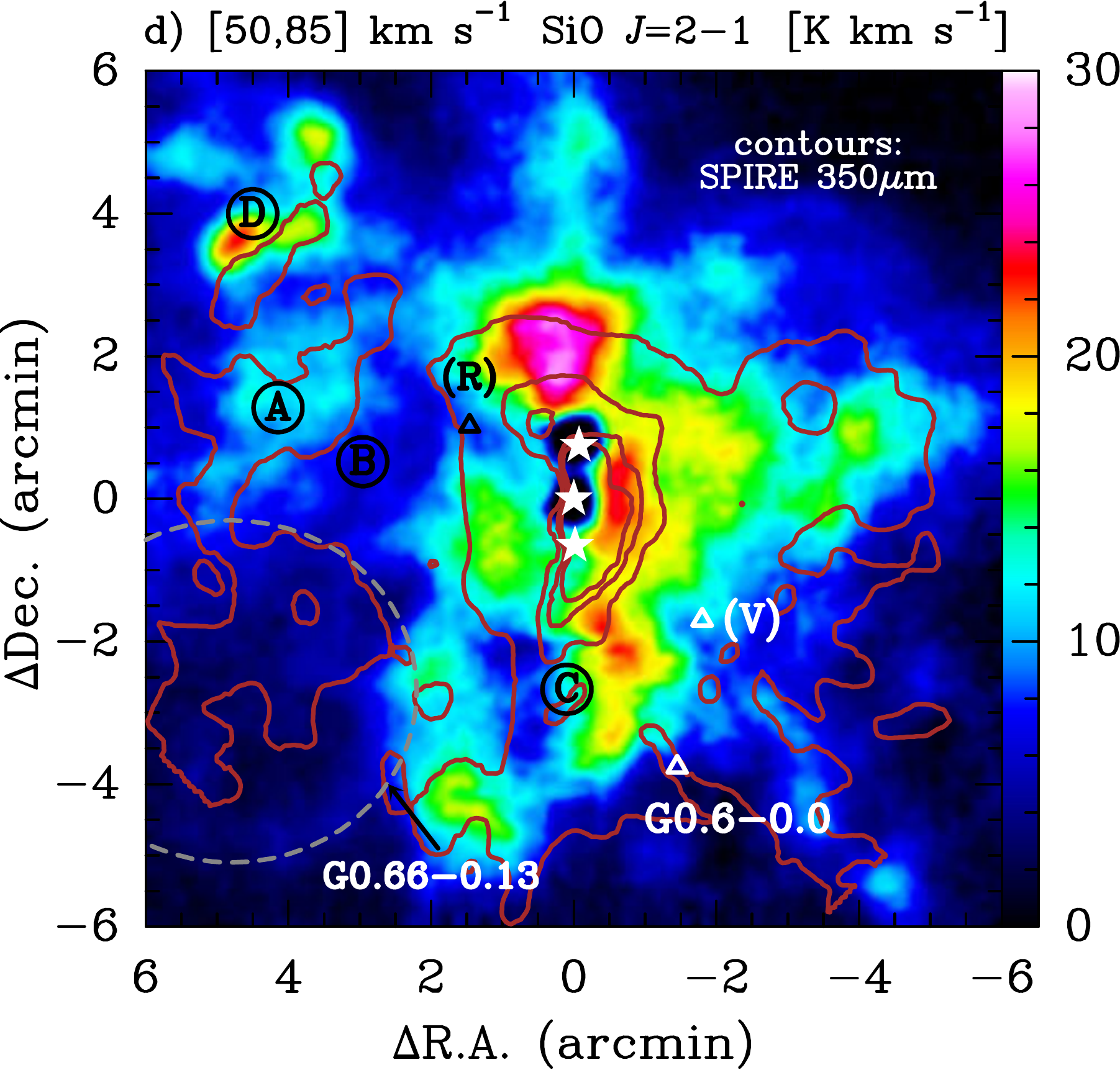}
\vspace{0.2cm}
\caption{IRAM\,30m maps of the \mbox{\ce{N2H+} (1$-$0)} and \mbox{SiO~(2$-$1)}  line intensities integrated
in the  \mbox{15$-$50\,\kms} and 50$-$85\,\kms ranges. 
Contours in panels $a$ and $c$ represent the 2015 hard X-ray emission integrated from 3 to 79\,keV  \citep[][]{Zhang_2015}. \mbox{Gray:~$19\cdot10^{-6}\;\rm{ph\,s^{-1}\,pixel^{-1}}$}. \mbox{Cyan:~$22\cdot10^{-6}\;\rm{ph\,s^{-1}\,pixel^{-1}}$}. \mbox{Red:~(24, 26, 28)\,$\cdot$\,$10^{-6}\;\rm{ph\,s^{-1}\,pixel^{-1}}$}. 
Dashed white lines delineate the regions with no X-ray observations.
Contours in panels $b$ and $d$ represent the 350\,$\upmu$m emission from 8 to 68 by 20 ($10^3$)\,MJy\,sr$^{-1}$.}
\label{fig:30mSiO}
\end{figure*}
%---------------------------------------------------------------

\begin{figure*}[ht]
\centering
\includegraphics[height=0.44\textwidth]{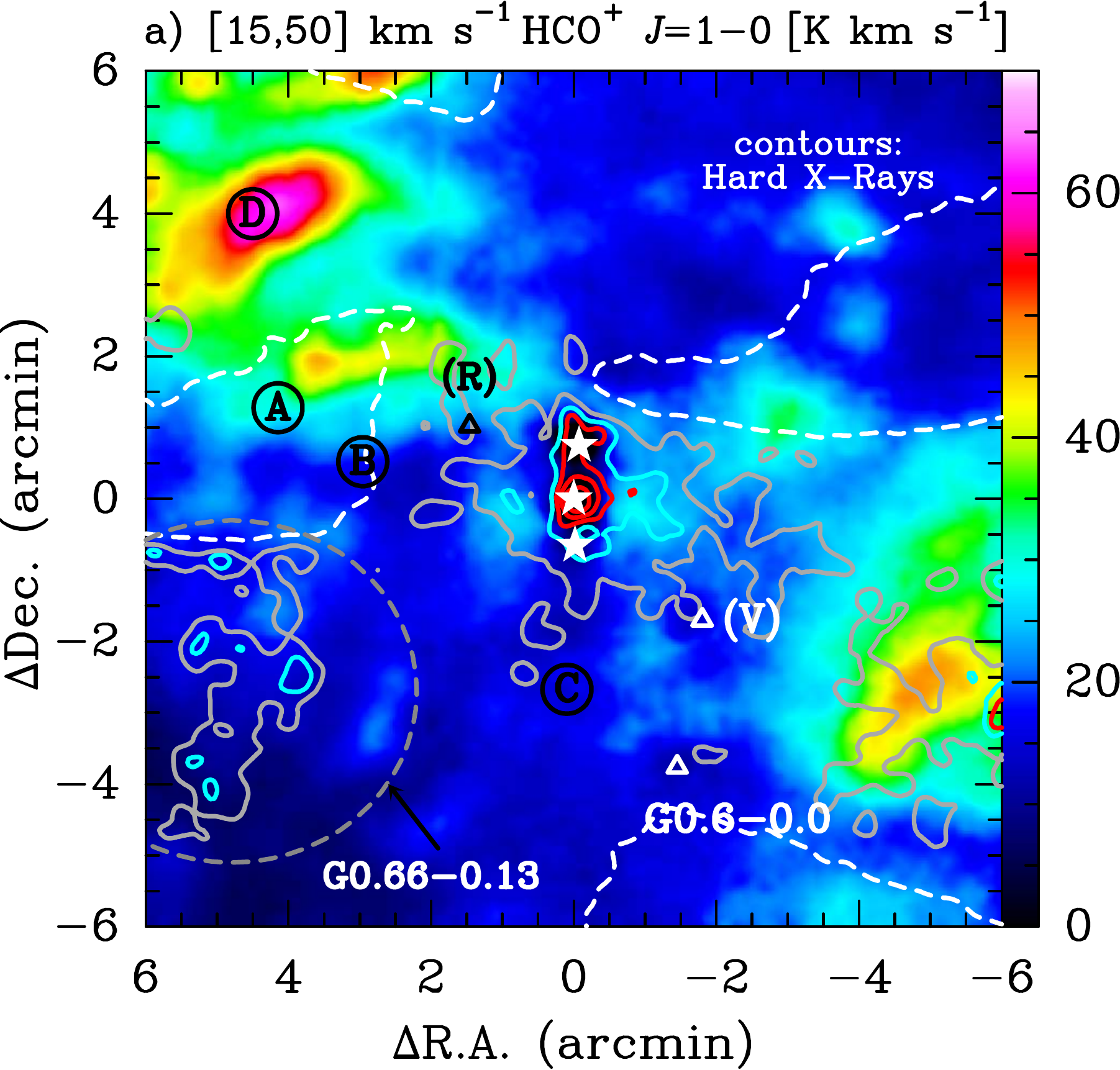}
\hspace{0.3cm}
\includegraphics[height=0.44\textwidth]{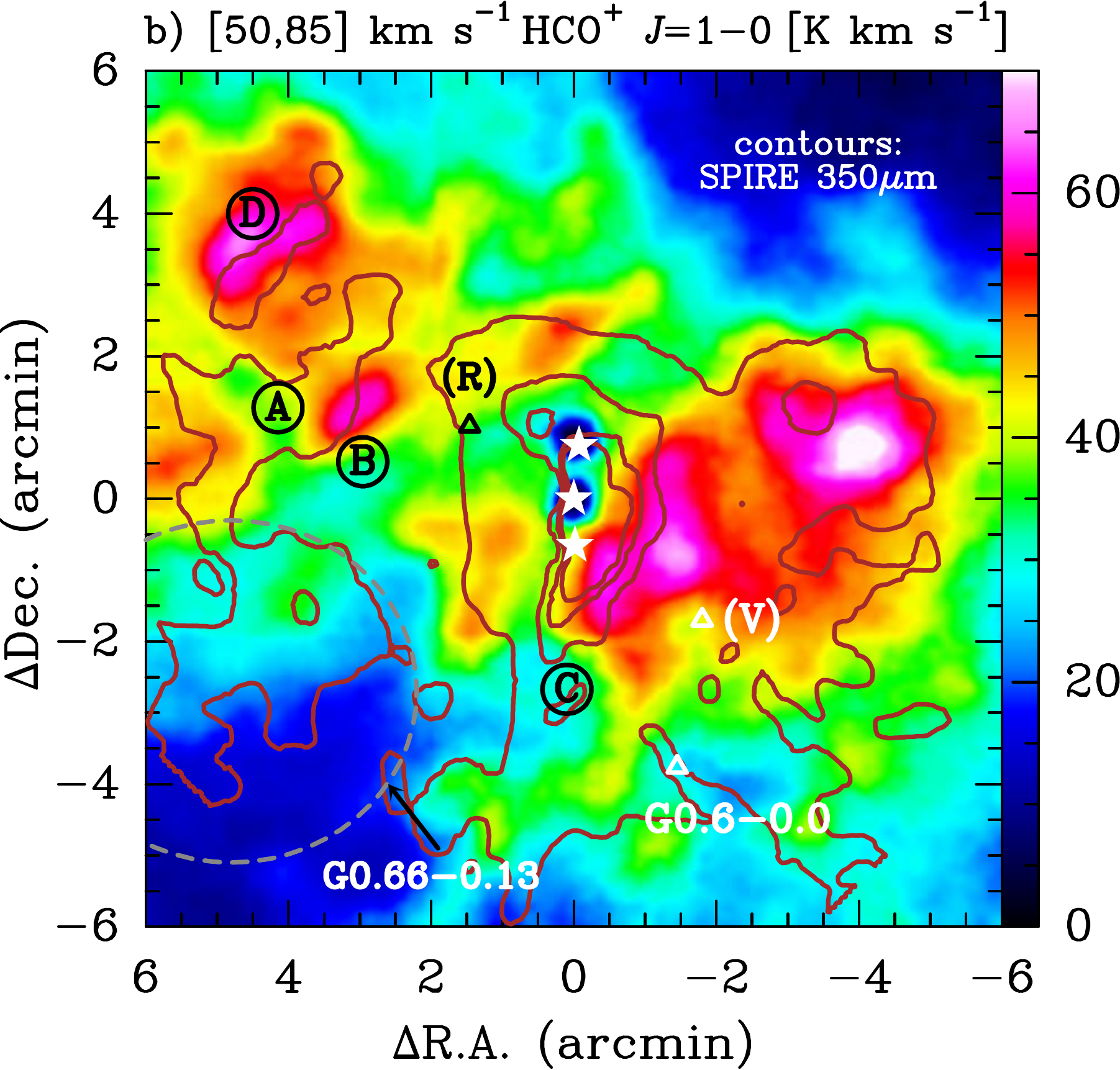} 

\vspace{0.2cm}
\includegraphics[height=0.44\textwidth]{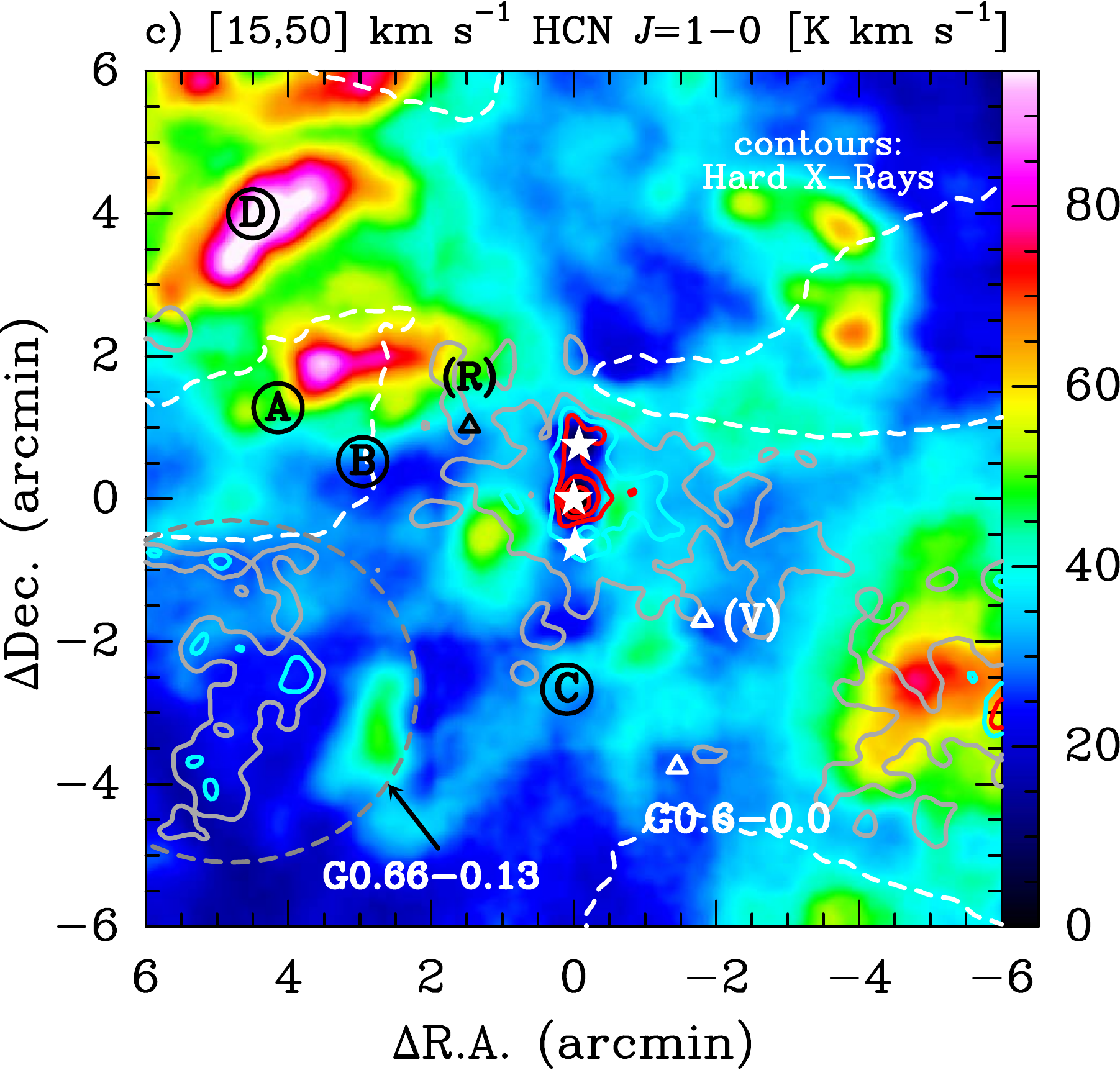}
\hspace{0.35cm}
\includegraphics[height=0.44\textwidth]{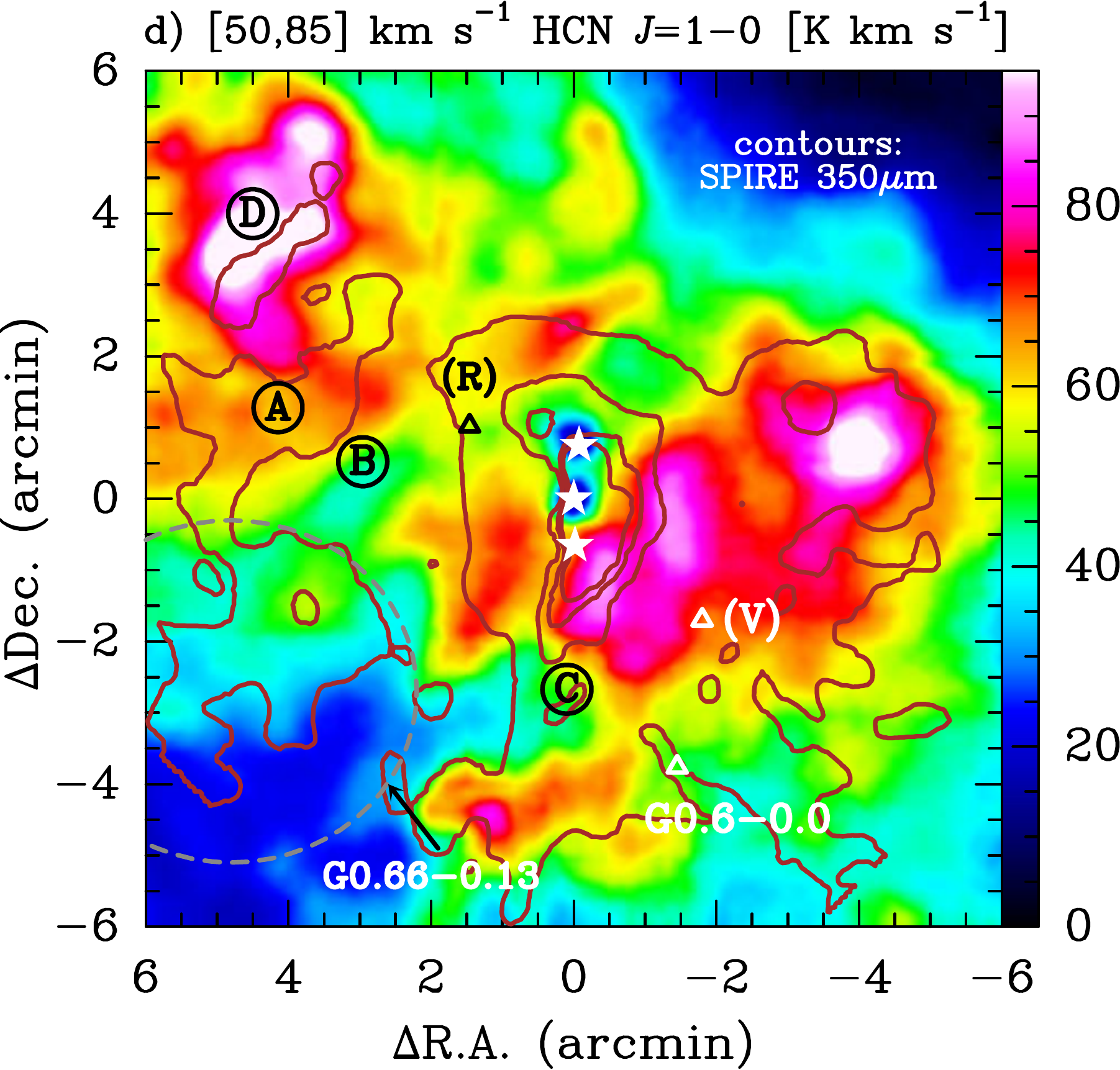}
\vspace{0.35cm}
\caption{IRAM\,30m maps of the \ce{HCO+}~(1$-$0) and \ce{HCN}~(1$-$0) line intensities integrated  in the 15$-$50\,\kms\;and 50$-$85\,\kms\;ranges.  Contours in panels $a$ and $c$ represent the hard X-Rays emission integrated from 3 to 79 keV, using the 2015 \textit{NuSTAR} data \citep[][]{Zhang_2015}. \mbox{Gray: $19\cdot10^{-6}\;\rm{ph\,s^{-1}\,pixel^{-1}}$}. \mbox{Cyan: $22\cdot10^{-6}\;\rm{ph\,s^{-1}\,pixel^{-1}}$}. \mbox{Red: (24, 26, 28)\,$\cdot$\,$10^{-6}\;\rm{ph\,s^{-1}\,pixel^{-1}}$}. 
Dashed white lines delineate the regions with no X-ray observations
Contours in panels $b$ and $d$ represent the 350\,$\upmu$m data from 8 to 68 by 20 ($\cdot10^3$)\,MJy\,sr$^{-1}$. }
\label{fig:30mHCN-HCO}
\end{figure*}
%---------------------------------------------------------------

\subsection{Spatial distribution of HCN, \ce{HCO+}, SiO, and \ce{N2H+}\label{ssec:maps30m}} 

\subsubsection{Integrated intensity line maps} \label{subsubsec:integratedmaps30m}

Figure~\ref{fig:30mtotal} shows the  \mbox{\ce{N2H+} (1$-$0)} (Fig.~\ref{fig:30mtotal}a), \mbox{\ce{HCO+} (1$-$0)} (Fig.~\ref{fig:30mtotal}b), \mbox{SiO (2$-$1)} (Fig.~\ref{fig:30mtotal}c), and \mbox{HCN (1$-$0)} (Fig.~\ref{fig:30mtotal}d)  integrated line intensity maps (over the entire
emission LSR velocity range \mbox{$-20$ to 120\,\kms}). 
Unexpectedly, the spatial distribution of the \mbox{SiO (2$-$1)} emission, usually
associated to warm shocked gas, resembles that of \mbox{\ce{N2H+} (1$-$0)}, considered as a tracer of quiescent cold gas  (see discussion in Sec.~\ref{subsubsec:SiO-N2Hp}). In Sgr~B2 their spatial distribution is extended and moderately correlated\footnote{We use the Spearman's correlation, instead of the usual Pearson's correlation, because the molecular emission is not normally distributed. Spearman's correlation is a non parametric statistic which measures the monotonic relationship. A $\rho$\,=\,0 does not imply that there is no relationship between the variables, just that it is not monotonic.}  ($\rho$\,=\,0.7). 

The \mbox{HCN (1$-$0)} and \mbox{\ce{HCO+} (1$-$0)} emission shows a similar spatial distribution. Their integrated line emission is strongly correlated ($\rho$\,=\,0.95). 
The average \mbox{HCN/H$^{13}$CN (1$-$0)} and \mbox{\ce{HCO+}\,/\,H$^{13}$CO$^+$ (1$-$0)} line intensity ratios in the mapped area are 7 and 13, respectively. These ratios are lower than the  $^{12}$C\,/\,$^{13}$C isotopic abundance ratio derived in Sgr~B2 \citep[$\sim$\,20$-$30; e.g.,][]{Langer_1990}.
Therefore,  the \mbox{HCN (1$-$0)} and \mbox{\ce{HCO+} (1$-$0)} emission is optically thick at large spatial scales \citep[for a detailed analysis in the entire CMZ see][]{Jones2012}. 

Owing to the  strong mm continuum emission from \mbox{Sgr~B2(M, N)} cores, these low lying molecular lines turn to \mbox{absorption} \mbox{(i.e., $T_{\rm ex}$\,$<$\,$T_{\rm cont}$)} at many LSR velocities. The absorptions are produced by molecular gas in translucent clouds
of the galactic spiral arms located in the line of sight toward the GC \citep[but not related to Sgr~B2, e.g.,][]{Greaves94} and by  gas in the Sgr~B2 envelope itself, that absorbs part of the line and continuum emission from the cores. These absorptions are responsible of the apparent lack of molecular emission toward the cores in the integrated line intensity maps of Figs.~\ref{fig:30mtotal}, \ref{fig:30mSiO}, and \ref{fig:30mHCN-HCO}.

\subsubsection{Sgr~B2 envelope gas velocity components} \label{subsubsec:velcomp}

\begin{figure}[t]
    %\centering
    \vspace{0.1cm}
    \hspace{-0.2cm}
    \includegraphics[width=0.47\textwidth]{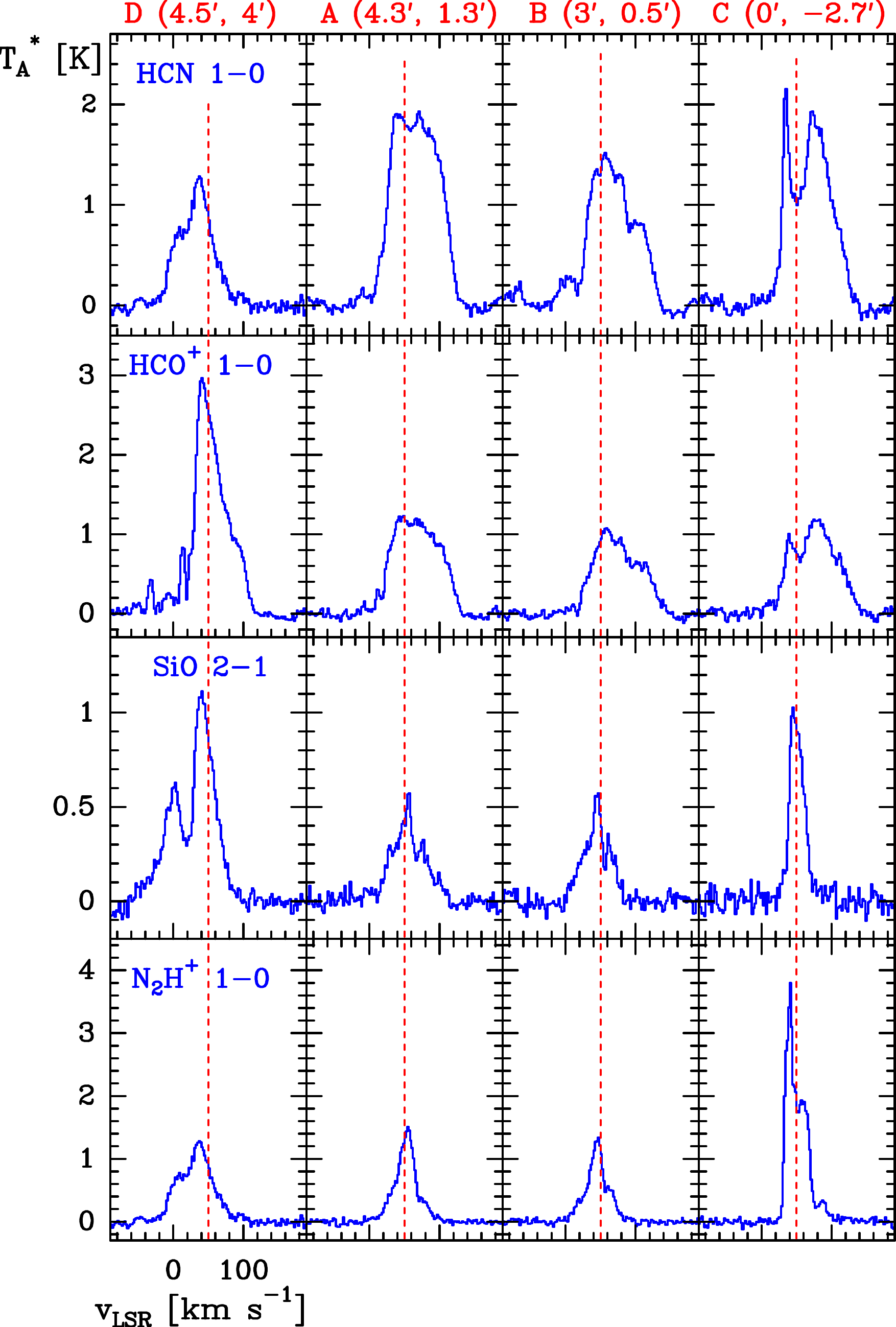}
    \vspace{0.1cm}
    \caption{HCN~(1$-$0), \ce{HCO+}~(1$-$0), \mbox{SiO~(2$-$1)}, and \mbox{\ce{N2H+}~(1$-$0)} line profiles  toward positions D, A, B, and C. The  vertical dashed line at $v_{\rm LSR}$\,=\,50\,\kms\;marks a representative
    velocity of  Sgr B2 envelope.} %\citep[][]{Thiel_2019}.}
    \label{fig:30mspectra}
    %\vspace{-0.2cm}
\end{figure}

The gas kinematics in the CMZ is complex and intricate. The average number of velocity components per observed line of sight in the CMZ is 1.6, and \mbox{$>$\,2} in Sgr~B2 \mbox{\citep{Henshaw16}}. In order to ease the discussion, here we split the wide range of emission from Sgr~B2 molecular cloud in two different velocity ranges \mbox{$\rm{v_{LSR}}$\,$=$\,$[15,~50]$\,\kms} and \mbox{[50,~85]\,\kms}. The  \mbox{$\rm{v_{LSR}}$\,$=$\,$[50,~85]$\,\kms}\;component (see Figs.~\ref{fig:30mSiO} and \ref{fig:30mHCN-HCO}) covers the main emission velocities of the Sgr~B2 molecular cloud \citep{Thiel_2019}. Figure \ref{fig:30mSiO} shows the spatial distribution of the \mbox{\ce{N2H+}~(1$-$0)} and \mbox{SiO~(2$-$1)} emission integrated in the \mbox{[15,~50]\,\kms} and \mbox{[50,~85]\,\kms}\;velocity ranges separately. Figure~\ref{fig:30mspectra} shows  \mbox{HCN~(1$-$0)}, \mbox{\ce{HCO+}~(1$-$0)}, \mbox{SiO~(2$-$1)}, and \mbox{\ce{N2H+}~(1$-$0)} line profiles toward the A, B, C, and D positions.

Line emission in the velocity range \mbox{[15,~50]~\kms}\;is typically associated to  the CMZ and it also includes gas in Sgr~B2 envelope \citep[e.g.,][]{deVicente97}.
\mbox{Figures~\ref{fig:30mSiO} and \ref{fig:30mHCN-HCO} a and c} compare the \mbox{\ce{N2H+} (1$-$0)}, \mbox{SiO (2$-$1)}, \mbox{HCN (1$-$0)}, and \mbox{\ce{HCO+} (1$-$0)} emission maps in the [15,~50]\,\kms\; range. In this velocity range we see two main emission structures. One of them, around position D, is located about 6\,arcmin ($\sim$\,15\,pc) northeast of Sgr~B2(M), and the other,  region \mbox{G0.66--0.13},  lays at the southeast of the main cores. This last feature seems to coincide with the hard X-ray continuum emission \citep{Zhang_2015}. A remarkable characteristic of the southeastern structure is the very similar spatial distribution of the \mbox{\ce{N2H+} (1$-$0)} and \mbox{SiO (2$-$1)} emission, and the  relative fainter  level of the \mbox{HCN (1$-$0)} and \mbox{\ce{HCO+} (1$-$0)} emission. The northern structure, however, shows bright emission in the \mbox{\ce{N2H+}}, \mbox{SiO}, \mbox{HCN} and \mbox{\ce{HCO+}} lines, as well as in the mid-$J$ CO lines. Interestingly, the  \mbox{SiO~(2$-$1)} emission in the  LSR velocity range \mbox{[15,~50]\,\kms} and the \mbox{mid-$J$~CO} emission are moderately correlated ($\rho$\,$\sim$\,0.8) along the east part of the map \mbox{($\Delta\rm{R.A.}\geq2'$)}. These specific regions will be discussed in Sect.~\ref{subsubsec:XDR_06-013} and Sect.~\ref{subsubsec:P}.

%---------------------------------------------------------------

%---------------------------------------------------------------

\begin{table}%[h]
\centering
\caption{Summary of spatial emission correlation  coefficients$^3$.} \label{tab:corr30m}
\resizebox{0.5\textwidth}{!}{
\begin{tabular}{llccc }
\hline \hline
          &           &   \multicolumn{3}{c}{Spearman's rank$^a$}    \\
\multicolumn{1}{c}{$y$}   & \multicolumn{1}{c}{$x$}            & $\rho$(tot.) & $\rho$(cores) & $\rho$(env.) \\
\hline
$I$(\NII) & $24\,\upmu\rm{m}$ [W m$^{-2}$ sr$^{-1}$] &  0.80 & 0.90 & 0.78 \\
$I$(\NII) & $70\,\upmu\rm{m}$ [W m$^{-2}$ sr$^{-1}$] &  0.81 & 0.80 & 0.81 \\
$I$(\ce{N2H+} 1$-$0) & $I$(SiO 2$-$1) &  0.70 & 0.60 & 0.70 \\
$I$(HCN 1$-$0) & $I$(\ce{HCO+} 1$-$0) &  0.95 & 0.96 & 0.95 \\
$I$(CO 7$-$6) & $I$(SiO 2$-$1) & 0.76 & 0.21 & 0.78 \\
$\sum I_{\rm{CO}}$ & $\sum I_{[\rm{C\textsc{i}}]}$ &  0.88 & 0.80 & 0.87 \\
\hline
\end{tabular}
}
\tablefoot{$^a$Total mapped area, around the cores, and in the envelope.}
\end{table}

\subsection{Spatial distribution of the dust continuum emission}
%---------------------------------------------------------------

\begin{figure*}
\centering
\includegraphics[height=0.44\textwidth]{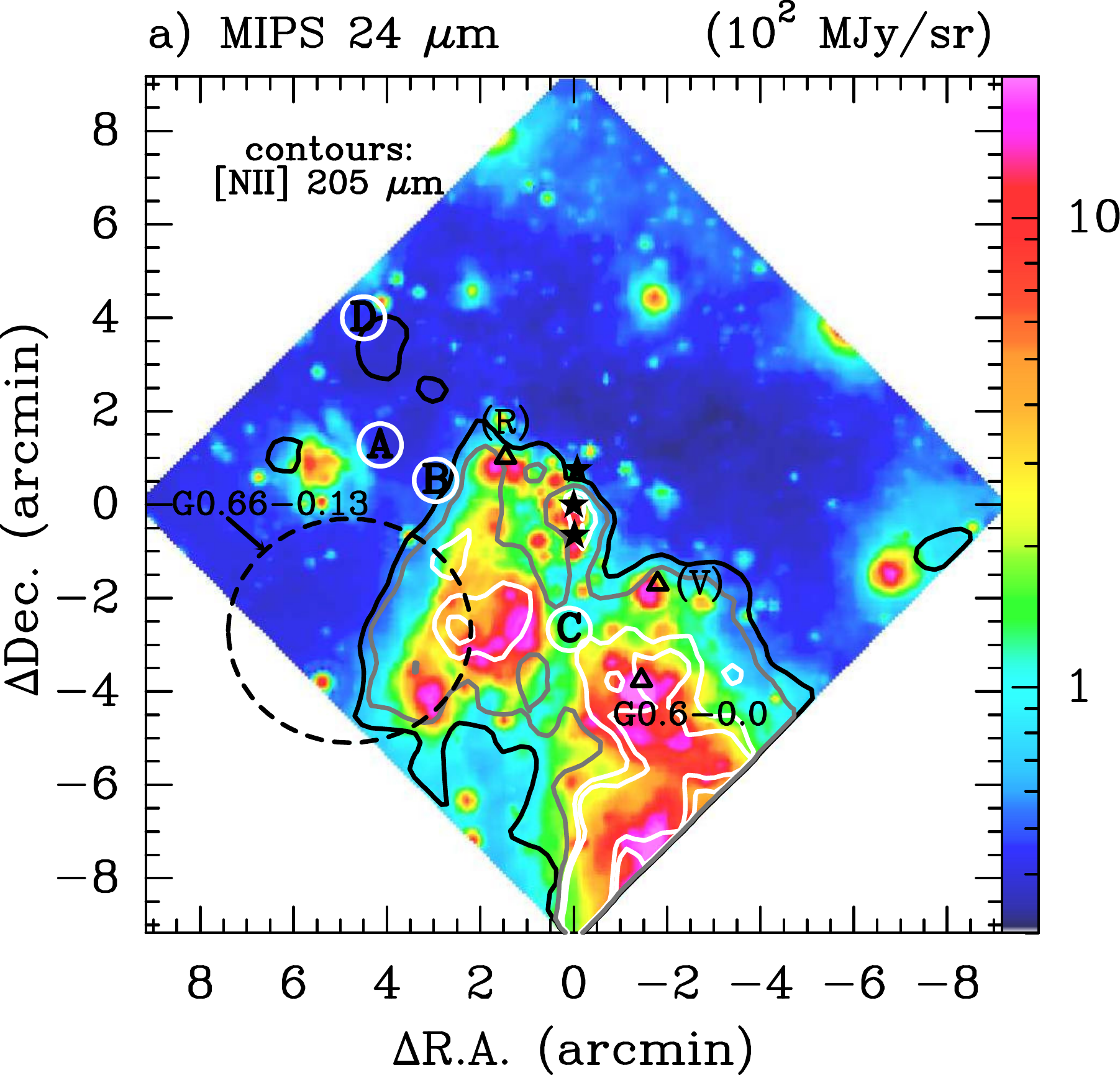}\hspace{0.7cm}
\includegraphics[height=0.44\textwidth]{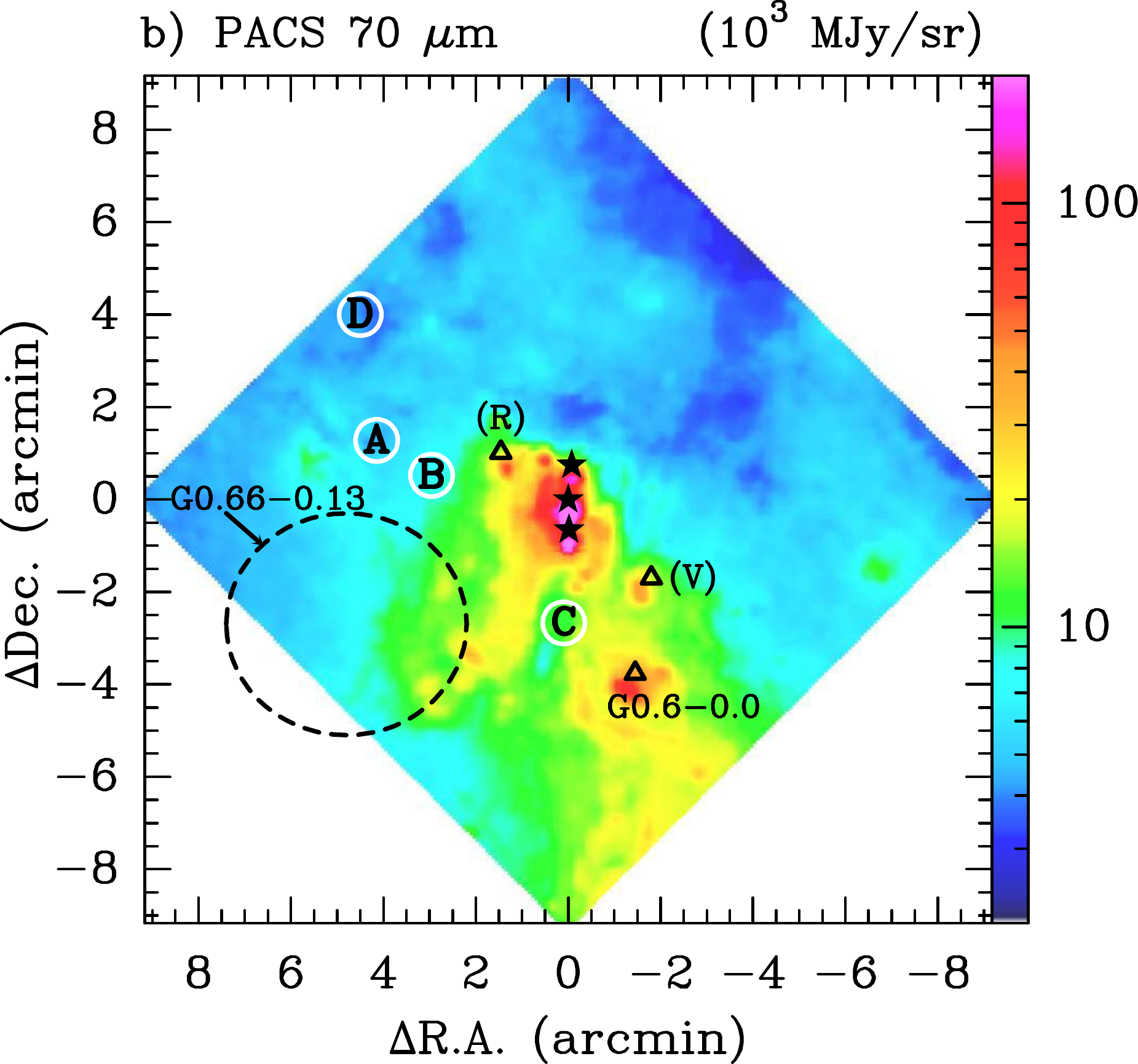} \\ \vspace{0.2cm}
\includegraphics[height=0.44\textwidth]{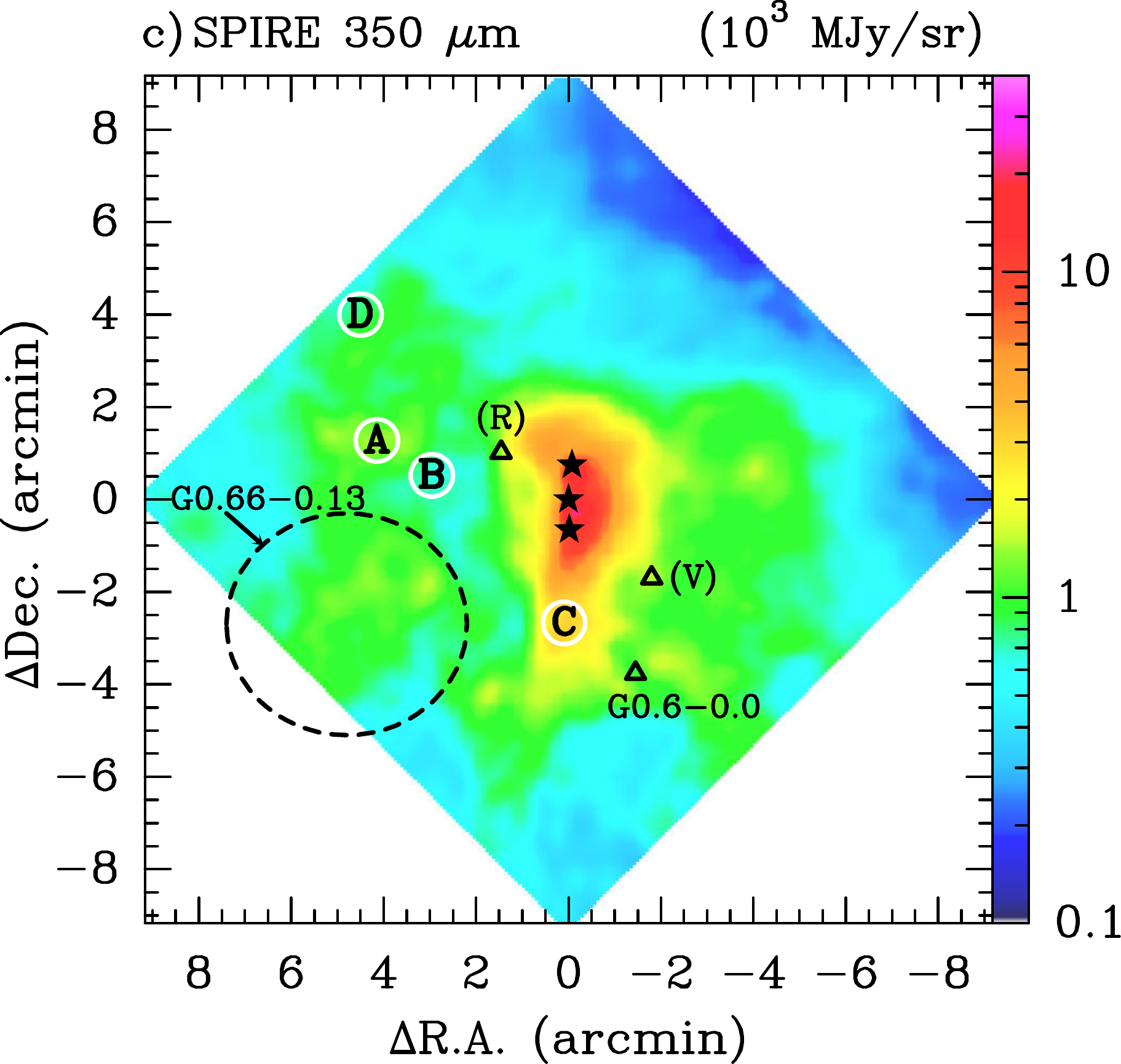} \hspace{0.5cm}
\includegraphics[height=0.44\textwidth]{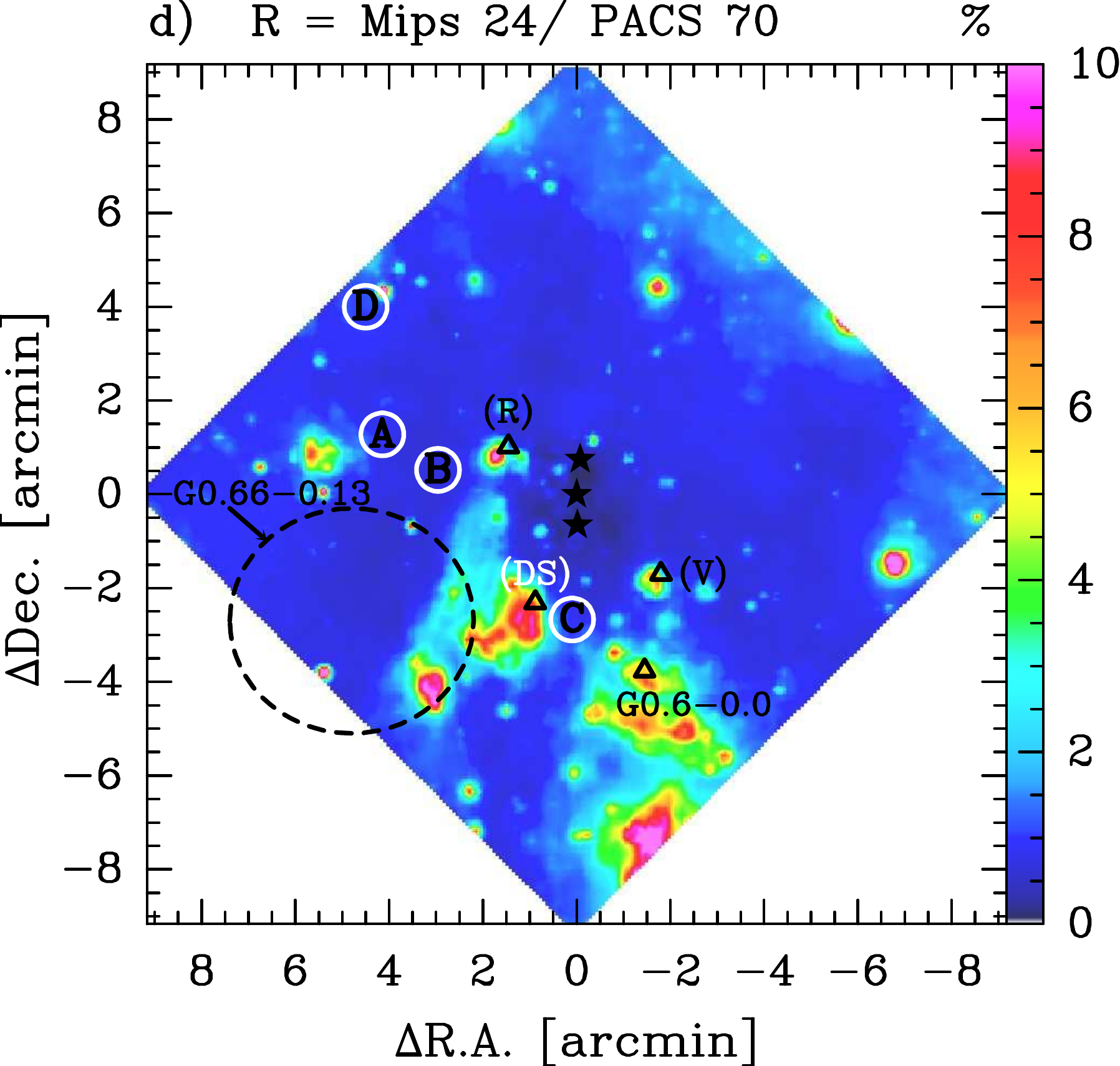}
\vspace{0.1cm}
\caption{Dust continuum emission at different wavelengths. ($a$)~MIPS 24\,$\upmu$m (VSGs), ($b$)~PACS 70\,$\upmu$m, ($c$)~SPIRE 350\,$\upmu$m,
and ($d$)~\mbox{$R_{\%}$\,=\,$I_{24}$\,/\,$I_{70}$\,$\times$\,100 ratio}. Black stars mark the location of the massive star-forming cores N, M, and S. Black triangles mark other \HII\;regions \citep{Mehringer1992,Mehringer1993, Meng2019}. The dashed  circle marks a specific X-ray irradiated region \citep{Zhang_2015}.} 
\label{fig:DUST}
\end{figure*}

Figure~\ref{fig:DUST} shows the different spatial distribution of the dust continuum emission 
at 24\,$\upmu$m ({Fig.~\ref{fig:DUST}a)}, 70\,$\upmu$m ({Fig.~\ref{fig:DUST}b)}, and 350\,$\upmu$m ({Fig.~\ref{fig:DUST}c)}. In the southern envelope, the \mbox{\NII\,205\,$\upmu$m} emission spatially correlates with the 24\,$\upmu$m dust emission (Fig.~\ref{fig:DUST}a) and also with that at 70\,$\upmu$m  (Fig.~\ref{fig:DUST}b).

%-----------------------------------------------------------------------------------------
%-----------------------------------------------------------------------------------------

In addition, there are other compact 24\,$\upmu$m emission sources that do not show bright \mbox{\NII\,205\,$\upmu$m} emission counterpart. In particular, those located in the northern part of the envelope. In UV-illuminated environments, the 70\,$\upmu$m continuum emission is often the sum of the  dust thermal emission from FUV-heated warm big grains (BGs) and that produced by hotter, stochastically heated,  very small grains  \citep[VSGs;][]{Desert1990}. The latter grains, VSGs, are typically present in the ionized gas and are heated by trapped Ly\,$\alpha$ radiation \citep{Salgado16}. In PDR-like environments, the emission from VSGs can contribute to the  observed  dust continuum emission at $\lambda$\,$<$\,100\,$\upmu$m, especially in the PACS\,70\,$\upmu$m images: \mbox{$I_{70}$(Obs.)\,$=$\,$I_{70}$(BGs)\,+\, $I_{70}$(VSGs)}. Since the SED of hot VSGs roughly peaks at $\sim$\,24\,$\upmu$m, this implies that \mbox{$I_{70}$(VSGs)\,$<$\,$I_{24}$(VSGs)}, and thus the observed $I_{24}$(VSGs) is an upper limit to $I_{70}$(VSGs). Hence, the maximum contribution \mbox{(in $\%$) of the VSGs emission} to the 70\,$\upmu$m continuum emission  in PDR environments ($R_{\%}$) is 
\mbox{$R_{\%}$\,$=$\,100$\cdot$$I_{24}$(VSGs)\,/\,$I_{70}$(Obs.)
$\geq$\,100$\cdot$$I_{70}$(VSGs)\,/\,$I_{70}$(Obs.)}. 
\mbox{Figure~\ref{fig:DUST}d}  shows a map of $R_{\%}$ at 6$''$ resolution. The  $R_{\%}$ ratio increases to \mbox{$\sim$\,10$\%$} in known \HII\;regions such as Sgr~B2(V) and (R), as well as toward \mbox{G0.6$-$0.0} (but not toward the main star-forming cores). Even taking into account intensity calibration uncertainties, or a foreground extinction correction of the 24\,$\upmu$m emission (increasing the  emission by up to a factor of $\sim$\,2.5) we conclude that the contribution of VSGs to the 70\,$\upmu$m emission is not large. 
Indeed, detailed dust emission  models of strongly irradiated PDRs, including multiple grain populations, predict that the VSG emission contribution to the FIR continuum emission is $\lesssim$\,5\,$\%$ \citep[][]{Arab_2012}. Hence, the reddish areas of 
\mbox{Fig.~\ref{fig:DUST}d} where $R_{\%}$\,$>$\,5\%, show the greatest  contribution
from hot VSGs present in ionized gas. Still, the 70\,$\upmu$m  emission in Sgr~B2 resembles that of 24\,$\upmu$m, and this implies that the \HII\,regions are surrounded by dusty PDR interfaces with the molecular gas. 

The continuum emission at longer submm wavelengths stems solely from the BGs population. This emission traces the dust-temperature weighted molecular gas column density, $N$(H$_2$). The 350~$\upmu$m emission (Fig.~\ref{fig:DUST}c), as well as at longer wavelengths, peaks toward the main star-forming cores and spreads out toward the east, west, and south into the extended envelope.

In the rest of the paper we consider that the observed \mbox{mid-$J$ CO}, \CI~(Figs.~\ref{fig:SPIRE}a, b, and c), and submm dust thermal emission (Fig.~\ref{fig:DUST}c) arise from Sgr~B2 envelope. That is, all emitting sources are physically associated. This seems justified because their spatial distribution  has a roughly spherical morphology around the main star-forming cores. In addition, any cloud not related to Sgr~B2, but located in the line of sight, will have lower
excitation conditions and column densities that are too low to dominate the observed submm line and continuum emission. On the other hand, the extended \NII\,205\,$\upmu$m emission  (Fig.~\ref{fig:SPIRE}d) and the mid-infrared emission from hot VSGs (Fig.~\ref{fig:DUST}a) do not show such a spherical distribution, and one might argue that they do not belong to the Sgr~B2 complex (but see Sect.~\ref{subsubsec:clumpymed}).

%%%%%%%%%%%%%%%%%%%%%%
\section{Analysis} \label{sect:Analysis}
%%%%%%%%%%%%%%%%%%%%%%

\subsection{Large-scale dust emission: SED fitting} \label{Sect:SED}
%---------------------------------------------

\begin{figure}[t]
    \hspace{-0.4cm}
    \includegraphics[width=0.5\textwidth]{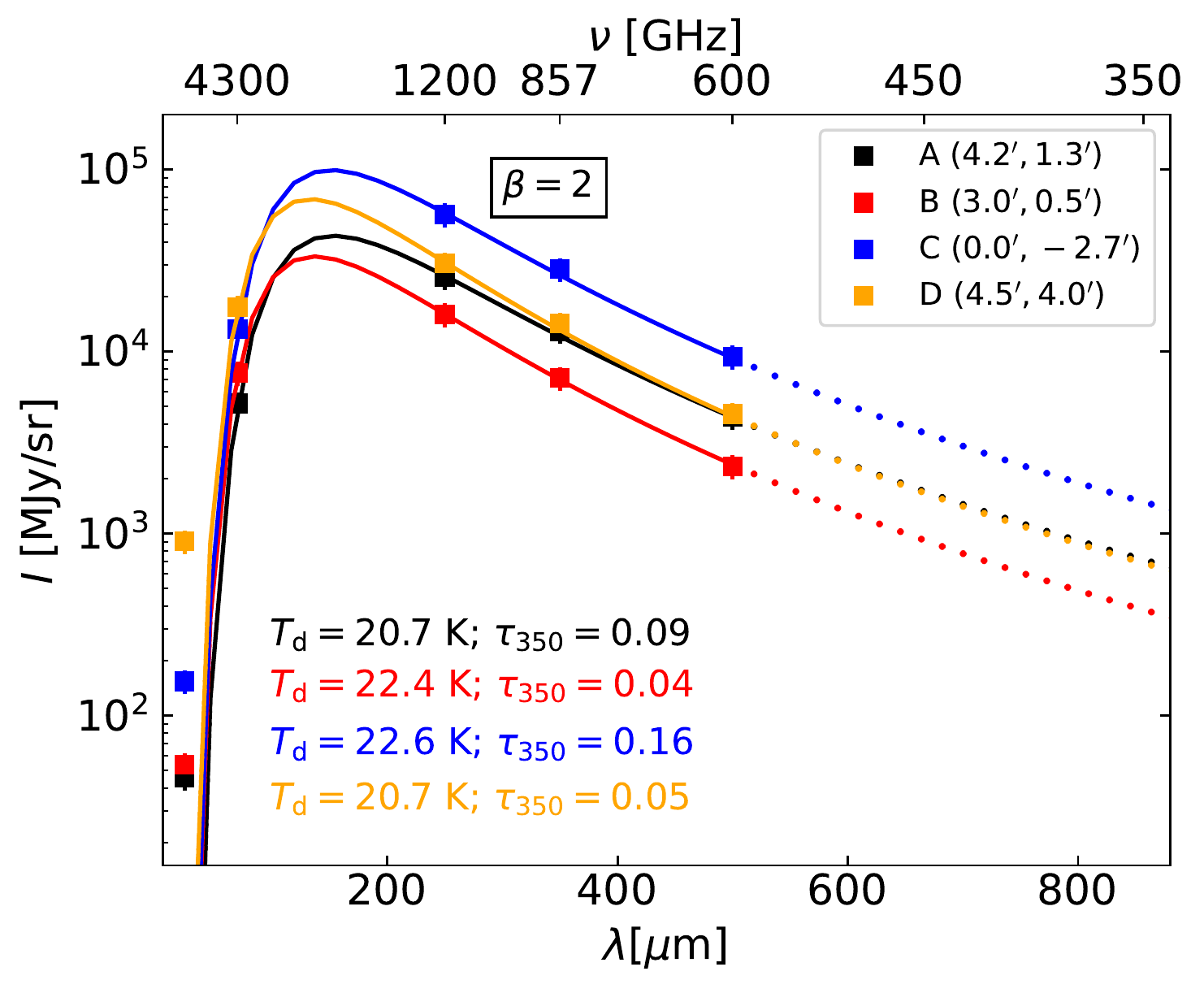}
    \vspace{0.2cm}
    \caption{Dust emission:  MIPS\,24\,$\upmu$m, HiGAL 70, 250, 350, and 500\,$\upmu$m  photometric data, all convolved to 42$''$ resolution (squares, with an intensity uncertainty of $\lesssim$\,15$\%$). Continuous  curves show the best-fit BG SEDs of representative positions A, B, C, and D. The 24\,$\upmu$m emission is produced by hotter VSGs.
    The area below the  continuous curves (40 to 500\,$\upmu\rm{m}$) is the integrated FIR intensity ($I\rm{_{FIR}}$).}
    \label{fig:SED_post}
\end{figure}

In this section we determine the effective dust temperature ($T\rm{_d}$), dust opacity ($\tau\rm{_{\nu}}$), and integrated FIR intensity ($I_{\rm FIR}$) of BGs  from a pixel-to-pixel fit to the dust SED. We fit, based on the Levenberg-Marquardt algorithm,
the continuum as a modified black body,
\begin{equation}
    I_\nu = B_\nu(T\rm{_d})\,(1-e^{-\tau_\nu}),
    \label{eq:mbb}
\end{equation}
where $B_\nu(T\rm{_d})$ is the intensity of a black body at a temperature $T\rm{_d}$ and frequency $\nu$. The dust opacity is parametrized as \mbox{$\tau_\nu$\,$=$\,$\tau\rm{_{\nu,\,ref}}(\nu$\,/\,$\nu_{\rm{ref}})^\beta$}, where $\beta$ is the grain-emissivity index, and we chose as reference frequency, $\nu_{\rm{ref}}$, 857\,GHz ($\simeq$\,350\,$\upmu$m). We let $T\rm{_d}$ and $\tau\rm{_{\nu,\,ref}}$ be free parameters and use a fixed grain-emissivity index of $\beta$\,$=$\,2, representative of silicate grains \citep[][and references therein]{Hirashita2007}. Depending on the observed wavelengths, previous studies report emissivity indexes in the range \mbox{$\beta$\,$\sim$\,1.1$-$2.5} in Sgr~B2 \citep{Dowell1999,Etxaluze13,Schmiedeke16,Arendt2019}.
We checked that results obtained by fixing $\beta$ or leaving it as a free parameter (giving close to 2) do not change $T_{\rm{d}}$ or $\tau_{350}$ significantly. 
Figure~\ref{fig:SED_post} shows the observed SED and best fit results of the four representative positions in Sgr~B2 envelope. 

\begin{figure*}[!ht]
    %\vspace{-0.2cm}
    \centering
    \includegraphics[height=0.435\textwidth]{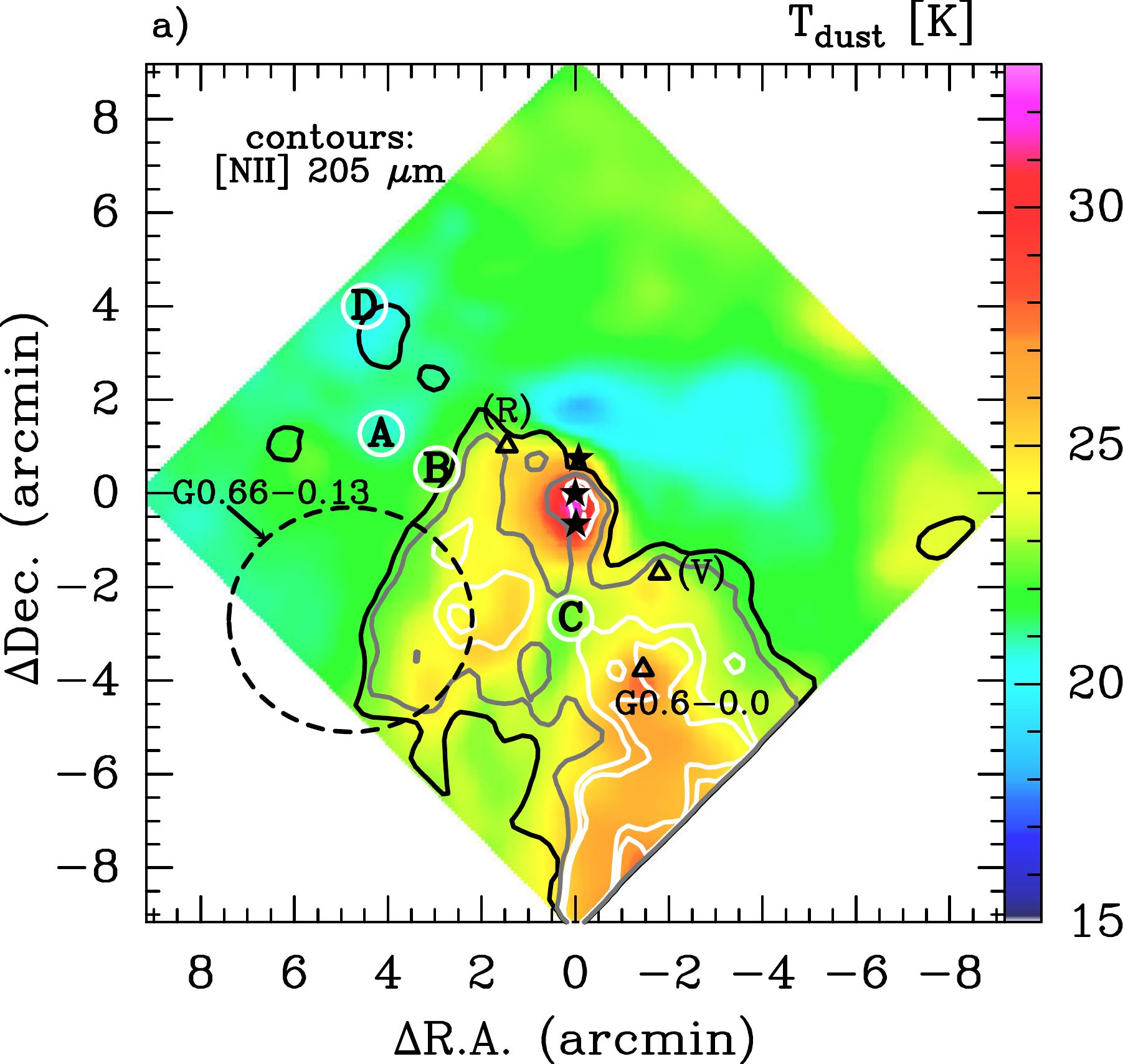}
    \hspace{0.4cm}
    \includegraphics[height=0.435\textwidth]{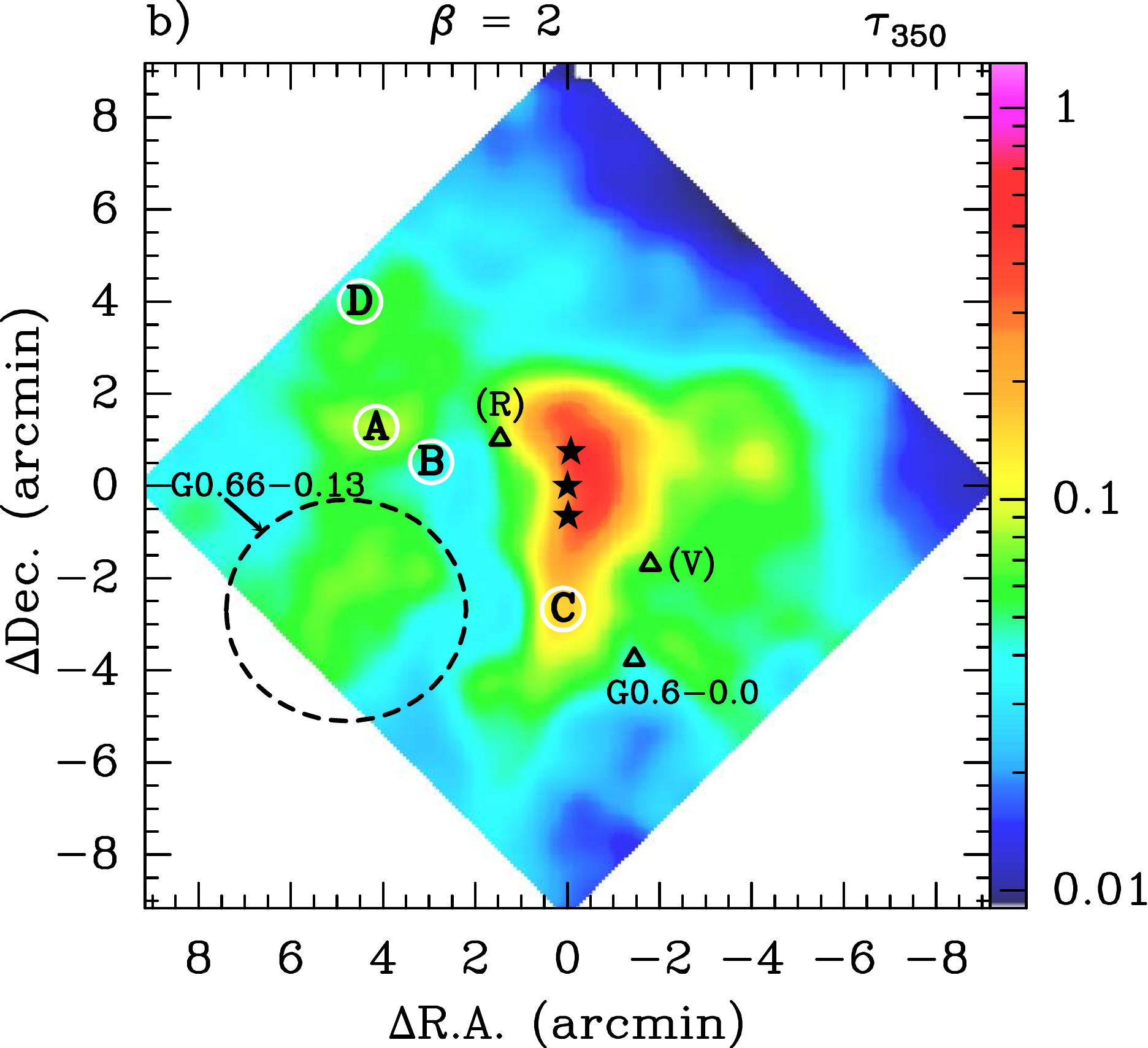}\\ \vspace{0.4cm}
    \hspace{-0.3cm}\includegraphics[height=0.435\textwidth]{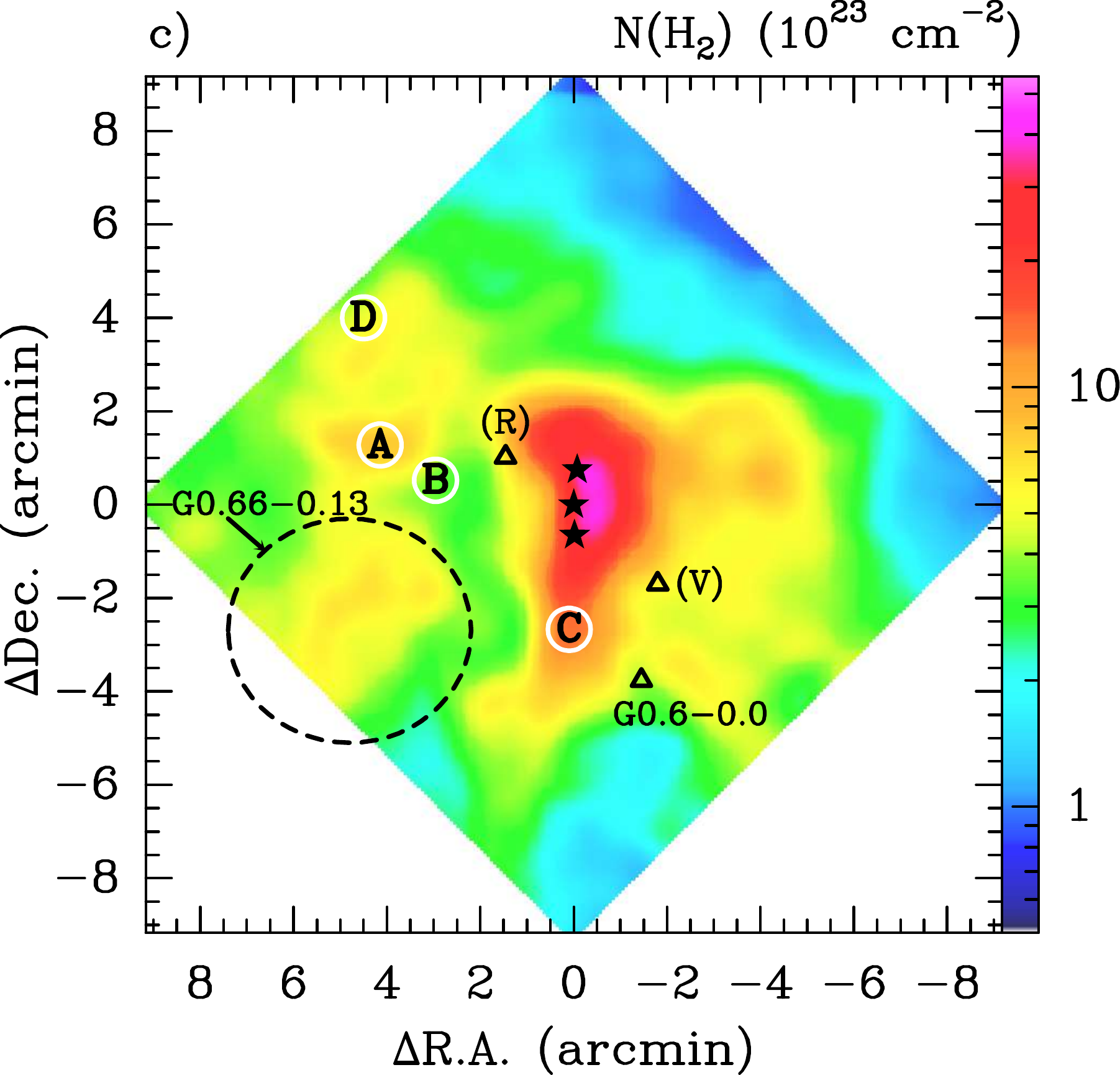}
    \hspace{0.45cm}
    \includegraphics[height=0.435\textwidth]{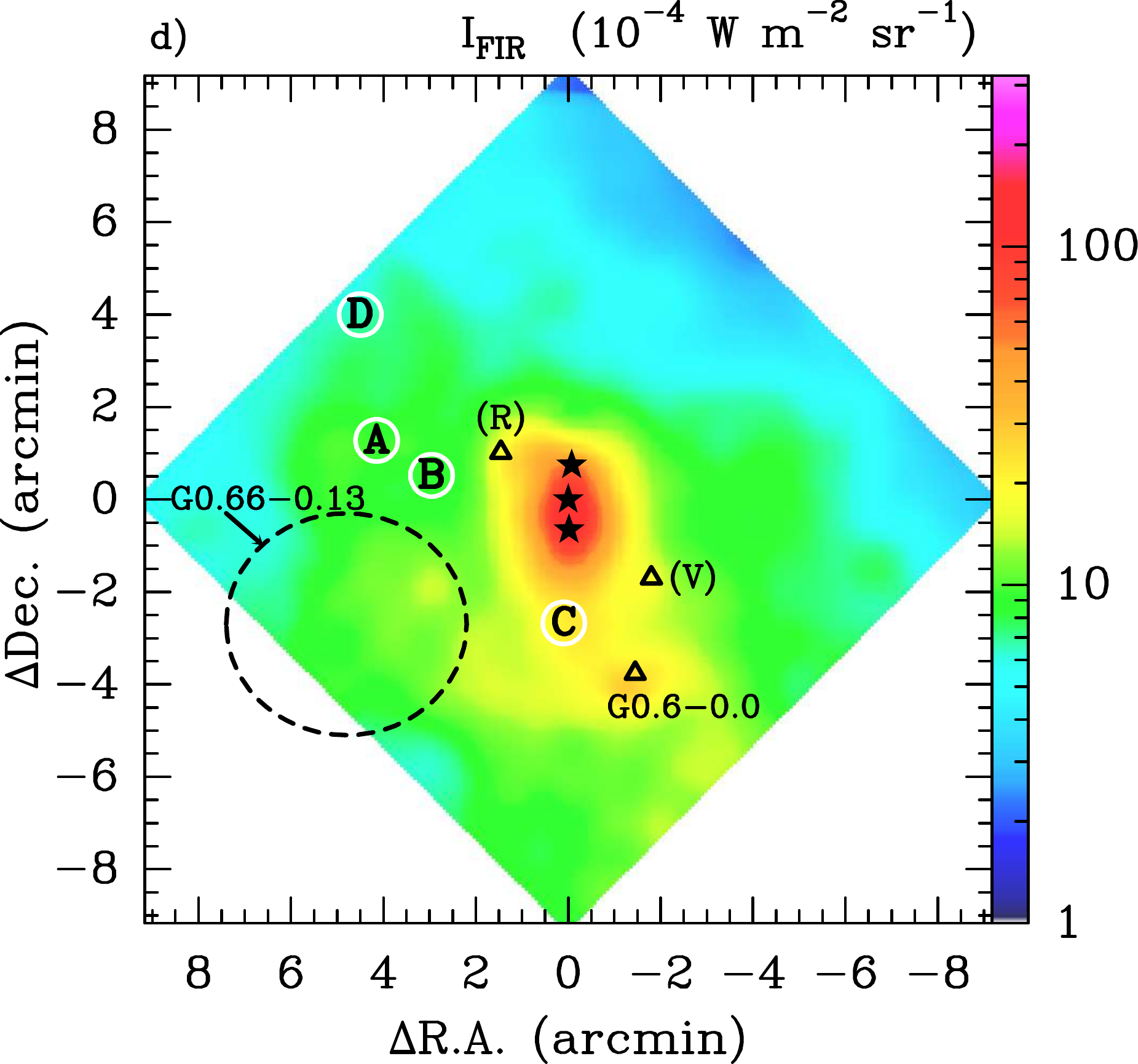}
    \vspace{0.2cm}
    \caption{Dust emission properties. a) Dust temperature derived from SED fist to the photometric data. Contours represent the \NII\,205\,$\rm{\upmu m}$ data. Black: $6\cdot10^{-8}$\,W\,m$^{-2}$\, sr$^{-1}$, gray: $8.5\cdot10^{-8}$\,W\,m$^{-2}$\, sr$^{-1}$, and red: $12\cdot10^{-8}$\,W\,m$^{-2}$\, sr$^{-1}$. b) Map of the dust opacity at 350\,$\upmu$m.
    c) Spatial distribution of the molecular gas column density $N$(H$_2$). 
    d) Far-IR surface brightness map integrated from  40 to 500\, $\upmu$m. 
    All these maps have a uniform angular resolution of 42$\arcsec$.}
    \label{fig:Td_Nh2_LFIR}
    %\vspace{-0.1cm}
\end{figure*}

Figure~\ref{fig:Td_Nh2_LFIR}a shows the resulting  dust temperature map in the region. The average $T\rm{_{d}}$ is  \mbox{$\simeq$\,23~K}, with a standard deviation\footnote{We define the  standard deviation or dispersion of the dust temperature  as: %\begin{equation}
\mbox{$\sigma$\,=\,$\sqrt{\frac{1}{\rm{N}}\sum_{i\rm{_{pixel}}=1}^{\rm{N}}\left (T_{\rm{d,}\mathit{\;i}}-T\rm{_{d,average}}\right )^2}$}} $\sigma$\,=2\,K. The maximum $T_{\rm{d}}$ value in the envelope is at  $21''$ ($\sim$\,0.8~pc) south of Sgr~B2(M), with $T\rm{_{d}}$\,$\simeq$\,32\,K. Far from the main star-forming cores, the warmest dust is located in and around the region \mbox{G0.6$-$0.0}, where the average dust temperature raises to about 28~K. We note that this analysis implicitly 
neglects the presence of $T\rm{_{d}}$ gradients along each line of sight.

%----------------------------------------------------------------------------
%----------------------------------------------------------------------------

%----------------------------------------------------------------------------

%----------------------------------------------------------------------------

From the SED fits we obtain a map of the dust opacity at 350\,$\upmu$m  (Fig.~\ref{fig:Td_Nh2_LFIR}b). At the angular resolution of the smoothed continuum
images (HPBW of 42$''$), we obtain \mbox{$\tau\rm{_{350\,\upmu m}}$\,$<$\,1} in the entire mapped region. In this approach, the dust continuum emission becomes optically thick at wavelengths shorter than \mbox{$\sim$190\,$\upmu$m} and \mbox{$\sim$220\,$\upmu$m} toward 
Sgr~B2(M) and (N), respectively. In the extended envelope, the 
continuum emission becomes  optically thick only below $\sim$50$-$100\,$\upmu$m (depending on the position).
 Therefore, we can safely  use the optically thin $350\,\upmu\rm{m}$ dust emission  to derive column densities and  cloud  masses.
We assume that most hydrogen is in molecular form, H$_2$, and derive the molecular gas column density, $N(\rm{H_2})$, as
\begin{equation}
    N(\rm{H_2})=\frac{\mathit{I}\rm{_{350}}\, R\rm{_{gd}}}{\mathit{B}\rm{_{350}}(\mathit{T}\rm{_d})\:\mathit{k}\rm{_{350}}\:\mu\: \mathit{m}\rm{_H}},
    \label{eq:Nh2}
\end{equation}
through the map, where $k\rm{_{350}}$\,=\,$2.38\,\rm{cm}^2\,g^{-1}$ is the dust absorption cross-section per dust mass at 350\,$\upmu$m \citep{Li_2001} and $I\rm{_{350}}$ is the intensity of the $350\,\upmu$m continuum emission.  We adopt the typical interstellar gas-to-dust mass ratio R$\rm{_{gd}}$\,=100 \citep{Hilde83}, a molecular weight per hydrogen molecule of \mbox{$\mu$\,=\,2.8} \citep{Kaufmfmann2008}, and $m\rm{_H}$ the hydrogen atom mass. Figure~\ref{fig:Td_Nh2_LFIR}c shows the spatial distribution of the \mbox{beam-averaged} \ce{H_2} column density  along each line of sight. 
The column densities toward  Sgr B2(M) and Sgr B2(N)
are \mbox{2.8$\cdot$10$^{24}$\,cm$^{-2}$} and \mbox{3.2$\cdot$10$^{24}$\,cm$^{-2}$}, and the dust temperatures are 30.6\,K and 24.7\,K, respectively. 
Our $N$(H$_2$) columns are consistent  with those derived from a more involved three dimensional dust radiative transfer model: 2.3$\cdot$10$^{24}$\,cm$^{-2}$ and 2.6$\cdot$10$^{24}$\,cm$^{-2}$ at a HPBW of 40$\arcsec$ \mbox{\citep[see model by][]{Schmiedeke16}}. Therefore, we are confident that our mass estimates are realistic within  $\sim$20$\%$.

We derive the molecular gas mass of the cloud as 
\begin{equation}
    M\rm{_{H{_2}}}=\mu\: \it{m}\rm{_H}\: \mathit{A}\rm{_{pixel}}\:\sum_{\rm{pixels}} \it{N}\rm{(H_2)},
    \label{eq:masadust}
\end{equation}
where $A\rm{_{pixel}}$ is the area  of each pixel %%%%\footnote{A$\rm{_{pixel}}=25\,\rm{arcsec}^2 = 5''\times5''$}
in cm$^2$. We obtain  \mbox{$M_{\rm H_2}$\,$\simeq$\,10$^7$\,\Ms} in the  mapped region. About $\sim$\,80\,\% of the mass  is in the  envelope, whereas $\sim$\,20\,\% is in the massive star-forming cores, that occupy about 1$\%$ of the  mapped area. Our derived  gas mass  agrees with previous estimations \citep{Goldsmith90,Schmiedeke16}. Interestingly, the estimated mass of the X-ray scattering gas in Sgr B2 %%%derived from broad-band X-ray observations (3$-$200\,keV), 
at a cloud's radius of about 30\,pc is 1.8$\cdot$10$^7$\,\Ms\ \citep{Revnivtsev2004}. \mbox{Assuming} standard interstellar grains, we derived the visual extinction  along each line of sight as \mbox{$A\rm{_V}$\,$\simeq$\,$N_{\rm{H}}$\,/\,1.9\,$\cdot$\,$10^{21}$\,mag} 
\mbox{\citep{Bohlin78}}. We obtain that $A_{\rm{V}}$ ranges from $\leq$\,100\,mag in the northwest edges of the map to $>$\,2\,$\cdot$\,10$^3$\,mag toward the main star-forming cores. These values include at least 25\,mag of extinction produced by foreground material not related to Sgr~B2 \mbox{\citep[e.g.,][]{Schultheis99}}.

We determined the FIR intensity, $I\rm{_{FIR}}$ (Fig.~\ref{fig:Td_Nh2_LFIR}d), by integrating each fitted SED from 40 to 500\,$\upmu$m \citep[][]{Sanders1996}. We computed the  FIR luminosity, $L\rm{_{FIR}}$, of a given area as
\begin{equation}
    L\rm{_{FIR}}=\it{L}(\rm{40-500}\,\upmu \rm{m}) =4\pi\, \rm{D}^2\, \it{I}\rm{_{FIR}}\,\Omega,
\end{equation}
where D\,$=$\,8.18\,kpc \citep{Gravity2019} and $\Omega$ is the solid angle subtended by the area of interest.  Figure~\ref{fig:Td_Nh2_LFIR}c shows the
resulting map of $I_{\rm{FIR}}$, where the more luminous regions are those around the massive star-forming cores and G0.6$-$0.0. The total $L_{\rm FIR}$ in the envelope is \mbox{$\simeq$\,2.6\,$\cdot$\,$10^7$\Ls} \mbox{(equivalent to a surface luminosity density of $2.7\cdot10^4\,\Ls$\,pc$^{-2}$)}, and it is \mbox{$\simeq$\,0.6}\,$\cdot$\,$10^7$\Ls \mbox{(or $5.0\cdot10^5\,\Ls$\,pc$^{-2}$)} in the cores. Thus, $\sim$\,80\% of the total FIR luminosity arises from the large-scale environment around the  star-forming cores. 

%--------------------------------------------------------------------------------------
\begin{table*}[t]
\centering
\caption{Summary of FIR- and submm-related magnitudes that
are spatially correlated in the mapped area of Sgr~B2.} \label{tab:corrfir}
\begin{tabular}{llccccc }%{@\vrule height 10pt depth 5pt width 0pt}}
\hline \hline
          &           & \multicolumn{2}{c}{} & \multicolumn{3}{c}{Spearman's rank$^b$}    \\
\multicolumn{1}{c}{$y$}   & \multicolumn{1}{c}{$x$}  & \multicolumn{1}{c}{slope$^a$}  & \multicolumn{1}{c}{intercept$^a$}          & $\rho$(total) & $\rho$(cores) & $\rho$(envelope) \\
\hline
$L_{\rm{FIR}}$\,/\,$M_{\rm{gas}}$ & $T_{\rm{dust}}$ [K]    & $\phantom{-}5.580\pm0.020$       & $-7.060\pm0.020$     & \phantom{-}0.94     & \phantom{-}0.92     & \phantom{-}0.96        \\
$L_{\rm{FIR}}$\,/\,$M_{\rm{gas}}$ & $24\,\upmu\rm{m}$ [MJy sr$^{-1}$]    & $\phantom{-}0.352\pm0.002$     & $-0.203\pm0.004$   & \phantom{-}0.78     & \phantom{-}0.94     & \phantom{-}0.79        \\
$I$(\CI)\,/\,$I_{\rm{FIR}}$  & $I_{\rm{FIR}}$      & $-0.721\pm0.003$    & $-6.529\pm0.009$   & -0.84    & -0.98    & -0.82       \\
$I$(\CI)\,/\,$I_{\rm{FIR}}$    & $T_{\rm{dust}}$     & $-4.910\pm0.030$      & $\phantom{-}2.340\pm0.040$      & -0.57    & -0.91    & -0.55       \\
$I$(\NII)\,/\,$I_{\rm{FIR}}$   & $I_{\rm{FIR}}$       & $-0.413\pm0.005$    & $-5.510\pm0.010$     & -0.23    & -0.79    & -0.12       \\

\hline
\end{tabular}
\tablefoot{$^a$$\rm{log_{10}}$($y$)\,=\,slope$\cdot$$\rm{log_{10}}$($x$)\,+\,intercept.
$^b$Total mapped area, around the cores, and in the envelope.}
\end{table*}
%--------------------------------------------------------------------------------------

\subsection{Large scale FUV and EUV radiation illumination}

We estimated the approximated flux of stellar FUV photons emitted by OB-type stars ($G_0$ in units of the Habing field) as
\begin{equation}
\label{eq-G0}  G_0\simeq\frac{1}{2}\frac{I_{\rm{FIR}}\,[\rm{W}\,m^{-2}\,sr^{-1}]}{1.3\cdot10^{-7}},
\end{equation} from \cite{Hollenbach_1999}. 
This relation assumes that the FIR continuum is emitted by \mbox{FUV-heated} dust grains in a \mbox{face-on} PDR. Values of $G_0$\,$\simeq$\,2\,$\cdot$\,10$^3$ in the envelope would correspond to $I_{\rm FIR}$ contours of \mbox{$\gtrsim$\,6\,$\cdot$\,10$^{-4}$}\,$\rm{W}\,m^{-2}\,sr^{-1}$ in  Fig.~\ref{fig:Td_Nh2_LFIR}d. Since embedded star-formation  (associated with cores of large dust column densities) also produces non-PDR dust continuum emission, this $I_{\rm FIR}$ map should be interpreted as a rough upper limit to $G_0$.

\begin{figure}[t]
    %\vspace{-0.2cm}
    \centering    \includegraphics[height=0.46\textwidth]{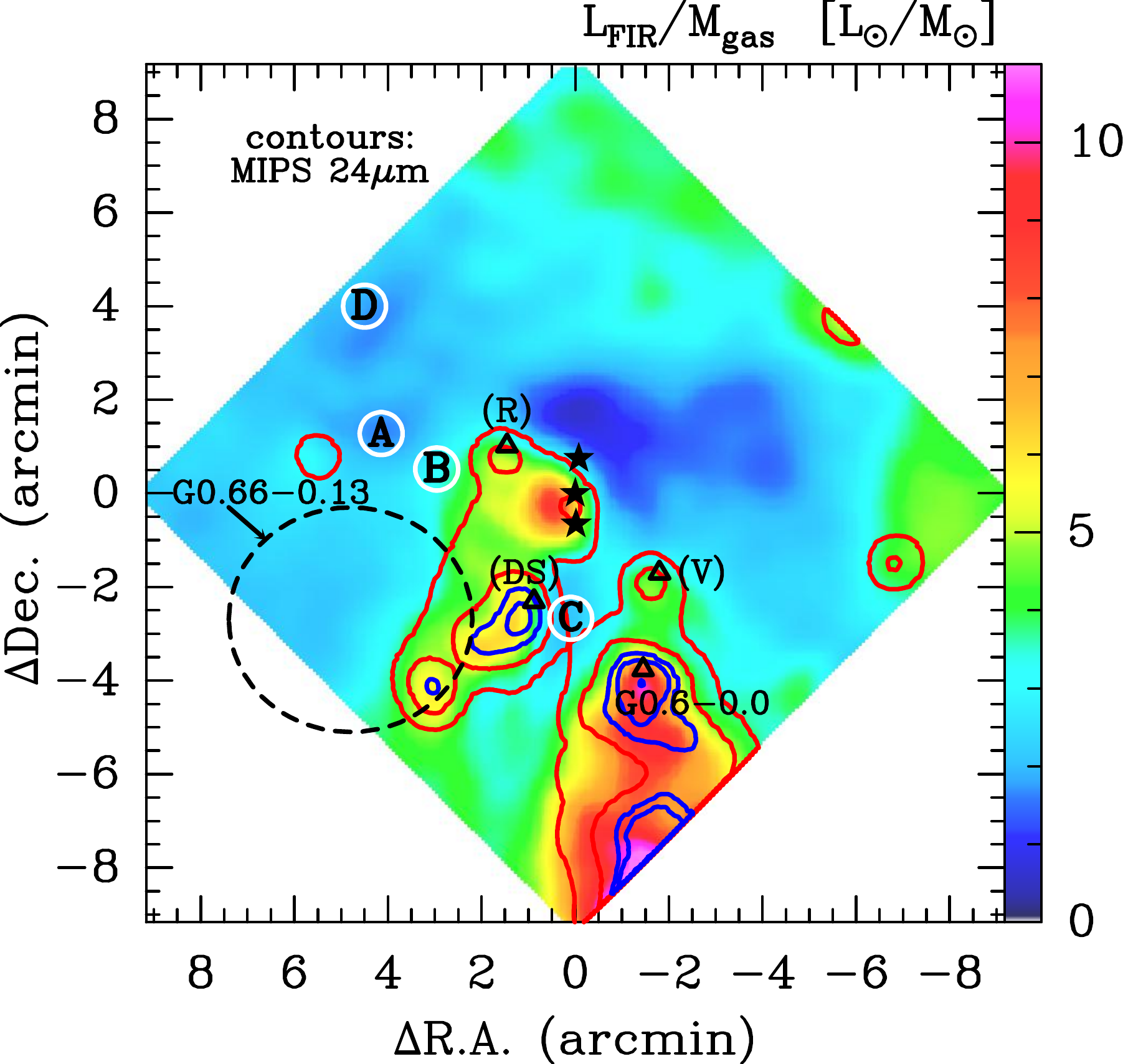}
 
    \caption{Map of the FIR-luminosity to gas-mass ratio \mbox{$L_{\rm FIR}$\,/\,$M_{\rm gas}$} in units of \mbox{$L_{\odot}$\,/\,$M_{\odot}$} per pixel (at an angular resolution of 42$''$).
Contours represent 24\,$\upmu$m emission levels from 3$\cdot$10$^4$\,MJy\,sr$^{-1}$ (red) to 200\,MJy\,sr$^{-1}$ (blue).}
    \label{fig:Lfir_Mgas}
    %\vspace{-0.1cm}
\end{figure}

Figure~\ref{fig:Lfir_Mgas} shows a map of the FIR-luminosity to gas-mass ratio (\mbox{$L_{\rm FIR}$\,/\,$M_{\rm H_2}$}) computed from the dust SED fits. In the
\mbox{\HII~regions} of high-mass star-forming clouds,  this ratio is expected to follow the ionization parameter \mbox{$U$\,$=$\,$Q({\rm H})\,/\,4\pi\,c\,n_e R^2$}, where $Q({\rm H})$ is the number of  EUV ionizing photons per second \citep[dominated by \mbox{Sgr\,B2(M and N)}, with about 10$^{50.3}$\,s$^{-1}$, e.g.,][]{Gaume1995}, $R$~is the distance from the ionizing star(s), and $n_e$ is the electron density. 
Our maps observationally support this association because the spatial distribution of the \mbox{$L_{\rm FIR}$\,/\,$M_{\rm H_2}$}  ratio roughly follows that of \NII\,205\,$\upmu$m  and is also very similar to the 24\,$\upmu$m hot dust emission \mbox{($\rho$\,$\simeq$\,0.8)}. We find that the spatial distribution of  \mbox{$L_{\rm FIR}$\,/\,$M_{\rm H_2}$} in Sgr~B2  is strongly correlated with $T_{\rm d}^{5.6\pm0.1}$ ($\rho$\,$\simeq$\,0.94).  Table~\ref{tab:corrfir} shows the parameters and \mbox{$\rho$ coefficients} of several FIR- and \mbox{submm-related} magnitudes that show a spatial correlation in the mapped area. The low effective dust temperatures  \mbox{($T_{\rm d}$\,$\simeq$\,20$-$25\,K)} explain the generally low \mbox{$L_{\rm FIR}$\,/\,$M_{\rm H_2}$\,$\lesssim$\,5\,\Ls\,\Ms$^{-1}$} values in Sgr~B2 envelope. Toward lines-of-sight of bright \NII\,205\,$\upmu$m and 24\,$\upmu$m dust emission, $T_{\rm d}$ increases above \mbox{$\simeq$\,25$-$30\,K}, leading to \mbox{$L_{\rm FIR}$\,/\,$M_{\rm H_2}$\,$\simeq$\,4$-$11\,\Ls\,\Ms$^{-1}$}. The increased $T_{\rm d}$ must be associated with warmer dust in the PDRs that border the extended ionized gas structures. Still, neither $T_{\rm d}$ nor $L_{\rm FIR}$\,/\,$M_{\rm gas}$ reach the high ratios observed toward much more compact and  denser \HII~regions in the immediate  environment of young massive stars  
\citep[\mbox{$L_{\rm FIR}$\,/\,$M_{\rm gas}$\,$\simeq$\,100\,\Ls\,\Ms$^{-1}$} at $\sim$\,0.2\,pc from  the Trapezium cluster in Orion, e.g.,][]{Goi15}. This confirms that, at large spatial scales  in Sgr~B2, the ionizing radiation has low $U$ values:  diluted radiation produced by distant massive stars  \citep[for photoionization models of
Sgr~B2, see][]{Goicoechea_2004}. Adopting their average value of $n_e$\,=\,240\,cm$^{-3}$ for the extended ionized gas component and \mbox{$Q$(H)\,$\simeq$\,10$^{50.3}$\,s$^{-1}$}, we obtain  $U$\,$\simeq$\,10$^{-3}$ at 6$'$ (15\,pc) from \mbox{Sgr\,B2(M)}. This  low ionization parameter is consistent with the  low \mbox{$L_{\rm FIR}$\,/\,$M_{\rm H_2}$} ratio and $T_{\rm d}$ values that we determine at large scales in Sgr~B2 \citep[see model
predictions for a broad range of $U$ parameters in][]{Abel09}. The spatial distribution of the ionizing stars with respect to the extended cloud and the clumpiness of the medium must determine the large-scale effects of this  radiation.

%----------------------------------------------------------------------------
%----------------------------------------------------------------------------

\subsection{Large-scale mid-$J$ CO and \CI: Gas physical conditions} \label{subsec:exccond}
%----------------------------------------
\begin{figure*}[!ht]
\centering
\hspace{-0.7cm}
\includegraphics[height=0.40\textwidth]{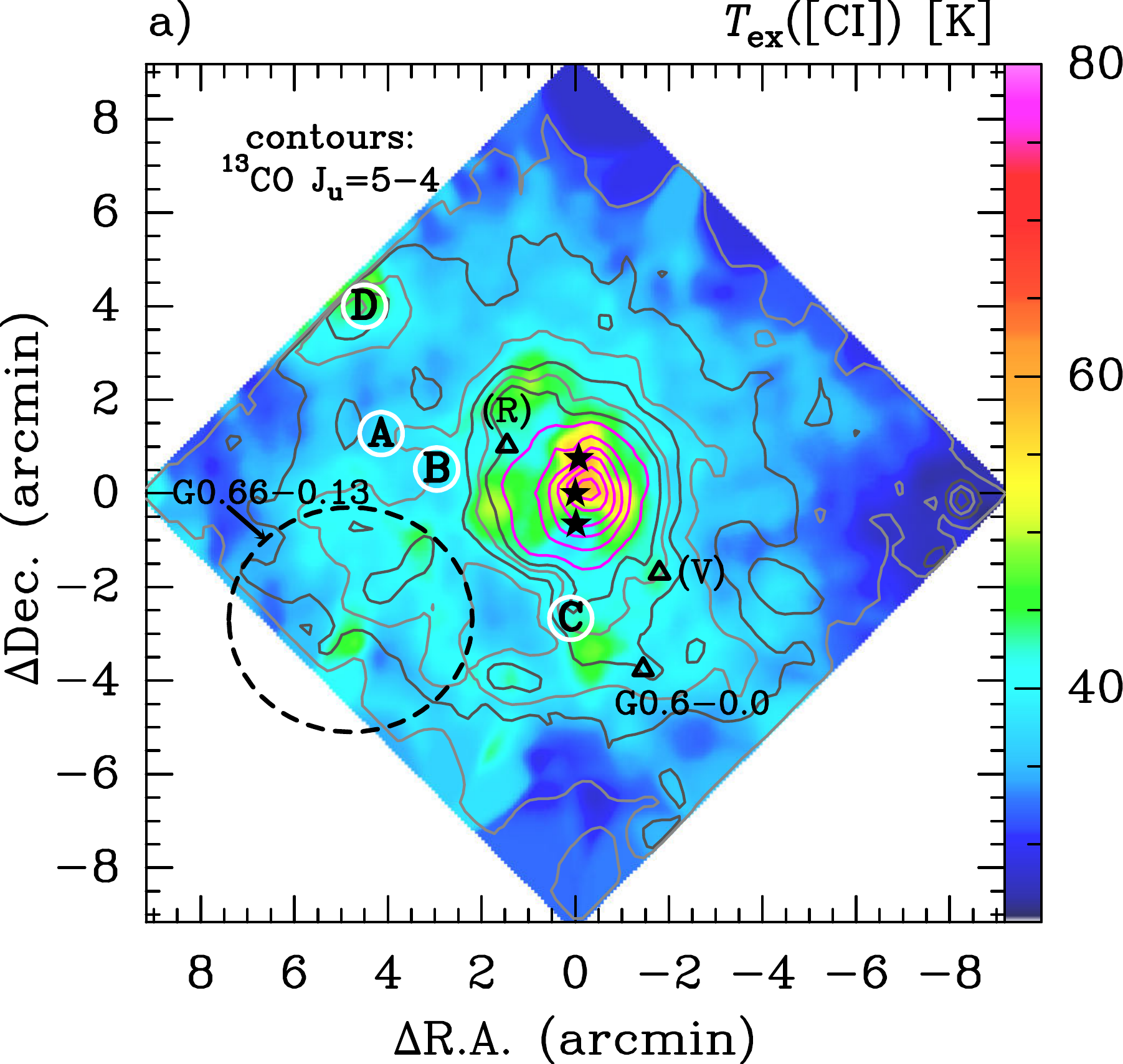} \hspace{1.1cm}
\includegraphics[height=0.40\textwidth]{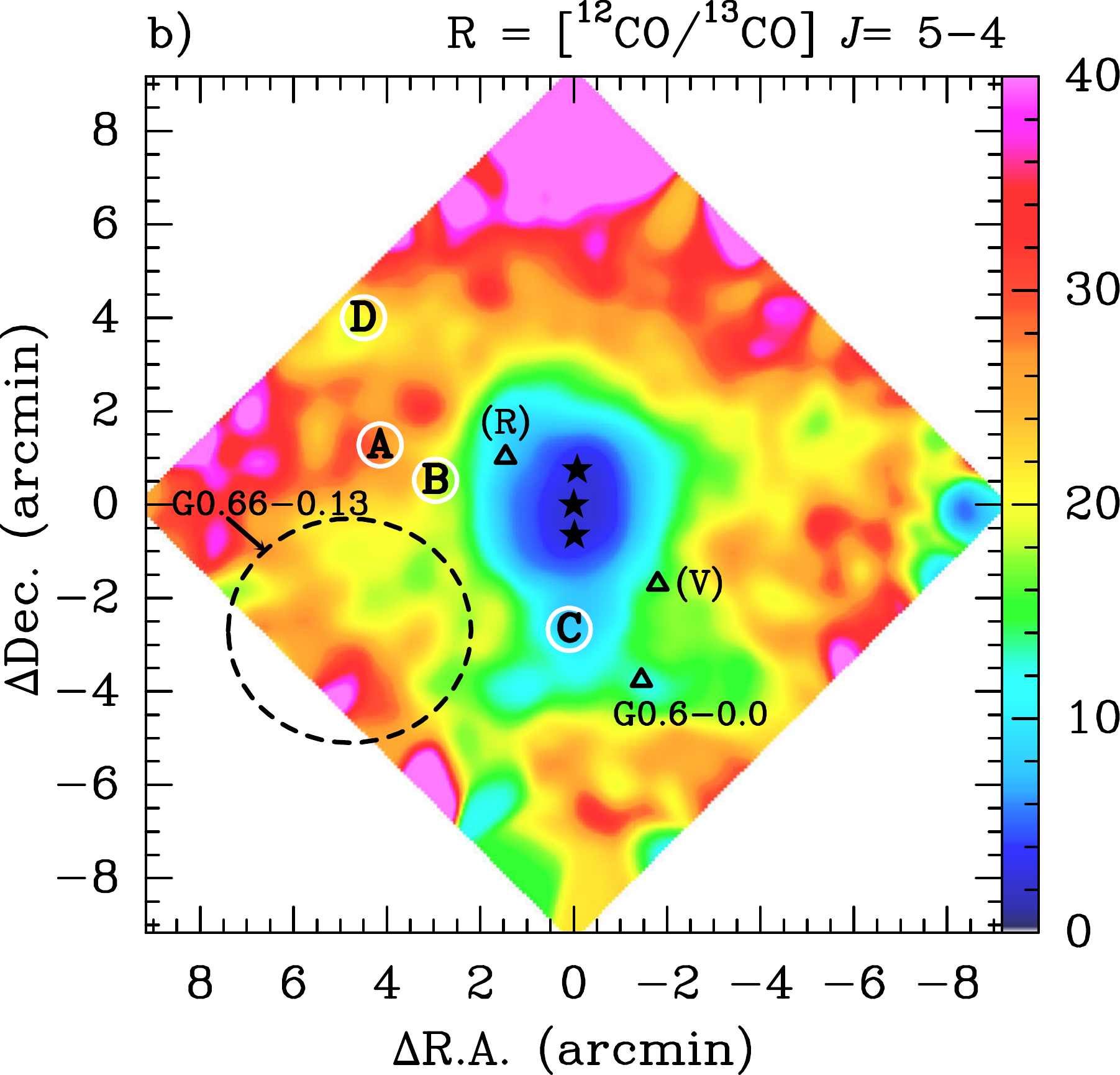}\\ \vspace{0.4cm} %\label{fig:Tex}
 %\hspace{0.2cm} %\label{fig:ratCO1213} 
\includegraphics[height=0.40\textwidth]{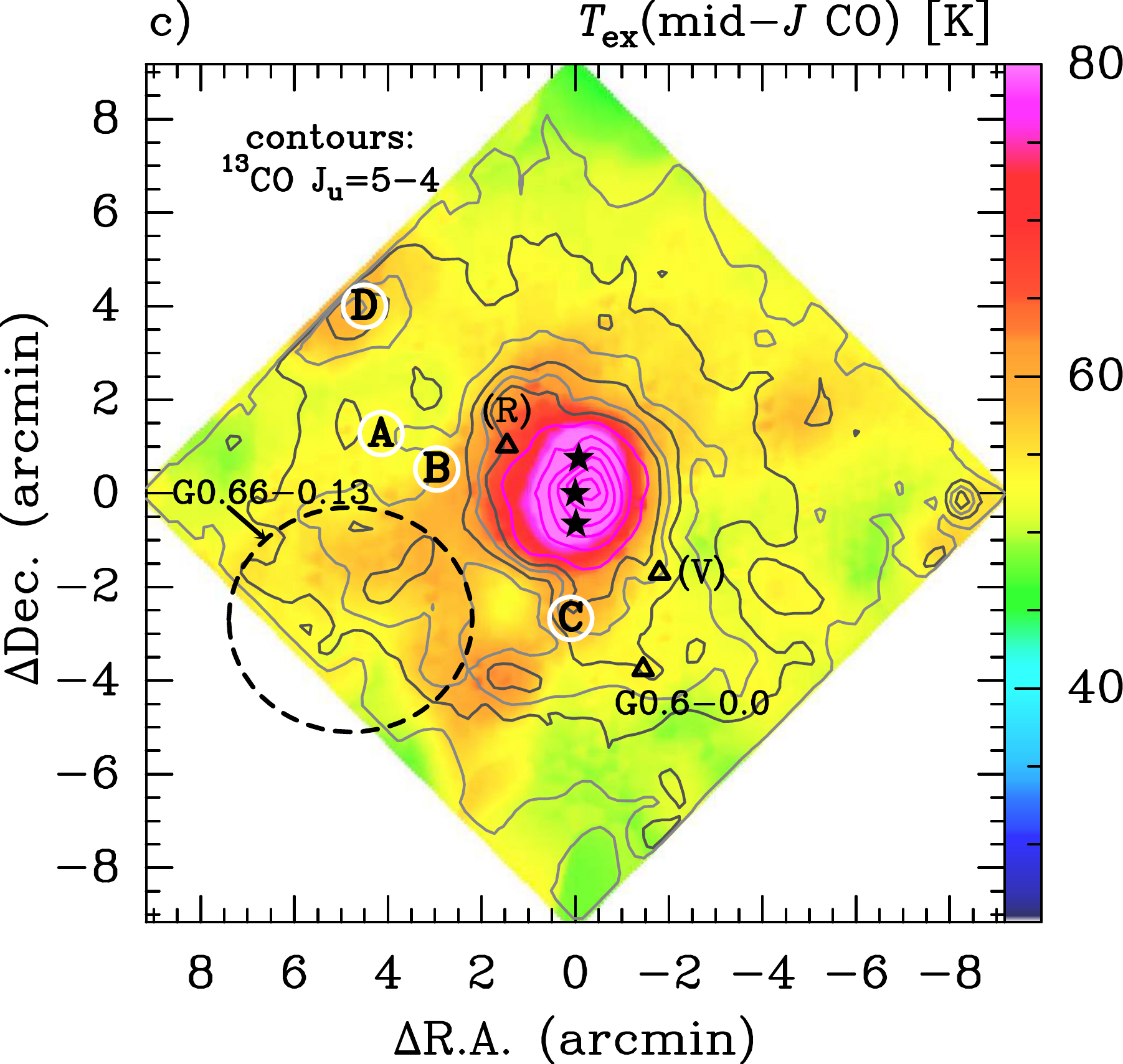} \hspace{0.5cm}
\includegraphics[height=0.40\textwidth]{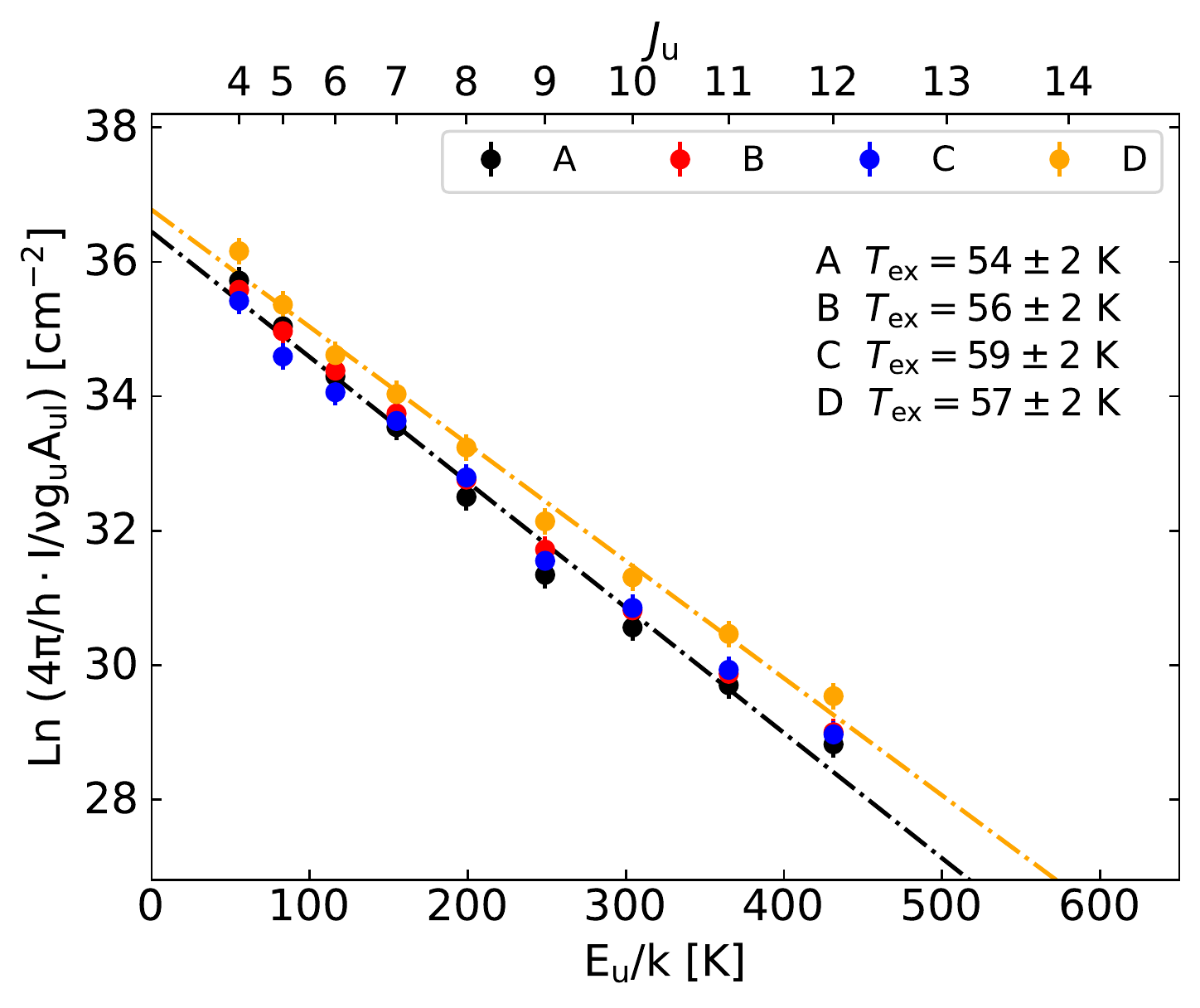}
\caption{Maps of $T_{\rm{ex}}$ for a) atomic carbon and c) mid-$J$ CO. Contours represent the $^{13}$CO $J$\,=\,5$-$4 emission from \mbox{$10^{-9}$\,W\,m$^{-2}$\,sr$^{-1}$} (gray) to \mbox{$2.8\cdot10^{-8}$ \,W\,m$^{-2}$\,sr$^{-1}$} (magenta) tracing  high $N$(\ce{H2}) regions. b) Line surface brightness $^{12}$CO\,/\,$^{13}$CO $J$\,=\,5$-$4 ratio map.
d) Population diagrams of $^{12}$CO lines observed with \mbox{SPIRE-FTS} toward positions A, B, C, and D in Sgr~B2 (see Table~\ref{table:TTNN}). Dashed curves refer to fits to positions A and D.}
\label{fig:Texrotdrtot}
%\vspace{-0.3cm}
\end{figure*}
%----------------------------------------

\subsubsection{Multi-position population diagrams} 
In this section we turn back to the properties of the neutral gas and derive the excitation temperature ($T\rm{_{ex}}$) and column density ($N$) of CO and atomic carbon. We first assume that the \mbox{mid-$J$~CO} rotational levels and \CI\;fine-structure levels are populated following a Boltzmann distribution at a single excitation temperature, $T_{\rm{ex}}$(mid-$J$ CO) and $T_{\rm{ex}}$(C), respectively. We analyze their line emission using the population  diagram technique to derive $T\rm{_{ex}}$ and $N$ \citep[see e.g.,][]{Goldsmith_1999}.
\begin{itemize}
\item[(i)] Excitation temperatures as lower limits to $T_{\rm k}$: 
\end{itemize}

Figure~\ref{fig:Texrotdrtot}a shows the spatial distribution of the derived
\mbox{excitation temperature} for atomic carbon. \mbox{$T_{\rm{ex}}$(C)}  ranges from 26\,K close to the map edges, to \mbox{$\sim$\,60\,K}, around  Sgr~B2(M). The average \mbox{$T\rm{_{ex}}$(C)} value in the mapped area is 36\,$\pm$\,5\,K.
If the gas density $n$(H$_2$) is on the order or lower than the critical density  of these fine-structure lines ($n_{\rm cr}$ of a few 10$^3$\,cm$^{-3}$, see Table~\ref{tab:linesall}), the population of these levels will not exactly be in local thermodynamic equilibrium (LTE) and $T\rm{_{ex}}$(C) will be a lower limit to the 
actual gas kinetic temperature~($T\rm{_k}$).

Before determining \mbox{$T_{\rm{ex}}$(mid-$J$~CO)}, we investigate whether the \mbox{mid-$J$~CO} emission is optically thin. \mbox{Figure~\ref{fig:Texrotdrtot}b} shows a map of the  \mbox{$I(\rm{CO})$\,/$\,I(\rm{^{13}CO})$\;$J$\,=\,5$-$4} line intensity ratio. This ratio ranges from $\sim$\,3 toward the main star-forming cores to $\gtrsim$\,40 at the map edges. 
For line intensity ratios significantly below the \mbox{$^{12}\rm{C}$\,/\,$^{13}\rm{C}$} isotopic abundance ratio in the GC \mbox{($\sim$\,20$-$30)} the CO emission is optically thick or self-absorbed. This is the case of the \mbox{CO (5$-$4)}  emission toward the N, M, S cores and their immediate surroundings ($\sim$\,5$-$10\,pc away). However, in the extended envelope (the spatial scales that dominate the emitted  \mbox{mid-$J$} CO line luminosity) the \mbox{CO (5$-$4)}, and thus all \mbox{higher $J$} lines, are optically thin.
HIFI high-spectral resolution observations of
these \mbox{mid-$J$} $^{12}$CO lines toward \mbox{Sgr~B2(M, N)} cores show pronounced
self-absorption features \citep{Neill2014,Indriolo17}. These are produced
by the less excited foreground  CO in the envelope that absorbs the intense, and optically thick, hot CO emission from the cores. However, we do not expect self-absorption to
dominate in the   envelope, as we proved that the
extended \mbox{mid-$J$ CO} emission is optically thin.

For optically thin line emission, the CO rotational population diagrams provide accurate values for \mbox{$T_{\rm{ex}}$(mid-$J$ CO)} and \mbox{$N$(mid-$J$ CO)}. Figure~\ref{fig:Texrotdrtot}c shows the spatial distribution of $T_{\rm{ex}}$(mid-$J$ CO) derived from fitting the CO rotational ladder at each position. 
Figure~\ref{fig:Texrotdrtot}d shows examples of the population diagrams at the four representative positions A, B, C, and D of the extended envelope and Table~\ref{table:TTNN} lists the fit parameters. The extended envelope displays \mbox{$T\rm{_{ex}}$(mid-$J$ CO)} in the range 45$-$65\,K. Again, $T_{\rm{ex}}$ is  a lower limit to the kinetic temperature of the CO gas component if \mbox{$n$(H$_2$)\,$\lesssim$\,$n_{\rm cr}$} (these \mbox{mid-$J$} lines have much higher critical densities, \mbox{$n_{\rm cr}$\,$\simeq$\,10$^5$\,$-$\,10$^6$\,cm$^{-3}$}, than the \CI\; lines). The excitation temperature \mbox{$T\rm{_{ex}}$(mid-$J$ CO)} toward the main star-forming cores  is higher, reaching \mbox{$>$\,100\,K}. \cite{Etxaluze13} previously analyzed their complete CO rotational ladder emission.

\begin{itemize}
\item[(ii)] Column densities of warm CO and atomic carbon: 
\end{itemize}

\begin{figure}[!ht]
\centering
\includegraphics[height=0.41\textwidth]{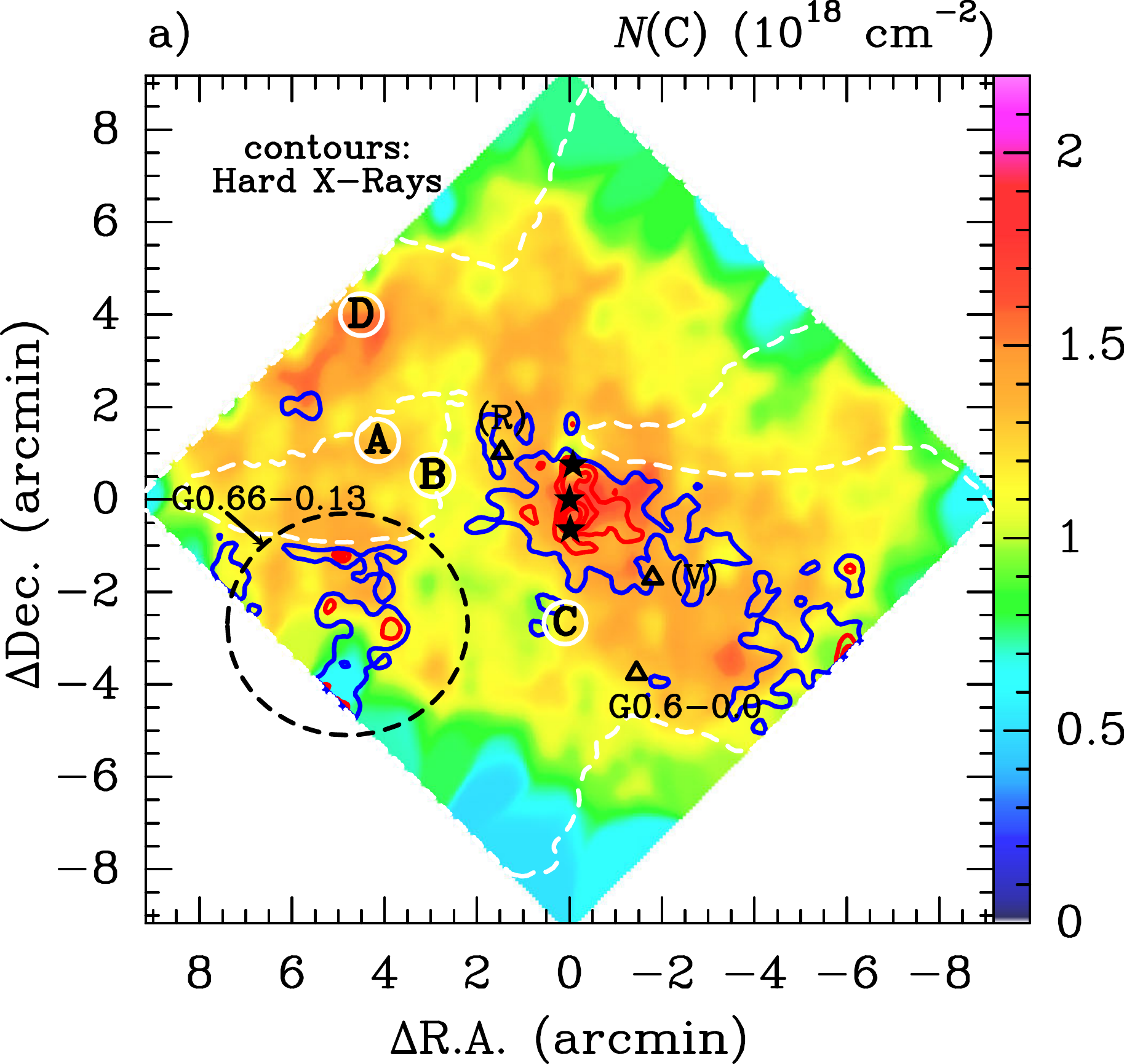}\\ \vspace{0.3cm}
\includegraphics[height=0.41\textwidth]{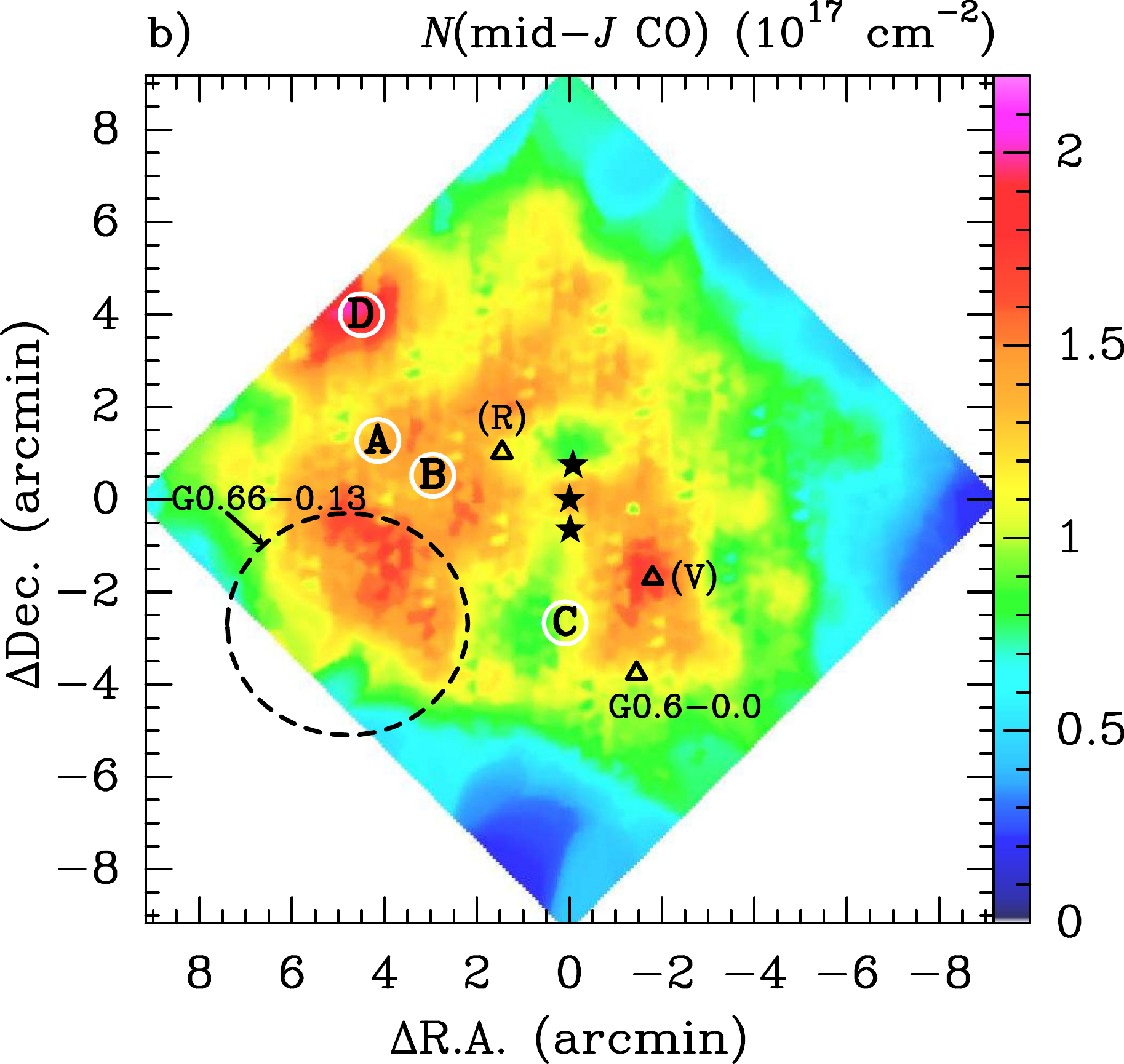} 
\caption{Maps of $N$(C) and $N$(warm CO)  derived from the population diagram analysis. a) Contours represent the hard X-Rays integrated from 3 to 79\,keV, using the 2015 \textit{NuSTAR} data \citep[][]{Zhang_2015}. Red: (22, 24, 26, 28)\,$\cdot$\,$10^{-6}\;\rm{ph\,s^{-1}\,pixel^{-1}}$. \mbox{Blue: $19\cdot10^{-6}\;\rm{ph\,s^{-1}\,pixel^{-1}}$}. The regions inside the white dashed contours are those of no exposures.}
\vspace{-0.3cm}
\label{fig:Ncoc}
\end{figure}

Figure~\ref{fig:Ncoc} shows the spatial distribution of the column density of atomic carbon and  \mbox{mid-$J$~CO} (hereafter warm CO) inferred from the population diagrams. 
\mbox{$N$(warm CO)} ranges from \mbox{$0.2\cdot10^{17}$\,cm$^{-2}$} to 
\mbox{2\,$\cdot$\,10$^{17}$\,cm$^{-2}$}  in the region around position D. Assuming a range of typical CO abundances with respect to H$_2$, \mbox{(1$-$10)\,$\cdot$\,$10^{-5}$}, these column densities imply that the warm CO is confined to a narrow gas layer of \mbox{$A_{\rm{V}}$\,$\approx$\,0.5$-$5\,mag} depth (i.e.,~it does not arise
from the entire column of material along each line of sight). The column densities of atomic carbon are higher, from \mbox{$5\cdot10^{17}$\,cm$^{-2}$} to \mbox{$1.7\cdot10^{18}$\,cm$^{-2}$}, and extend from northeast to southwest, through the center. The spatial distribution of the $N$(C) and $N$(warm~CO) column density maps, as well as the total emission intensity ($\sum I_{[\rm{C\textsc{i}}]}$ and $\sum I_{\rm{CO}}$; see Table~\ref{tab:corr30m}) are moderately correlated, with $\rho$\,$\gtrsim$\,0.8.

%------------------------------------------------------------------------
\begin{table}[!htb]
%\vspace{-0.1cm}
\caption{Warm CO and C excitation temperatures and column densities derived from population diagrams toward the reference positions. \label{table:TTNN}} 
%\vspace{-0.3cm}
    \centering
    %\hspace{-0.5cm}
    \resizebox{0.49\textwidth}{!}{
    \begin{tabular}{cccccc}
    \hline\hline
        Position & $T\rm{_{ex}}$(mid-$J$ CO) & $N$(mid-$J$ CO) & $T\rm{_{ex}}$(C) & $N$(C) \\
                 &    [K] & [10$^{17}$ cm$^{-2}$] & [K] & [10$^{17}$ cm$^{-2}$] \\
                 \hline
               A & 54 $\pm$ 2 & 1.3 $\pm$ 0.3 & 38 $\pm$ 10  & 13 $\pm$ 4  \\
               B & 56 $\pm$ 2 & 1.4 $\pm$ 0.2 & 38 $\pm$ 10  & 12 $\pm$ 3  \\
               C & 59 $\pm$ 2 & 1.1 $\pm$ 0.1 & 42 $\pm$ 12  & 12 $\pm$ 4  \\
               D & 57 $\pm$ 2 & 2.0 $\pm$ 0.3 & 45 $\pm$ 14  & 15 $\pm$ 5 \\
               \hline
    \end{tabular}
    }
    \tablefoot{$\Delta T_{\rm{ex}}$(C) and $\Delta N$(C) are due to observational uncertainties.}
\end{table}
%------------------------------------------------------------------------

%------------------------------------------------------------------------
%------------------------------------------------------------------------

\subsubsection{Gas physical conditions from non-LTE models}  \label{subsubsec:LVG}

To determine the actual  gas temperature ($T\rm{_k}$) and density ($n\rm{_{H_2}}$),
we first tried to reproduce the observed mid-$J$ CO spectral line energy distributions (SLEDs) and \CI\ fine-structure lines with single-slab non-LTE excitation and radiative transfer models. We used the RADEX code \citep{vanderTak_2007}, by means of the ndRadex tool \citep{ndRADEX}, to generate a large number of models covering a wide range of conditions. As input\footnote{We considered \mbox{CO\,--H$_2$}, \mbox{C\,--H$_2$}, and  \mbox{C\,--$e$} inelastic collision rate coefficients of \cite{Yang10}, \cite{Schroder91}, and \cite{Johnson87}, respectively. The model assumes that the H$_2$ ortho-to-para ratio is thermalized to $T_{\rm k}$. We checked that \mbox{C\,--$e$} collisions do not start to contribute to the \CI~excitation unless the  ionization fraction (the electron abundance $x_{\rm e}$\,=\,$n_{\rm e}$\,/\,$n_{\rm H}$) is unrealistically high, $x_{\rm e}$\,$>$\,10$^{-2}$.} parameters we considered: $n$(H$_2$) (from 10 to 10$^6$\,cm$^{-2}$), $T\rm{_k}$ (from 10$^2$ to 10$^6$\,K), and the gas column density ($N$). We compare the observed CO and \CI\, line intensities to those predicted by the models. Table~\ref{table:radex} summarizes the best fit model parameters for the four reference positions in the Sgr~B2 envelope. The best fit is obtained by finding the minimum root mean square (rms) value\footnote{We define the rms value of each model fit as \mbox{$\rm{rms^2}$\,=\,$\frac{1}{\rm{N}}\sum_{\it{i}=\rm{1}}^{\rm{N}}\rm{\it{log}}_{10}\left (\it{I}^{i}\rm{_{obs}}/\it{I}^{i}\rm{_{mod}} \right )^2$}, where N is the number of lines, $\it{I^{i}}\rm{_{obs}}$ is the observed intensity, and $\it{I^{i}}\rm{_{mod}}$ that of the model.}. We find that the best fit models have similar $N$(C) and $N$(CO) columns than those derived from the population  diagrams.  This agrees with optically thin line emission in the extended envelope.

%-----------------------------------------------------------------------------------
\begin{table}[t]
\caption{Physical conditions obtained from single-slab non-LTE  models. \label{table:radex}} 
    %\centering
    \hspace{-0.2cm}
    \resizebox{0.5\textwidth}{!}{
    \begin{tabular}{cccccccc@{\vrule height 10pt depth 5pt width 0pt}}
    \hline\hline
        Pos. & $T\rm{_{k,\,CO}}$ & $n\rm{_{H_{2}}}$ &   $P_{\rm th}$(CO)     & $T\rm{_{k,\,C}}$ & $n\rm{_{H_{2}}}$  & $P_{\rm th}$(C) \\
             &    [K]          &  [cm$^{-3}$]       &  [K\, cm$^{-3}$] &     [K]        & [cm$^{-3}$] & [K\, cm$^{-3}$]\\
                 \hline
           A & 			   255 & 1.7$\cdot$10$^4$   & 8.8$\cdot$10$^6$ & 51 & 0.4$\cdot$10$^4$ & 3.6$\cdot$10$^5$   \\
           B & 			   153 & 3.8$\cdot$10$^4$   & 1.2$\cdot$10$^7$ & 51 & 0.3$\cdot$10$^4$ & 3.0$\cdot$10$^5$  \\
           C &             245 & 2.8$\cdot$10$^4$   & 1.4$\cdot$10$^7$ & 61 & 0.3$\cdot$10$^4$ & 3.8$\cdot$10$^5$  \\
           D &             265 & 2.0$\cdot$10$^4$   & 1.1$\cdot$10$^7$ & 81 & 0.2$\cdot$10$^4$ & 3.6$\cdot$10$^5$ \\
               \hline
    \end{tabular}
    }
    %\tablefoot{}
    \vspace{-0.3cm}
\end{table}
%-----------------------------------------------------------------------------------

\begin{figure}[t]
    %\vspace{0.3cm}
    \centering
    \includegraphics[height=0.41\textwidth]{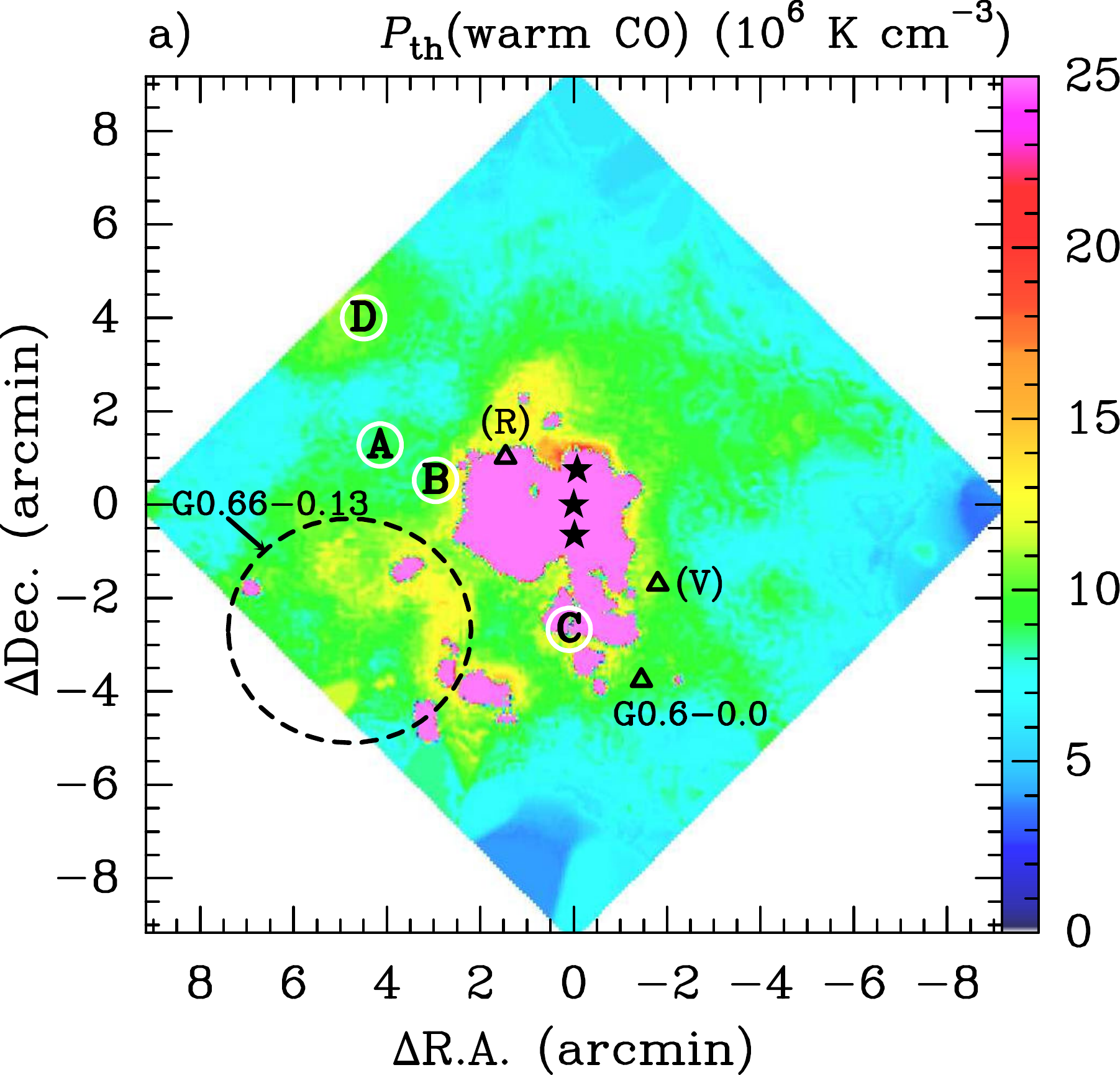}\\
    \vspace{0.3cm}       
    \includegraphics[height=0.41\textwidth]{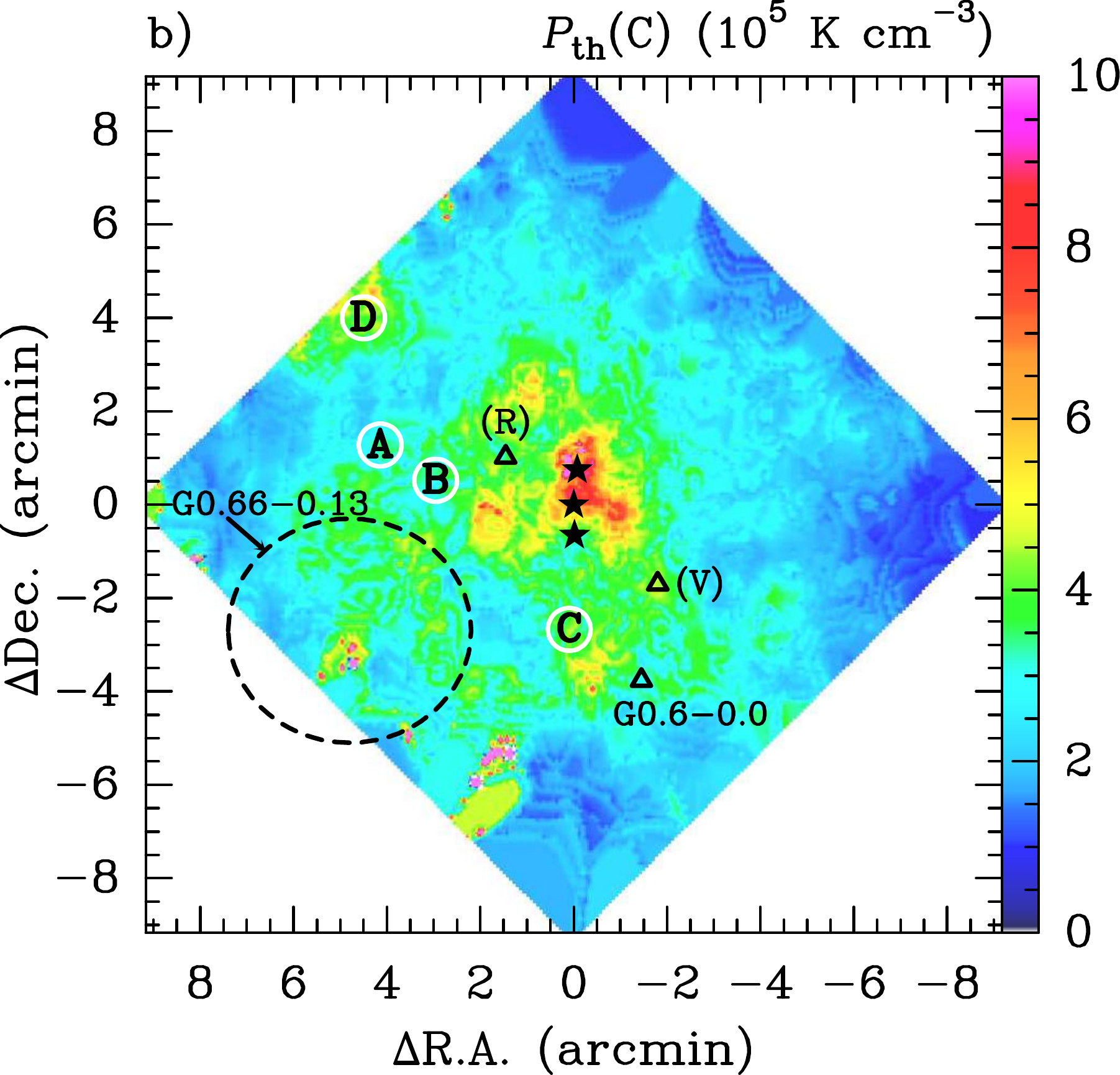}
    \caption{Maps of  $P_{\rm th}$/$k$ derived from single-slab non-LTE
    excitation models applied to \mbox{mid-$J$ CO} (upper panel) and \CI\ lines (lower panel).}
    \label{fig:PthCO} 
    %\vspace{-0.3cm}
\end{figure}

The first remarkable result  is that the \mbox{mid-$J$ CO} and \CI\;lines seem to trace two different gas components with different physical conditions (lower $T_{\rm k}$ and density for the \mbox{\CI-emitting} gas).
From the fit results we admit that the best combination of  $T_{\rm{k}}$ and $n_{\rm{H_2}}$  is slightly degenerated (in the sense that temperature and density are often anticorrelated). However, the product of these two magnitudes, the gas thermal  pressure \mbox{$P_{\rm th}/k$\,=\,$n_{\rm H}$\,$T_{\rm k}$}  (with \mbox{$n_{\rm{H}}$\,=\,$n(\rm{H})+2\it{n}(\rm{H_2})\approx2\it{n}(\rm{H_2})$} in GC molecular clouds) is more accurately constrained. \mbox{Figure~\ref{fig:PthCO}a} shows the spatial distribution of the resulting $P\rm{_{th}}$(warm CO). Following the definition of the specific region around the main star-forming cores (N, M, and S) and those of the extended envelope \citep[see ][]{Schmiedeke16}, we derive an average gas thermal pressure \mbox{$P\rm{_{th}(warm\;CO)}$\,=\,(7.8\,$\pm$\,1.3)$\cdot$10$^6$\,K\,cm$^{-3}$} in the envelope, and \mbox{$P\rm{_{th}}$(warm CO)\,=\,(1.1\,$\pm$\,0.9)\,$\cdot$\,$10^8$\,K\,cm$^{-3}$} around the main cores. Since we do not take into account self-absorption of the CO core emission by the envelope, our $P\rm{_{th}}$ determination toward \mbox{Sgr~B2(N, M, and S)} cores is likely a lower limit. We note that the regions of  high thermal pressure extend further east and south into the envelope than the typical definition of the main star-forming cores. These are regions of denser and warmer molecular gas, likely permeated by the strong FUV  field that emerges from tens of young  OB stars inside the cores and that escapes the region. Although not very different morphologically, the thermal pressure map obtained from  the \CI\;lines (Fig.~\ref{fig:PthCO}b) reveals a  much lower and more uniform gas pressure component: \mbox{$P\rm{_{th}(C)}$\,=\,$(2.8\pm 1.5)\cdot10^5$\,K\,cm$^{-3}$} in the envelope.

  %-----------------------------------------------------------------------------------
%-----------------------------------------------------------------------------------

  %---------------------------------------------------
\begin{figure*}[!h]
    \centering
    \hspace{-0.1cm} \includegraphics[width=0.505\textwidth]{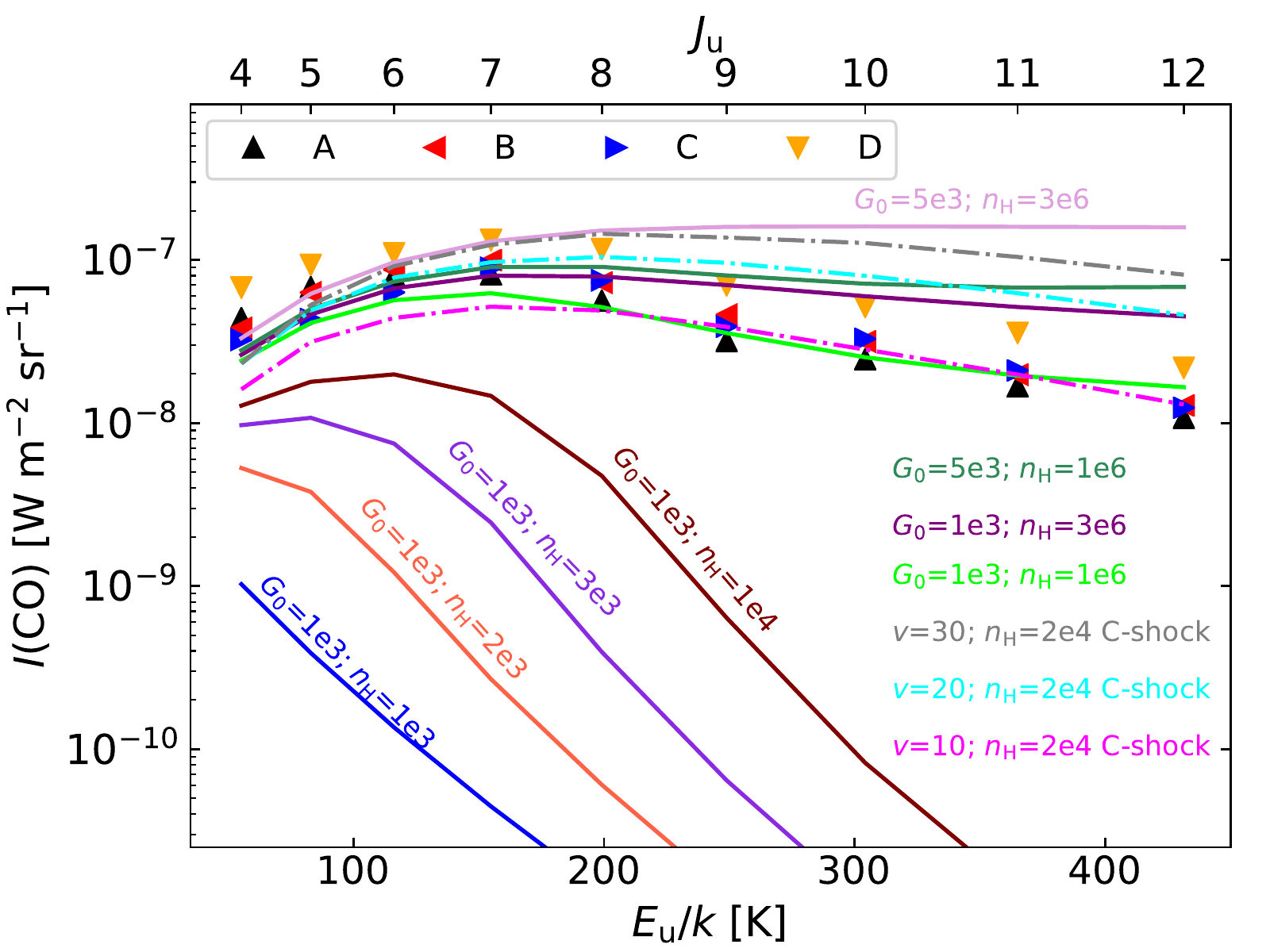} \hspace{-0.45cm}    \includegraphics[width=0.505\textwidth]{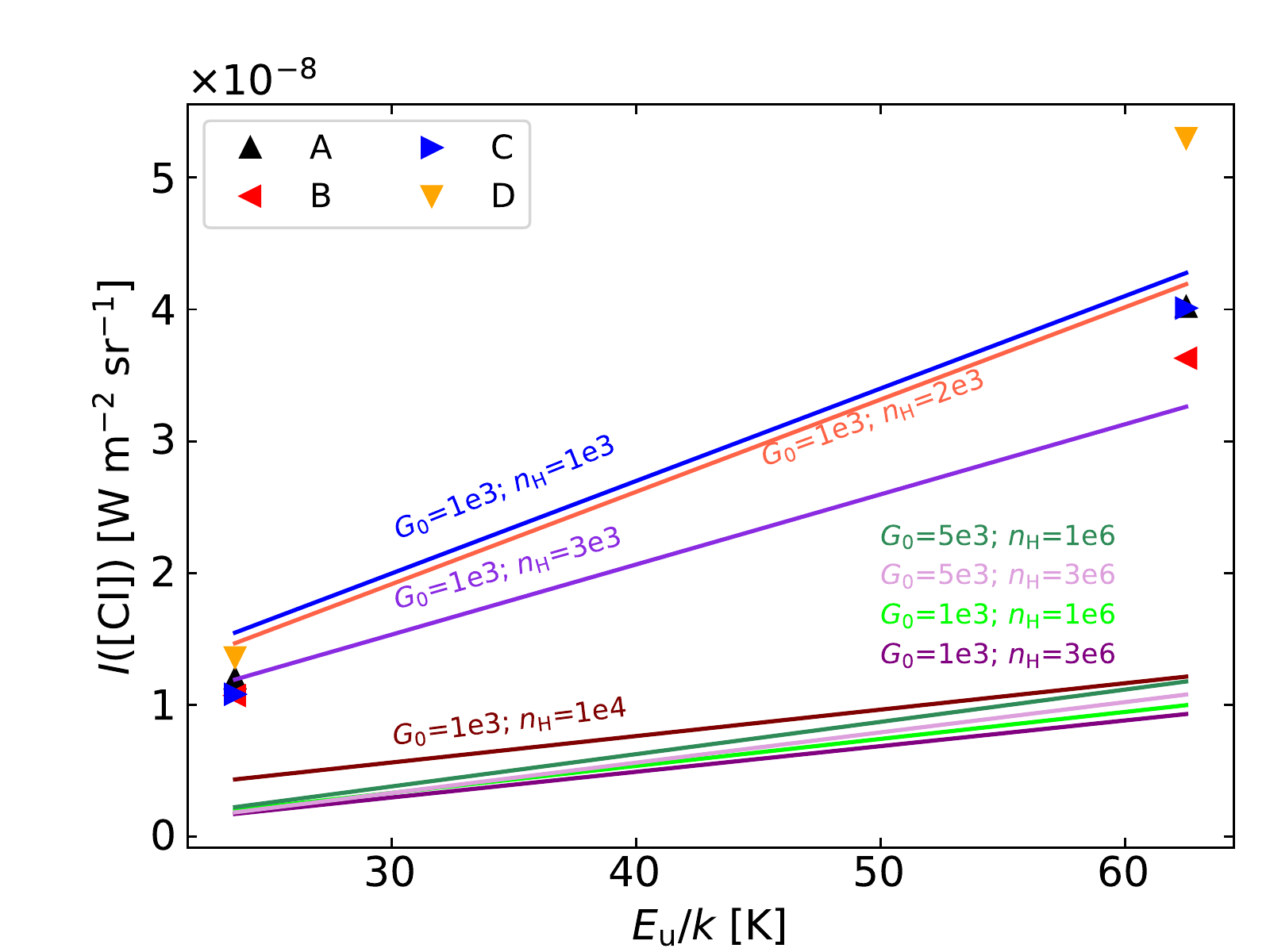}
    \caption{Observed \mbox{mid-$J$ CO} ($left panel$) and \CI\ line intensities ($right panel$) toward positions A, B, C, and D (shown as triangles). Observed CO and \CI~lines have an  intensity uncertainty of $\sim$20$\%$.  Continuous curves show predictions from PDR models with different values of $G_0$ and $n_{\rm H}$ (in units of cm$^{-3}$). Dashed curves in the left panel show  $C$-type shock model predictions \citep[from][]{Flower2010}.} 
    \label{fig:COsleds}
\end{figure*}
%---------------------------------------------------

\subsection{\mbox{PDR  and shock models applied to the Sgr~B2 envelope}} \label{sec:PDRmod}
In order to go beyond the previous simple single-slab analysis and to guide our interpretation of such a complex region, in this section we compare the observed molecular and atomic carbon line intensities with the prediction of specific PDR and shock models. These models take into account the gas temperature gradients expected in these environments.

\subsubsection{FUV and CR irradiation models} 

We use the Meudon PDR  \citep{LePetit_2006} to model the Sgr~B2 envelope as a constant density cloud illuminated by FUV radiation
fields ($G_0$) of different strengths, and adopting enhanced (compared to that in disk GMCs) H$_2$ cosmic-ray ionization rates \citep[\mbox{$\zeta_{\rm{CR}}$\,$\lesssim$\,$8\cdot 10^{-15}$\,s$^{-1}$} for the LSR velocity range \mbox{greater than $+40$\,km\,s$^{-1}$} and  associated with the dense molecular cloud, see][]{Indriolo15}. We note that \cite{Armijos2020} estimate a much lower X-ray ionization rate of $\zeta_{\rm X}$\,$\simeq$\,10$^{-19}$\,s$^{-1}$ in the envelope. In order to constrain the range of physical conditions that better reproduce the observed mid-$J$ CO lines toward A, B, C, and D positions, we ran several PDR models with different $G_0$ values and gas densities $n_{\rm H}$. Since the observed CO and \CI\ emission are very extended, we did not use any filling factor to correct the line intensity predictions.

Figure~\ref{fig:COsleds} (left panel) and Table~\ref{tab:rmsmodels} show that the best PDR \mbox{models fitting} the observed \mbox{mid-$J$ CO}  line intensities are those with \mbox{$G_0$\,$\simeq$\,10$^3$}, \mbox{$\zeta_{\rm{CR}}$\,=\,$2\cdot10^{-15}$\,s$^{-1}$}, and \mbox{$n_{\rm H}$\,$\simeq$\,$(1-3)\cdot10^6$\,cm$^{-3}$}. While PDR models of lower gas density can still crudely reproduce the observed intensities of the \mbox{lowest-$J$ CO} lines, only high density PDR models reproduce the line intensities beyond $J_{\rm u}$\,$>$\,7 (i.e., beyond the SLED peak, see Fig.~\ref{fig:COsleds}~left). In these high density models, the mid-$J$~CO emission arises from a narrow layer of warm molecular gas that occupies only \mbox{$\Delta A_{\rm{V}}\lesssim2$\,mag}. This agrees with  the low $N$(warm CO) compared to the much larger total $N$(H$_2$) columns derived  toward each position  (equivalent to $A_{\rm V}$ of a few tens of magnitudes). This points to the presence of spatially unresolved clumps or ``thin'' sheets or filaments of dense  molecular gas, exposed to strong FUV fields and distributed uniformly enough throughout the envelope so as to create relatively uniform emission; perhaps similar to the small-scale filamentary structures detected by ALMA at the edges of nearby PDRs \citep[e.g.,][]{Goico16}. Molecular globules and pillars are also  readily seen in strongly illuminated  clouds at large distances from the exciting stars \citep[][]{Dent09,Gahm13,Goico20}. All these overdense  gas structures are the consequence of  stellar feedback (UV radiation, winds, and shells), mediated by the presence of magnetic fields of different strengths and orientations.

The origin of the \CI\,492,\,809\,GHz lines in GMCs has been historically difficult to constrain because the observed \CI\;emission is generally correlated to that of \mbox{low-$J$ $^{13}$CO} lines \citep[e.g.,][]{Tauber1995}. Thus, the \CI\,emission is often more spatially extended than the predictions of dense PDR models where the emission is confined to a narrow layer in between the \ce{C+} and \mbox{low-$J$ CO} emission layers \citep[e.g.,][]{Tielens_1985a}. In Sgr~B2,  the \CI\;492, 809\,GHz line emission is widespread and relatively uniform at pc scales (Fig.~\ref{fig:SPIRE}b). 
However, the same dense PDR models that fit the  \mbox{mid-$J$ CO} emission greatly underestimate the intensity of the widespread \CI\,492,\,809\,GHz emission (Fig.~\ref{fig:COsleds}~right). The only way to reproduce their extended emission  is adding a lower density\footnote{The \CI-emitting gas in Sgr~B2 is much denser than the warm \mbox{($T_{\rm k}$\,$\simeq$\,200$-$500\,K)} and  diffuse clouds ($n_{\rm H}$\,$\lesssim$\,100\,cm$^{-3}$)  invoked to explain the presence of absorption lines from excited metastable levels of H$_3^+$ \citep[e.g.,][]{Goto11}. These diffuse clouds in the CMZ typically absorb at negative LSR velocities and have higher $\zeta_{\rm{CR}}$ rates \citep[\mbox{$\gtrsim 10^{-14}\,\rm{s}^{-1}$}, e.g.,][]{Indriolo15,LePetit16}. The different $\zeta_{\rm{CR}}$ values likely imply that $\zeta_{\rm{CR}}$ decreases for increasing depths inside dense clouds.}, \mbox{$n_{\rm H}$\,$\simeq$\,(1$-$2)\,$\cdot$\,10$^3$\,cm$^{-3}$}, component illuminated by the same FUV field, \mbox{$G_0$\,$\simeq$\,10$^3$}, but extending to \mbox{$\Delta A_{\rm V}$\,$\simeq$\,20\,mag} (to match the derived $N$(C) columns and observed \CI\ line intensities). In this component, most of the gas-phase carbon is in atomic form, with \mbox{$N$(C)/$N$(CO)\,$\simeq$\,10}. The gas temperatures predicted by the PDR model are similar to the \mbox{$T_{\rm ex}$(C)}   inferred from the population diagrams (Fig.~\ref{fig:Texrotdrtot}). However, we note that at the $n_{\rm H}$  and $T_{\rm k}$ of the \CI-emitting gas, the predicted \mbox{mid-$J$ CO} line emission is negligible.

\begin{figure*}[h]
\centering
\includegraphics[width=0.99\textwidth]{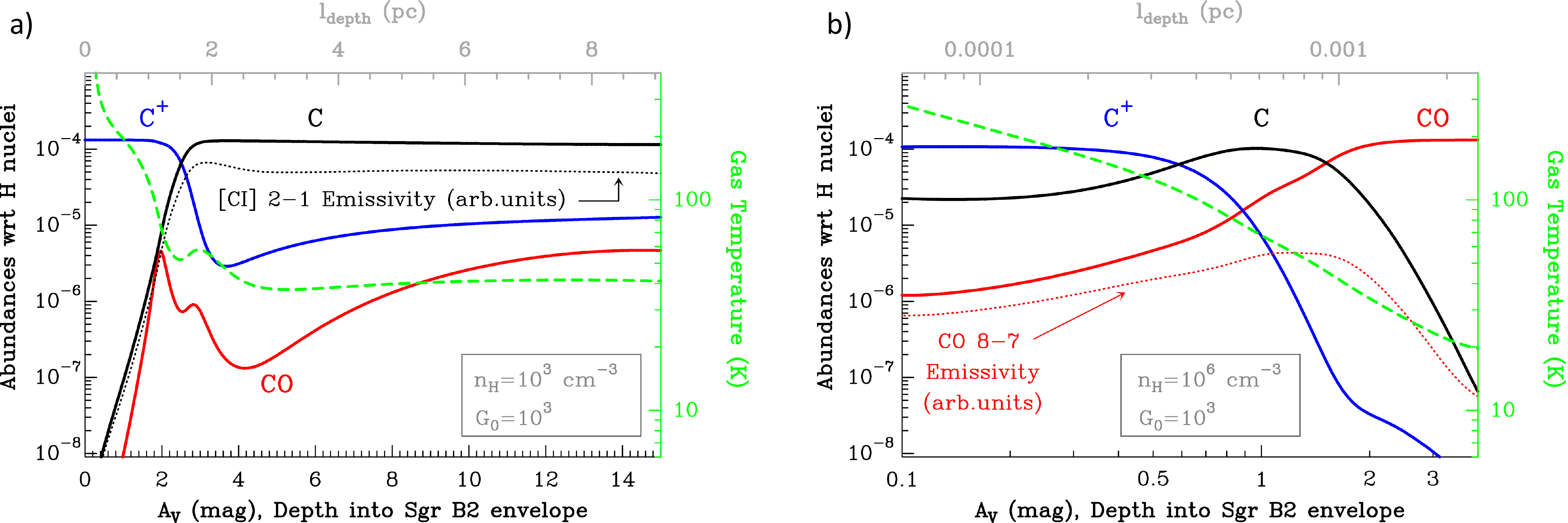}
    \caption{Abundance, line emissivity, and $T_{\rm k}$ profiles of PDR models with 
    \mbox{$G_0$\,$=$\,10$^3$} and \mbox{$\zeta_{\rm{CR}}$\,$=$\,2$\cdot$10$^{-15}$\,s$^{-1}$}, 
    consistent with the observed \CI~emission (\mbox{$n_{\rm H}$\,=\,10$^3$\,cm$^{-3}$}; \textit{left}) and the \mbox{mid-$J$ CO emission} (\mbox{$n_{\rm H}$\,=\,10$^6$\,cm$^{-3}$}; \textit{right}). We note the different spatial scales (upper axis).}
    \label{fig:PDR_struct}
 \end{figure*}

Figure~\ref{fig:PDR_struct} shows the depth-dependent C$^+$, C, and CO abundance profiles predicted by the low-density (Fig.~\ref{fig:PDR_struct}a) and \mbox{high-density} (Fig.~\ref{fig:PDR_struct}b) PDR models with $G_0$\,=\,10$^3$. The dotted curves show the predicted \CI\,809\,GHz and CO\,(8$-$7) line emissivities (in arbitrary units): ``extended'' along the line of sight for \CI\;and ``narrow''  for the \mbox{mid-$J$\,CO} lines. Most of the \CI\,emission  in \mbox{Fig.~\ref{fig:PDR_struct}a} arises from cloud depths where the gas temperature is controlled by $\zeta_{\rm CR}$, in the range
\mbox{$T_{\rm k}$\,$\simeq$\,40$-$60\,K} for \mbox{$\zeta_{\rm{CR}}$\,$=$\,(2$-$6)\,$\cdot$\,10$^{-15}$\,s$^{-1}$}, and results in \mbox{$x_e$\,$\gtrsim$\,5$\cdot$\,10$^{-5}$} and 
\mbox{$N$(C)\,/\,$N$(C$^+$)\,$\gtrsim$\,4}. In the \mbox{high-density} model of \mbox{Fig.~\ref{fig:PDR_struct}b}, the mid-$J$\,CO emission arises from denser gas  where $T_{\rm k}$ is set by $G_0$. This produces a steep temperature gradient in the mid-$J$\,CO-emitting layers from a few hundred K  to about 40\,K.

%-----------------------------------------------------------------------
%-----------------------------------------------------------------------

%-------------------------------------------------------------------------
\begin{table}[t] %[htpb]
\centering
\caption{Rms$^5$ of the best PDR and shock models for \mbox{mid-$J$ CO} lines.\label{tab:rmsmodels}}
\begin{tabular}{lccccc @{\vrule height 5pt depth 10pt width 0pt}}
\hline
\hline
%Model                & A     & B     & C     & D     \\
\multicolumn{1}{c}{PDR Model}                                                                                       & \multicolumn{4}{c}{rms\,\,/\,\,Position}              \\
$G_0$ {[}Habing{]},\;$n_{\rm{H}}$ {[}cm$^{-3}${]}   & A     & B     & C                         & D          \\
\hline
$G_0$\,=\,$1\cdot10^3$, $n_{\rm{H}}$\,=\,$1\cdot10^6$   & \textbf{0.15} & \textbf{0.15} & \textbf{0.11} & 0.32 \\
$G_0$\,=\,$1\cdot10^3$, $n_{\rm{H}}$\,=\,$3\cdot10^6$   &    0.33       &  0.27         &   0.26        & \textbf{0.24} \\
$G_0$\,=\,$5\cdot10^3$, $n_{\rm{H}}$\,=\,$1\cdot10^6$   &    0.41       &  0.34         &   0.34        & 0.27 \\
\hline
\multicolumn{1}{c}{$C$-type shock model}                                                                                      & \multicolumn{4}{c}{}              \\
$\rm{v_s}$ {[}\kms{]},\;$n_{\rm{H}}$ {[}cm$^{-3}${]}   &  \multicolumn{4}{c}{}  \\
\hline
$\rm{v_s}$\,=\,20, $n_{\rm{H}}$\,=\,$2\cdot10^4$         & 0.39  & 0.32  & 0.32  & \textbf{0.25} \\
$\rm{v_s}$\,=\,10, $n_{\rm{H}}$\,=\,$2\cdot10^4$        & 0.22  & 0.22  & 0.16  & 0.39  \\  \hline
\end{tabular}
%}
\tablefoot{Models in boldface text produce the best line intensity fits.}
\vspace{-0.5cm}
\end{table}
%\clearpage
%-------------------------------------------------------------------------

\subsubsection{Shock models of the mid-$J$ CO emission}

We also compared the line intensity and shape of  \mbox{mid-$J$ CO} SLEDs with  predictions of molecular shock models reported by \cite{Flower2010}. The first conclusion is that high-velocity  \mbox{$J$-type} shock models  (\mbox{v$_{\rm s}$\,=\,40\,\kms}) produce stronger CO emission and shifted to \mbox{higher-$J$ CO} lines. These models neither reproduce the observed \mbox{mid-$J$ CO} line intensities nor the global shape of the CO SLED. For \mbox{nondissociative} \mbox{$C$-type shocks}, and  as v$_{\rm s}$ and $n_{\rm H}$ decreases, the peak of the CO SLED occurs at lower $J$ numbers. Models with 
\mbox{$n_{\rm H}$\,=\,2\,$\cdot$\,10$^4$\,cm$^{-3}$} and \mbox{v$_{\rm s}$\,=\,10$-$20\,\kms} produce a CO SLED reasonably similar to the observed in Sgr~B2 envelope. However, only for position D, a \mbox{$C$-type} shock model with \mbox{$\rm{v_s}$\,=\,20\,\kms} and \mbox{$n_{\rm H}$\,=\,2\,$\cdot$\,10$^4$\,cm$^{-3}$} results in a rms comparable to that of the best  PDR model   (see Table~\ref{tab:rmsmodels} and left panel of Fig.~\ref{fig:COsleds}). Although not shown in \mbox{Fig.~\ref{fig:COsleds}},  \mbox{$J$-type} and $C$-type  shock models  produce little \CI\ emission \citep[][]{Neufeld1989,Hollenbach1989} compared to the intensities observed in Sgr~B2.

\subsubsection{Enhanced ionization-rates, extended \ce{N2H+} emission}\label{subsubsec:CRs_N2H+}

The \ce{N2H+}~(1$-$0) emission in Sgr~B2 is extended at pc scales. Considering previous observations of nearby star-forming clouds of the Galactic disk, this is an unexpected result. In these clouds, the \ce{N2H+} emission traces cold and dense gas shielded from FUV radiation \citep{Caselli2002}. Hence, the  \mbox{\ce{N2H+}} emission is confined to the dense filamentary ``bones''  of these clouds and to cold prestellar cores inside them  \citep[e.g.,][]{Sanhueza2012,Kirk2013,Pety17,Hacar17,Hacar18}.

The main chemical pathway to the formation of \ce{N2H+} is the reaction of N$_2$ with H$_3^+$ \citep[e.g.,][]{Herbst1973}:
\begin{equation}
    \centering
    \rm{N_2 + H_3^+ \rightarrow N_2H^+ + H_2}.
    \label{eq:n2hpform}
\end{equation}
The ion H$_3^+$ forms after ionization of H$_2$ by cosmic rays, so any enhancement of the ionization rate (including X-rays) will increase the  H$_3^+$ production rate \citep{Indriolo2012}. The cosmic-ray ionization rate in  Sgr~B2  is \mbox{$\zeta\rm{_{CR}}$\,$\simeq$\,(1$-$10)\,$\cdot$\,$10^{-15}\, \rm{s}^{-1}$} \citep{Zhang_2015,Indriolo15}, a factor of 10$-$100 higher than the average cosmic-ray ionization rate in clouds of the Galactic disk, where the main \ce{N2H+} destruction route is reacting  with CO. As the ionization rate increases, so does the abundance of helium ions, ionized by cosmic-ray particles. Hence, destruction of CO by reactions with He$^+$ (in addition to reactions with H$_{3}^{+}$ to form \ce{HCO+})  becomes increasingly important. In this regime, reactions of He$^+$ with N$_2$ also enhance the formation of \ce{N2+}, which further reacts with H$_2$ and increases the \ce{N2H+} production. Because of the higher ionization fractions $x_{\rm e}$ at high $\zeta\rm{_{CR}}$,  the destruction of both \ce{N2H+} and \ce{HCO+}  becomes dominated by dissociative recombinations with electrons. This leads to a decrease of the \ce{HCO+}/\ce{N2H+} column density ratio as $\zeta\rm{_{CR}}$ increases \citep[see also][]{Ceccarelli2014}. This reasoning applies as long as H$_{3}^{+}$ destruction is dominated by reactions with CO \citep[so called chemistry in the ``low ionization phase'' (LIP);][]{PdF92,LeBourlot93}. In the LIP, and  as $\zeta\rm{_{CR}}$ increases, protonation reactions with H$_{3}^{+}$ dominate the formation of increasingly more abundant molecular ions such as \ce{N2H+}.
 
%-----------------------------------------------------------------------
\begin{figure}[t]
    \centering   
    \includegraphics[height=0.3\textwidth]{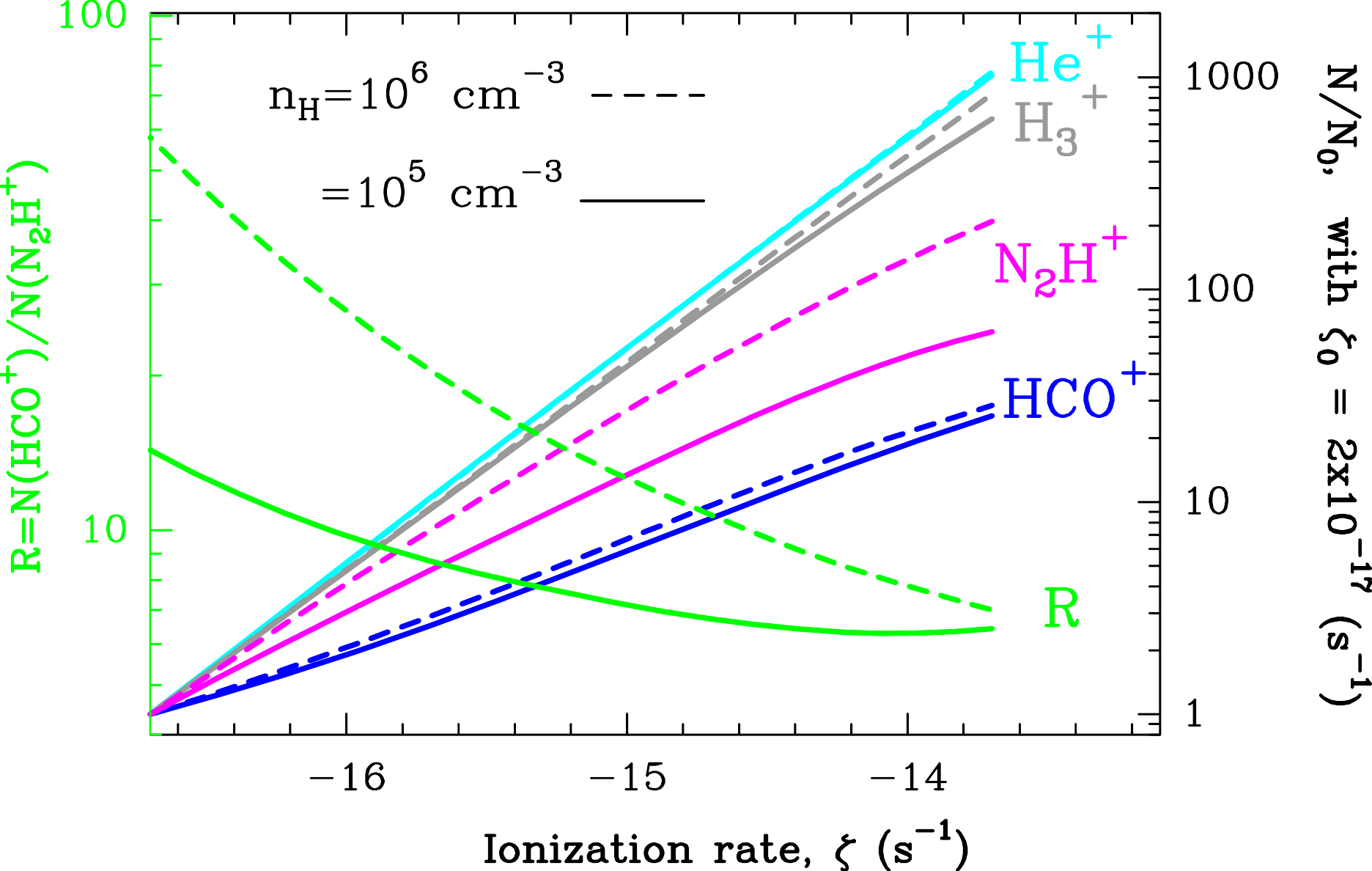}
    \vspace{0.4cm}
    \caption{Model predictions for different values of $\zeta\rm{_{CR}}$
    (and fixed $G_0$\,=\,10$^3$, $n_{\rm H}$\,=\,10$^5$, 10$^6$\,cm$^{-3}$, and 
    $A_{V, \rm{tot}}$\,=\,15\,mag). Continuous curves show the column density enhancement
    $N/N_0$ with respect to the  models   with \mbox{$\zeta_{\rm{CR,0}}$\,$=$\,2$\cdot$10$^{-17}$\,s$^{-1}$} 
    (roughly the ionization rate in disk GMCs). }
    \label{fig:PDR_ion}
\end{figure}
%-----------------------------------------------------------------------

In the ``high ionization phase'' (HIP\footnote{Our best models
of the \CI~emission belong to the HIP category. They lead to
\mbox{$x_e$\,$\gtrsim$\,5$\cdot$\,10$^{-5}$} and high \mbox{$N$(C)/$N$(CO)}
and \mbox{$N$(C)/$N$(C$^+$)} ratios.}), however, the abundance of electrons increases to the point where 
H$_{3}^{+}$ destruction by electron recombinations becomes  comparable to  destruction by  CO
\citep[e.g.,][]{LeBourlot93}. In this regime, the production of molecular ions from  H$_{3}^{+}$ becomes less important and the predicted \ce{HCO+} or \ce{N2H+} abundances are much lower. Hence, the moderately high  $G_0$ and $\zeta\rm{_{CR}}$ values in Sgr\,B2 imply that 
\ce{N2H+} arises from dense gas ($n_{\rm H}$\,$>$10$^4$\,cm$^{-3}$) at relatively large $A_V$ 
(so that $x_e$ does not reach the high values of the HIP).
\mbox{Figure~\ref{fig:PDR_ion}} show models of this kind for  Sgr\,B2 envelope (\mbox{$G_0$\,=\,10$^3$}, \mbox{$n_{\rm H}$\,=\,10$^5$ and 10$^6$\,cm$^{-3}$}, and increasing  $\zeta\rm{_{CR}}$).
For \mbox{$\zeta_{\rm{CR}}$\,$=$\,(2$-$6)\,$\cdot$\,10$^{-15}$\,s$^{-1}$}, these models predict $x_e$ of several 10$^{-8}$ to several 10$^{-7}$ at \mbox{$A_{V}$\,=\,15\,mag}.
This chemistry is markedly  different  to that of cold molecular clouds in the disk of the Galaxy, with lower $\zeta_{\rm{CR}}$ and thus lower electron and H$_{3}^{+}$ abundances.  There, \ce{N2H+} is only abundant where CO  freezes-out at high extinction depths and cold temperatures  \citep[\mbox{$T_{\rm{d}}$\,$\simeq$\,$T_{\rm{k}}$\,$<$\,20\,K}; see e.g.,][]{Caselli2002,Bergin_2007}.

%%%%%%%%%%%%%%%%%%%%%%
\section{Discussion} \label{sect:discussion}
%%%%%%%%%%%%%%%%%%%%%%

%%\subsection{Observational tracers of different gas heating mechanisms in the GC} 
\subsection{Tracers of different gas heating mechanisms} 
\label{sec:heatingmec}
%--------------------------------------------
In this section we discuss how to observationally distinguish the main
 heating mechanisms of the warm molecular gas in Sgr~B2.

\subsubsection{Radiation versus shocks and the CO/FIR intensity ratio.} \label{subsec:lumrat}

%----------------------------------------------------------------------------
\begin{table*}[t]
%\vspace{0.4cm}
\caption{Luminosities in Sgr B2 (envelope, main cores, and total).  \label{table:MLB2}}
    \centering
    \resizebox{1\textwidth}{!}{
    \begin{tabular}{ccccccccc@{\vrule height 8pt depth 4pt width 0pt}}
    \hline\hline
        Region &  $L\rm{_{FIR}}^a$ & $L\rm{_{\rm{CO}}}$ &  \LCI    &  \LNII & $L\rm{_{\rm{H_{2}O,\,752}}}$ &  $L\rm{_{CO}}$\,/\,$L\rm{_{FIR}}$ & \LCI\,/\,$L\rm{_{FIR}}$  & \LNII\,/\,$L\rm{_{FIR}}$ \\
               &  [\Ls]            & [\Ls]              &  [\Ls]  &  [\Ls]  &  [\Ls]                                & ($10^{-4}$)                   & ($10^{-4}$)          & ($10^{-4}$) \\ 
                 \hline
               
               Envelope   &  ($2.6\pm0.5$)$\cdot 10^{7}$ & ($9\pm2$)$\cdot$10$^3$ & ($13 \pm 3$)$\cdot$10$^{2}$  &   ($17\pm3$)$\cdot$10$^2$       & 56\,$\pm$\,11 &  $4\pm1$   & $0.5\pm0.1$  &  $0.7\pm0.2$ \\
               Main star- &  \multirow{2}{*}{($0.6\pm0.1$)$\cdot$10$^{7}$} & \multirow{2}{*}{($1.2\pm0.2$)$\cdot$10$^3$} & \multirow{2}{*}{($0.9\pm0.2$)$\cdot$10$^{2}$} & \multirow{2}{*}{($1.0\pm0.2$)$\cdot$10$^2$}    & \multirow{2}{*}{$37\pm7$} & \multirow{2}{*}{$2.0\pm0.7$}  &  \multirow{2}{*}{$0.15\pm0.06$} & \multirow{2}{*}{$0.17\pm0.07$}  \\
               forming cores & & & & & & &  & \\
               \hline
               Total &  ($3.3\pm0.7$)$\cdot$10$^{7}$ & ($11\pm2$)$\cdot$10$^3$ & ($14\pm3$)$\cdot$10$^{2}$ & ($18\pm4$)$\cdot$10$^2$     & $93\pm20$ & $3\pm1$   & $0.4\pm0.1$ &  $0.6\pm0.2$ \\
               \hline
    \end{tabular}
    }
    \tablefoot{The area around the main star-forming cores ($\sim$12\,pc$^2$) corresponds to the ``moderate density'' region of  \cite{Schmiedeke16}. The total region is the entire mapped region, that is a rhombus with a 43.8\,pc diagonal
    (area of $\sim$960\,pc$^2$). The envelope region  ($\sim$948\,pc$^2$) is the difference between the main star-forming cores and the total mapped area. 
    The $M\rm{_{H_2}}$ masses we compute in the envelope, main star-forming cores, and entire region are $8.9\cdot10^6$\,\Ms, $2.3\cdot10^6$\,\Ms, and 1.1\,$\cdot$\,$10^7$\,\Ms, respectively.
     $^{(a)}$ $L\rm{_{FIR}}$ from 40 to 500\,$\upmu$m  and adopting D\,=\,8.18\,kpc.}
\end{table*}

%----------------------------------------------------------------------------

\begin{figure}[t]
    %\vspace{-0.3cm}
    \centering       
    \includegraphics[height=0.49\textwidth]{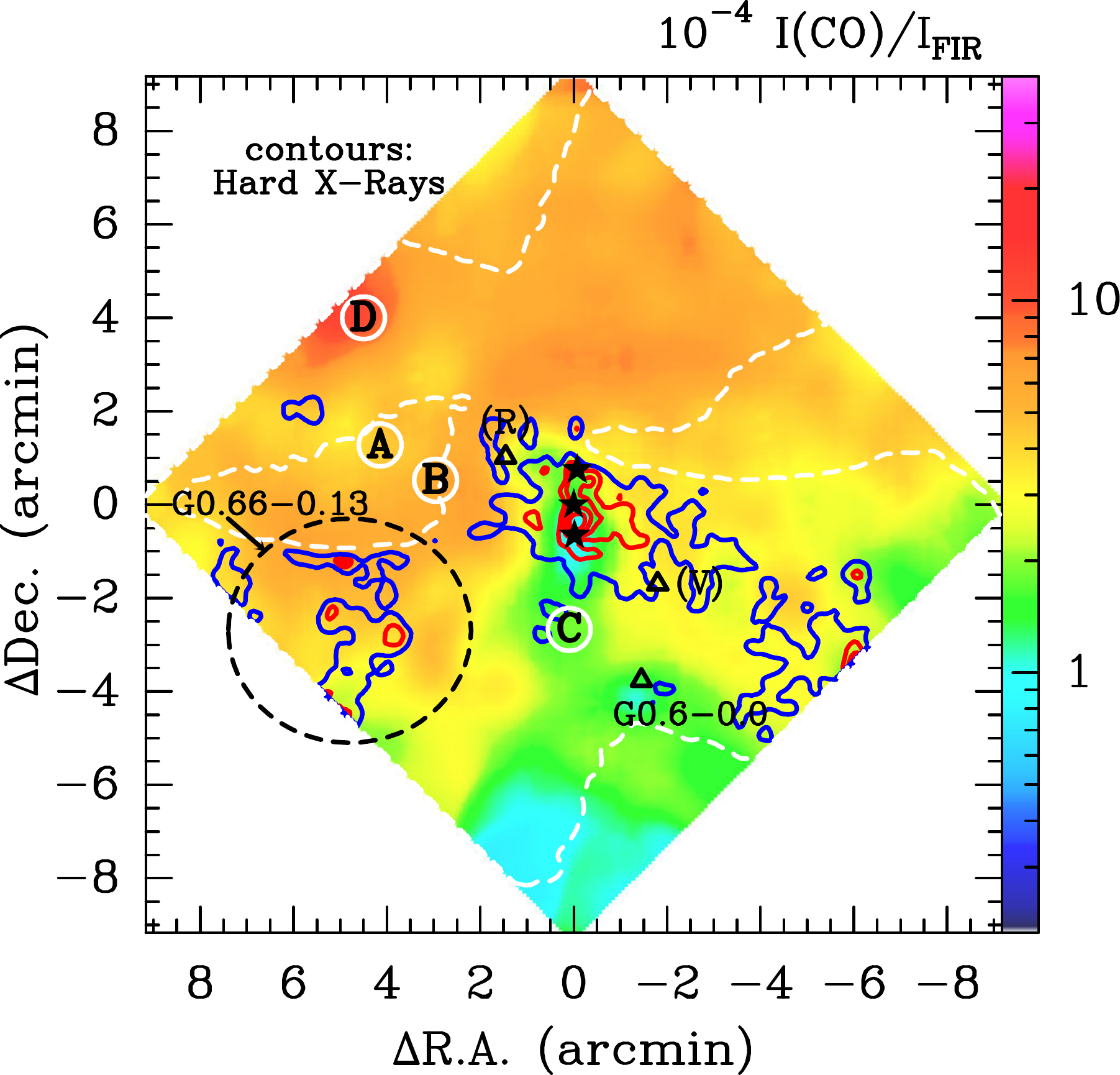}
    \caption{CO to FIR intensity ratio.  Contours represent the hard X-ray emission using the 2015 \textit{NuSTAR} data (S. Zhang priv. comm.; no data inside the areas enclosed
    by the white dashed lines). Red: (22, 24, 26, 28)\,$\cdot$\,$10^{-6}\;\rm{ph\,s^{-1}\,pixel^{-1}}$. \mbox{Blue: $19\cdot10^{-6}\;\rm{ph\,s^{-1}\,pixel^{-1}}$}.}
    \label{fig:Lcofir}
    %\vspace{-0.3cm}
\end{figure}

One possible observational diagnostic to discriminate between radiative and mechanical heating is the total CO to FIR luminosity ratio\footnote{We  add the CO line
intensities from $J_{\rm u}$\,$\geq$\,4 only, but we note that in warm gas, the low-$J$ CO lines have a very modest contribution to the total CO surface brightness in W\,m$^{-2}$\,sr$^{-1}$ units.}. Dust grains are more effectively  heated in \mbox{FUV-irradiated} environments (producing strong FIR continuum emission). On the other hand,  shocks typically heat the molecular gas to higher temperatures, but not the grains. Therefore, we expect higher  \mbox{$I\rm{_{CO}}/\mathit{I}\rm{_{FIR}}$} intensity ratios in  shocks than in PDRs \mbox{\citep[e.g.,][]{Meijerink2013}}. \mbox{Figure~\ref{fig:Lcofir}} shows a map of the spatial distribution of the \mbox{$I\rm{_{CO}}/\mathit{I}\rm{_{FIR}}$} intensity ratio in Sgr~B2. The derived ratios range from about \mbox{$9\cdot10^{-5}$} toward 25$''$ south of Sgr~B2(M), to  1.2\,$\cdot$\,$10^{-3}$ toward position D. The average ratio in  the mapped area is \mbox{$I\rm{_{CO}}/\mathit{I}\rm{_{FIR}}$\,=\,$(4\pm1)$\,$\cdot$\,$10^{-4}$} (see Table~\ref{table:MLB2}). The envelope of Sgr~B2 shows a wide range of \mbox{$I\rm{_{CO}}/\mathit{I}\rm{_{FIR}}$} ratios (see \mbox{Fig.~\ref{fig:Lcofir}}). At first order we see two differentiated regions, one with the low \mbox{$I\rm{_{CO}}/\mathit{I}\rm{_{FIR}}$} ratios in greenish and blueish, that includes the main star-forming cores and \mbox{G0.6$-$0.0} in the south,
and coincides with the brightest \NII\,205\,$\upmu$m emitting regions.
 The second region includes most of the envelope, where the ratio is $\geq$\,4\,$\cdot$\,$10^{-4}$, in orangish, and its maximum value peaks toward position D.

%-----------------------------------------------------------------------
%-----------------------------------------------------------------------

 %-----------------------------------------------------------------------
\begin{figure}[ht]
    \centering    
    \includegraphics[height=0.46\textwidth]{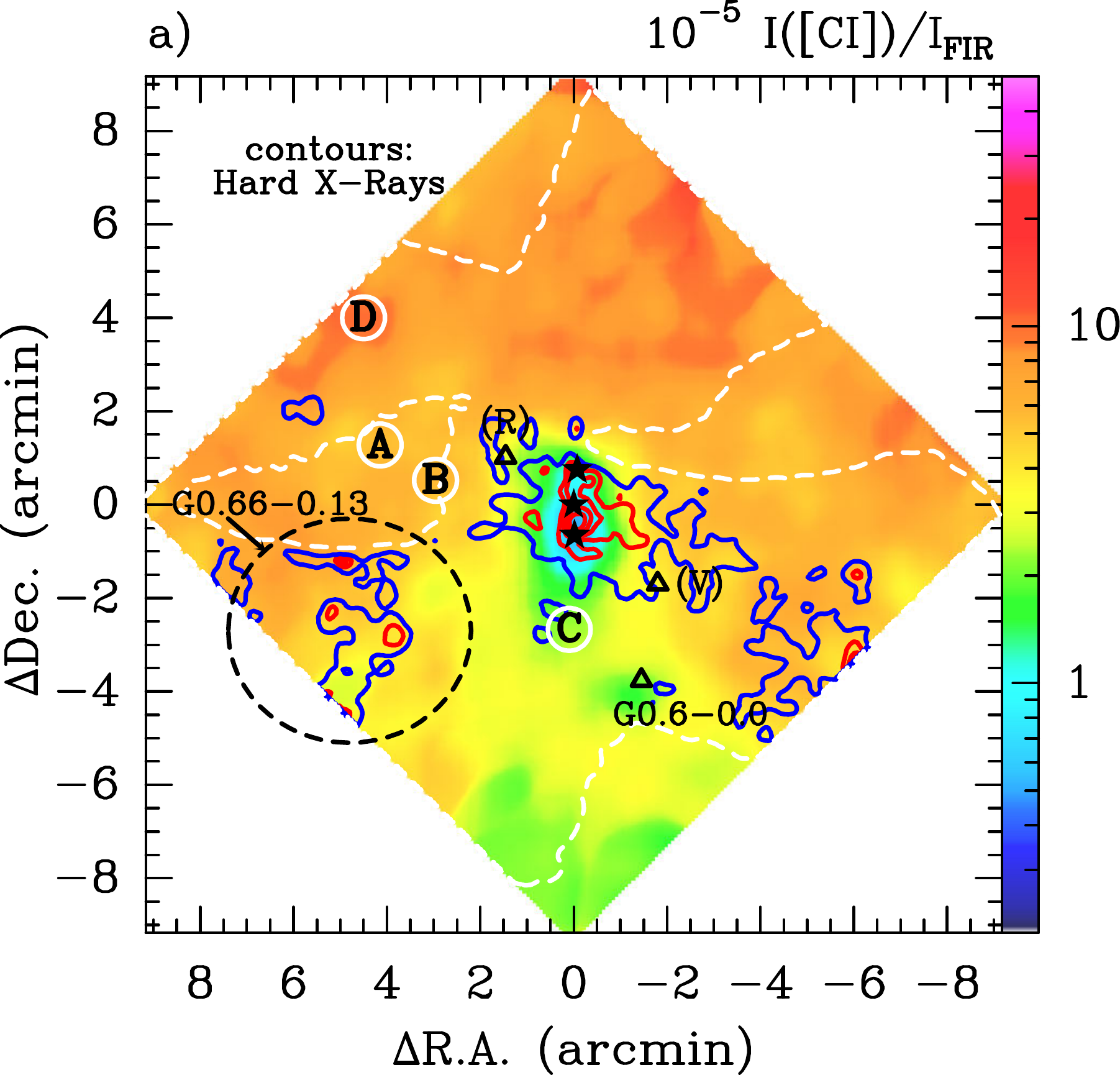}\\ \vspace{0.3cm}
\includegraphics[height=0.46\textwidth]{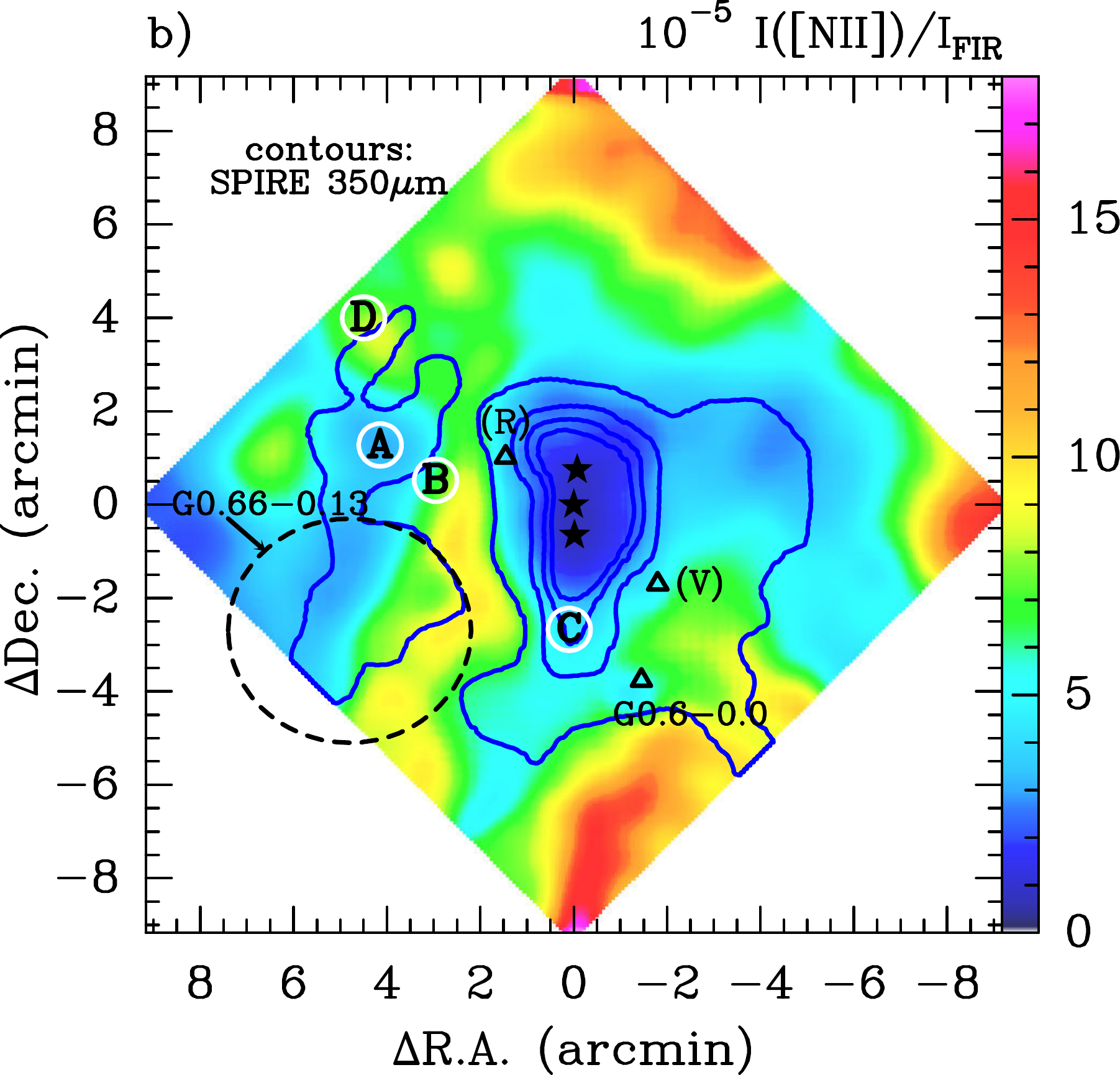} \vspace{0.2cm}
\caption{Spatial distribution of the \CI\,($492+809$ GHz) and \NII\,205\,$\upmu$m to FIR intensity ratios in the Sgr B2. a) Contours represent the hard X-ray emission integrated from 3 to 79 keV, using the 2015 \textit{NuSTAR} data (same as in Fig.~\ref{fig:Lcofir}).  b) Contours represent the SPIRE~350\,$\upmu$m dust emission. 
}
    \label{fig:Lcifir}
\end{figure}
%-----------------------------------------------------------------------

The \mbox{$I\rm{_{CO}}/\mathit{I}\rm{_{FIR}}$} intensity ratio has been observationally measured in prototypical PDRs and shocked gas regions of the Galaxy and in extragalactic sources. It is $\gtrsim$\,$6$\,$\cdot$\,$10^{-4}$ toward \mbox{Orion BN/KL} outflows  \citep{Goicopacs15}, $\simeq$\,$3\cdot10^{-4}$ toward the strongly illuminated PDR the Orion Bar \citep{Joblin18}, and $\simeq$\,$10^{-3}$ toward Sgr~A* and the circumnuclear disk  
\citep[the CND; where \mbox{FUV-irradiated} shocks likely dominate the hot molecular gas heating;][]{Goicoechea2013,Goicoechea2018a}. In a representative sample of local (U)LIRGs (the HerCULES sample), the \mbox{$L\rm{_{CO}}/\mathit{L}\rm{_{FIR}}$} ratio is remarkably constant, \mbox{$\sim$\,(1.5$-$3.5)\,$\cdot$\,10$^{-4}$}, for all $L\rm{_{FIR}}$ and redshifts. This excludes NGC~6240, which shows a much higher ratio,  \mbox{$1.1$\,$\cdot$\,$10^{-3}$} \citep{Rosenberg2015}. NGC~6240 is a luminous merging galaxy in which mechanical  widespread shocks, in addition to UV radiation, explains the observed \mbox{high-$J$} CO lines \citep{Meijerink2013,Rosenberg2015}.
Comparing these intensity ratios with those inferred in Sgr~B2 (see Table~\ref{table:MLB2}), we conclude that the lower \mbox{$I\rm{_{CO}}/\mathit{I}\rm{_{FIR}}$} values measured close to the main  star-forming cores and toward many locations in the southern part of the envelope resemble more those observed in  PDRs. The higher  values in the north of the envelope resemble more those seen in 
shocks.

\subsubsection{The \CI\,/\,FIR and \NII\,/\,FIR ratios and their ``deficit"}

Figure~\ref{fig:Lcifir}a shows the spatial distribution of the  \mbox{$I\rm{_{[CI]}}$\,/\,$\mathit{I}\rm{_{FIR}}$} intensity ratio over Sgr~B2. 
The \mbox{$I\rm{_{[CI]}}$\,/\,$\mathit{I}\rm{_{FIR}}$} ratio decreases from $\sim$\,10$^{-4}$ in the envelope, to several 10$^{-6}$ toward the bright FIR star-forming cores (see \mbox{Table~\ref{table:MLB2}}). 
The decrease of \mbox{$I\rm{_{[CI]}}/\mathit{I}\rm{_{FIR}}$} for increasing FIR intensity reminds of the so called C$^+$/FIR  ``deficit'' seen in observations of the \CII\,158\,$\upmu$m line toward local galaxies hosting vigorous star-formation \citep[e.g.,][]{Luhman98,Malhotra01}. This extragalactic deficit may apply to other FIR fine-structure lines (\OI, [\OIII], \NII, [\NIII], ...) that trace different phases of the ISM \citep[][]{Carpio11,Herrera-Camus18}. In Sgr\,B2 we find that the \mbox{$I\rm{_{[CI]}}$\,/\,$\mathit{I}\rm{_{FIR}}$} intensity ratio is  anticorrelated with the FIR emission ($\rho$\,=\,$-$0.84; see Table~\ref{tab:corrfir}). The \CI\;emission in Sgr~B2 only changes by a factor of $\sim$\,4 whereas $I_{\rm FIR}$ changes by more than two orders of magnitude. Hence the variations of the  \mbox{$I\rm{_{[CI]}}$\,/\,$\mathit{I}\rm{_{FIR}}$} are dominated by variations of $I_{\rm FIR}$, either because $G_0$ reaches high values in the envelope (leading to strong FIR emission of 
\mbox{FUV-heated}  dust) or because the dust column density is very large toward  star-forming cores; producing an excess of non-PDR, internally heated,  FIR dust emission.

 Figure~\ref{fig:Lcifir}b  shows the spatial distribution of the  
\mbox{$I_{\rm [NII]205}$/$I_{\rm FIR}$} intensity ratio. The \NII\,205\,$\upmu$m line traces  extended  low-density ionized gas. The spatial distribution of the \mbox{$I_{\rm [CI]}$/$I_{\rm FIR}$}  and  \mbox{$I_{\rm [NII]205}$/$I_{\rm FIR}$}  intensity ratios are different. Plotted against the FIR emission, the \mbox{$I_{\rm [NII]205}$/$I_{\rm FIR}$} ratio globally decreases for increasing $I_{\rm FIR}$ but the anticorrelation is only apparent in the region around the main star-forming cores (with \mbox{$\rho$\,=\,$-0.79$}, \mbox{Table~\ref{tab:corrfir}}). As we have seen before, the \NII\,205\,$\upmu$m emission is correlated with the 24\,$\upmu$m emission dominated by hot VSGs (see Table~\ref{tab:corrfir}). Hence, while we do not see the same behavior of the \mbox{$I_{\rm [NII]205}$/$I_{\rm FIR}$} and  \mbox{$I\rm{_{[CI]492,809}}/\mathit{I}\rm{_{FIR}}$} intensities ratios, both magnitudes always show the lowest values toward the massive cores of  embedded high-mass star formation (very large column densities of dust).

\subsection{Sgr B2 as a template for SB and (U)LIRG emission}\label{subsec:extragtemplate}
The spectrum of Sgr~B2 main cores has been previously compared to that of prototypical nearby galaxies such as Arp~220, M82, and NGC~253 \citep[e.g.,][]{Gonzalez-Alfonso2004,Goicoechea_2004,Martin_2006,Etxaluze13,Kamenetzky14}.  Arp~220 is a merger and a SB template of very dusty ULIRGs. The SB galaxy M82 is considered as an "extragalactic PDR" \citep{Martin2009}.  NGC~253 is an isolated spiral galaxy (in an earlier stage of evolution than M82) in which widespread shocks seem to dominate a significant fraction of the  molecular gas heating \citep{Perez-Beau2018}. NGC\,1068  hosts a prototypical CND and an active galactic nuclei \citep[AGN, e.g.,][]{Aladro15}. IC~342 is a \mbox{face-on} spiral galaxy with a nucleus likely similar to  our GC \citep[e.g.,][]{Rolling2016}. 
In this  section we discuss how the spatially resolved line emission from Sgr~B2 cores and envelope compares to the generally unresolved emission from these iconic galaxies.

\subsubsection{Widespread and correlated SiO and \ce{N2H+} emission}\label{subsubsec:SiO-N2Hp}

In Sgr B2, the \mbox{SiO (2$-$1)} and \mbox{\ce{N2H+} (1$-$0)} emission is extended at parsec scales, and is moderately correlated ($\rho$\,=\,0.7). As the SiO abundance is specifically enhanced by  sputtering of silicate grains, bright SiO emission often traces shocked molecular gas \citep{MartinPintado1992,Mikami1992,Bergin_2007,Jimenez-Serra2010}. In addition, observations of  GC clouds show a correlation between the SiO emission and the intensity of the 6.4\,keV Fe$^0$ line \citep{MartinPintado2000}. The origin of the Fe K$\alpha$ line may be related to the presence of strong shocks, but also to an enhanced flux of LECRs \citep{Valinia2000}.
The first  extragalactic detection of SiO and \ce{N2H+} were toward the nuclei of the nearby spiral galaxies NGC~253, M82, Maffei~2, IC~342 and in NGC~6946 only \ce{N2H+} \citep{Mauersberger91}. These authors already  suggested that SiO and \ce{N2H+} could coexist in the warm molecular gas of NGC~253 nucleus.

%---------------------------------------------------------------------------------------
\begin{table}[t]
    \caption{Line intensity ratios in Sgr~B2 and other sources.}
    %\centering
    \hspace{-0.3cm}
    \begin{tabular}{lccc @{\vrule height 8pt depth 3pt width 0pt}}
    \hline\hline
        %\hspace{1mm}
        Source & $\frac{I(\rm{N_2H^+)}}{\textit{I}(\rm{SiO})}$ & $\frac{I(\rm{HCO^+})}{\textit{I}(\rm{N_2H^+})}$ & $\frac{I(\rm{HCO^+})}{\textit{I}(\rm{HCN})}$ \\
        %\hspace{1mm}  \\
        \hline
         Sgr B2 entire area & \phantom{0}2.1 & \phantom{0}1.9 & 0.6\\
         Sgr B2 envelope & \phantom{0}2.0 & \phantom{0}1.9 & 0.6 \\
         Sgr B2 main cores & \phantom{0}2.4 & \phantom{0}1.6 & 0.6 \\
         \hline
         Arp 220$^a$ (ULIRG SB) & \phantom{0}1.8 & \phantom{0}2\phantom{.0} & 0.4\\
         NGC 253$^a$ (SB) & \phantom{0}2.2 & \phantom{0}5.4  & 0.9\\
         NGC 1068$^a$ (AGN+SB) & \phantom{0}3.3 & \phantom{0}7.2 & 0.6 \\
%         M83$^a$ (SB) & \phantom{0}4.9 & \phantom{0}8\phantom{.0} & 0.9 \\
         M82$^a$ (SB) & \phantom{0}5.5 & 20.5 & 1.4\\
         IC 342$^b$ & \phantom{0}4\phantom{.0} & \phantom{0}3\phantom{.0} & 0.7 \\

        \hline

         CMZ$^c$ & \phantom{0}1.9 & \phantom{0}4\phantom{.0} & 0.6\\
         Sgr A$^c$ & \phantom{0}2.8 & \phantom{0}3.4 & 0.6 \\
         Sgr C$^c$ & \phantom{0}4\phantom{.0} & \phantom{0}5\phantom{.0} & 0.6 \\
         Orion~B$^d$ & $>$10 & 24\phantom{.0} & 1.1 \\
         \hline

          \\Reference positions in Sgr B2:  & & \\\hline
          Sgr B2 A  v$\rm{_{LSR}}$\,=\,$[-20, 120]$ \kms& \phantom{0}2.1 & \phantom{0}1.9 & 0.6 \\
          Sgr B2 B & \phantom{0}2.2 & \phantom{0}1.8 & 0.7 \\
          Sgr B2 C & \phantom{0}3.6 & \phantom{0}0.7 & 0.6 \\
          Sgr B2 D & \phantom{0}1.2 & \phantom{0}2.2 & 0.6 \\
          G0.6--0.0 & \phantom{0}3.0 & \phantom{0}1.6 & 0.6 \\
          Sgr B2(DS) & \phantom{0}5.0 & \phantom{0}1.5 & 0.6 \\
          G0.66--0.13 & \phantom{0}3.6 & \phantom{0}0.5 & 0.6 \\  \hline
          
          Sgr B2 v$\rm{_{LSR}}=[15, 50]$ \kms & \phantom{0}2.1 & \phantom{0}1.1 & 0.6 \\
          Sgr B2, position A  & \phantom{0}2.2 & \phantom{0}1.5 & 0.6 \\
          Sgr B2, position B & \phantom{0}2.2 & \phantom{0}0.7 & 0.6 \\
          Sgr B2, position C & \phantom{0}4.0 & \phantom{0}0.3 & 0.5 \\
          Sgr B2, position D & \phantom{0}1.4 & \phantom{0}1.7 & 0.6 \\
          G0.6--0.0 & \phantom{0}4.5 & \phantom{0}0.9 & 0.4 \\
          Sgr B2(DS) & \phantom{0}4.8 & \phantom{0}0.7 & 0.5 \\
          G0.66--0.13 & \phantom{0}3.2 & \phantom{0}0.3 & 0.5 \\ \hline
          
          Sgr B2 v$\rm{_{LSR}}$\,=\,$[50, 85]$ \kms & \phantom{0}2.2 & \phantom{0}2.1 & 0.7 \\
          Sgr B2, position A  & \phantom{0}2.3 & \phantom{0}1.4 & 0.6 \\
          Sgr B2, position B & \phantom{0}2.1 & \phantom{0}2.4 & 0.7 \\
          Sgr B2, position C & \phantom{0}3.2 & \phantom{0}0.8 & 0.6 \\
          Sgr B2, position D & \phantom{0}1.0 & \phantom{0}4.3 & 0.6 \\
          G0.6--0.0 & \phantom{0}3.4 & \phantom{0}1.2 & 0.6 \\
          Sgr B2(DS) & \phantom{0}3.2 & \phantom{0}1.5 & 0.7 \\
          G0.66--0.13 & \phantom{0}2.0 & \phantom{0}2.7 & 0.6 \\
         \hline
    \end{tabular}
    \vspace{0.1cm}
    \tablefoot{First three rows refer to whole mapped area, the envelope, and the main star-forming cores (as in Table~\ref{table:MLB2}). The second part of the table refers to the whole mapped area and toward  positions A, B, C, and D, in the velocity ranges [15, 50]\,\kms, and [50, 85]\,\kms. $^a$\cite{Aladro15}. $^b$\cite{Takano19}. $^c$\cite{Jones2012}.
    $^d$\cite{Pety17}   \label{tab:ratios30mSgrb2}}
    \vspace{-0.2cm}
\end{table}
%---------------------------------------------------------------------------------------

If we assume a roughly uniform ionization rate in Sgr~B2, our maps show that the \mbox{\ce{N2H+}\,(1$-$0)\,/\,SiO\,(2$-$1)} intensity ratio decreases in regions where we suspect that the heating of molecular gas  is driven or at least affected by shocks. Table~\ref{tab:ratios30mSgrb2} lists the \mbox{\ce{N2H+}\,(1$-$0)\,/\,SiO\,(2$-$1)} intensity toward  different environments of Sgr~B2 and in other representative clouds of the GC compared to their observed values in prototypical nearby galaxies.  In Sgr~B2, the \mbox{\ce{N2H+}\,(1$-$0)\,/\,SiO\,(2$-$1)} intensity ratio is relatively uniform \mbox{($\sim$\,2$-$4)}, and only toward the reference position D the ratio decreases to $\sim$\,1. This region is also a maximum in the \mbox{$I$(CO)\,/\,$I_{\rm{FIR}}$} intensity ratio \mbox{($\simeq$\,2\,$\cdot$\,10$^{-3}$)}, the emission lines  show broad and intricate profiles, and \mbox{$C$-type} shock models are able to reproduce the observed \mbox{mid-$J$ CO} SLED.
 
On the other hand, the highest \mbox{\ce{N2H+}\,(1$-$0)\,/\,SiO\,(2$-$1)} intensity ratios  ($\sim$\,4.5$-$5) are observed toward \mbox{G0.6--0.0} and Sgr~B2(DS), where the FIR emission is strong (equivalent to an upper limit $G_{0}$ value of several~10$^4$). We suspect that the intermediate ratios between  2 and 3  imply the concerted contribution of both radiative and mechanical heating. The \mbox{\ce{N2H+}(1$-$0)\,/\,SiO\,(2$-$1)}  intensity ratios observed toward the prototypical galaxies listed in  Table~\ref{tab:ratios30mSgrb2}  imply that  shocks are globally important in  ULIRG Arp~220, whereas SB M82 is dominated by widespread \mbox{FUV radiation}. NGC~253 and NGC~1068 show intermediate values, comparable with regions of the Sgr~B2 where UV radiation and widespread shocks drive the heating of the extended molecular gas.

\subsubsection{\ce{HCO+}/\,\ce{N2H+} and \ce{HCO+}/\,\ce{HCN} intensity ratios}

Our chemical models predict that the \ce{HCO+}/\,\ce{N2H+} abundance ratio decreases for increasing CR ionization rates $\zeta_{\rm{CR}}$ (Fig.~\ref{fig:PDR_ion}). The average value of the \mbox{\ce{HCO+}/\,\ce{N2H+}~(1$-$0)} integrated line intensity ratio in Sgr~B2 is indeed low $\sim$\,2, much lower than that observed at (comparable) large spatial scales in disk GMCs  \citep[e.g., $\sim$\,24 in Orion~B,][]{Pety17}. The third column in \mbox{Table~\ref{tab:ratios30mSgrb2}} shows that the \ce{HCO+}/\,\ce{N2H+}~(1$-$0) intensity ratio observed  in the CMZ is in the range 3$-$5 \citep{Jones2012}. These values are slightly higher than those found in Sgr~B2 which either indicates that the ionization rate is higher in Sgr~B2 or that the opacity of the 
\mbox{\ce{HCO+}\,(1$-$0)} line is higher because of the particularly large column densities of molecular material in Sgr~B2. We checked that the \mbox{H$^{13}$CO$^+$/\,\ce{N2H+} (1$-$0)} line ratio multiplied by the GC isotopic ratio $^{12}$C\,/\,$^{13}$C\,$=$\,30 follows the same behavior and has a typical value of $\sim$\,5 in the Sgr~B2 envelope. Adopting a ratio of $\sim$\,2 to 5 for a GMC of enhanced $\zeta_{\rm{CR}}$, the \mbox{\ce{HCO+}/\,\ce{N2H+} (1$-$0)} line intensity ratios observed in Arp~220, NGC~253, and IC~342 are closer to those of Sgr~B2 \citep{Aladro15,Takano19}, while those of NGC~1068  are slightly higher. The starburst M82 shows a much higher \mbox{\ce{HCO+}/\,\ce{N2H+}~(1$-$0)}  intensity ratio, closer to the values observed in disk GMCs such as Orion~B, and thus are compatible with molecular clouds of lower CR ionization rates.

Turning to the 	\mbox{\ce{HCO+}/\,HCN~(1$-$0)} intensity ratio, the average value in Sgr~B2 is 0.6, similar to the intensity ratio observed in many locations of the CMZ \citep{Jones2012} but smaller than the 1.1  average value in Orion~B \citep[over large spatial scales;][]{Pety17} or at much smaller spatial scales in prototypical   PDRs such as the Orion Bar  ($\sim$\,1.3; \mbox{S. Cuadrado}, priv.comm.).  As in PDRs, pure SB galaxies such as M82 show \mbox{\ce{HCO+}/\,HCN~(1$-$0)} line intensity ratios \mbox{above 1}. This is a probe of the dominant role of FUV radiation in the gas heating and chemistry of these regions. ULIRGs hosting larger amounts of dense molecular gas, enhanced star-formation rates, and AGNs, often show \mbox{\ce{HCO+}/\,HCN~(1$-$0)} intensity ratios lower \mbox{than 1} \citep[0.6 in NGC\,1068 and 0.4 in Arp\,220;][]{Aladro15}, thus similar to the observed values in Sgr~B2. We conclude that in terms of the  \ce{HCO+}/\,\ce{N2H+} and \ce{HCO+}/\,\ce{HCN} line intensity ratios, Sgr~B2 is closer to the values observed in ULIRG Arp~220.

\subsubsection{Submm emission lines in Sgr~B2 and in nearby galaxies}

The relative abundances of certain molecular species \mbox{(SiO, \ce{N2H+}, ...)} shed light on the presence of  UV radiation, CRs, and shocks (if one understands their chemistry). 
In addition, FIR/submm fine-structure lines and \mbox{mid-$J$ CO} lines directly measure the cooling of the warm neutral gas in star-forming regions. Compared to \mbox{low-$J$ CO} lines, \mbox{mid-$J$ CO} lines have much higher $n_{\rm cr}$ and their excitation require higher gas thermal pressures. The detection of  extended \mbox{mid-$J$ CO} emission in Sgr~B2 reveals the presence of a widespread component of high thermal pressure molecular gas far from the massive star-forming cores. The luminosity emitted by the \mbox{mid-$J$ CO} lines in the extended envelope of Sgr~B2 is nearly a factor of ten higher than that emitted from the cores (\mbox{Table~\ref{table:MLB2}}). This suggests that the large-scale cloud environment can dominate much of the extragalactic submm line emission.

 \cite{Kamenetzky14} studied the complete CO rotational ladder  toward a sample of nearby galaxies,
including the prototypical galaxies Arp\,220, NGC\,253, and M82.
Using their published CO SLEDs we obtain the following  excitation temperatures \mbox{$T_{\rm ex}$(mid-$J$ CO)}: 
69\,$\pm$\,2\,K,  66\,$\pm$\,4\,K, and 64\,$\pm$\,4\,K, respectively. These values are  similar
to those of Sgr~B2 envelope (Fig.~\ref{fig:Texrotdrtot}c) but lower than
 toward the massive star-forming cores 
\citep[\mbox{$T_{\rm ex}$\,$\gtrsim$\,150\,K};][]{Etxaluze13}. \cite{Kamenetzky14} fitted their extragalactic CO line fluxes with two gas components (``cold'' and ``warm''). The gas thermal pressure  of their warm component, \mbox{$P_{\rm{ th}}/k$\,=\,$n_{\rm{H}}\,T_{\rm K}$}, is \mbox{$\simeq$\,1.6\,$\cdot$\,10$^7$\,K\,cm$^{-3}$} in Arp\,220, 
\mbox{1.3\,$\cdot$\,10$^7$\,K\,cm$^{-3}$} in NGC\,253, and \mbox{4.0\,$\cdot$\,10$^6$\,K\,cm$^{-3}$} in M82. The pressures inferred in Arp\,220 and NGC\,253  are similar to the ones we infer  in the extended envelope of Sgr~B2 from the  \mbox{mid-$J$ CO} lines.  In addition, the thermal pressure of their cold gas component (dominated by the \mbox{lowest-$J$ CO} emission) is several \mbox{10$^5$\,K\,cm$^{-3}$}, very much alike that we derive from the extended  \CI-emitting  component in Sgr~B2. As a summary, \mbox{Table~\ref{tab:ratiosSPIRESgrb}} compiles the line ratios \mbox{CO 10$-$9\,/\,4$-$3} (a proxy of high-pressure molecular gas), \mbox{\NII\,205\,$\upmu$m\,/\,\CI\,809\,GHz} (that seems to measure the contribution of ionizing EUV photons), and \mbox{\CI\,809\,GHz\,/\,CO 7$-$6} in Sgr~B2 and in other galactic sources. Again, Sgr~B2 shows more similarities with Arp\,220 and NGC\,253 in these line ratios, and only the \mbox{\CI\,809\,GHz\,/\,CO (7$-$6)} is markedly different in NGC\,1068 (\mbox{AGN\,$+$\,SB}).

%---------------------------------------------------------------------------------
\begin{table}[t]
    \caption{Molecular line intensity ratios in Sgr~B2 and in other sources.}
    \hspace{-0.3cm}
    \resizebox{0.5\textwidth}{!}{
    \begin{tabular}{lccc @{\vrule height 8pt depth 3pt width 0pt}}
    \hline\hline

        Source & $\frac{\rm{CO\,(10-9)}}{\rm{CO\,(4-3)}}$ & $\frac{[\rm{C}\,{\textsc i}]\,809\,\rm{GHz}}{\rm{CO\,(7-6)}}$ & $\frac{[\rm{N}\,{\textsc{ii}}]\,205\,\upmu\rm{m}}{[\rm{C}\,{\textsc i}]\,809\,\rm{GHz}}$ \\

        \hline
         Sgr B2 entire area & $0.7\pm0.3$ & $0.5\pm0.2$ & $1.7\pm0.7$ \\
         Sgr B2 envelope & $0.6\pm0.2$ & $0.5\pm0.2$ & $1.7\pm0.7$ \\
         Sgr B2 main cores & $3\pm1$ & $0.4\pm0.2$ & $1.3\pm0.5$ \\

         \hline
         Arp 220$^a$ (ULIRG SB) & \phantom{0}1.3$^b$ & 0.5 & 2.2\\
         NGC 253$^a$ (SB) & 1.4 & 0.7  & 1.7\\
         NGC 1068$^a$ (AGN+SB) & 1.3 & 1.5 & 6.4 \\
         M82$^a$ (SB) & 1.2 & 0.6 & 5.0\\
         IC 342$^c$ & 0.3 & 0.5 & 1.3 \\

        \hline
         Sgr\,A$^*$/CND$\,^{d}$ & 3.5 & 0.4 & 4 \\

         \hline

          \\Reference positions in Sgr B2:  & & \\\hline
          Sgr B2, position A & $0.6\pm0.2$ & $0.5\pm0.2$ & $0.8\pm0.3$ \\
          Sgr B2, position B & $0.8\pm0.3$ & $0.4\pm0.1$ & $1.4\pm0.6$ \\
          Sgr B2, position C & $1.0\pm0.4$ & $0.4\pm0.2$ & $2\pm1$ \\
          Sgr B2, position D & $0.8\pm0.3$ & $0.4\pm0.1$ & $1.0\pm0.4$ \\
         \hline
    \end{tabular}
    }
    \tablefoot{$^a$\cite{Kamenetzky14}. $^b$\cite{Rangwala2011}. $^c$\cite{Rigopoulou2013} $^d$\cite{Goicoechea2013}.  \label{tab:ratiosSPIRESgrb} }
    \vspace{-0.4cm}
\end{table}
%---------------------------------------------------------------------------------

\subsection{Discussion of specific regions in Sgr~B2 complex}

We conclude our discussion by zooming again to specific regions of the Sgr~B2 complex
that show peculiar conditions.

\vspace{-0.2cm}
\subsubsection{G0.6$-$0.0: \HII~regions in a clumpy medium} \label{subsubsec:clumpymed}

The southern and eastern regions of the envelope show bright \NII\,205\,$\upmu$m emission  far from the  massive star-forming cores (Fig.~\ref{fig:SPIRE}d).
Radio-continuum observations  previously revealed  the presence of
a diffuse ionized gas component \citep{Mehringer1992,Mehringer1993} and of FIR fine-structure lines at low-angular resolution \citep{Goicoechea_2004}. \citet{Mehringer1992} observed the H110$\alpha$ radio recombination line emission and studied the kinematics of the ionized gas in the Sgr~B region. The H110$\alpha$ line peaks at v$_{\rm{LSR}}\,\simeq$\,55\,\kms\,\,
toward G0.6$-$0.0, which is very similar to one of the emission peaks of the SiO~(2$-$1) and \ce{N2H+}~(1$-$0) lines in the same line of sight (\mbox{v$_{\rm{LSR}}\simeq$\,51\,\kms}). In addition, the emission from H radio recombination lines  toward \HII~regions such as \mbox{Sgr~B2(V, R, AA)} peak at analogous  velocities than at least two of the molecular lines (sometimes \ce{N2H+} and SiO, sometimes HCN and \ce{HCO+}). Therefore, the ionized and the surrounding molecular gas seem related. The similar velocities in the mapped area suggest that most of the emission sources are physically associated. The origin of the large-scale ionizing radiation is likely the large population of young OB stars in Sgr~B2 main cores plus a minor contribution from the less numerous massive stars in the envelope \citep[e.g., at least four late-type O stars in G0.6$-$0.0;][]{Mehringer1992}. 
Hence, their natal molecular  cores need to be inhomogeneous and clumpy enough so that radiation escapes, diffuses, and  illuminates several cloud surfaces along the line of sight \mbox{\citep[e.g.,][]{Tauber90,Goicoechea_2004}}. This would lead to dilute ionizing radiation at large scales and it is a very plausible scenario to  explain the low ionization parameter $U$  and the low $L_{\rm FIR}$/$M_{\rm gas}$  ratios inferred in Sgr~B2 envelope.   

\vspace{-0.2cm}
\subsubsection{Sgr B2(DS): Peculiar \HII\,region, $\gamma$-rays, and electrons}

The region Sgr~B2(DS) is located east of position C \citep[e.g.,][]{Ginsburg2018}. Our CO and \CI~maps (Figs.~\ref{fig:SPIRE} and \ref{fig:30mtotal}) reveal an emission cavity. However, the \NII\,205\,$\upmu$m map (Fig.~\ref{fig:SPIRE}d) shows an extended secondary emission peak in this region, as well as in the 24~$\upmu$m dust emission (Fig.~\ref{fig:DUST}a). In this particular region, the dust is warmer ($T_{\rm{d}}$>24\,K)  than the average $T_{\rm d}$ in the  envelope, and the 24\,$\upmu$m\,/\,70\,$\upmu$m ratio peaks at \mbox{$R_{\%}$\,$\simeq$\,8\,\%} 
(\mbox{Fig.~\ref{fig:DUST}d}) suggesting enhanced column densities of ionized gas. In addition, $L_{\rm{FIR}}$\,/\,$M_{\rm{gas}}$ is higher than 5\,$L_\odot$/$M_\odot$, which implies that the flux of ionizing photons is higher  than the average in the envelope.

The \textit{Fermi} LAT catalog  \citep{Fermi2012} shows a \mbox{$\gamma$-ray} emission source, \mbox{2FGL~J1747.3-2825c}, associated to Sgr~B2 \citep{YusefZadeh2013}. This source is located in Sgr~B2(DS) (see e.g., Fig.~\ref{fig:SPIRE}a). Recent observations  with the VLA  reveal an extended \HII\,region that displays a combination of thermal and nonthermal radio-continuum emission \citep{Meng2019}. The morphology of the thermal emission  is clumpy and concentrated, whereas the nonthermal emission is extended and diffuse.  Relativistic electrons can be the origin of the nonthermal component \citep{Padovani2019}.
Hence, Sgr~B2(DS) may be an example of local CR accelerator, having higher CR fluxes
than other regions in the envelope, and perhaps altering the C$^+$, C, and CO abundances \citep[see models by e.g.,][]{Bisbas2017}. We conclude that perhaps only in this region,  accelerated electrons and high-energy radiation produced by relativistic electrons  \citep[\mbox{$\gamma$-} and \mbox{X-rays}; e.g.,][]{Padovani2020} contribute to heat the dust, ionize the gas,  and clear the molecular gas around Sgr~B2(DS).

\subsubsection{G0.66--0.13: CRs, X-rays, and shocks}\label{subsubsec:XDR_06-013}

The spatial distribution of the \mbox{SiO (2$-$1)} and \mbox{\ce{N2H+} (1$-$0)} emission between 15 and 50 \kms~reveals a very bright extended region in the southeast of Sgr~B2 (shown in \mbox{Figs.~\ref{fig:30mSiO}a and c)}. The \mbox{HCN (1$-$0)} and \mbox{\ce{HCO+} (1$-$0)} emission, however, is surprisingly fainter. This region, bright in hard X-ray continuum up to well above 10\,keV, was reported as a new cloud feature by \cite{Zhang_2015}. The \mbox{6.4 keV Fe K$\alpha$} line flux toward this area shows  fast variability, that reached a maximum in 2012 and rapidly lessened within one year. These authors concluded that the best possible scenario to explain the Fe~K$\alpha$ flux variability is that \mbox{G0.66--0.13} is a X-ray reflection nebula
affected by past flaring activity from Sgr~A$^*$ \citep[see also][]{Terrier2018}. Alternatively, \cite{Zhang_2015} proposed that if the X-ray emission has actually reached a constant background level, then the emission can be explained as a result of LECRp bombardment.

The spatial distribution of the mid-$J$~CO line emission in \mbox{G0.66--0.13} follows  that of \mbox{SiO (2$-$1)}. Therefore, shocks likely contribute to compress and heat the gas. The \mbox{\ce{N2H+}/\,SiO} and \ce{HCO+}/\,\ce{N2H+} intensity ratios are 2.1 and 0.4, respectively (integrated in the $+$15 to $+$50\,\kms\;velocity range). The low \ce{HCO+}/\,\ce{N2H+} ratio suggests that the ionization rate is higher than the average rate in Sgr~B2. The spatial correlation of the \mbox{SiO (2$-$1)} with \mbox{\ce{N2H+} (1$-$0)} emission  suggests a common origin, probably related to the presence of shocks and enhanced ionization rates. While the  X-ray ionization rate seems too low  to affect the  gas chemistry \citep[$\zeta_{\rm X}$\,$\simeq$\,10$^{-19}$\,s$^{-1}$,][]{Armijos2020},  $\zeta_{\rm CR}$ is high enough \citep{Zhang_2015} to dominate the ionization in Sgr~B2.

The hard X-ray continuum and the \mbox{6.4 keV Fe K$\alpha$} line emission toward \mbox{G0.66--0.13} seem associated with the SiO~(2$-$1) emission at \mbox{15$-$25\,\kms}~velocities \citep[see Fig.~\ref{fig:30mSiO} and \ref{fig:30mHCN-HCO}, and also][]{Armijos2020}. Sources able to ultimately drive shocks and generate  (hard)  \mbox{X-ray emission} would possibly explain a common origin for the enhanced SiO and \mbox{Fe K$\alpha$ emission} \citep{MartinPintado2000}. However, even if shocks with \mbox{$\rm{v_s}$\,$\simeq$\,10$-$20\,\kms} exist in the envelope, it is unlikely that they  generate \mbox{X-rays} or UV radiation  \citep{Lehmann2020}. More generally, \cite{Tsuboi2015} suggest the presence of up to 6 shells in the region, one of them as a promising candidate for an ongoing cloud-cloud collision. This particular site, their Shell~3, overlaps with the \mbox{X-ray-emitting} region of \mbox{G0.66--0.13}. Our interpretation is that   shocks and enhanced CR fluxes  (ultimately leading to enhanced X-ray emission) coexist in  \mbox{G0.66--0.13}. The shocked gas in Shell 3 is likely the origin of the bright \mbox{SiO (2$-$1)} and  mid-$J$~CO emission we observe (Figs.~\ref{fig:SPIRE}a and \ref{fig:30mtotal}). An alternative scenario is that the shocked molecular gas and the enhanced $\zeta_{\rm CR}$ rates are produced by nearby supernovae remnants (SNRs). %%These are all sources of stellar feedback.

\subsubsection{G0.76--0.06: A peculiar shocked region}\label{subsubsec:P}
The region around  reference position D (\mbox{G0.76--0.06}) is located about 15\,pc away from Sgr~B2(M) core. This region shows bright submm- and mm-wave emission lines (see \mbox{Figs.~\ref{fig:SPIRE} and \ref{fig:30mtotal})}. Most observational tracers point to shocks driven by the intricate gas dynamics in this region. The \mbox{SiO~(2$-$1)}, \mbox{\ce{N2H+}}, \mbox{HCN}, and \mbox{\ce{HCO+}~(1$-$0)} line-profiles toward position D show more than three different velocity components in the LSR velocity range between $-50$\,\kms\; and $100$\,\kms. The derived column density of warm~CO toward position D, \mbox{(2.0$\pm$0.4)\,$\cdot$\,$10^{17}$\,cm$^{-2}$}, is the largest in the map. In addition, the \mbox{mid-$J$~CO} line intensities can be satisfactorily reproduced by \mbox{$C$-type} shock models (\mbox{Table~\ref{tab:rmsmodels}}). 

In this region the gas heating is likely driven by shocks and we see that large column densities of warm molecular gas coexists with colder dust \mbox{($T_{\rm{d}}$\,$\simeq$\,20$-$25\,K;} Fig.~\ref{fig:Td_Nh2_LFIR}a). Indeed, the $I_{\rm{CO}}$\,/\,$I_{\rm{FIR}}$ luminosity ratio attains its maximum value,  $\gtrsim$\,10$^{-3}$. Such high ratios are similar to those observed in protostellar outflows \mbox{\citep[e.g.,][]{Goicoechea12,Goicopacs15}}. Furthermore, the detection of very bright \mbox{SiO (2$-$1)} emission supports the presence of shocks in \mbox{G0.76--0.06}. This region is also a secondary peak in HNCO \citep{Henshaw16}, a chemical species that is particularly  abundant  in  shocks \mbox{\citep[e.g.,][]{Rodriguez-Fernandez2010}}.

Given the bright mid-$J$~CO and \mbox{SiO (2$-$1)} emission and broad line profiles  \citep[see also][]{Miyazaki2000}, one plausible scenario is that gas located in  the $x_1$ and $x_2$ orbits collides  in this area producing multiple shocks \citep{Binney1991}. 
An alternative  scenario invokes that \mbox{G0.76--0.06} falls close to the rims of two overlapping expanding shells, likely produced by SN relics (Shells~1 and 2 in Fig.~\ref{fig:30mtotal}), and the rim of Shell~3 \citep{Tsuboi2015}. 
%In either case, we conclude that the molecular gas heating in this region of Sgr~B2 envelope is dominated by the accumulation of multiple shocks.
 Higher angular resolution observations will  determine whether these shocks  trigger star formation.

%%%%%%%%%%%%%%%%%%%%%%%%%%%%%%%%%%
\section{Summary and conclusions} \label{sect:conclusions}
%%%%%%%%%%%%%%%%%%%%%%%%%%%%%%%%%%

We presented 168\,arcmin$^2$ ($\sim$1000\,pc$^2$)  spectral-images of the GC starburst region Sgr~B2 taken with \mbox{\textit{Herschel}/SPIRE-FTS}. We detected  widespread and \mbox{extended} emission from  $^{12}$CO and $^{13}$CO rotational ladders (\mbox{$J$\,=\,4$-$3} to \mbox{$J$\,=\,12$-$11} lines), \mbox{$p$-H$_2$O $2_{1,1}-2_{0,2}$}, \mbox{\CI\,492,\,809\,GHz}, and \mbox{\NII\,205\,$\upmu$m} lines far from the main star-forming cores. We also presented \mbox{velocity-resolved} maps of the \mbox{SiO\,(2$-$1)}, \ce{N2H+}, \ce{HCO+}, and \mbox{HCN\,(1$-$0)} emission taken  with the IRAM\,30m telescope. We  analyzed  this dataset in tandem with \textit{Spitzer} and \textit{Herschel}  images of the dust emission as well as \textit{NuSTAR} observations of the hard X-ray continuum. Rather than focusing on the massive star-forming cores, we studied the more extended Sgr~B2 envelope (the cloud \mbox{environment}). \mbox{Although} line fluxes at any specific position in the envelope are smaller than in the  cores, the larger area of the envelope means that emission lines integrated over the entire region will be dominated by the envelope and not by the cores. As a result, it is the extended environment, rather than the star-forming cores themselves, that may serve as a useful template when making comparisons to other galaxies where the telescope beams will encompass such extended  regions. We summarize our results  as follows:

--~Stellar EUV and FUV photons propagate and permeate the Sgr~B2 complex.
Owing to the clumpiness of the cloud that surrounds the main star-forming cores Sgr~B2(N and M), a fraction of the intense radiation field emitted by young massive stars  escapes to the cloud environment. The observed FIR luminosities are
equivalent to an upper limit of  $G_0$\,$\approx$10$^3$ far from these cores. The presence of less prominent local \HII~regions (more young massive stars) throughout  the envelope, especially in the southern region and \mbox{G0.6--0.0}, adds more UV radiation to the region. The detection of very extended \NII\,205\,$\upmu$m emission, tracing a  widespread component of diffuse ionized gas of low ionization  parameter $U$ \mbox{(i.e., diluted EUV}  radiation emitted far from the ionized gas) agrees with this scenario. The inferred  low ionization  parameters, $U$\,$\simeq$\,10$^{-3}$, are consistent with the observed low  \mbox{$L_{\rm FIR}$\,/\,$M_{\rm gas}$\,$\simeq$\,4$-$11\,\Ls\,\Ms$^{-1}$} luminosity-to-mass ratios, a proxy of $U$, and with the cold dust temperatures  in the region,  with an average of \mbox{$T_{\rm d}$\,$\simeq$\,23\,K}. The extended \NII\,205\,$\upmu$m  emission correlates with the 24\,$\upmu$m emission from hot VSGs (predominantly arising from dust in the ionized gas) and also with the 70\,$\upmu$m emission, arising from bigger grains in the surrounding neutral PDRs.

--~Spectral-images of the \mbox{mid-$J$~CO} lines (up to \mbox{$J$\,=\,12$-$11}) reveal the presence of a widespread component of  molecular gas at high thermal-pressure   
(\mbox{$P_{\rm th}$/$k\,\simeq$\,10$^7$\,K\,cm$^{-3}$}). The observed \mbox{mid-$J$~CO} SLEDs can  be explained  by PDR models if the emission arises from  clumps or thin  ($\Delta A_{\rm{V}}$\,$\sim$\,2\,mag) sheets of dense molecular gas ($n_{\rm{H}}$\,$\simeq$\,10$^6$\,cm$^{-3}$). PDR models with lower gas densities fail to reproduce the observed CO line intensities beyond \mbox{$J_{\rm u}$\,$\geq$\,7}. The origin of this thin component of \mbox{FUV-illuminated} dense  gas can be compressed gas layers  at the interfaces of colliding clouds and expanding shells \citep[e.g.,][]{Fukui2020}, or be  the result of over-pressurized photo-evaporating PDR cloud edges \citep[triggering  gas compression, e.g.,][]{Hosokawa06,Goico16,Bron18}.
The compression of molecular gas can also be produced by shocks triggered by  expanding bubbles associated with SNRs or winds from massive stars  in the region  \citep[e.g.,][]{Dwarkadas2011,Pabst20}. The presence of bubbles (i.e., of stellar \mbox{feedback}) and  cloud-cloud collisions  has been invoked in the past  \cite[e.g.,][]{MartinPintado1999,Tsuboi2015}. The correlation between the CO~(7$-$6) and SiO~(2$-$1) emission suggests that shocks play a role in compressing the gas at many positions.

--~The \mbox{\CI\,492,\,809\,GHz}  emission reveals a different component of very extended but lower density gas (\mbox{$n_{\rm{H}}$\,$\simeq$\,10$^3$\,cm$^{-3}$}).
These lines stem from large column densities of neutral gas at lower
 pressures  ($P_{\rm th}$/$k$\,$\simeq$\,10$^5$\,K\,cm$^{-3}$) than those of the
\mbox{mid-$J$~CO} emitting gas. The \CI\;intensities can be explained by chemical models with an enhanced  $\zeta_{\rm CR}$ of
\mbox{(2$-$6)\,$\cdot$\,$10^{-15}$\,s$^{-1}$}, a factor of $\simeq$\,10$-$100 times higher than in disk GMCs. Cosmic rays dominate the heating 
of this  gas component at \mbox{$T_{\rm k}$\,$\simeq$\,40$-$60\,K}
and produce significantly larger column densities of C than of C$^+$ and CO.

--~The enhanced cosmic-ray ionization rate in the dense molecular gas 
of  Sgr~B2 leads to enhanced column densities of \ce{N2H+} (through more abundant H$_{3}^{+}$, \ce{He+}, and N$_2^+$). This leads to extended and widespread  \mbox{\ce{N2H+}} emission (i.e., not only confined to cold and dense cores and filaments as in disk GMCs). Observationally, the higher ionization fraction of GC clouds is likely responsible of the  much lower \mbox{\ce{HCO+}/\,\ce{N2H+}\,(1$-$0)} line intensity ratios in Sgr~B2 ($\sim$\,2) compared to GMCs of the Galactic disk \mbox{\citep[$\sim$\,25 in Orion~B;][]{Pety17}}
or  pure starburst galaxies \citep[$\sim$\,20 in M82;][]{Aladro15}. In addition, we find that
 the spatial distribution of the \ce{HCO+} and \mbox{HCN (1$-$0)} emission in Sgr~B2  is strongly correlated and shows \mbox{\ce{HCO+}/\,HCN~(1$-$0)\,$<$\,1}  intensity ratios. This ratio differs from those measured in prototypical PDRs such as the Orion Bar, in Orion~B at large spatial scales \citep{Pety17}, or in M82, all showing \mbox{\ce{HCO+}/\,HCN~(1$-$0)\,$>$\,1}. Therefore, in addition to the role of widespread stellar FUV radiation, Sgr~B2 is affected by other mechanisms, notably higher cosmic-ray ionization rates and shocks, that contribute to the gas heating and produce its characteristic chemistry. The spatial correlation of the  \mbox{\ce{N2H+} (1$-$0)} and  \mbox{SiO (2$-$1)} emission implies that  both high ionization rates and shocks coexist
 at many positions.

--~Our  spectral-images unveil the complexity and more extreme conditions of the ISM of galactic nuclei compared to disk GMCs. Most, but not all, high-mass star formation  takes place in three massive cores that radiate $\sim$\,20\,\%~of the  FIR luminosity of the mapped region. However, the extended  cloud environment  dominates the  \mbox{H$_2$O\,752\,GHz ($\sim$\,60\,\%)}, \mbox{mid-$J$\,CO ($\sim$\,91\,\%)}, \mbox{\CI~($\sim$\,93\,\%)}, and \mbox{\NII\,205\,$\upmu$m~($\sim$\,95\,\%)} luminosities.

--~The \mbox{\ce{N2H+}/\,SiO},  \mbox{\ce{HCO+}/\,\ce{N2H+}}, and  \mbox{\ce{HCO+}/\,HCN} line intensity ratios, the CO SLEDs, and the submm spectra of Sgr~B2 envelope resemble more those observed toward the ULIRG galaxy Arp\,220, a massive burst of embedded star formation triggered by the collision of two smaller galaxies, and possibly hosting thousands of embedded Sgr~B2-like starbursts \citep[][]{Goicoechea_2004,Gonzalez-Alfonso2004,Rangwala2011}. However, pure starburst galaxies such as M82 show markedly different observational characteristics. A major difference is the much higher \mbox{$L_{\rm FIR}$\,/\,$M_{\rm H_2}$\,$>$\,80\,$L_{\odot}$\,$M_{\odot}^{-1}$} values in Arp~220 and in other galaxies hosting vigorous star formation \citep{Carpio11}. In Sgr~B2, \mbox{$L_{\rm FIR}$\,/\,$M_{\rm H_2}$} and $T_{\rm d}^{6}$ are strongly correlated (see Table~\ref{tab:corrfir}), but grains heated by dilute \mbox{stellar} UV photons are relatively cold at large scales. However, ULIRGs produce larger quantities of warmer dust because of the less diluted stellar radiation (more stars per molecular mass) and perhaps also due to non-PDR dust emission \mbox{(e.g., AGN activity)}. Extrapolating the observational scaling we derive  in Sgr~B2 (Table~\ref{tab:corrfir}), we  obtain  $T_{\rm d}$\,$\simeq$\,40\,K for \mbox{$L_{\rm FIR}$\,/\,$M_{\rm H_2}$\,$\simeq$\,80\,$L_{\odot}$\,$M_{\odot}^{-1}$}. These are precisely the values inferred in the extended component of Arp~220 \citep{Dwek2020} and thus suggest similar physical processes at place. Similarly to the C$^+$/\,FIR deficit observed in local galaxies hosting very active star-formation modes, we find a clear trend and a  pronounced decrease of \mbox{\CI\,/\,FIR} with increasing FIR intensity, with the lowest values seen toward the main massive star-forming cores.
The observed \mbox{\NII\,205\,$\upmu$m\,/\,FIR} ratio globally decreases with the FIR intensity too. This suggests that different FIR fine-structure lines tracing different ISM phases can show deficits. In Sgr~B2 these are mostly dominated by an excess of FIR dust emission toward the embedded star-forming cores.

%%OUTLOOK 
Mapping the key FIR and submm gas cooling lines of the ISM (\NII, \OI, \CII, \CI, and \mbox{mid-$J$ CO} lines) over larger spatial  scales: in the plane of the Milky Way and in entire nearby galaxies, is clearly needed to constrain the global properties of the ISM phases, and to study star-formation and  feedback. These are  fundamental ingredients to understand the evolution of galaxies.

\begin{acknowledgements}
This paper is dedicated to the memory of Bruce Swinyard, for his mentorship and his major contributions to the SPIRE-FTS performance and calibration. We are grateful to our  anonymous referee for constructive comments and very useful suggestions that improved the clarity of this paper.
We warmly thank 
S.~Zhang for sharing her \textit{NuSTAR} continuum X-ray mosaic image of Sgr~B2. We also thank J.~P.~Fonfria for his help setting up our models in different computers.
We thank the Spanish MICIU for funding support under grants AYA2017-85111-P
and PID2019-106110GB-I00. 
\end{acknowledgements}


\bibliographystyle{aa}
\bibliography{references}

\begin{thebibliography}{154}
\expandafter\ifx\csname natexlab\endcsname\relax\def\natexlab#1{#1}\fi

\bibitem[{{Abel} {et~al.}(2009){Abel}, {Dudley}, {Fischer}, {Satyapal}, \& {van
  Hoof}}]{Abel09}
{Abel}, N.~P., {Dudley}, C., {Fischer}, J., {Satyapal}, S., \& {van Hoof},
  P.~A.~M. 2009, \apj, 701, 1147

\bibitem[{{Aladro} {et~al.}(2015){Aladro}, {Mart{\'{\i}}n}, {Riquelme},
  {Henkel}, {Mauersberger}, {Mart{\'{\i}}n-Pintado}, {Wei{\ss}}, {Lefevre},
  {Kramer}, {Requena-Torres}, \& {Armijos-Abenda{\~n}o}}]{Aladro15}
{Aladro}, R., {Mart{\'{\i}}n}, S., {Riquelme}, D., {et~al.} 2015, \aap, 579,
  A101

\bibitem[{{Arab} {et~al.}(2012){Arab}, {Abergel}, {Habart}, {Bernard-Salas},
  {Ayasso}, {Dassas}, {Martin}, \& {White}}]{Arab_2012}
{Arab}, H., {Abergel}, A., {Habart}, E., {et~al.} 2012, \aap, 541, A19

\bibitem[{{Arendt} {et~al.}(2019){Arendt}, {Staguhn}, {Dwek}, {Morris},
  {Yusef-Zadeh}, {Benford}, {Kov{\'a}cs}, \& {Gonzalez-Quiles}}]{Arendt2019}
{Arendt}, R.~G., {Staguhn}, J., {Dwek}, E., {et~al.} 2019, \apj, 885, 71

\bibitem[{{Armijos-Abenda{\~n}o}
  {et~al.}(2020{\natexlab{a}}){Armijos-Abenda{\~n}o}, {Banda-Barrag{\'a}n},
  {Mart{\'\i}n-Pintado}, {D{\'e}nes}, {Federrath}, \&
  {Requena-Torres}}]{Armijos2020b}
{Armijos-Abenda{\~n}o}, J., {Banda-Barrag{\'a}n}, W.~E., {Mart{\'\i}n-Pintado},
  J., {et~al.} 2020{\natexlab{a}}, \mnras

\bibitem[{{Armijos-Abenda{\~n}o}
  {et~al.}(2020{\natexlab{b}}){Armijos-Abenda{\~n}o}, {Mart{\'\i}n-Pintado},
  {L{\'o}pez}, {Llerena}, {Harada}, {Requena-Torres}, {Mart{\'\i}n}, {Rivilla},
  {Riquelme}, \& {Aldas}}]{Armijos2020}
{Armijos-Abenda{\~n}o}, J., {Mart{\'\i}n-Pintado}, J., {L{\'o}pez}, E.,
  {et~al.} 2020{\natexlab{b}}, \apj, 895, 57

\bibitem[{{Benson} \& {Johnston}(1984)}]{Benson1984}
{Benson}, J.~M. \& {Johnston}, K.~J. 1984, \apj, 277, 181

\bibitem[{{Bergin} \& {Tafalla}(2007)}]{Bergin_2007}
{Bergin}, E.~A. \& {Tafalla}, M. 2007, \araa, 45, 339

\bibitem[{{Bernard} {et~al.}(2010){Bernard}, {Paradis}, {Marshall}, {Montier},
  {Lagache}, {Paladini}, {Veneziani}, {Brunt}, {Mottram}, {Martin},
  {Ristorcelli}, {Noriega-Crespo}, {Compi{\`e}gne}, {Flagey}, {Anderson},
  {Popescu}, {Tuffs}, {Reach}, {White}, {Benedettini}, {Calzoletti},
  {Digiorgio}, {Faustini}, {Juvela}, {Joblin}, {Joncas}, {Mivilles-Deschenes},
  {Olmi}, {Traficante}, {Piacentini}, {Zavagno}, \& {Molinari}}]{Bernard10}
{Bernard}, J.~P., {Paradis}, D., {Marshall}, D.~J., {et~al.} 2010, \aap, 518,
  L88

\bibitem[{{Binney} {et~al.}(1991){Binney}, {Gerhard}, {Stark}, {Bally}, \&
  {Uchida}}]{Binney1991}
{Binney}, J., {Gerhard}, O.~E., {Stark}, A.~A., {Bally}, J., \& {Uchida}, K.~I.
  1991, \mnras, 252, 210

\bibitem[{{Bisbas} {et~al.}(2017){Bisbas}, {van Dishoeck}, {Papadopoulos},
  {Sz{\H{u}}cs}, {Bialy}, \& {Zhang}}]{Bisbas2017}
{Bisbas}, T.~G., {van Dishoeck}, E.~F., {Papadopoulos}, P.~P., {et~al.} 2017,
  \apj, 839, 90

\bibitem[{{Bohlin} {et~al.}(1978){Bohlin}, {Savage}, \& {Drake}}]{Bohlin78}
{Bohlin}, R.~C., {Savage}, B.~D., \& {Drake}, J.~F. 1978, \apj, 224, 132

\bibitem[{{Bron} {et~al.}(2018){Bron}, {Ag{\'u}ndez}, {Goicoechea}, \&
  {Cernicharo}}]{Bron18}
{Bron}, E., {Ag{\'u}ndez}, M., {Goicoechea}, J.~R., \& {Cernicharo}, J. 2018,
  ArXiv e-prints

\bibitem[{{Carey} {et~al.}(2009){Carey}, {Noriega-Crespo}, {Mizuno}, {Shenoy},
  {Paladini}, {Kraemer}, {Price}, {Flagey}, {Ryan}, {Ingalls}, {Kuchar},
  {Pinheiro Gon{\c{c}}alves}, {Indebetouw}, {Billot}, {Marleau}, {Padgett},
  {Rebull}, {Bressert}, {Ali}, {Molinari}, {Martin}, {Berriman}, {Boulanger},
  {Latter}, {Miville-Deschenes}, {Shipman}, \& {Testi}}]{Carey09}
{Carey}, S.~J., {Noriega-Crespo}, A., {Mizuno}, D.~R., {et~al.} 2009, \pasp,
  121, 76

\bibitem[{{Caselli} {et~al.}(2002){Caselli}, {Benson}, {Myers}, \&
  {Tafalla}}]{Caselli2002}
{Caselli}, P., {Benson}, P.~J., {Myers}, P.~C., \& {Tafalla}, M. 2002, \apj,
  572, 238

\bibitem[{{Ceccarelli} {et~al.}(2014){Ceccarelli}, {Dominik},
  {L{\'o}pez-Sepulcre}, {Kama}, {Padovani}, {Caux}, \&
  {Caselli}}]{Ceccarelli2014}
{Ceccarelli}, C., {Dominik}, C., {L{\'o}pez-Sepulcre}, A., {et~al.} 2014,
  \apjl, 790, L1

\bibitem[{{Cernicharo} {et~al.}(2006){Cernicharo}, {Goicoechea}, {Pardo}, \&
  {Asensio-Ramos}}]{Cernicharo_2006}
{Cernicharo}, J., {Goicoechea}, J.~R., {Pardo}, J.~R., \& {Asensio-Ramos}, A.
  2006, \apj, 642, 940

\bibitem[{{Clark} {et~al.}(2018){Clark}, {Lohr}, {Patrick}, {Najarro}, {Dong},
  \& {Figer}}]{Clark2018}
{Clark}, J.~S., {Lohr}, M.~E., {Patrick}, L.~R., {et~al.} 2018, \aap, 618, A2

\bibitem[{{de Vicente} {et~al.}(1997){de Vicente}, {Martin-Pintado}, \&
  {Wilson}}]{deVicente97}
{de Vicente}, P., {Martin-Pintado}, J., \& {Wilson}, T.~L. 1997, \aap, 320, 957

\bibitem[{{Dent} {et~al.}(2009){Dent}, {Hovey}, {Dewdney}, {Burgess}, {Willis},
  {Lightfoot}, {Jenness}, {Leech}, {Matthews}, {Heyer}, \& {Poulton}}]{Dent09}
{Dent}, W.~R.~F., {Hovey}, G.~J., {Dewdney}, P.~E., {et~al.} 2009, \mnras, 395,
  1805

\bibitem[{{Desert} {et~al.}(1990){Desert}, {Boulanger}, \&
  {Puget}}]{Desert1990}
{Desert}, F.~X., {Boulanger}, F., \& {Puget}, J.~L. 1990, \aap, 500, 313

\bibitem[{{Dowell} {et~al.}(1999){Dowell}, {Lis}, {Serabyn}, {Gardner},
  {Kovacs}, \& {Yamashita}}]{Dowell1999}
{Dowell}, C.~D., {Lis}, D.~C., {Serabyn}, E., {et~al.} 1999, in Astronomical
  Society of the Pacific Conference Series, Vol. 186, The Central Parsecs of
  the Galaxy, ed. H.~{Falcke}, A.~{Cotera}, W.~J. {Duschl}, F.~{Melia}, \&
  M.~J. {Rieke}, 453

\bibitem[{{Dwarkadas}(2011)}]{Dwarkadas2011}
{Dwarkadas}, V.~V. 2011, \memsai, 82, 781

\bibitem[{{Dwek} \& {Arendt}(2020)}]{Dwek2020}
{Dwek}, E. \& {Arendt}, R.~G. 2020, \apj, 901, 36

\bibitem[{{Etxaluze} {et~al.}(2013){Etxaluze}, {Goicoechea}, {Cernicharo},
  {Polehampton}, {Noriega-Crespo}, {Molinari}, {Swinyard}, {Wu}, \&
  {Bally}}]{Etxaluze13}
{Etxaluze}, M., {Goicoechea}, J.~R., {Cernicharo}, J., {et~al.} 2013, \aap,
  556, A137

\bibitem[{{Fischer} {et~al.}(2010){Fischer}, {Sturm}, {Gonz{\'a}lez-Alfonso},
  {Graci{\'a}-Carpio}, {Hailey-Dunsheath}, {Poglitsch}, {Contursi}, {Lutz},
  {Genzel}, {Sternberg}, {Verma}, \& {Tacconi}}]{Fischer2010}
{Fischer}, J., {Sturm}, E., {Gonz{\'a}lez-Alfonso}, E., {et~al.} 2010, \aap,
  518, L41

\bibitem[{{Flower} \& {Pineau Des For{\^e}ts}(2010)}]{Flower2010}
{Flower}, D.~R. \& {Pineau Des For{\^e}ts}, G. 2010, \mnras, 406, 1745

\bibitem[{{Fukui} {et~al.}(2020){Fukui}, {Habe}, {Inoue}, {Enokiya}, \&
  {Tachihara}}]{Fukui2020}
{Fukui}, Y., {Habe}, A., {Inoue}, T., {Enokiya}, R., \& {Tachihara}, K. 2020,
  arXiv e-prints, arXiv:2009.05077

\bibitem[{{Gahm} {et~al.}(2013){Gahm}, {Persson}, {M{\"a}kel{\"a}}, \&
  {Haikala}}]{Gahm13}
{Gahm}, G.~F., {Persson}, C.~M., {M{\"a}kel{\"a}}, M.~M., \& {Haikala}, L.~K.
  2013, \aap, 555, A57

\bibitem[{{Gaume} {et~al.}(1995){Gaume}, {Claussen}, {de Pree}, {Goss}, \&
  {Mehringer}}]{Gaume1995}
{Gaume}, R.~A., {Claussen}, M.~J., {de Pree}, C.~G., {Goss}, W.~M., \&
  {Mehringer}, D.~M. 1995, \apj, 449, 663

\bibitem[{{Gerin} {et~al.}(2016){Gerin}, {Neufeld}, \& {Goicoechea}}]{Gerin16}
{Gerin}, M., {Neufeld}, D.~A., \& {Goicoechea}, J.~R. 2016, \araa, 54, 181

\bibitem[{{Ginsburg} {et~al.}(2018){Ginsburg}, {Bally}, {Barnes}, {Bastian},
  {Battersby}, {Beuther}, {Brogan}, {Contreras}, {Corby}, {Darling}, {De Pree},
  {Galv{\'a}n-Madrid}, {Garay}, {Henshaw}, {Hunter}, {Kruijssen}, {Longmore},
  {Lu}, {Meng}, {Mills}, {Ott}, {Pineda}, {S{\'a}nchez-Monge}, {Schilke},
  {Schmiedeke}, {Walker}, \& {Wilner}}]{Ginsburg2018}
{Ginsburg}, A., {Bally}, J., {Barnes}, A., {et~al.} 2018, \apj, 853, 171

\bibitem[{{Ginsburg} \& {Kruijssen}(2018)}]{Ginsburg2018b}
{Ginsburg}, A. \& {Kruijssen}, J.~M.~D. 2018, \apjl, 864, L17

\bibitem[{{Goicoechea} {et~al.}(2012){Goicoechea}, {Cernicharo}, {Karska},
  {Herczeg}, {Polehampton}, {Wampfler}, {Kristensen}, {van Dishoeck},
  {Etxaluze}, {Bern{\'e}}, \& {Visser}}]{Goicoechea12}
{Goicoechea}, J.~R., {Cernicharo}, J., {Karska}, A., {et~al.} 2012, \aap, 548,
  A77

\bibitem[{{Goicoechea} {et~al.}(2015{\natexlab{a}}){Goicoechea},
  {Chavarr{\'{\i}}a}, {Cernicharo}, {Neufeld}, {Vavrek}, {Bergin}, {Cuadrado},
  {Encrenaz}, {Etxaluze}, {Melnick}, \& {Polehampton}}]{Goicopacs15}
{Goicoechea}, J.~R., {Chavarr{\'{\i}}a}, L., {Cernicharo}, J., {et~al.}
  2015{\natexlab{a}}, \apj, 799, 102

\bibitem[{{Goicoechea} {et~al.}(2013){Goicoechea}, {Etxaluze}, {Cernicharo},
  {Gerin}, {Neufeld}, {Contursi}, {Bell}, {De Luca}, {Encrenaz}, {Indriolo},
  {Lis}, {Polehampton}, \& {Sonnentrucker}}]{Goicoechea2013}
{Goicoechea}, J.~R., {Etxaluze}, M., {Cernicharo}, J., {et~al.} 2013, \apjl,
  769, L13

\bibitem[{{Goicoechea} {et~al.}(2020){Goicoechea}, {Pabst}, {Kabanovic},
  {Santa-Maria}, {Marcelino}, {Tielens}, {Hacar}, {Bern{\'e}}, {Buchbender},
  {Cuadrado}, {Higgins}, {Kramer}, {Stutzki}, {Suri}, {Teyssier}, \&
  {Wolfire}}]{Goico20}
{Goicoechea}, J.~R., {Pabst}, C.~H.~M., {Kabanovic}, S., {et~al.} 2020, \aap,
  639, A1

\bibitem[{{Goicoechea} {et~al.}(2016){Goicoechea}, {Pety}, {Cuadrado},
  {Cernicharo}, {Chapillon}, {Fuente}, {Gerin}, {Joblin}, {Marcelino}, \&
  {Pilleri}}]{Goico16}
{Goicoechea}, J.~R., {Pety}, J., {Cuadrado}, S., {et~al.} 2016, \nat, 537, 207

\bibitem[{{Goicoechea} {et~al.}(2004){Goicoechea},
  {Rodr{\'{\i}}guez-Fern{\'a}ndez}, \& {Cernicharo}}]{Goicoechea_2004}
{Goicoechea}, J.~R., {Rodr{\'{\i}}guez-Fern{\'a}ndez}, N.~J., \& {Cernicharo},
  J. 2004, \apj, 600, 214

\bibitem[{{Goicoechea} {et~al.}(2019){Goicoechea}, {Santa-Maria}, {Bron},
  {Teyssier}, {Marcelino}, {Cernicharo}, \& {Cuadrado}}]{Goicoechea2019}
{Goicoechea}, J.~R., {Santa-Maria}, M.~G., {Bron}, E., {et~al.} 2019, \aap,
  622, A91

\bibitem[{{Goicoechea} {et~al.}(2018){Goicoechea}, {Santa-Maria}, {Teyssier},
  {Cernicharo}, {Gerin}, \& {Pety}}]{Goicoechea2018a}
{Goicoechea}, J.~R., {Santa-Maria}, M.~G., {Teyssier}, D., {et~al.} 2018, \aap,
  616, L1

\bibitem[{{Goicoechea} {et~al.}(2015{\natexlab{b}}){Goicoechea}, {Teyssier},
  {Etxaluze}, {Goldsmith}, {Ossenkopf}, {Gerin}, {Bergin}, {Black},
  {Cernicharo}, {Cuadrado}, {Encrenaz}, {Falgarone}, {Fuente}, {Hacar}, {Lis},
  {Marcelino}, {Melnick}, {M{\"u}ller}, {Persson}, {Pety}, {R{\"o}llig},
  {Schilke}, {Simon}, {Snell}, \& {Stutzki}}]{Goi15}
{Goicoechea}, J.~R., {Teyssier}, D., {Etxaluze}, M., {et~al.}
  2015{\natexlab{b}}, \apj, 812, 75

\bibitem[{{Goldsmith} \& {Langer}(1999)}]{Goldsmith_1999}
{Goldsmith}, P.~F. \& {Langer}, W.~D. 1999, \apj, 517, 209

\bibitem[{{Goldsmith} {et~al.}(1990){Goldsmith}, {Lis}, {Hills}, \&
  {Lasenby}}]{Goldsmith90}
{Goldsmith}, P.~F., {Lis}, D.~C., {Hills}, R., \& {Lasenby}, J. 1990, \apj,
  350, 186

\bibitem[{{Goldsmith} {et~al.}(1992){Goldsmith}, {Lis}, {Lester}, \&
  {Harvey}}]{Goldsmith1992}
{Goldsmith}, P.~F., {Lis}, D.~C., {Lester}, D.~F., \& {Harvey}, P.~M. 1992,
  \apj, 389, 338

\bibitem[{{Gonz{\'a}lez-Alfonso} {et~al.}(2004){Gonz{\'a}lez-Alfonso}, {Smith},
  {Fischer}, \& {Cernicharo}}]{Gonzalez-Alfonso2004}
{Gonz{\'a}lez-Alfonso}, E., {Smith}, H.~A., {Fischer}, J., \& {Cernicharo}, J.
  2004, \apj, 613, 247

\bibitem[{{Goto} {et~al.}(2011){Goto}, {Usuda}, {Geballe}, {Indriolo},
  {McCall}, {Henning}, \& {Oka}}]{Goto11}
{Goto}, M., {Usuda}, T., {Geballe}, T.~R., {et~al.} 2011, \pasj, 63, L13

\bibitem[{{Graci{\'a}-Carpio} {et~al.}(2011){Graci{\'a}-Carpio}, {Sturm},
  {Hailey-Dunsheath}, {Fischer}, {Contursi}, {Poglitsch}, {Genzel},
  {Gonz{\'a}lez-Alfonso}, {Sternberg}, {Verma}, {Christopher}, {Davies},
  {Feuchtgruber}, {de Jong}, {Lutz}, \& {Tacconi}}]{Carpio11}
{Graci{\'a}-Carpio}, J., {Sturm}, E., {Hailey-Dunsheath}, S., {et~al.} 2011,
  \apjl, 728, L7

\bibitem[{{Gravity Collaboration} {et~al.}(2019){Gravity Collaboration},
  {Abuter}, {Amorim}, {Baub{\"o}ck}, {Berger}, {Bonnet}, {Brandner},
  {Cl{\'e}net}, {Coud{\'e} Du Foresto}, {de Zeeuw}, {Dexter}, {Duvert},
  {Eckart}, {Eisenhauer}, {F{\"o}rster Schreiber}, {Garcia}, {Gao}, {Gendron},
  {Genzel}, {Gerhard}, {Gillessen}, {Habibi}, {Haubois}, {Henning}, {Hippler},
  {Horrobin}, {Jim{\'e}nez-Rosales}, {Jocou}, {Kervella}, {Lacour},
  {Lapeyr{\`e}re}, {Le Bouquin}, {L{\'e}na}, {Ott}, {Paumard}, {Perraut},
  {Perrin}, {Pfuhl}, {Rabien}, {Rodriguez Coira}, {Rousset}, {Scheithauer},
  {Sternberg}, {Straub}, {Straubmeier}, {Sturm}, {Tacconi}, {Vincent}, {von
  Fellenberg}, {Waisberg}, {Widmann}, {Wieprecht}, {Wiezorrek}, {Woillez}, \&
  {Yazici}}]{Gravity2019}
{Gravity Collaboration}, {Abuter}, R., {Amorim}, A., {et~al.} 2019, \aap, 625,
  L10

\bibitem[{{Greaves} \& {Williams}(1994)}]{Greaves94}
{Greaves}, J.~S. \& {Williams}, P.~G. 1994, \aap, 290, 259

\bibitem[{{Griffin} {et~al.}(2010){Griffin}, {Abergel}, {Abreu}, {Ade},
  {Andr{\'e}}, {Augueres}, {Babbedge}, {Bae}, {Baillie}, {Baluteau}, {Barlow},
  {Bendo}, {Benielli}, {Bock}, {Bonhomme}, {Brisbin}, {Brockley-Blatt},
  {Caldwell}, {Cara}, {Castro-Rodriguez}, {Cerulli}, {Chanial}, {Chen},
  {Clark}, {Clements}, {Clerc}, {Coker}, {Communal}, {Conversi}, {Cox},
  {Crumb}, {Cunningham}, {Daly}, {Davis}, {de Antoni}, {Delderfield}, {Devin},
  {di Giorgio}, {Didschuns}, {Dohlen}, {Donati}, {Dowell}, {Dowell}, {Duband},
  {Dumaye}, {Emery}, {Ferlet}, {Ferrand}, {Fontignie}, {Fox}, {Franceschini},
  {Frerking}, {Fulton}, {Garcia}, {Gastaud}, {Gear}, {Glenn}, {Goizel},
  {Griffin}, {Grundy}, {Guest}, {Guillemet}, {Hargrave}, {Harwit}, {Hastings},
  {Hatziminaoglou}, {Herman}, {Hinde}, {Hristov}, {Huang}, {Imhof}, {Isaak},
  {Israelsson}, {Ivison}, {Jennings}, {Kiernan}, {King}, {Lange}, {Latter},
  {Laurent}, {Laurent}, {Leeks}, {Lellouch}, {Levenson}, {Li}, {Li},
  {Lilienthal}, {Lim}, {Liu}, {Lu}, {Madden}, {Mainetti}, {Marliani}, {McKay},
  {Mercier}, {Molinari}, {Morris}, {Moseley}, {Mulder}, {Mur}, {Naylor},
  {Nguyen}, {O'Halloran}, {Oliver}, {Olofsson}, {Olofsson}, {Orfei}, {Page},
  {Pain}, {Panuzzo}, {Papageorgiou}, {Parks}, {Parr-Burman}, {Pearce},
  {Pearson}, {P{\'e}rez-Fournon}, {Pinsard}, {Pisano}, {Podosek}, {Pohlen},
  {Polehampton}, {Pouliquen}, {Rigopoulou}, {Rizzo}, {Roseboom}, {Roussel},
  {Rowan-Robinson}, {Rownd}, {Saraceno}, {Sauvage}, {Savage}, {Savini},
  {Sawyer}, {Scharmberg}, {Schmitt}, {Schneider}, {Schulz}, {Schwartz},
  {Shafer}, {Shupe}, {Sibthorpe}, {Sidher}, {Smith}, {Smith}, {Smith},
  {Spencer}, {Stobie}, {Sudiwala}, {Sukhatme}, {Surace}, {Stevens}, {Swinyard},
  {Trichas}, {Tourette}, {Triou}, {Tseng}, {Tucker}, {Turner}, {Vaccari},
  {Valtchanov}, {Vigroux}, {Virique}, {Voellmer}, {Walker}, {Ward}, {Waskett},
  {Weilert}, {Wesson}, {White}, {Whitehouse}, {Wilson}, {Winter}, {Woodcraft},
  {Wright}, {Xu}, {Zavagno}, {Zemcov}, {Zhang}, \& {Zonca}}]{Griffin10}
{Griffin}, M.~J., {Abergel}, A., {Abreu}, A., {et~al.} 2010, \aap, 518, L3

\bibitem[{{Hacar} {et~al.}(2017){Hacar}, {Alves}, {Tafalla}, \&
  {Goicoechea}}]{Hacar17}
{Hacar}, A., {Alves}, J., {Tafalla}, M., \& {Goicoechea}, J.~R. 2017, \aap,
  602, L2

\bibitem[{{Hacar} {et~al.}(2018){Hacar}, {Tafalla}, {Forbrich}, {Alves},
  {Meingast}, {Grossschedl}, \& {Teixeira}}]{Hacar18}
{Hacar}, A., {Tafalla}, M., {Forbrich}, J., {et~al.} 2018, \aap, 610, A77

\bibitem[{{Hasegawa} {et~al.}(1994){Hasegawa}, {Sato}, {Whiteoak}, \&
  {Miyawaki}}]{Hasegawa1994}
{Hasegawa}, T., {Sato}, F., {Whiteoak}, J.~B., \& {Miyawaki}, R. 1994, \apjl,
  429, L77

\bibitem[{{Hatchfield} {et~al.}(2020){Hatchfield}, {Battersby}, {Keto},
  {Walker}, {Barnes}, {Callanan}, {Ginsburg}, {Henshaw}, {Kauffmann},
  {Kruijssen}, {Longmore}, {Lu}, {Mills}, {Pillai}, {Zhang}, {Bally},
  {Butterfield}, {Contreras}, {Ho}, {Ott}, {Patel}, \&
  {Tolls}}]{Hatchfield2020}
{Hatchfield}, H.~P., {Battersby}, C., {Keto}, E., {et~al.} 2020, arXiv
  e-prints, arXiv:2009.05052

\bibitem[{{Henshaw} {et~al.}(2016){Henshaw}, {Longmore}, {Kruijssen}, {Davies},
  {Bally}, {Barnes}, {Battersby}, {Burton}, {Cunningham}, {Dale}, {Ginsburg},
  {Immer}, {Jones}, {Kendrew}, {Mills}, {Molinari}, {Moore}, {Ott}, {Pillai},
  {Rathborne}, {Schilke}, {Schmiedeke}, {Testi}, {Walker}, {Walsh}, \&
  {Zhang}}]{Henshaw16}
{Henshaw}, J.~D., {Longmore}, S.~N., {Kruijssen}, J.~M.~D., {et~al.} 2016,
  \mnras, 457, 2675

\bibitem[{{Herbst} \& {Klemperer}(1973)}]{Herbst1973}
{Herbst}, E. \& {Klemperer}, W. 1973, \apj, 185, 505

\bibitem[{{Herrera-Camus} {et~al.}(2018){Herrera-Camus}, {Sturm},
  {Graci{\'a}-Carpio}, {Lutz}, {Contursi}, {Veilleux}, {Fischer},
  {Gonz{\'a}lez-Alfonso}, {Poglitsch}, {Tacconi}, {Genzel}, {Maiolino},
  {Sternberg}, {Davies}, \& {Verma}}]{Herrera-Camus18}
{Herrera-Camus}, R., {Sturm}, E., {Graci{\'a}-Carpio}, J., {et~al.} 2018, \apj,
  861, 95

\bibitem[{{HESS Collaboration} {et~al.}(2016){HESS Collaboration},
  {Abramowski}, {Aharonian}, {Benkhali}, {Akhperjanian}, {Ang{\"u}ner},
  {Backes}, {Balzer}, {Becherini}, {Tjus}, {Berge}, {Bernhard}, {Bernl{\"o}hr},
  {Birsin}, {Blackwell}, {B{\"o}ttcher}, {Boisson}, {Bolmont}, {Bordas},
  {Bregeon}, {Brun}, {Brun}, {Bryan}, {Bulik}, {Carr}, {Casanova},
  {Chakraborty}, {Chalme-Calvet}, {Chaves}, {Chen}, {Chr{\'e}tien},
  {Colafrancesco}, {Cologna}, {Conrad}, {Couturier}, {Cui}, {Davids},
  {Degrange}, {Deil}, {Dewilt}, {Djannati-Ata{\"\i}}, {Domainko}, {Donath},
  {Drury}, {Dubus}, {Dutson}, {Dyks}, {Dyrda}, {Edwards}, {Egberts}, {Eger},
  {Ernenwein}, {Espigat}, {Farnier}, {Fegan}, {Feinstein}, {Fernandes},
  {Fernandez}, {Fiasson}, {Fontaine}, {F{\"o}rster}, {F{\"u}{\ss}ling},
  {Gabici}, {Gajdus}, {Gallant}, {Garrigoux}, {Giavitto}, {Giebels},
  {Glicenstein}, {Gottschall}, {Goyal}, {Grondin}, {Grudzi{\'n}ska}, {Hadasch},
  {H{\"a}ffner}, {Hahn}, {Hawkes}, {Heinzelmann}, {Henri}, {Hermann}, {Hervet},
  {Hillert}, {Hinton}, {Hofmann}, {Hofverberg}, {Hoischen}, {Holler}, {Horns},
  {Ivascenko}, {Jacholkowska}, {Jamrozy}, {Janiak}, {Jankowsky},
  {Jung-Richardt}, {Kastendieck}, {Katarzy{\'n}ski}, {Katz}, {Kerszberg},
  {Kh{\'e}lifi}, {Kieffer}, {Klepser}, {Klochkov}, {Klu{\'z}niak}, {Kolitzus},
  {Komin}, {Kosack}, {Krakau}, {Krayzel}, {Kr{\"u}ger}, {Laffon}, {Lamanna},
  {Lau}, {Lefaucheur}, {Lefranc}, {Lemi{\'e}re}, {Lemoine-Goumard}, {Lenain},
  {Lohse}, {Lopatin}, {Lu}, {Lui}, {Marandon}, {Marcowith}, {Mariaud}, {Marx},
  {Maurin}, {Maxted}, {Mayer}, {Meintjes}, {Menzler}, {Meyer}, {Mitchell},
  {Moderski}, {Mohamed}, {Mor{\r{a}}}, {Moulin}, {Murach}, {de Naurois},
  {Niemiec}, {Oakes}, {Odaka}, {{\"O}ttl}, {Ohm}, {Opitz}, {Ostrowski}, {Oya},
  {Panter}, {Parsons}, {Arribas}, {Pekeur}, {Pelletier}, {Petrucci}, {Peyaud},
  {Pita}, {Poon}, {Prokoph}, {P{\"u}hlhofer}, {Punch}, {Quirrenbach}, {Raab},
  {Reichardt}, {Reimer}, {Reimer}, {Renaud}, {de Los Reyes}, {Rieger},
  {Romoli}, {Rosier-Lees}, {Rowell}, {Rudak}, {Rulten}, {Sahakian}, {Salek},
  {Sanchez}, {Santangelo}, {Sasaki}, {Schlickeiser}, {Sch{\"u}ssler}, {Schulz},
  {Schwanke}, {Schwemmer}, {Seyffert}, {Simoni}, {Sol}, {Spanier}, {Spengler},
  {Spies}, {Stawarz}, {Steenkamp}, {Stegmann}, {Stinzing}, {Stycz}, {Sushch},
  {Tavernet}, {Tavernier}, {Taylor}, {Terrier}, {Tluczykont}, {Trichard},
  {Tuffs}, {Valerius}, {van der Walt}, {van Eldik}, {van Soelen},
  {Vasileiadis}, {Veh}, {Venter}, {Viana}, {Vincent}, {Vink}, {Voisin},
  {V{\"o}lk}, {Vuillaume}, {Wagner}, {Wagner}, {Wagner}, {Weidinger},
  {Weitzel}, {White}, {Wierzcholska}, {Willmann}, {W{\"o}rnlein}, {Wouters},
  {Yang}, {Zabalza}, {Zaborov}, {Zacharias}, {Zdziarski}, {Zech}, {Zefi}, \&
  {{\.Z}ywucka}}]{Hess16}
{HESS Collaboration}, {Abramowski}, A., {Aharonian}, F., {et~al.} 2016, \nat,
  531, 476

\bibitem[{{Hildebrand}(1983)}]{Hilde83}
{Hildebrand}, R.~H. 1983, \qjras, 24, 267

\bibitem[{{Hirashita} {et~al.}(2007){Hirashita}, {Hibi}, \&
  {Shibai}}]{Hirashita2007}
{Hirashita}, H., {Hibi}, Y., \& {Shibai}, H. 2007, \mnras, 379, 974

\bibitem[{{Hollenbach} \& {McKee}(1989)}]{Hollenbach1989}
{Hollenbach}, D. \& {McKee}, C.~F. 1989, \apj, 342, 306

\bibitem[{{Hollenbach} \& {Tielens}(1999)}]{Hollenbach_1999}
{Hollenbach}, D.~J. \& {Tielens}, A.~G.~G.~M. 1999, Rev. Mod. Phys., 71, 173

\bibitem[{{Hosokawa} \& {Inutsuka}(2006)}]{Hosokawa06}
{Hosokawa}, T. \& {Inutsuka}, S.-i. 2006, \apj, 646, 240

\bibitem[{{Hüttemeister} {et~al.}(1993){Hüttemeister}, {Wilson}, {Henkel}, \&
  {Mauersberger}}]{Huttemeister1993}
{Hüttemeister}, S., {Wilson}, T.~L., {Henkel}, C., \& {Mauersberger}, R. 1993,
  \aap, 276, 445

\bibitem[{{Hüttemeister} {et~al.}(1995){Hüttemeister}, {Wilson},
  {Mauersberger}, {Lemme}, {Dahmen}, \& {Henkel}}]{Huttemeister1995}
{Hüttemeister}, S., {Wilson}, T.~L., {Mauersberger}, R., {et~al.} 1995, \aap,
  294, 667

\bibitem[{{Indriolo} {et~al.}(2017){Indriolo}, {Bergin}, {Goicoechea},
  {Cernicharo}, {Gerin}, {Gusdorf}, {Lis}, \& {Schilke}}]{Indriolo17}
{Indriolo}, N., {Bergin}, E.~A., {Goicoechea}, J.~R., {et~al.} 2017, \apj, 836,
  117

\bibitem[{{Indriolo} \& {McCall}(2012)}]{Indriolo2012}
{Indriolo}, N. \& {McCall}, B.~J. 2012, \apj, 745, 91

\bibitem[{{Indriolo} {et~al.}(2015){Indriolo}, {Neufeld}, {Gerin}, {Schilke},
  {Benz}, {Winkel}, {Menten}, {Chambers}, {Black}, {Bruderer}, {Falgarone},
  {Godard}, {Goicoechea}, {Gupta}, {Lis}, {Ossenkopf}, {Persson},
  {Sonnentrucker}, {van der Tak}, {van Dishoeck}, {Wolfire}, \&
  {Wyrowski}}]{Indriolo15}
{Indriolo}, N., {Neufeld}, D.~A., {Gerin}, M., {et~al.} 2015, \apj, 800, 40

\bibitem[{{Jim{\'e}nez-Serra} {et~al.}(2010){Jim{\'e}nez-Serra}, {Caselli},
  {Tan}, {Hernand ez}, {Fontani}, {Butler}, \& {van Loo}}]{Jimenez-Serra2010}
{Jim{\'e}nez-Serra}, I., {Caselli}, P., {Tan}, J.~C., {et~al.} 2010, \mnras,
  406, 187

\bibitem[{{Joblin} {et~al.}(2018){Joblin}, {Bron}, {Pinto}, {Pilleri}, {Le
  Petit}, {Gerin}, {Le Bourlot}, {Fuente}, {Berne}, {Goicoechea}, {Habart},
  {K{\"o}hler}, {Teyssier}, {Nagy}, {Montillaud}, {Vastel}, {Cernicharo},
  {R{\"o}llig}, {Ossenkopf-Okada}, \& {Bergin}}]{Joblin18}
{Joblin}, C., {Bron}, E., {Pinto}, C., {et~al.} 2018, \aap, 615, A129

\bibitem[{{Johnson} {et~al.}(1987){Johnson}, {Burke}, \&
  {Kingston}}]{Johnson87}
{Johnson}, C.~T., {Burke}, P.~G., \& {Kingston}, A.~E. 1987, Journal of Physics
  B Atomic Molecular Physics, 20, 2553

\bibitem[{{Jones} {et~al.}(2012){Jones}, {Burton}, {Cunningham},
  {Requena-Torres}, {Menten}, {Schilke}, {Belloche}, {Leurini},
  {Mart{\'\i}n-Pintado}, {Ott}, \& {Walsh}}]{Jones2012}
{Jones}, P.~A., {Burton}, M.~G., {Cunningham}, M.~R., {et~al.} 2012, \mnras,
  419, 2961

\bibitem[{{Kamenetzky} {et~al.}(2014){Kamenetzky}, {Rangwala}, {Glenn},
  {Maloney}, \& {Conley}}]{Kamenetzky14}
{Kamenetzky}, J., {Rangwala}, N., {Glenn}, J., {Maloney}, P.~R., \& {Conley},
  A. 2014, \apj, 795, 174

\bibitem[{{Kauffmann} {et~al.}(2008){Kauffmann}, {Bertoldi}, {Bourke}, {Evans},
  \& {Lee}}]{Kaufmfmann2008}
{Kauffmann}, J., {Bertoldi}, F., {Bourke}, T.~L., {Evans}, N.~J., I., \& {Lee},
  C.~W. 2008, \aap, 487, 993

\bibitem[{{Kirk} {et~al.}(2013){Kirk}, {Myers}, {Bourke}, {Gutermuth},
  {Hedden}, \& {Wilson}}]{Kirk2013}
{Kirk}, H., {Myers}, P.~C., {Bourke}, T.~L., {et~al.} 2013, \apj, 766, 115

\bibitem[{{Koyama} {et~al.}(1996){Koyama}, {Maeda}, {Sonobe}, {Takeshima},
  {Tanaka}, \& {Yamauchi}}]{Koyama1996}
{Koyama}, K., {Maeda}, Y., {Sonobe}, T., {et~al.} 1996, \pasj, 48, 249

\bibitem[{{Langer} {et~al.}(2015){Langer}, {Goldsmith}, {Pineda}, {Velusamy},
  {Requena-Torres}, \& {Wiesemeyer}}]{Langer2015}
{Langer}, W.~D., {Goldsmith}, P.~F., {Pineda}, J.~L., {et~al.} 2015, \aap, 576,
  A1

\bibitem[{{Langer} \& {Penzias}(1990)}]{Langer_1990}
{Langer}, W.~D. \& {Penzias}, A.~A. 1990, \apj, 357, 477

\bibitem[{{Le Bourlot} {et~al.}(1993){Le Bourlot}, {Pineau des Forets},
  {Roueff}, \& {Schilke}}]{LeBourlot93}
{Le Bourlot}, J., {Pineau des Forets}, G., {Roueff}, E., \& {Schilke}, P. 1993,
  \apjl, 416, L87

\bibitem[{{Le Petit} {et~al.}(2006){Le Petit}, {Nehm{\'e}}, {Le Bourlot}, \&
  {Roueff}}]{LePetit_2006}
{Le Petit}, F., {Nehm{\'e}}, C., {Le Bourlot}, J., \& {Roueff}, E. 2006, \apjs,
  164, 506

\bibitem[{{Le Petit} {et~al.}(2016){Le Petit}, {Ruaud}, {Bron}, {Godard},
  {Roueff}, {Languignon}, \& {Le Bourlot}}]{LePetit16}
{Le Petit}, F., {Ruaud}, M., {Bron}, E., {et~al.} 2016, \aap, 585, A105

\bibitem[{{Lee} {et~al.}(2019){Lee}, {Madden}, {Le Petit}, {Gusdorf},
  {Lesaffre}, {Wu}, {Lebouteiller}, {Galliano}, \& {Chevance}}]{Lee2019}
{Lee}, M.~Y., {Madden}, S.~C., {Le Petit}, F., {et~al.} 2019, \aap, 628, A113

\bibitem[{{Lehmann} {et~al.}(2020){Lehmann}, {Godard}, {Pineau des For{\^e}ts},
  \& {Falgarone}}]{Lehmann2020}
{Lehmann}, A., {Godard}, B., {Pineau des For{\^e}ts}, G., \& {Falgarone}, E.
  2020, arXiv e-prints, arXiv:2010.01042

\bibitem[{{Li} \& {Draine}(2001)}]{Li_2001}
{Li}, A. \& {Draine}, B.~T. 2001, \apj, 554, 778

\bibitem[{{Lis} \& {Goldsmith}(1990)}]{Lis1990}
{Lis}, D.~C. \& {Goldsmith}, P.~F. 1990, \apj, 356, 195

\bibitem[{{Luhman} {et~al.}(1998){Luhman}, {Satyapal}, {Fischer}, {Wolfire},
  {Cox}, {Lord}, {Smith}, {Stacey}, \& {Unger}}]{Luhman98}
{Luhman}, M.~L., {Satyapal}, S., {Fischer}, J., {et~al.} 1998, \apjl, 504, L11

\bibitem[{{Maeda} {et~al.}(2002){Maeda}, {Baganoff}, {Feigelson}, {Morris},
  {Bautz}, {Brandt}, {Burrows}, {Doty}, {Garmire}, {Pravdo}, {Ricker}, \&
  {Townsley}}]{Maeda2002}
{Maeda}, Y., {Baganoff}, F.~K., {Feigelson}, E.~D., {et~al.} 2002, \apj, 570,
  671

\bibitem[{{Malhotra} {et~al.}(2001){Malhotra}, {Kaufman}, {Hollenbach},
  {Helou}, {Rubin}, {Brauher}, {Dale}, {Lu}, {Lord}, {Stacey}, {Contursi},
  {Hunter}, \& {Dinerstein}}]{Malhotra01}
{Malhotra}, S., {Kaufman}, M.~J., {Hollenbach}, D., {et~al.} 2001, \apj, 561,
  766

\bibitem[{{Martin} \& {Downes}(1972)}]{Martin1972}
{Martin}, A.~H.~M. \& {Downes}, D. 1972, \aplett, 11, 219

\bibitem[{{Mart{\'\i}n} {et~al.}(2009){Mart{\'\i}n}, {Mart{\'\i}n-Pintado}, \&
  {Viti}}]{Martin2009}
{Mart{\'\i}n}, S., {Mart{\'\i}n-Pintado}, J., \& {Viti}, S. 2009, \apj, 706,
  1323

\bibitem[{{Mart{\'{\i}}n} {et~al.}(2006){Mart{\'{\i}}n}, {Mauersberger},
  {Mart{\'{\i}}n-Pintado}, {Henkel}, \& {Garc{\'{\i}}a-Burillo}}]{Martin_2006}
{Mart{\'{\i}}n}, S., {Mauersberger}, R., {Mart{\'{\i}}n-Pintado}, J., {Henkel},
  C., \& {Garc{\'{\i}}a-Burillo}, S. 2006, \apjs, 164, 450

\bibitem[{{Martin-Pintado} {et~al.}(1992){Martin-Pintado}, {Bachiller}, \&
  {Fuente}}]{MartinPintado1992}
{Martin-Pintado}, J., {Bachiller}, R., \& {Fuente}, A. 1992, \aap, 254, 315

\bibitem[{{Mart{\'\i}n-Pintado} {et~al.}(2000){Mart{\'\i}n-Pintado}, {de
  Vicente}, {Rodr{\'\i}guez-Fern{\'a}ndez}, {Fuente}, \&
  {Planesas}}]{MartinPintado2000}
{Mart{\'\i}n-Pintado}, J., {de Vicente}, P., {Rodr{\'\i}guez-Fern{\'a}ndez},
  N.~J., {Fuente}, A., \& {Planesas}, P. 2000, \aap, 356, L5

\bibitem[{{Mart{\'\i}n-Pintado} {et~al.}(1999){Mart{\'\i}n-Pintado}, {Gaume},
  {Rodr{\'\i}guez-Fern{\'a}ndez}, {de Vicente}, \&
  {Wilson}}]{MartinPintado1999}
{Mart{\'\i}n-Pintado}, J., {Gaume}, R.~A., {Rodr{\'\i}guez-Fern{\'a}ndez}, N.,
  {de Vicente}, P., \& {Wilson}, T.~L. 1999, \apj, 519, 667

\bibitem[{{Mashian} {et~al.}(2015){Mashian}, {Sturm}, {Sternberg}, {Janssen},
  {Hailey-Dunsheath}, {Fischer}, {Contursi}, {Gonz{\'a}lez-Alfonso},
  {Graci{\'a}-Carpio}, {Poglitsch}, {Veilleux}, {Davies}, {Genzel}, {Lutz},
  {Tacconi}, {Verma}, {Wei{\ss}}, {Polisensky}, \& {Nikola}}]{Mashian2015}
{Mashian}, N., {Sturm}, E., {Sternberg}, A., {et~al.} 2015, \apj, 802, 81

\bibitem[{{Mauersberger} \& {Henkel}(1991)}]{Mauersberger91}
{Mauersberger}, R. \& {Henkel}, C. 1991, \aap, 245, 457

\bibitem[{{Mehringer} {et~al.}(1993){Mehringer}, {Palmer}, {Goss}, \&
  {Yusef-Zadeh}}]{Mehringer1993}
{Mehringer}, D.~M., {Palmer}, P., {Goss}, W.~M., \& {Yusef-Zadeh}, F. 1993,
  \apj, 412, 684

\bibitem[{{Mehringer} {et~al.}(1992){Mehringer}, {Yusef-Zadeh}, {Palmer}, \&
  {Goss}}]{Mehringer1992}
{Mehringer}, D.~M., {Yusef-Zadeh}, F., {Palmer}, P., \& {Goss}, W.~M. 1992,
  \apj, 401, 168

\bibitem[{{Meijerink} {et~al.}(2013){Meijerink}, {Kristensen}, {Wei{\ss}}, {van
  der Werf}, {Walter}, {Spaans}, {Loenen}, {Fischer}, {Israel}, {Isaak},
  {Papadopoulos}, {Aalto}, {Armus}, {Charmand aris}, {Dasyra}, {Diaz-Santos},
  {Evans}, {Gao}, {Gonz{\'a}lez-Alfonso}, {G{\"u}sten}, {Henkel}, {Kramer},
  {Lord}, {Mart{\'\i}n-Pintado}, {Naylor}, {Sanders}, {Smith}, {Spinoglio},
  {Stacey}, {Veilleux}, \& {Wiedner}}]{Meijerink2013}
{Meijerink}, R., {Kristensen}, L.~E., {Wei{\ss}}, A., {et~al.} 2013, \apjl,
  762, L16

\bibitem[{{Meng} {et~al.}(2019){Meng}, {S{\'a}nchez-Monge}, {Schilke},
  {Padovani}, {Marcowith}, {Ginsburg}, {Schmiedeke}, {Schw{\"o}rer}, {DePree},
  {Veena}, \& {M{\"o}ller}}]{Meng2019}
{Meng}, F., {S{\'a}nchez-Monge}, {\'A}., {Schilke}, P., {et~al.} 2019, \aap,
  630, A73

\bibitem[{{Mikami} {et~al.}(1992){Mikami}, {Umemoto}, {Yamamoto}, \&
  {Saito}}]{Mikami1992}
{Mikami}, H., {Umemoto}, T., {Yamamoto}, S., \& {Saito}, S. 1992, \apjl, 392,
  L87

\bibitem[{{Miyazaki} \& {Tsuboi}(2000)}]{Miyazaki2000}
{Miyazaki}, A. \& {Tsuboi}, M. 2000, \apj, 536, 357

\bibitem[{{Molinari} {et~al.}(2011){Molinari}, {Bally}, {Noriega-Crespo},
  {Compi{\`e}gne}, {Bernard}, {Paradis}, {Martin}, {Testi}, {Barlow}, {Moore},
  {Plume}, {Swinyard}, {Zavagno}, {Calzoletti}, {Di Giorgio}, {Elia},
  {Faustini}, {Natoli}, {Pestalozzi}, {Pezzuto}, {Piacentini}, {Polenta},
  {Polychroni}, {Schisano}, {Traficante}, {Veneziani}, {Battersby}, {Burton},
  {Carey}, {Fukui}, {Li}, {Lord}, {Morgan}, {Motte}, {Schuller},
  {Stringfellow}, {Tan}, {Thompson}, {Ward-Thompson}, {White}, \&
  {Umana}}]{Molinari11}
{Molinari}, S., {Bally}, J., {Noriega-Crespo}, A., {et~al.} 2011, \apjl, 735,
  L33

\bibitem[{{Molinari} {et~al.}(2010){Molinari}, {Swinyard}, {Bally}, {Barlow},
  {Bernard}, {Martin}, {Moore}, {Noriega-Crespo}, {Plume}, {Testi}, {Zavagno},
  {Abergel}, {Ali}, {Andr{\'e}}, {Baluteau}, {Benedettini}, {Bern{\'e}},
  {Billot}, {Blommaert}, {Bontemps}, {Boulanger}, {Brand}, {Brunt}, {Burton},
  {Campeggio}, {Carey}, {Caselli}, {Cesaroni}, {Cernicharo}, {Chakrabarti},
  {Chrysostomou}, {Codella}, {Cohen}, {Compiegne}, {Davis}, {de Bernardis}, {de
  Gasperis}, {Di Francesco}, {di Giorgio}, {Elia}, {Faustini}, {Fischera},
  {Fukui}, {Fuller}, {Ganga}, {Garcia-Lario}, {Giard}, {Giardino}, {Glenn},
  {Goldsmith}, {Griffin}, {Hoare}, {Huang}, {Jiang}, {Joblin}, {Joncas},
  {Juvela}, {Kirk}, {Lagache}, {Li}, {Lim}, {Lord}, {Lucas}, {Maiolo},
  {Marengo}, {Marshall}, {Masi}, {Massi}, {Matsuura}, {Meny}, {Minier},
  {Miville-Desch{\^e}nes}, {Montier}, {Motte}, {M{\"u}ller}, {Natoli}, {Neves},
  {Olmi}, {Paladini}, {Paradis}, {Pestalozzi}, {Pezzuto}, {Piacentini},
  {Pomar{\`e}s}, {Popescu}, {Reach}, {Richer}, {Ristorcelli}, {Roy}, {Royer},
  {Russeil}, {Saraceno}, {Sauvage}, {Schilke}, {Schneider-Bontemps},
  {Schuller}, {Schultz}, {Shepherd}, {Sibthorpe}, {Smith}, {Smith},
  {Spinoglio}, {Stamatellos}, {Strafella}, {Stringfellow}, {Sturm}, {Taylor},
  {Thompson}, {Tuffs}, {Umana}, {Valenziano}, {Vavrek}, {Viti}, {Waelkens},
  {Ward-Thompson}, {White}, {Wyrowski}, {Yorke}, \& {Zhang}}]{Molinari2010}
{Molinari}, S., {Swinyard}, B., {Bally}, J., {et~al.} 2010, \pasp, 122, 314

\bibitem[{{Morris} \& {Serabyn}(1996)}]{Morris1996}
{Morris}, M. \& {Serabyn}, E. 1996, \araa, 34, 645

\bibitem[{{Murakami} {et~al.}(2000){Murakami}, {Koyama}, {Sakano}, {Tsujimoto},
  \& {Maeda}}]{Murakami2000}
{Murakami}, H., {Koyama}, K., {Sakano}, M., {Tsujimoto}, M., \& {Maeda}, Y.
  2000, \apj, 534, 283

\bibitem[{{Neill} {et~al.}(2014){Neill}, {Bergin}, {Lis}, {Schilke},
  {Crockett}, {Favre}, {Emprechtinger}, {Comito}, {Qin}, {Anderson},
  {Burkhardt}, {Chen}, {Harris}, {Lord}, {McGuire}, {McNeill}, {Monje},
  {Phillips}, {Steber}, {Vasyunina}, \& {Yu}}]{Neill2014}
{Neill}, J.~L., {Bergin}, E.~A., {Lis}, D.~C., {et~al.} 2014, \apj, 789, 8

\bibitem[{{Neufeld} \& {Dalgarno}(1989)}]{Neufeld1989}
{Neufeld}, D.~A. \& {Dalgarno}, A. 1989, \apj, 344, 251

\bibitem[{{Nolan} {et~al.}(2012){Nolan}, {Abdo}, {Ackermann}, {Ajello},
  {Allafort}, {Antolini}, {Atwood}, {Axelsson}, {Baldini}, {Ballet},
  {Barbiellini}, {Bastieri}, {Bechtol}, {Belfiore}, {Bellazzini}, {Berenji},
  {Bignami}, {Blandford}, {Bloom}, {Bonamente}, {Bonnell}, {Borgland},
  {Bottacini}, {Bouvier}, {Brandt}, {Bregeon}, {Brigida}, {Bruel}, {Buehler},
  {Burnett}, {Buson}, {Caliandro}, {Cameron}, {Campana}, {Ca{\~n}adas},
  {Cannon}, {Caraveo}, {Casandjian}, {Cavazzuti}, {Ceccanti}, {Cecchi},
  {{\c{C}}elik}, {Charles}, {Chekhtman}, {Cheung}, {Chiang}, {Chipaux},
  {Ciprini}, {Claus}, {Cohen-Tanugi}, {Cominsky}, {Conrad}, {Corbet}, {Cutini},
  {D'Ammando}, {Davis}, {de Angelis}, {DeCesar}, {DeKlotz}, {De Luca}, {den
  Hartog}, {de Palma}, {Dermer}, {Digel}, {Silva}, {Drell}, {Drlica-Wagner},
  {Dubois}, {Dumora}, {Enoto}, {Escande}, {Fabiani}, {Falletti}, {Favuzzi},
  {Fegan}, {Ferrara}, {Focke}, {Fortin}, {Frailis}, {Fukazawa}, {Funk},
  {Fusco}, {Gargano}, {Gasparrini}, {Gehrels}, {Germani}, {Giebels},
  {Giglietto}, {Giommi}, {Giordano}, {Giroletti}, {Glanzman}, {Godfrey},
  {Grenier}, {Grondin}, {Grove}, {Guillemot}, {Guiriec}, {Gustafsson},
  {Hadasch}, {Hanabata}, {Harding}, {Hayashida}, {Hays}, {Hill}, {Horan},
  {Hou}, {Hughes}, {Iafrate}, {Itoh}, {J{\'o}hannesson}, {Johnson}, {Johnson},
  {Johnson}, {Johnson}, {Kamae}, {Katagiri}, {Kataoka}, {Katsuta}, {Kawai},
  {Kerr}, {Kn{\"o}dlseder}, {Kocevski}, {Kuss}, {Lande}, {Landriu},
  {Latronico}, {Lemoine-Goumard}, {Lionetto}, {Llena Garde}, {Longo},
  {Loparco}, {Lott}, {Lovellette}, {Lubrano}, {Madejski}, {Marelli}, {Massaro},
  {Mazziotta}, {McConville}, {McEnery}, {Mehault}, {Michelson}, {Minuti},
  {Mitthumsiri}, {Mizuno}, {Moiseev}, {Mongelli}, {Monte}, {Monzani},
  {Morselli}, {Moskalenko}, {Murgia}, {Nakamori}, {Naumann-Godo}, {Norris},
  {Nuss}, {Nymark}, {Ohno}, {Ohsugi}, {Okumura}, {Omodei}, {Orlando}, {Ormes},
  {Ozaki}, {Paneque}, {Panetta}, {Parent}, {Perkins}, {Pesce-Rollins},
  {Pierbattista}, {Pinchera}, {Piron}, {Pivato}, {Porter}, {Racusin},
  {Rain{\`o}}, {Rando}, {Razzano}, {Razzaque}, {Reimer}, {Reimer}, {Reposeur},
  {Ritz}, {Rochester}, {Romani}, {Roth}, {Rousseau}, {Ryde}, {Sadrozinski},
  {Salvetti}, {Sanchez}, {Saz Parkinson}, {Sbarra}, {Scargle}, {Schalk},
  {Sgr{\`o}}, {Shaw}, {Shrader}, {Siskind}, {Smith}, {Spandre}, {Spinelli},
  {Stephens}, {Strickman}, {Suson}, {Tajima}, {Takahashi}, {Takahashi},
  {Tanaka}, {Thayer}, {Thayer}, {Thompson}, {Tibaldo}, {Tibolla}, {Tinebra},
  {Tinivella}, {Torres}, {Tosti}, {Troja}, {Uchiyama}, {Vandenbroucke}, {Van
  Etten}, {Van Klaveren}, {Vasileiou}, {Vianello}, {Vitale}, {Waite},
  {Wallace}, {Wang}, {Werner}, {Winer}, {Wood}, {Wood}, {Wood}, {Yang}, \&
  {Zimmer}}]{Fermi2012}
{Nolan}, P.~L., {Abdo}, A.~A., {Ackermann}, M., {et~al.} 2012, \apjs, 199, 31

\bibitem[{{Ott}(2010)}]{Ott10}
{Ott}, S. 2010, in Astronomical Society of the Pacific Conference Series, Vol.
  434, Astronomical Data Analysis Software and Systems XIX, ed. Y.~{Mizumoto},
  K.-I. {Morita}, \& M.~{Ohishi}, 139

\bibitem[{{Pabst} {et~al.}(2019){Pabst}, {Higgins}, {Goicoechea}, {Teyssier},
  {Berne}, {Chambers}, {Wolfire}, {Suri}, {Guesten}, {Stutzki}, {Graf},
  {Risacher}, \& {Tielens}}]{Pabst2019}
{Pabst}, C., {Higgins}, R., {Goicoechea}, J.~R., {et~al.} 2019, \nat, 565, 618

\bibitem[{{Pabst} {et~al.}(2020){Pabst}, {Goicoechea}, {Teyssier}, {Bern{\'e}},
  {Higgins}, {Chambers}, {Kabanovic}, {G{\"u}sten}, {Stutzki}, \&
  {Tielens}}]{Pabst20}
{Pabst}, C.~H.~M., {Goicoechea}, J.~R., {Teyssier}, D., {et~al.} 2020, \aap,
  639, A2

\bibitem[{{Padovani} {et~al.}(2020){Padovani}, {Ivlev}, {Galli}, {Offner},
  {Indriolo}, {Rodgers-Lee}, {Marcowith}, {Girichidis}, {Bykov}, \&
  {Kruijssen}}]{Padovani2020}
{Padovani}, M., {Ivlev}, A.~V., {Galli}, D., {et~al.} 2020, \ssr, 216, 29

\bibitem[{{Padovani} {et~al.}(2019){Padovani}, {Marcowith},
  {S{\'a}nchez-Monge}, {Meng}, \& {Schilke}}]{Padovani2019}
{Padovani}, M., {Marcowith}, A., {S{\'a}nchez-Monge}, {\'A}., {Meng}, F., \&
  {Schilke}, P. 2019, \aap, 630, A72

\bibitem[{{P{\'e}rez-Beaupuits} {et~al.}(2018){P{\'e}rez-Beaupuits},
  {G{\"u}sten}, {Harris}, {Requena-Torres}, {Menten}, {Wei{\ss}},
  {Polehampton}, \& {van der Wiel}}]{Perez-Beau2018}
{P{\'e}rez-Beaupuits}, J.~P., {G{\"u}sten}, R., {Harris}, A., {et~al.} 2018,
  \apj, 860, 23

\bibitem[{{Pety} {et~al.}(2017){Pety}, {Guzm{\'a}n}, {Orkisz}, {Liszt},
  {Gerin}, {Bron}, {Bardeau}, {Goicoechea}, {Gratier}, {Le Petit}, {Levrier},
  {{\"O}berg}, {Roueff}, \& {Sievers}}]{Pety17}
{Pety}, J., {Guzm{\'a}n}, V.~V., {Orkisz}, J.~H., {et~al.} 2017, \aap, 599, A98

\bibitem[{{Pilbratt} {et~al.}(2010){Pilbratt}, {Riedinger}, {Passvogel},
  {Crone}, {Doyle}, {Gageur}, {Heras}, {Jewell}, {Metcalfe}, {Ott}, \&
  {Schmidt}}]{Pilbratt10}
{Pilbratt}, G.~L., {Riedinger}, J.~R., {Passvogel}, T., {et~al.} 2010, \aap,
  518, L1

\bibitem[{{Pineau des Forets} {et~al.}(1992){Pineau des Forets}, {Roueff}, \&
  {Flower}}]{PdF92}
{Pineau des Forets}, G., {Roueff}, E., \& {Flower}, D.~R. 1992, \mnras, 258,
  45P

\bibitem[{{Rangwala} {et~al.}(2011){Rangwala}, {Maloney}, {Glenn}, {Wilson},
  {Rykala}, {Isaak}, {Baes}, {Bendo}, {Boselli}, {Bradford}, {Clements},
  {Cooray}, {Fulton}, {Imhof}, {Kamenetzky}, {Madden}, {Mentuch}, {Sacchi},
  {Sauvage}, {Schirm}, {Smith}, {Spinoglio}, \& {Wolfire}}]{Rangwala2011}
{Rangwala}, N., {Maloney}, P.~R., {Glenn}, J., {et~al.} 2011, \apj, 743, 94

\bibitem[{{Revnivtsev} {et~al.}(2004){Revnivtsev}, {Churazov}, {Sazonov},
  {Sunyaev}, {Lutovinov}, {Gilfanov}, {Vikhlinin}, {Shtykovsky}, \&
  {Pavlinsky}}]{Revnivtsev2004}
{Revnivtsev}, M.~G., {Churazov}, E.~M., {Sazonov}, S.~Y., {et~al.} 2004, \aap,
  425, L49

\bibitem[{{Rigopoulou} {et~al.}(2013){Rigopoulou}, {Hurley}, {Swinyard},
  {Virdee}, {Croxall}, {Hopwood}, {Lim}, {Magdis}, {Pearson}, {Pellegrini},
  {Polehampton}, \& {Smith}}]{Rigopoulou2013}
{Rigopoulou}, D., {Hurley}, P.~D., {Swinyard}, B.~M., {et~al.} 2013, \mnras,
  434, 2051

\bibitem[{{Rodr{\'\i}guez-Fern{\'a}ndez}
  {et~al.}(2010){Rodr{\'\i}guez-Fern{\'a}ndez}, {Tafalla}, {Gueth}, \&
  {Bachiller}}]{Rodriguez-Fernandez2010}
{Rodr{\'\i}guez-Fern{\'a}ndez}, N.~J., {Tafalla}, M., {Gueth}, F., \&
  {Bachiller}, R. 2010, \aap, 516, A98

\bibitem[{{R{\"o}llig} {et~al.}(2016){R{\"o}llig}, {Simon}, {G{\"u}sten},
  {Stutzki}, {Israel}, \& {Jacobs}}]{Rolling2016}
{R{\"o}llig}, M., {Simon}, R., {G{\"u}sten}, R., {et~al.} 2016, \aap, 591, A33

\bibitem[{{Rosenberg} {et~al.}(2015){Rosenberg}, {van der Werf}, {Aalto},
  {Armus}, {Charmandaris}, {D{\'\i}az-Santos}, {Evans}, {Fischer}, {Gao},
  {Gonz{\'a}lez-Alfonso}, {Greve}, {Harris}, {Henkel}, {Israel}, {Isaak},
  {Kramer}, {Meijerink}, {Naylor}, {Sanders}, {Smith}, {Spaans}, {Spinoglio},
  {Stacey}, {Veenendaal}, {Veilleux}, {Walter}, {Wei{\ss}}, {Wiedner}, {van der
  Wiel}, \& {Xilouris}}]{Rosenberg2015}
{Rosenberg}, M.~J.~F., {van der Werf}, P.~P., {Aalto}, S., {et~al.} 2015, \apj,
  801, 72

\bibitem[{{Salgado} {et~al.}(2016){Salgado}, {Bern{\'e}}, {Adams}, {Herter},
  {Keller}, \& {Tielens}}]{Salgado16}
{Salgado}, F., {Bern{\'e}}, O., {Adams}, J.~D., {et~al.} 2016, \apj, 830, 118

\bibitem[{{Sanders} \& {Mirabel}(1996)}]{Sanders1996}
{Sanders}, D.~B. \& {Mirabel}, I.~F. 1996, \araa, 34, 749

\bibitem[{{Sanhueza} {et~al.}(2012){Sanhueza}, {Jackson}, {Foster}, {Garay},
  {Silva}, \& {Finn}}]{Sanhueza2012}
{Sanhueza}, P., {Jackson}, J.~M., {Foster}, J.~B., {et~al.} 2012, \apj, 756, 60

\bibitem[{{Sato} {et~al.}(2000){Sato}, {Hasegawa}, {Whiteoak}, \&
  {Miyawaki}}]{Sato2000}
{Sato}, F., {Hasegawa}, T., {Whiteoak}, J.~B., \& {Miyawaki}, R. 2000, \apj,
  535, 857

\bibitem[{{Schinnerer} {et~al.}(2013){Schinnerer}, {Meidt}, {Pety}, {Hughes},
  {Colombo}, {Garc{\'\i}a-Burillo}, {Schuster}, {Dumas}, {Dobbs}, {Leroy},
  {Kramer}, {Thompson}, \& {Regan}}]{Schinnerer2013}
{Schinnerer}, E., {Meidt}, S.~E., {Pety}, J., {et~al.} 2013, \apj, 779, 42

\bibitem[{{Schirm} {et~al.}(2014){Schirm}, {Wilson}, {Parkin}, {Kamenetzky},
  {Glenn}, {Rangwala}, {Spinoglio}, {Pereira-Santaella}, {Baes}, {Barlow},
  {Clements}, {Cooray}, {De Looze}, {Karczewski}, {Madden}, {R{\'e}my-Ruyer},
  \& {Wu}}]{Schirm2014}
{Schirm}, M. R.~P., {Wilson}, C.~D., {Parkin}, T.~J., {et~al.} 2014, \apj, 781,
  101

\bibitem[{{Schmiedeke} {et~al.}(2016){Schmiedeke}, {Schilke}, {M{\"o}ller},
  {S{\'a}nchez-Monge}, {Bergin}, {Comito}, {Csengeri}, {Lis}, {Molinari},
  {Qin}, \& {Rolffs}}]{Schmiedeke16}
{Schmiedeke}, A., {Schilke}, P., {M{\"o}ller}, T., {et~al.} 2016, \aap, 588,
  A143

\bibitem[{{Schroder} {et~al.}(1991){Schroder}, {Staemmler}, {Smith}, {Flower},
  \& {Jaquet}}]{Schroder91}
{Schroder}, K., {Staemmler}, V., {Smith}, M.~D., {Flower}, D.~R., \& {Jaquet},
  R. 1991, Journal of Physics B Atomic Molecular Physics, 24, 2487

\bibitem[{{Schultheis} {et~al.}(1999){Schultheis}, {Ganesh}, {Simon}, {Omont},
  {Alard}, {Borsenberger}, {Copet}, {Epchtein}, {Fouqu{\'e}}, \&
  {Habing}}]{Schultheis99}
{Schultheis}, M., {Ganesh}, S., {Simon}, G., {et~al.} 1999, \aap, 349, L69

\bibitem[{{Swinyard} {et~al.}(2014){Swinyard}, {Polehampton}, {Hopwood},
  {Valtchanov}, {Lu}, {Fulton}, {Benielli}, {Imhof}, {Marchili}, {Baluteau},
  {Bendo}, {Ferlet}, {Griffin}, {Lim}, {Makiwa}, {Naylor}, {Orton},
  {Papageorgiou}, {Pearson}, {Schulz}, {Sidher}, {Spencer}, {van der Wiel}, \&
  {Wu}}]{Swinyard14}
{Swinyard}, B.~M., {Polehampton}, E.~T., {Hopwood}, R., {et~al.} 2014, \mnras,
  440, 3658

\bibitem[{{Takano} {et~al.}(2019){Takano}, {Nakajima}, \& {Kohno}}]{Takano19}
{Takano}, S., {Nakajima}, T., \& {Kohno}, K. 2019, \pasj, 71, S20

\bibitem[{Taniguchi(2019)}]{ndRADEX}
Taniguchi, A. 2019, astropenguin/ndradex

\bibitem[{{Tauber} \& {Goldsmith}(1990)}]{Tauber90}
{Tauber}, J.~A. \& {Goldsmith}, P.~F. 1990, \apjl, 356, L63

\bibitem[{{Tauber} {et~al.}(1995){Tauber}, {Lis}, {Keene}, {Schilke}, \&
  {Buettgenbach}}]{Tauber1995}
{Tauber}, J.~A., {Lis}, D.~C., {Keene}, J., {Schilke}, P., \& {Buettgenbach},
  T.~H. 1995, \aap, 297, 567

\bibitem[{{Tayal}(2011)}]{Tayal11}
{Tayal}, S.~S. 2011, \apjs, 195, 12

\bibitem[{{Terrier} {et~al.}(2018){Terrier}, {Clavel}, {Soldi}, {Goldwurm},
  {Ponti}, {Morris}, \& {Chuard}}]{Terrier2018}
{Terrier}, R., {Clavel}, M., {Soldi}, S., {et~al.} 2018, \aap, 612, A102

\bibitem[{{Terrier} {et~al.}(2010){Terrier}, {Ponti}, {B{\'e}langer},
  {Decourchelle}, {Tatischeff}, {Goldwurm}, {Trap}, {Morris}, \&
  {Warwick}}]{Terrier2010}
{Terrier}, R., {Ponti}, G., {B{\'e}langer}, G., {et~al.} 2010, \apj, 719, 143

\bibitem[{{Thiel} {et~al.}(2019){Thiel}, {Belloche}, {Menten}, {Giannetti},
  {Wiesemeyer}, {Winkel}, {Gratier}, {M{\"u}ller}, {Colombo}, \&
  {Garrod}}]{Thiel_2019}
{Thiel}, V., {Belloche}, A., {Menten}, K.~M., {et~al.} 2019, \aap, 623, A68

\bibitem[{{Tielens} \& {Hollenbach}(1985)}]{Tielens_1985a}
{Tielens}, A.~G.~G.~M. \& {Hollenbach}, D. 1985, \apj, 291, 722

\bibitem[{{Tsuboi} {et~al.}(2015){Tsuboi}, {Miyazaki}, \&
  {Uehara}}]{Tsuboi2015}
{Tsuboi}, M., {Miyazaki}, A., \& {Uehara}, K. 2015, \pasj, 67, 109

\bibitem[{{Valinia} {et~al.}(2000){Valinia}, {Tatischeff}, {Arnaud}, {Ebisawa},
  \& {Ramaty}}]{Valinia2000}
{Valinia}, A., {Tatischeff}, V., {Arnaud}, K., {Ebisawa}, K., \& {Ramaty}, R.
  2000, \apj, 543, 733

\bibitem[{{van der Tak} {et~al.}(2007){van der Tak}, {Black}, {Sch{\"o}ier},
  {Jansen}, \& {van Dishoeck}}]{vanderTak_2007}
{van der Tak}, F.~F.~S., {Black}, J.~H., {Sch{\"o}ier}, F.~L., {Jansen}, D.~J.,
  \& {van Dishoeck}, E.~F. 2007, \aap, 468, 627

\bibitem[{{van der Werf} {et~al.}(2010){van der Werf}, {Isaak}, {Meijerink},
  {Spaans}, {Rykala}, {Fulton}, {Loenen}, {Walter}, {Wei{\ss}}, {Armus},
  {Fischer}, {Israel}, {Harris}, {Veilleux}, {Henkel}, {Savini}, {Lord},
  {Smith}, {Gonz{\'a}lez-Alfonso}, {Naylor}, {Aalto}, {Charmand aris},
  {Dasyra}, {Evans}, {Gao}, {Greve}, {G{\"u}sten}, {Kramer},
  {Mart{\'\i}n-Pintado}, {Mazzarella}, {Papadopoulos}, {Sanders}, {Spinoglio},
  {Stacey}, {Vlahakis}, {Wiedner}, \& {Xilouris}}]{vanderwerf2010}
{van der Werf}, P.~P., {Isaak}, K.~G., {Meijerink}, R., {et~al.} 2010, \aap,
  518, L42

\bibitem[{{Whiteoak} \& {Gardner}(1983)}]{Whiteoak1983}
{Whiteoak}, J.~B. \& {Gardner}, F.~F. 1983, \mnras, 205, 27P

\bibitem[{{Wu} {et~al.}(2018){Wu}, {Bron}, {Onaka}, {Le Petit}, {Galliano},
  {Languignon}, {Nakamura}, \& {Okada}}]{Wu18}
{Wu}, R., {Bron}, E., {Onaka}, T., {et~al.} 2018, \aap, 618, A53

\bibitem[{{Yang} {et~al.}(2010){Yang}, {Stancil}, {Balakrishnan}, \&
  {Forrey}}]{Yang10}
{Yang}, B., {Stancil}, P.~C., {Balakrishnan}, N., \& {Forrey}, R.~C. 2010,
  \apj, 718, 1062

\bibitem[{{Yusef-Zadeh} {et~al.}(2009){Yusef-Zadeh}, {Hewitt}, {Arendt},
  {Whitney}, {Rieke}, {Wardle}, {Hinz}, {Stolovy}, {Lang}, {Burton}, \&
  {Ramirez}}]{YusefYadeh2009}
{Yusef-Zadeh}, F., {Hewitt}, J.~W., {Arendt}, R.~G., {et~al.} 2009, \apj, 702,
  178

\bibitem[{{Yusef-Zadeh} {et~al.}(2013){Yusef-Zadeh}, {Hewitt}, {Wardle},
  {Tatischeff}, {Roberts}, {Cotton}, {Uchiyama}, {Nobukawa}, {Tsuru}, {Heinke},
  \& {Royster}}]{YusefZadeh2013}
{Yusef-Zadeh}, F., {Hewitt}, J.~W., {Wardle}, M., {et~al.} 2013, \apj, 762, 33

\bibitem[{{Zhang} {et~al.}(2015){Zhang}, {Hailey}, {Mori}, {Clavel}, {Terrier},
  {Ponti}, {Goldwurm}, {Bauer}, {Boggs}, {Christensen}, {Craig}, {Harrison},
  {Hong}, {Nynka}, {Soldi}, {Stern}, {Tomsick}, \& {Zhang}}]{Zhang_2015}
{Zhang}, S., {Hailey}, C.~J., {Mori}, K., {et~al.} 2015, \apj, 815, 132

\end{thebibliography}

\begin{appendix}\label{Sect:Appendix}

\section{Complementary SPIRE-FTS surface brightness maps of Sgr~B2}

\begin{figure*}[!ht]
\centering
\includegraphics[height=0.4\textwidth]{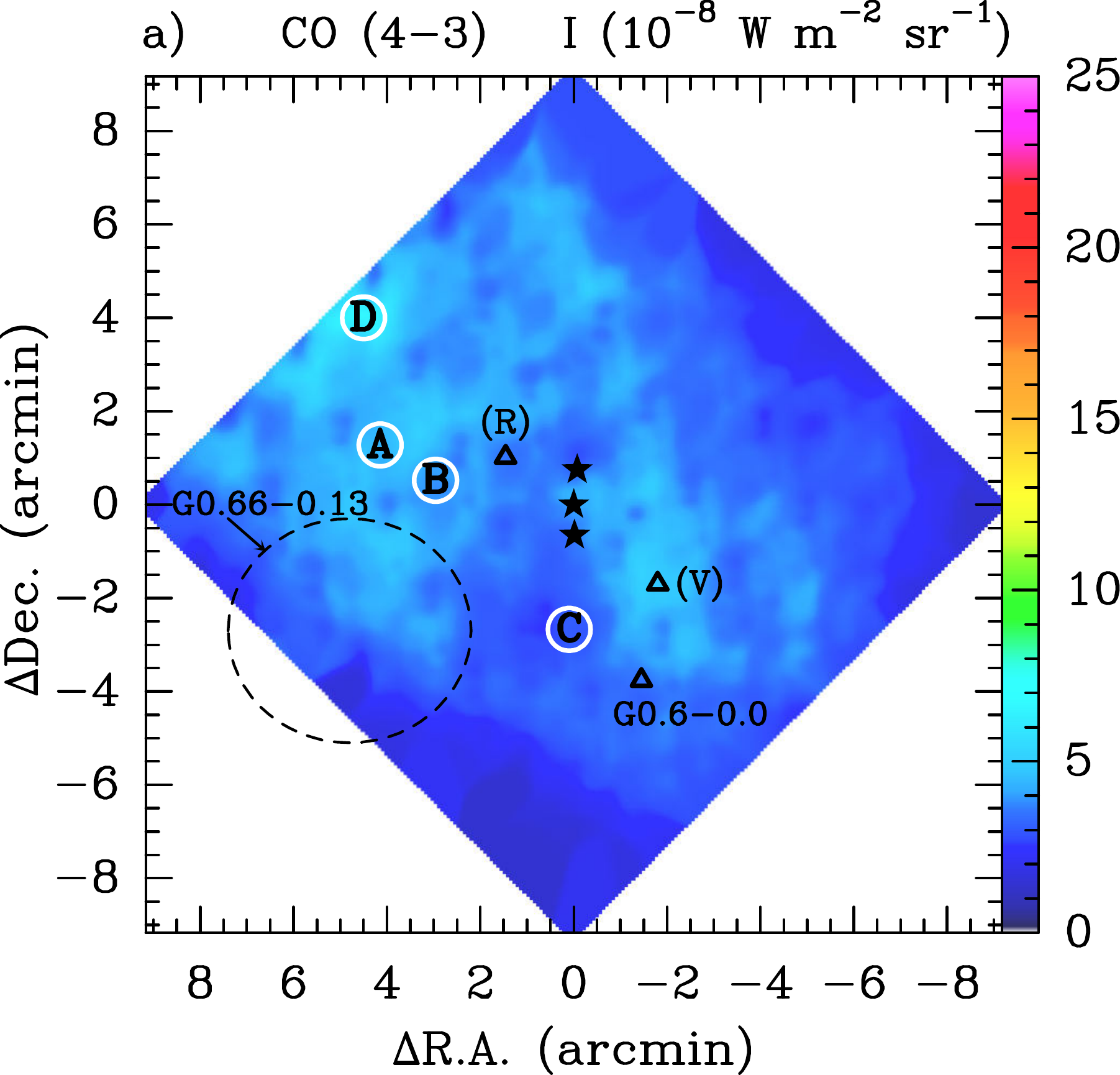}  \hspace{0.8cm}
\includegraphics[height=0.4\textwidth]{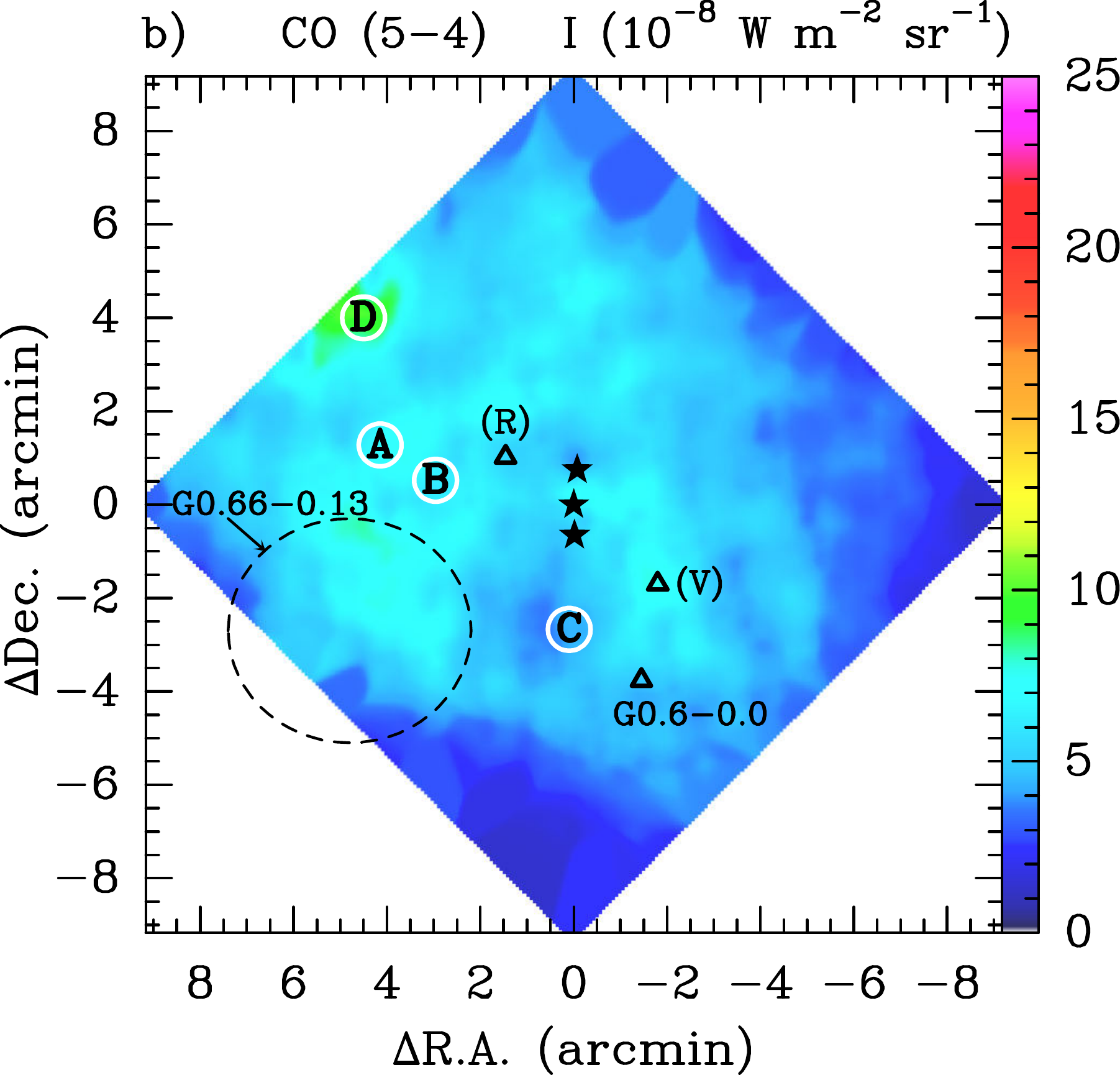} \\ \vspace{0.4cm}
\includegraphics[height=0.4\textwidth]{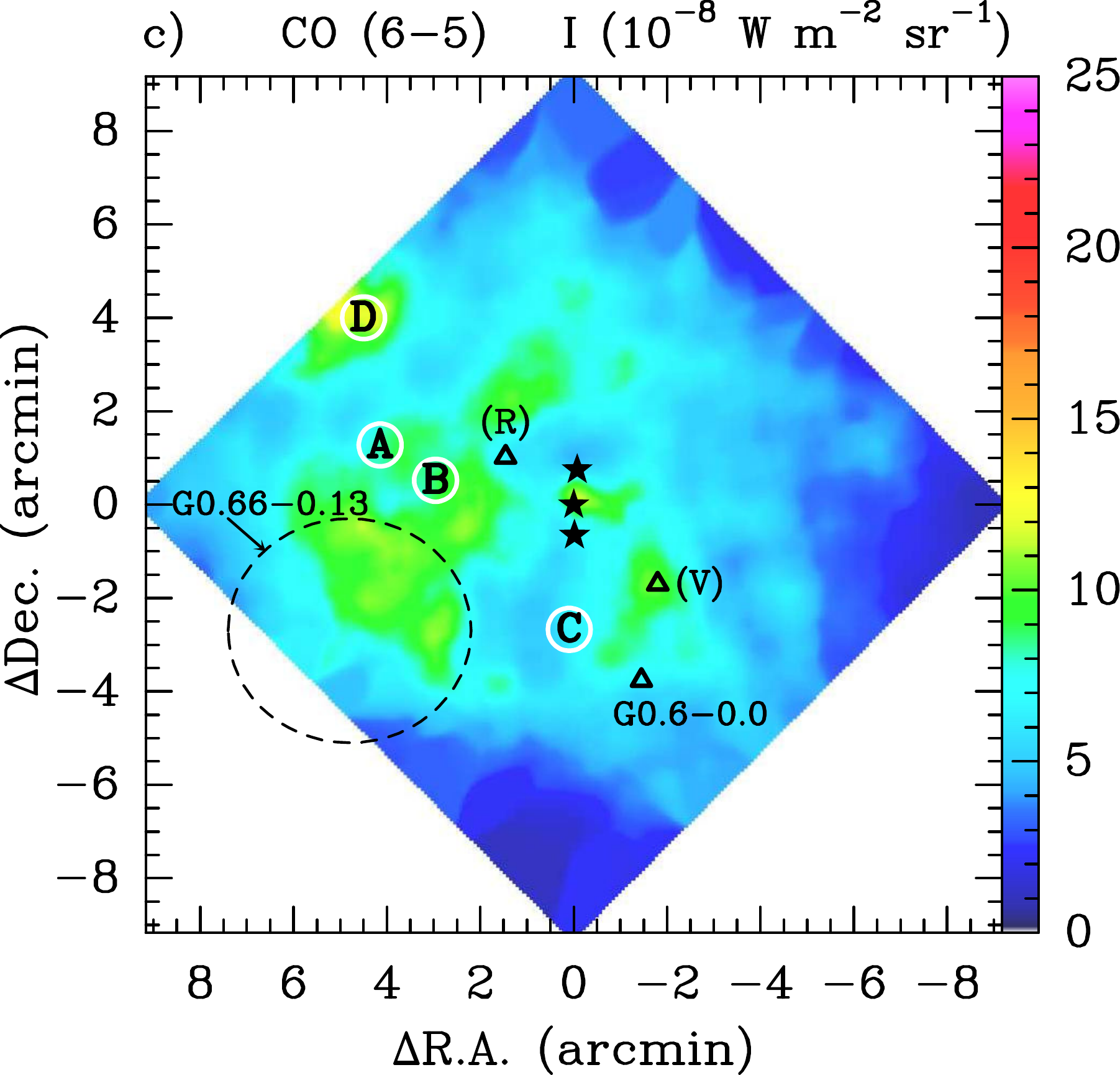} \hspace{0.8cm}
\includegraphics[height=0.4\textwidth]{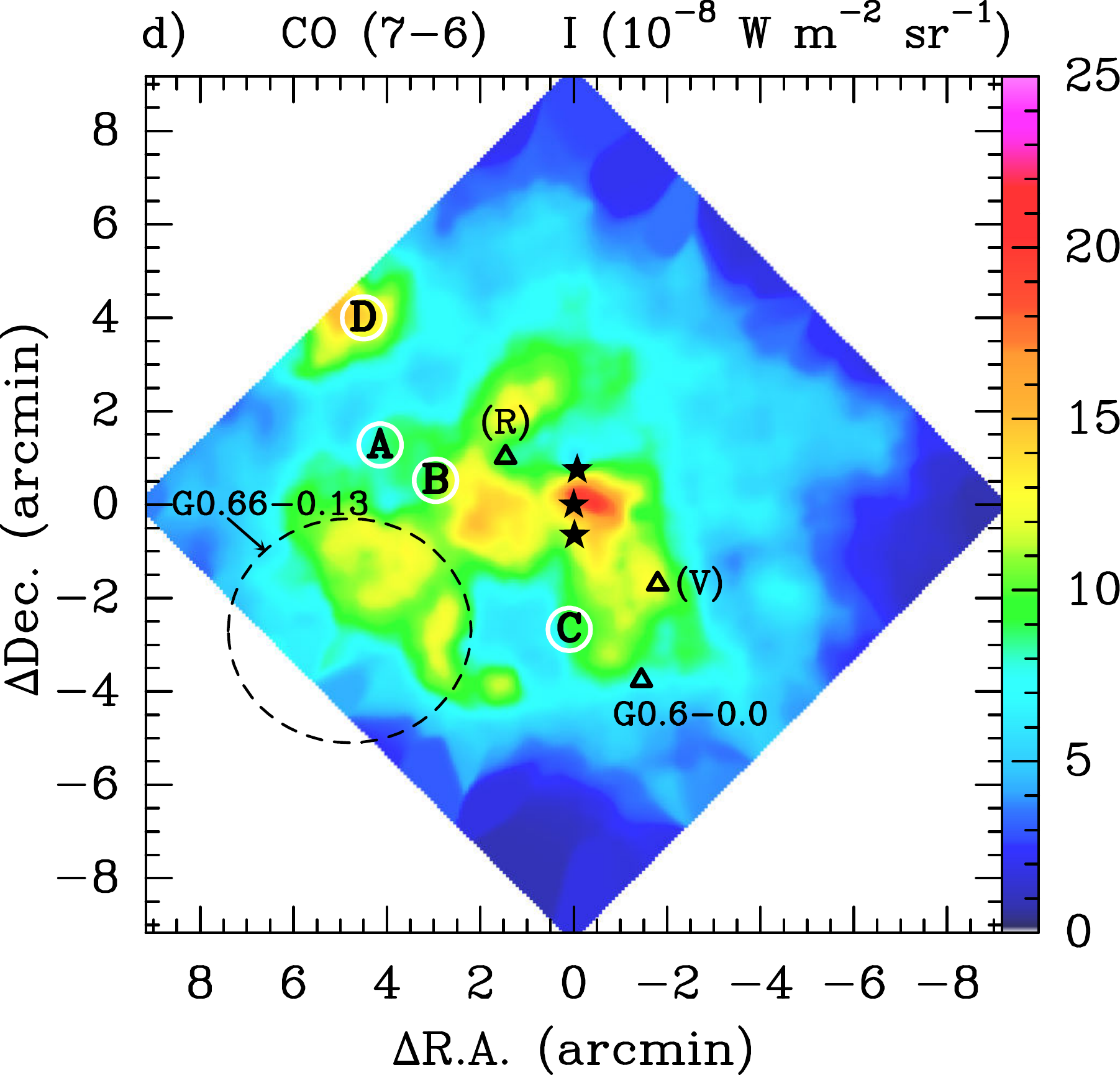} \\ \vspace{0.4cm}
\includegraphics[height=0.4\textwidth]{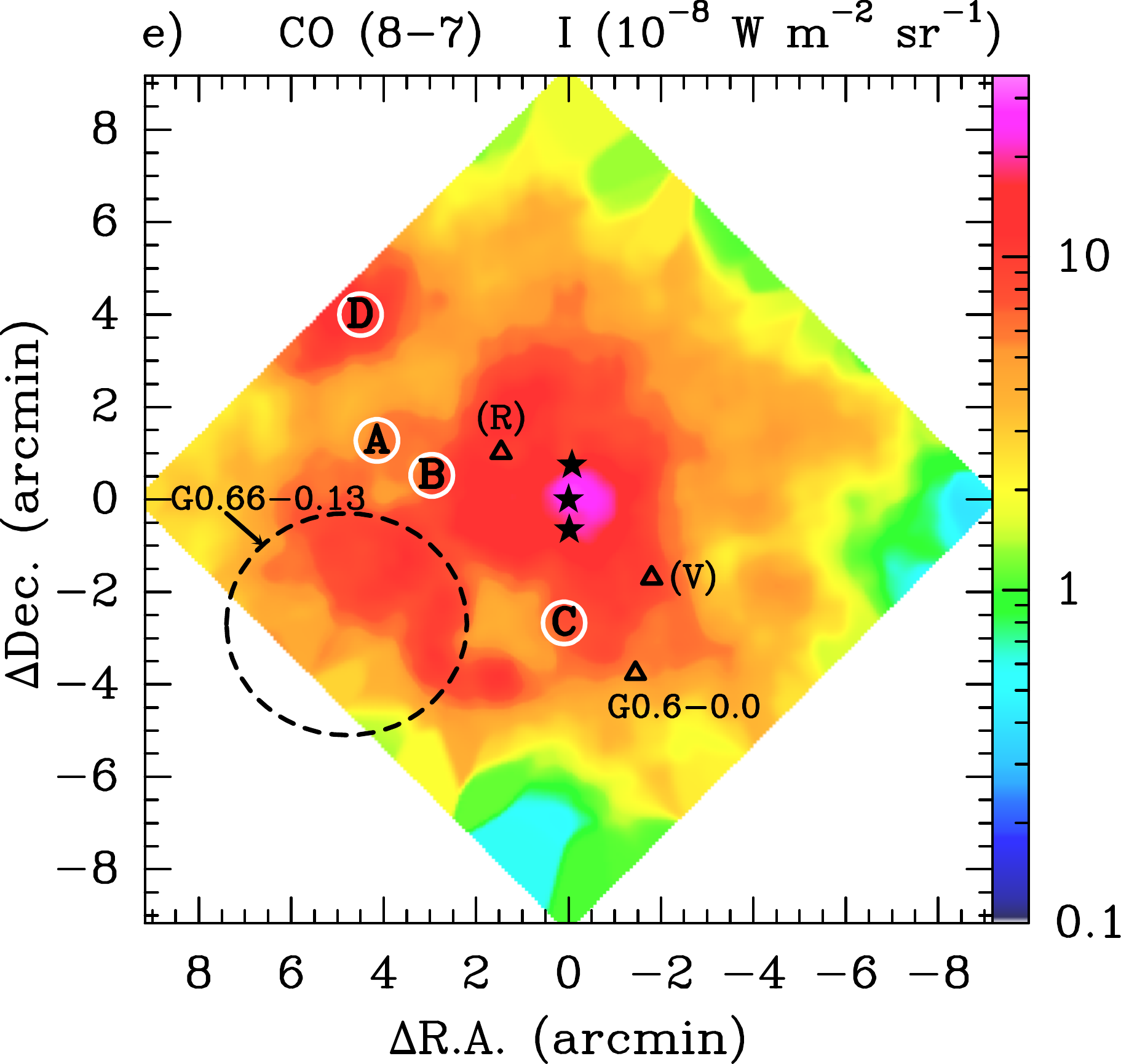} \hspace{0.8cm}
\includegraphics[height=0.4\textwidth]{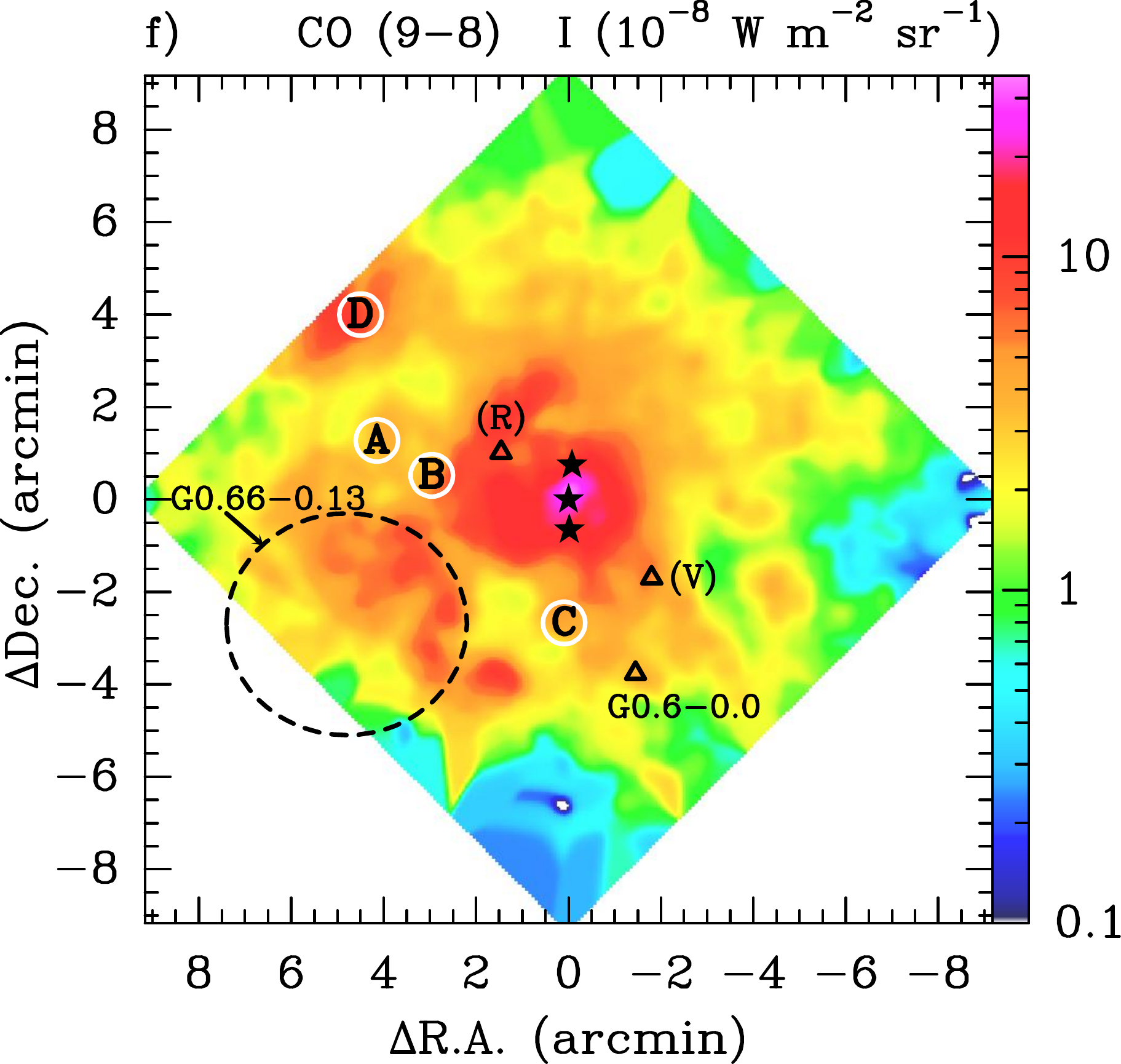}\\
\caption{SPIRE-FTS surface brightness maps of the mid-$J$ CO line emission. }
\label{fig:apendSPIRECO}
\end{figure*}

\begin{figure*}[!ht]
\centering
\includegraphics[height=0.4\textwidth]{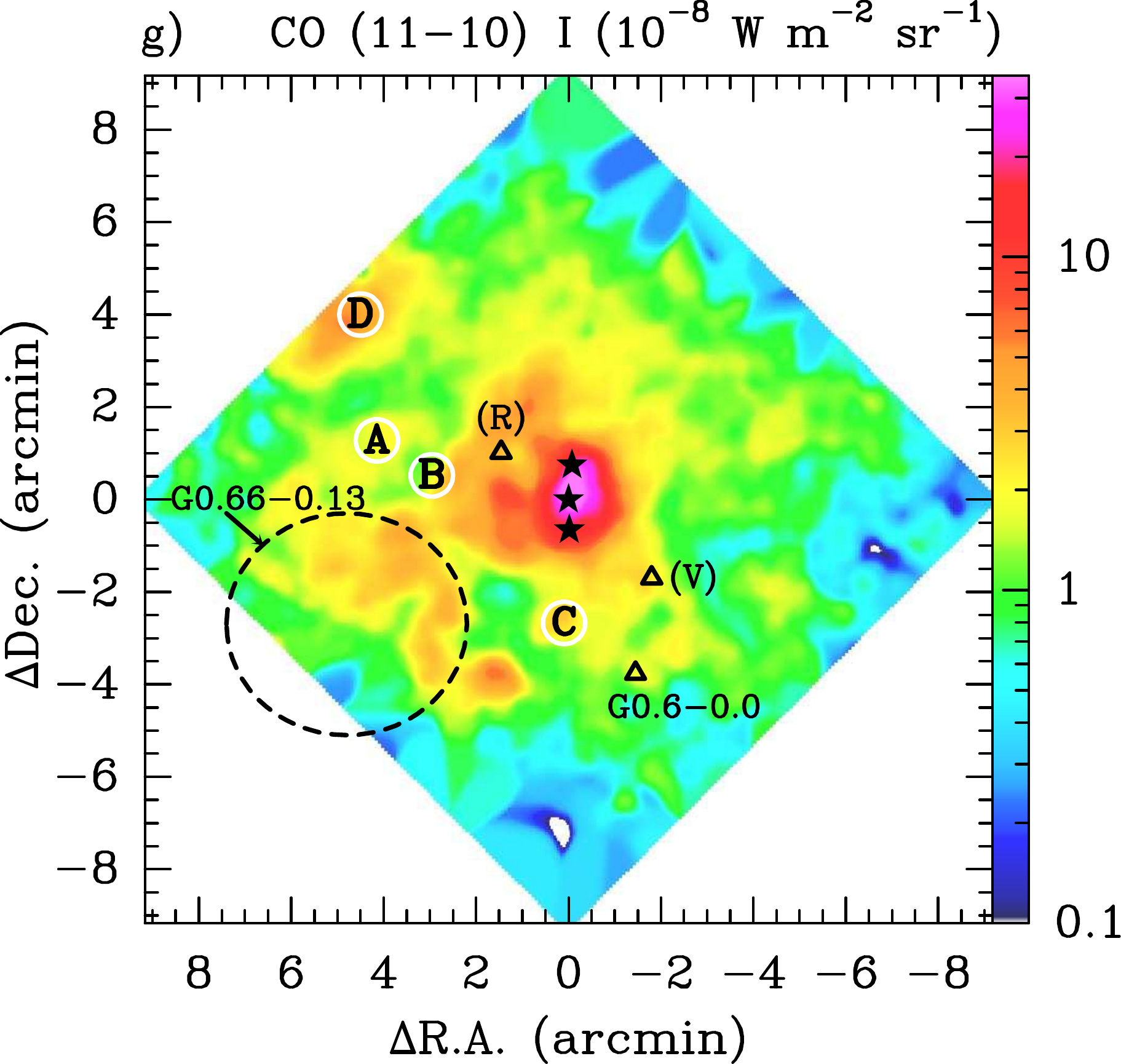}\hspace{0.8cm}
\includegraphics[height=0.4\textwidth]{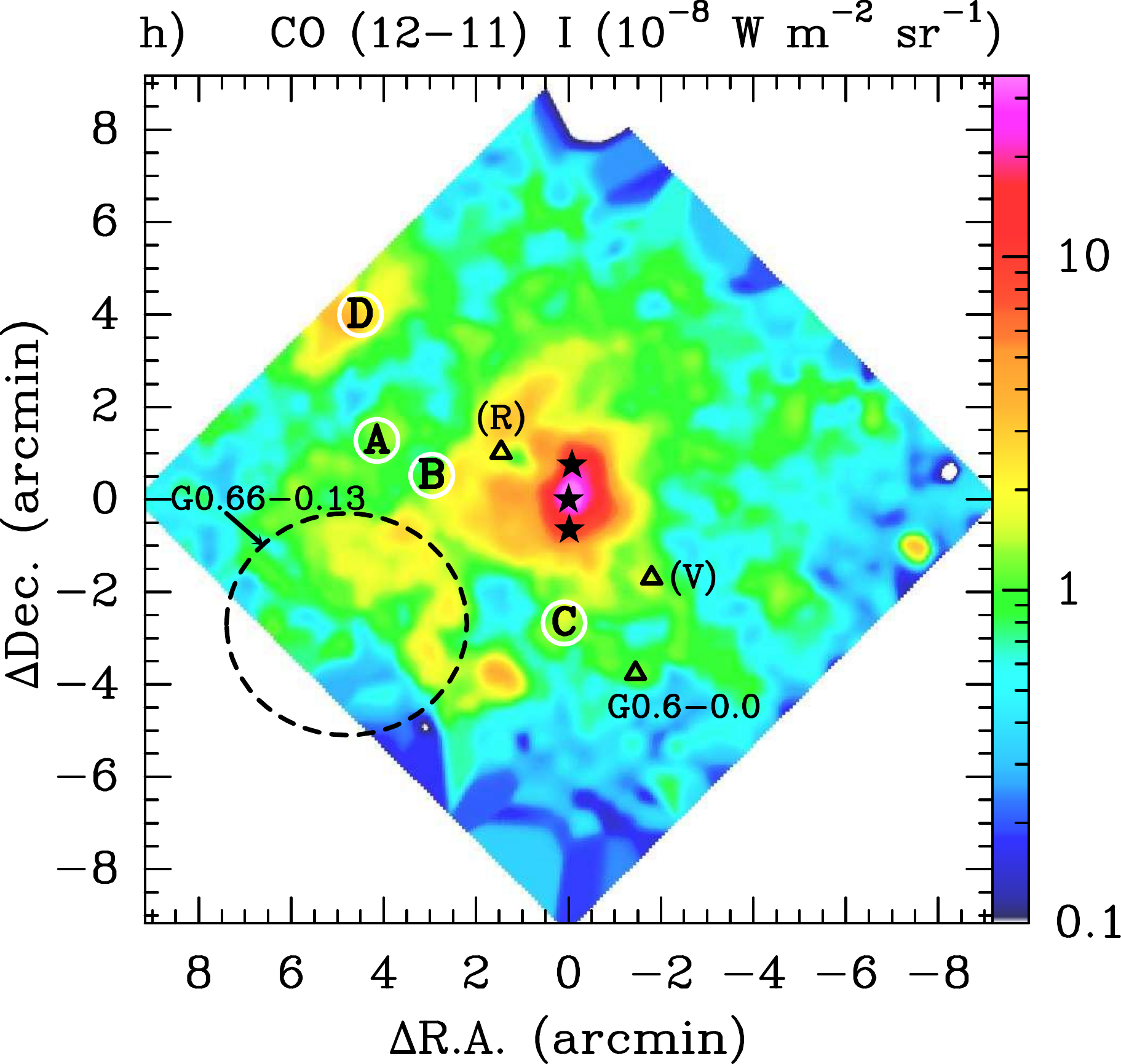}\\ \vspace{0.4cm}
\hspace{-0.4cm}\includegraphics[height=0.4\textwidth]{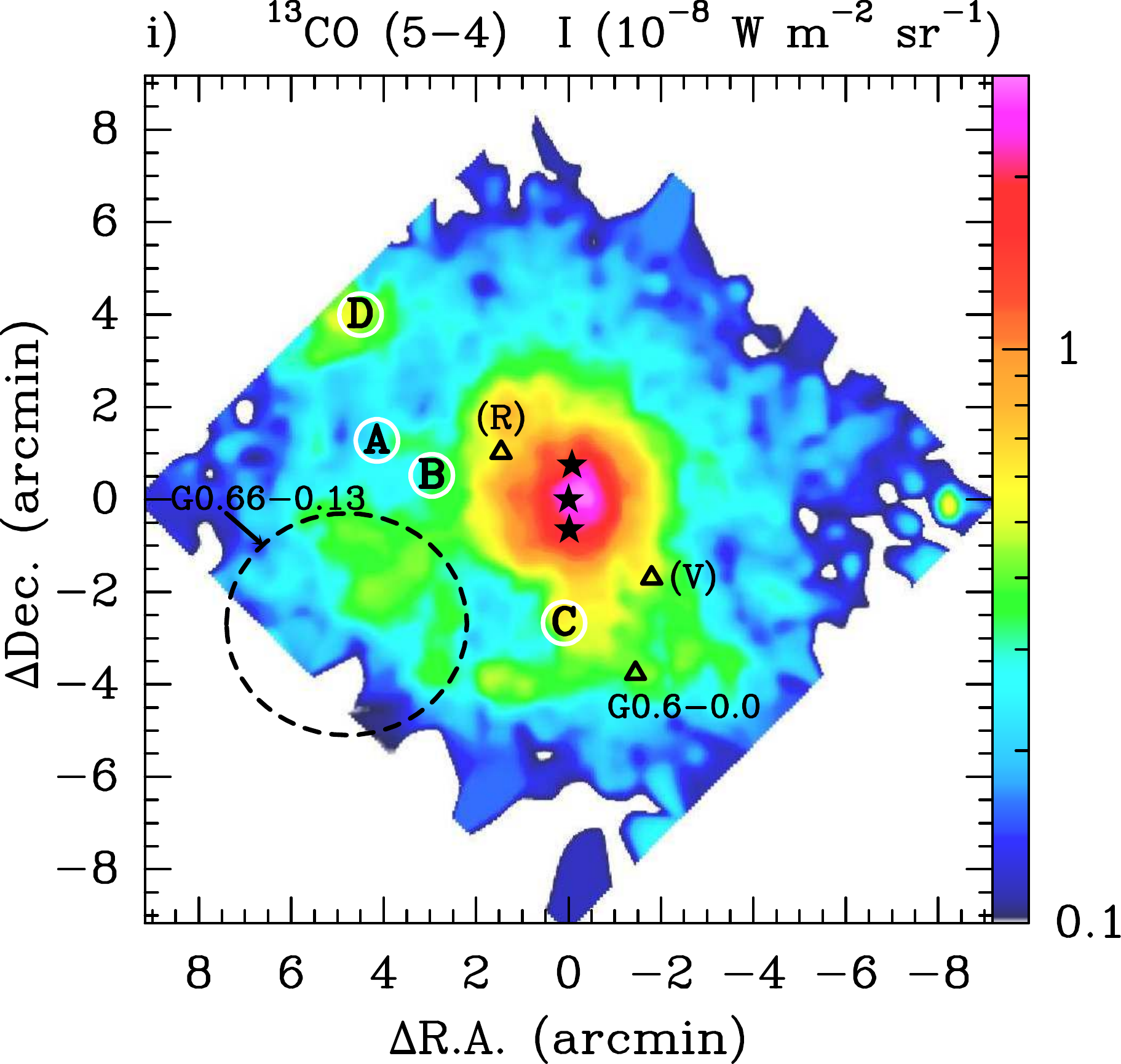} \hspace{0.9cm}
\includegraphics[height=0.4\textwidth]{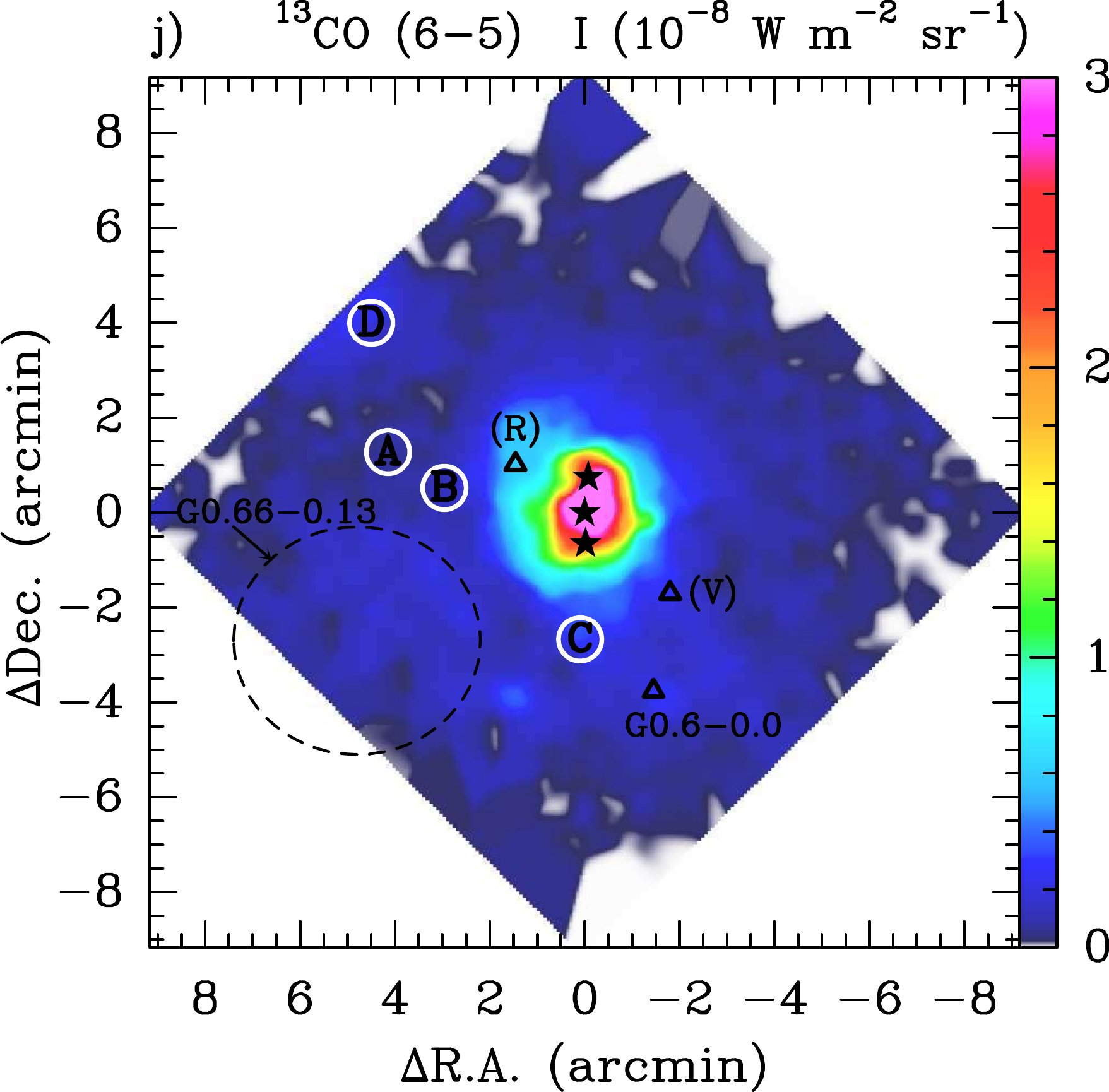}\\ \vspace{0.4cm}
\includegraphics[height=0.4\textwidth]{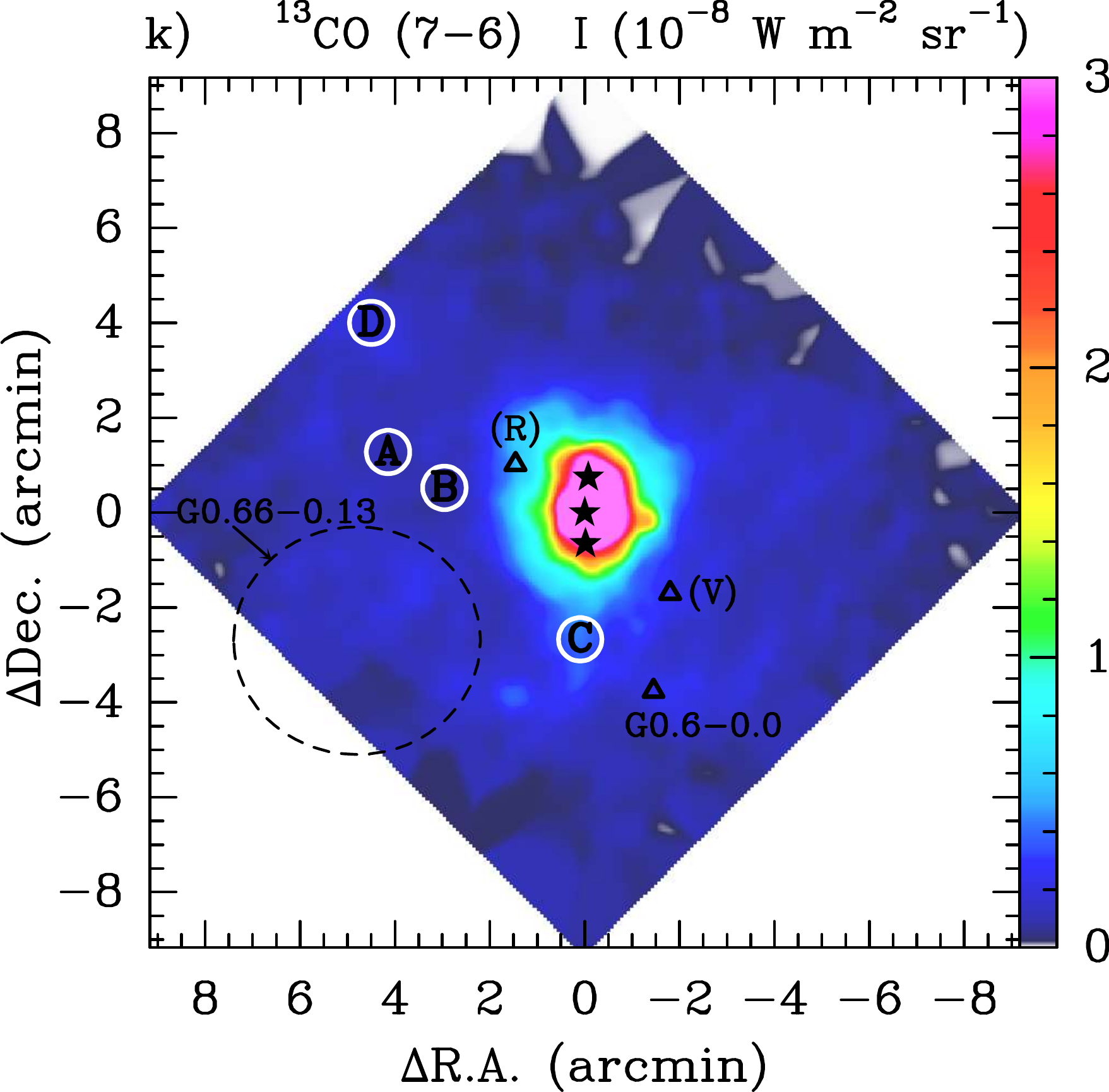} \hspace{0.8cm}
\includegraphics[height=0.4\textwidth]{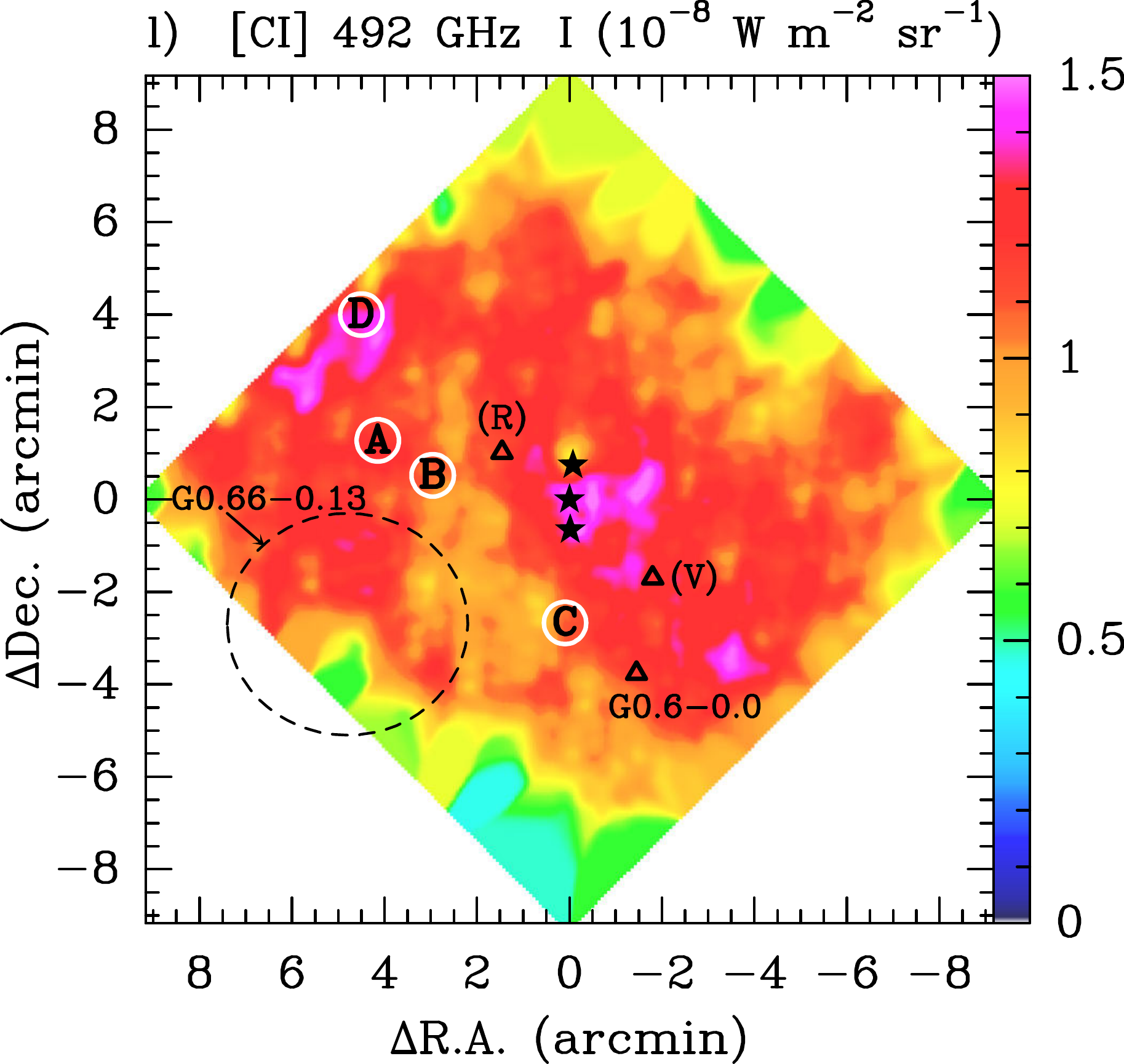}
\caption{SPIRE-FTS surface brightness maps of CO, $^{13}$CO, and \CI~line emission.}
\label{fig:apendSPIRE13COCI}
\end{figure*}

\end{appendix}

\end{document}